%% file: paper.tex
\documentclass[preprint, 10pt, 3p, twocolumn]{elsarticle}

\usepackage{amssymb}
\usepackage{lineno}

\usepackage[binary-units=true]{siunitx}
\usepackage{wrapfig}
\usepackage{xcolor}
\usepackage{subcaption}
\usepackage{pgf}
\usepackage{adjustbox}
\usepackage{amsmath}

\graphicspath{{./figures/}}

\newcommand{\cf}{cf.\ }

\newcommand{\BigO}[1]{$\mathcal{O}(#1)$}

\definecolor{revcolor}{HTML}{F57C00}

\usepackage[ruled]{algorithm2e} % For algorithms

\SetAlFnt{\small}
\SetAlCapFnt{\small}
\SetAlCapNameFnt{\small}
\SetAlCapHSkip{0pt}
\IncMargin{-\parindent}

\SetStartEndCondition{ }{}{}%
\SetKwProg{Fn}{def}{\string:}{}
\SetKwFunction{Range}{range}%%
\SetKw{KwTo}{in}
\SetKwFor{For}{for}{\string:}{}%
\SetKwIF{If}{ElseIf}{Else}{if}{:}{elif}{else:}{}%
\SetKwFor{While}{while}{:}{}%
\SetKwSwitch{Switch}{Case}{Other}{switch}{:}{case}{else}{}{}%
\AlgoDontDisplayBlockMarkers\SetAlgoNoEnd\SetAlgoNoLine%

\journal{Computer-Aided Design}

\begin{document}

\begin{frontmatter}



\title{Fast Exact Booleans for Iterated CSG using Octree-Embedded BSPs}


\author[rwth-address]{Julius Nehring-Wirxel}
\ead{nehring-wirxel@cs.rwth-aachen.de}
\author[rwth-address]{Philip Trettner}
\ead{trettner@cs.rwth-aachen.de}
\author[rwth-address]{Leif Kobbelt}
\ead{kobbelt@cs.rwth-aachen.de}

\address[rwth-address]{
RWTH Aachen,
Lehrstuhl für Informatik 8,
Ahornstraße 55,
52074 Aachen,
Germany 
}

\begin{abstract}
\input{paper-00-abstract.tex}
\end{abstract}

\begin{keyword}
    Plane-based Geometry \sep
    CSG \sep 
    Mesh Booleans \sep
    BSP \sep
    Octree \sep
    Integer Arithmetic
\end{keyword}

\end{frontmatter}

% \linenumbers

\input{paper-01-intro.tex}
\input{paper-02-related-work.tex}
\input{paper-03-method.tex}
\input{paper-04-evaluation.tex}
\input{paper-05-applications.tex}
\input{paper-06-future-work.tex}
\input{paper-07-conclusion.tex}
\section*{Funding}
This work was supported by the Gottfried-Wilhelm-Leibniz Programme of the Deutsche Forschungsgemeinschaft DFG, by the Excellence Initiative of the German federal and state government and by the European Regional Development Fund within the HDV-Mess project under the funding code EFRE-0500038.



%%       \citet{<label>} ==> Jones et al. [21]
%%       \citep{<label>} ==> [21]
%%

\bibliographystyle{elsarticle-num-names} 
\bibliography{paper}





% \end{thebibliography}
\end{document}

%% file: paper-00-abstract.tex
We present octree-embedded BSPs, a volumetric mesh data structure suited for performing a sequence of Boolean operations (iterated CSG) efficiently.
At its core, our data structure leverages a plane-based geometry representation and integer arithmetics to guarantee unconditionally robust operations.
These typically present considerable performance challenges which we overcome by using custom-tailored fixed-precision operations and an efficient algorithm for cutting a convex mesh against a plane.
Consequently, BSP Booleans and mesh extraction are formulated in terms of mesh cutting.
The octree is used as a global acceleration structure to keep modifications local and bound the BSP complexity.
With our optimizations, we can perform up to \num{2.5} million mesh-plane cuts per second on a single core, which creates roughly 40--50 million output BSP nodes for CSG.
We demonstrate our system in two iterated CSG settings: sweep volumes and a milling simulation.

%% file: paper-01-intro.tex
% !TeX root = ./paper.tex

\section{Introduction}

Mesh Booleans or \textit{Constructive Solid Geometry} (CSG) are popular and intuitive ways to model and design objects.
\emph{Iterated CSG} is the task of applying a sequence of CSG operations on some initial object.
For typical use-cases such as \textit{carving} or \textit{sculpting} each operation is simple but the resulting object turns more complex the more operations are applied.
This setting is also common for simulations of manufacturing processes such as CNC milling or 3D printing.

Iterated CSG is typically quite challenging since the result of each iteration is used as input for the next operation.
Therefore the algorithm is required to be unconditionally robust. 
Otherwise the simulation degenerates quickly.
When the intermediate result grows more complex and is subject to many local modifications, it is unacceptable to use methods where the computation time scales with the complexity of the workpiece.
Instead, the runtime of each operation should only depend on the complexity of the actually modified region.
These two problems are especially severe in the aforementioned manufacturing simulations as many small, potentially aligned or almost aligned operations are performed.

While there is extensive literature on how to quickly perform exact Booleans on large input meshes \citep{Bernstein09, Campen10,Zhou16,Sheng18,douze2015quickcsg,Magalhaes17}, the work on iterated CSG is mostly limited to inexact \textit{voxel} or \textit{dexel} based methods with their own set of scaling and discretization issues \citep{lynn2017voxel,sun2018geometric}.
In fact, we found that all modern \textit{exact} approaches are not well suited for iterated CSG tasks (\cf Section~\ref{sec:app}).
To overcome the stability issues, many exact approaches use arbitrary precision computations \citep{Zhou16,Barki15,Sheng18}.
However, these methods produce meshes with rounded vertex positions after an operation has finished.
Extending the lifetime of the arbitrary precision numbers across many iterations is usually not feasible as the quantization complexity increases rapidly after each iteration.
Consequently, current solutions suffer from either stability issues or poor performance.

Our proposed solution is based on Binary Space Partition (BSP) trees.
BSPs have already been used to perform robust CSG operations with a plane-based geometry representation \cite{Bernstein09, Campen10}.
However, they require conversion from mesh to BSP (import) and BSP to mesh (extraction), both with their own robustness issues.
While fast for small trees, BSP-based Booleans, in the worst-case, scale at least with \BigO{n^2} for $n$ nodes, even if the output complexity is only \BigO{n} \citep{lysenko2008improved}.
Extracting a polygonal mesh from a BSP is conceptually simple: start with a cube large enough to contain the result and recursively perform mesh-plane cuts until the leaves of the BSP are reached.
This results in a set of convex meshes from which those belonging to \textbf{out} nodes are discarded.
With infinite precision, each intermediate mesh is perfectly convex. 
However, rounding caused by fixed-precision floating point arithmetics often leads to small non-convexities.
This greatly complicates further cutting steps with epsilon tolerances and topology-repair steps \cite{Wang13}.
These problems are basically guaranteed to happen if the BSP is constructed from a triangle mesh: normally, three planes intersect in a single point and it is unlikely that a fourth plane contains the same point.
In a triangle mesh, the average number of faces around a vertex is six, meaning that, by construction, six planes should intersect in a single point, something extraordinarily unlikely when working with floating point values.
A stable extraction can still be achieved by either accepting these inaccuracies and formulate a topologically robust cutting \citep{Wang13} or by using plane-based geometry and exact predicates \citep{Bernstein09,Campen10}.
We use fixed-precision integer arithmetic (Section~\ref{sec:method:int}) to provide a fast and reliable numerical foundation which enables us to formulate a simple, yet exact and efficient algorithm for extracting the mesh surface of a BSP tree (Section~\ref{sec:method:cut} and \ref{sec:method:extract}).

Even if the numerical substrate is fast, BSP-based CSG still suffers from scaling problems.
Therefore, we superimpose a global octree structure and store a BSP in each octree cell.
These octree-embedded BSPs should not be seen as a temporary constructs such as in \cite{Campen10} but instead as a volumetric data structure that persists through the entire sequence of CSG operations.
While the octree introduces some overhead as the cell borders split triangles, it imposes strict limits on the BSP complexity and thus mitigates the scaling issues.
Additionally, it provides localization of operations and a set of efficient culling operations to limit the CSG computation to the affected regions.
An octree can be seen as a special BSP tree, a fact that we exploit to formulate efficient merge and split operations.
Still, the special structure of an octree guarantees good worst-case behavior that cannot be provided by more general BSPs.
Conceptually, our octree-embedded BSPs can be seen as a single gigantic BSP where the upper levels follow an octree structure to provide spatial partitioning and only the lower levels adapt to the geometry.

Finally, we added various improvements to the classical mesh import, BSP Booleans, and mesh extraction algorithms.
Importing a mesh into our data structure is fast with our exact computations as we can first build the octree and then locally construct BSPs using a small number of triangles for each octree cell.
With floating point computation this tends to produce cracks at octree cell borders.
\subsection{Contribution}

In summary, we present

\begin{itemize}
    \item octree-embedded BSPs as a data structure for performing high-performance iterated CSG operations.
    \item exact and efficient CSG based on BSP Booleans using custom-tailored \num{256} bit integer arithmetic.
    \item a BSP to mesh conversion using an efficient algorithm to cut planes against convex half-edge meshes.
\end{itemize}

%% file: paper-02-related-work.tex
% !TEX root paper.tex

\section{Related Work}
\label{sec:related-work}

There is extensive literature for Booleans on meshes \citep{naylor1990merging,Hachenberger07,Pavic10,Bernstein09,Campen10,Mei13,Barki15}.
The trilemma for different approaches are essentially performance, accuracy, and algorithmic stability.
Usually, only two of these can be fulfilled in a satisfying manner.
The main causes for these problems are rounding issues with fixed size floating point numbers, poor performance of exact calculation schemes, and high algorithmic complexity due to an immense number of topological configurations.

Binary Space Partition based approaches were kick-started by \citet{naylor1990merging}.
They introduced a procedure to merge two BSPs into a single one that represents a Boolean of the two inputs.
To prune empty nodes from the resulting BSP, a polygon cutting algorithm is part of the merge procedure.
Additionally, the conversion from and to a mesh requires the same polygon cutting algorithm.
Unfortunately, polygon cutting tends to be highly unstable when using fixed length floating point arithmetics \citep{lysenko2008improved}.
\citet{lysenko2008improved} present an improved BSP CSG algorithm by eliminating the polygon cutting from the merge procedure and replacing it by a satisfiability check for a linear program.
The constraints of the program are the same as the half-spaces from the current subtree to the root of the BSP.
If there is no solution to the linear program, the subtree is empty.

\citet{Bernstein09} go a different route to deal with instability.
They propose using exact filtered predicates \citep{shewchuk97a} in the domain of BSP merging.
To avoid having to increase the resolution when creating new intersection vertices they use a purely plane-based representation for convex polygons which is a subset of the class of Nef polygons \citep{nef1978}.
Plane-based modelling itself was already proposed by \citep{sugihara90}.
Vertices are defined as the intersection of three non-coplanar planes, convex polygons consist of a supporting plane and a list of planes determining the boundary of the polygon.
Unfortunately the BSP merging procedure scales poorly with the size of the input BSP \citep{lysenko2008improved}.
To mitigate the scaling problems \citet{Campen10} embed \textit{Bernstein's} BSP merging algorithm into an octree that localizes CSG operations.
Only small, overlapping parts of the input meshes are converted to BSPs, limiting the complexity of each single BSP merge.
The result is converted back to the original mesh. 
They also give guarantees for exactness of the operations as well as the conversion from and to mesh representation.
However, these guarantees force them to limit the precision of the input mesh.

Since the conversion between BSP and mesh is not trivial, \citet{Wang13} propose an efficient algorithm to extract a mesh from a BSP.
They do not use exact calculations but instead achieve stability by making sure that their mesh data structure is always in a sound state topologically.

To avoid conversions to and from plane-based representations, many approaches perform CSG operations directly on the mesh.
Some avoid floating point issues by approximating a seam where the two input meshes intersect \citep{Pavic10, Wang11, Schmidt16}.
Others present methods that perform well on many meshes but rely on careful tuning of epsilon values and may still fail \cite{Feito13,Anguita15}.
\citet{Barki15} first compute the intersections of all triangles of the input meshes using exact rational number representations.
They then determine which triangles belong to the result of the Boolean using an ordering of triangles around their shared edges.
\citet{Sheng18} pursue a similar approach: 
they first cut intersecting input triangles, but instead of using exact rationals to represent the new primitives, they use plane based vertex, edge, and face representations similar to \citep{Bernstein09}.
\citet{Magalhaes17} perform exact calculations on volume meshes and additionally use symbolic perturbation to avoid many special cases.
Using actual vertex perturbation, \citet{douze2015quickcsg} introduce a very fast but unstable algorithm that executes successfully on a large number of meshes.

Besides the purely mesh or plane-based methods, \citet{Hachenberger07} introduced a Nef-polyhedral-based approach that is implemented in the CGAL library.
It is both exact and has good stability, but is quite slow and its internal data-structures are very memory consuming.

In addition, there are a few alternative approaches:
\citet{Zhou16} combine exact calculations with a winding number based approach.
They derive the winding number of vertices from one mesh in respect to another input mesh topologically and use that information to determine which parts of the mesh to keep.
Notably, they provide a good implementation of their method in \texttt{libigl} \cite{libigl}.
Recently, \citet{CLSA20} improved the performance of this approach by deriving lazy floating-point predicates to replace the exact rational arithmetic used by \cite{Zhou16}.
Similarly, \citet{barill2018fast} introduce a general approach to determine winding numbers as a measurement of \textit{insideness} which can be used to perform Boolean operations.

Beside the mesh or plane based approaches, voxel-based approaches are also often used for CSG operations \citep{jense1989voxel,jang2000voxel,lynn2017voxel,hattab2019interactive}.
By principle they suffer from aliasing effects and limited resolution.
Additionally, the machining industry often uses a \textit{dexel}-based representation \citep{benouamer1997bridging,sun2018geometric}.
Dexel, also called \textit{multi-layer-z-map} are essentially a 2-dimensional regular grid with a list of height values that describe a transition from \textit{inside} to \textit{outside} or vice versa.
While this representation is often extended into three dimensions, it still suffers from aliasing problems similar to voxel-based approaches.

Our proposed method builds upon \citet{Bernstein09} and \citet{Campen10}.
Instead of using filtered floating-point numbers, we show how the required geometric predicates can be formulated and implemented using fixed-width integers.
By limiting the largest intermediate value to 128, 192, or 256 bit, we present a set of extremely practical speed-precision tradeoffs.
Furthermore, we show that instead of storing vertices as the intersection of three planes, a representation as 4D homogeneous coordinates has several advantages.
In particular, the most expensive predicate does not involve a $4 \times 4$ determinant anymore but rather a cheaper 4D dot product.
Algorithmically, we present a fast plane-against-convex-mesh cutting algorithm that allows us to implement exact BSP booleans efficiently using mesh extraction.
To the best of our knowledge, this results in the fastest exact BSP booleans to date and simultaneously provides high quality surface meshes and a method for reducing BSP redundancies.
Finally, our octree and the one from \citet{Campen10} serve subtly different purposes:
Theirs is a temporary acceleration structure used to find and localize the intersecting surface.
Our octree-embedded BSPs are a persistent data structure that are kept across iterations and could be considered an ``imposed structure'' on the upper levels of a global BSP.

%% file: paper-03-method.tex
% !TEX root paper.tex

\section{Method}
\label{sec:method}

% -----------------------------------------------------------------------------------------------------
\subsection{Overview}

Our method builds on the ideas of \cite{naylor1990merging, Bernstein09, Campen10}.
Geometry is represented as a BSP tree where each \emph{leaf} node has a label (\textbf{in} and \textbf{out} for simplicity but the algorithms are trivially extended to a multi-material setting).
Each \emph{non-leaf} BSP-node contains a plane that splits the domain into a positive and a negative half-space.
Thus, BSP trees represent a hierarchy of convex polyhedral cells and the object geometry is the union of all \textbf{in}-labeled leaf cells.
Instead of using floating point numbers, we base our entire system on integers and guarantee exactness of all results by analyzing and bounding the required number of bits (Section~\ref{sec:method:int}).
By using fixed-precision instead of arbitrary-precision arithmetic we can cater to all practical scenarios with minimal performance overhead.
Our system uses a plane-based geometry representation where each plane equation has integer coefficients.
While all input points must be integer, the resulting vertices are represented by the intersection of three planes (as 4D homogeneous coordinates) and can thus lie at fractional positions.
Because CSG operations cannot introduce new planes, all possible positions are representable using combinations of input planes.

Equipped with fast and exact predicates, we present a practical and efficient algorithm for converting BSP trees into meshes (Section~\ref{sec:method:extract}).
Instead of using convex Nef polyhedra \citep{nef1978,Bernstein09}, our boundary extraction algorithm operates on convex polyhedra stored in a half-edge data structure with a fast plane-mesh cutting procedure as its core operation.
This algorithm significantly outperforms the state of the art such as \citep{cork2013,Hachenberger07,Zhou16,douze2015quickcsg} while staying completely exact.

\begin{figure}
    \centering
    \fontsize{7pt}{11pt}\selectfont
    \def\svgwidth{\columnwidth}
    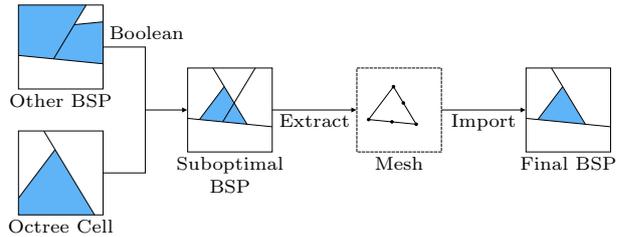
    \caption{
        Performing Booleans on an octree-embedded BSP data structure and another BSP can lead to suboptimal resulting BSPs.
        Extracting and reimporting the BSP's boundary usually improves the BSP quality.
    }
    \label{fig:bsp:workflow}
\end{figure}

Boundary extraction is typically a slow, error-prone, and often overlooked step.
However, based on our fast and exact extraction scheme, other essential BSP operations can be implemented.
Section~\ref{sec:method:Booleans} describes how performing CSG on BSPs is straightforward and faster than the state of the art when using our boundary extraction as a building block.
Removing redundancy from BSPs can be implemented by extracting the represented surface and converting it back to a BSP without any loss (Section~\ref{sec:method:simpl}).
A typical workflow of our method is depicted in Figure~\ref{fig:bsp:workflow}.

Section~\ref{sec:method:octree} presents our octree-embedded BSPs, a global octree that stores a BSP in each cell and locally adapts to the BSP complexity.
When performing iterated CSG, we maintain the octree across iterations and perform octree-BSP merges in each step.

The final module of the system is a pair of algorithms to import triangular meshes into our data structure: from polygon soup to BSP and from triangle soup to octree-embedded BSP.

% -----------------------------------------------------------------------------------------------------
\subsection{Arithmetic Foundation}
\label{sec:method:int}
Floating point computation often introduces round-off errors that tend to violate geometric invariants and threaten algorithmic stability.
Using arbitrary-precision libraries such as \texttt{gmp} \cite{Granlund12} typically incur prohibitive performance overhead.
Instead, we base our computation on fixed-precision integer arithmetic and exact representations.
We start by introducing our required operations.
Then, given a maximum number of bits that we are willing to use for intermediate results, we work backwards to determine the largest range of input values that guarantees overflow-free computation.

Input meshes are given with integer vertex positions and (non-normalized) integer normals.
Inner BSP nodes store planes $p$ in plane equation form with integer coefficients $a_p, b_p, c_p, d_p$ where $(a_p, b_p, c_p) = n_p$ is the plane normal and  $d_p = -v^T n_p$ is the offset from the origin (for any point $v$ that lies on the plane).
While input vertices are integer, the result of CSG operations can contain vertices at fractional positions. 
Thus, in a plane-based geometry representation, we do not actually compute vertex positions but treat each vertex as the intersection of three planes.
\citet{Bernstein09} describe working with planes and intersection of planes in the context of BSP booleans:
Given two planes $p$ and $q$, 
\begin{linenomath*}\begin{equation}
    n_p \times n_q = \vec{0}
\end{equation}\end{linenomath*}
tests if two planes are coplanar.
Three planes $p$, $q$, $r$ intersect in a unique point if the determinant 
\begin{linenomath*}\begin{equation}
    |n_p ~ n_q ~ n_r| \neq 0.
\end{equation}\end{linenomath*}
BSP extraction and mesh cutting requires another operation: given a point (as the intersection of planes $p$, $q$, $r$) and a plane $s$, classify if the point lies on the positive side, on the negative side, or exactly on $s$.
This can be computed using the sign of a $4 \times 4$ times a $3 \times 3$ determinant:
\begin{linenomath*}\begin{equation}
    \text{sign}\left|p~q~r~s\right| \cdot \text{sign}\left|n_p~n_q~n_r\right|.
\end{equation}\end{linenomath*}
Note that these tests can be performed with integer arithmetic exclusively.

In this formulation, vertex classification requires computing an expensive $4 \times 4$ determinant.
We derive a different but equivalent way to classify vertices based on homogeneous coordinates.
While our version basically results in the same formula, we offer an interpretation that immediately clarifies why a large part of the computation does not depend on $s$ and can therefore be precomputed and reused.
Given three planes $p$, $q$, and $r$, their intersection point can be found by solving
\begin{linenomath*}\begin{equation}
    \begin{pmatrix}
        n_p^T \\
        n_q^T \\
        n_r^T     
    \end{pmatrix} x =
    \begin{pmatrix}
        a_p & b_p & c_p \\
        a_q & b_q & c_q \\
        a_r & b_r & c_r
    \end{pmatrix} x =
    \begin{pmatrix}
        -d_p \\
        -d_q \\
        -d_r
    \end{pmatrix}.
\end{equation}\end{linenomath*}
Using Cramer's rule, $x = \left(|A_1|,  |A_2|, |A_3|\right) / |A|$, where $A = \left(n_p, n_q, n_r\right)^T$ and $A_i$ is $A$ with the $i$-th column replaced by $\left(d_p, d_q, d_r\right)$.
Written in 4D homogeneous coordinates, $x = \left(|A_1|,  |A_2|, |A_3|, |A|\right)$, or equivalently
\begin{linenomath*}\begin{equation}
    \label{eq:homocoords}
    \text{intersect}(p, q, r) = \begin{pmatrix}
        x_1 \\
        x_2 \\
        x_3 \\
        x_4
    \end{pmatrix}
    = \begin{vmatrix}
        \mathbf{e_1} & \mathbf{e_2} & \mathbf{e_3} & \mathbf{e_4} \\
        a_p & b_p & c_p & d_p \\
        a_q & b_q & c_q & d_q \\
        a_r & b_r & c_r & d_r 
    \end{vmatrix},
\end{equation}\end{linenomath*}
where $\mathbf{e_i}$ are the canonical unit vectors.
We classify points $x$ relative to plane $s$ by computing the sign of the distance to the plane $n_s^T x + d_s$.
When given in homogeneous coordinates, we can avoid fractional results by multiplying with $x_4$ and accounting for its sign:
\begin{linenomath*}\begin{equation}
    \text{classify}(x, s) = \text{sign}(x^T s) \cdot \text{sign}(x_4)
\end{equation}\end{linenomath*}
In contrast to \citet{Bernstein09}, we store vertex positions as easily interpretable homogeneous 4D integer coordinates.
The costly $4 \times 4$ determinant for classifying vertices is split into the construction of the intersection points (pre-compute four $3 \times 3$ determinants) and the actual classification (a 4D dot product).
In our mesh cutting algorithm, classification is performed more frequently than intersection construction, thus resulting in a significant overall performance gain.
(See Table~\ref{tab:math} in the Evaluation for benchmarks.)

When exact computation is required, typical approaches use either arbitrary-precision libraries (like \texttt{gmp}) or filtered floating-point predicates (like \citet{shewchuk97a}).
We found these either too slow or having insufficient resolution.
Instead, we use fixed-precision integers and analyze the operand bounds of each operation to choose appropriate bit widths.
Modern \texttt{x64} CPUs allow highly efficient implementation of wide integer addition and multiplication.
Chaining the \texttt{adc} instruction (add with carry) $k$ times implements $64\cdot k$ bit integer addition.
Multiplication is efficient because the \texttt{mul} instruction multiplies two \num{64} bit integers and stores the \num{128} bit result in two registers.
This can be used as a building block for a simple long multiplication scheme.
For very large numbers, this scheme is inefficient as it scales with \BigO{n \cdot m} (given $n$ and $m$ bit inputs) but we found it well suited for integers consisting of only a few \num{64} bit blocks.

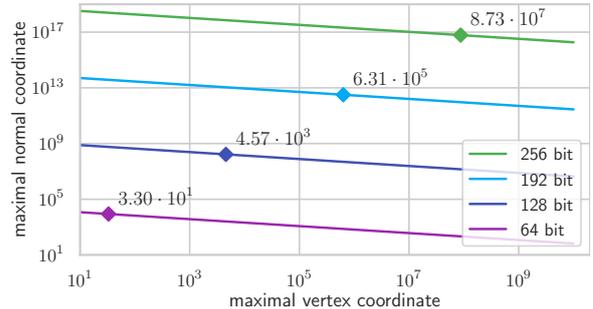
\begin{figure}
    \centering    
    \resizebox{\linewidth}{!}{\input{figures/size-tradeoff.pgf}}
    \caption{
        Given fixed-precision integer arithmetic, different tradeoffs between maximal vertex and normal coordinates exist.
        Below these curves, all our computations are guaranteed exact and overflow-free.
        The annotated points mark the maximal vertex coordinate that can be safely used when computing normals from vertex positions via edge cross-products.
        While these points are useful when importing a triangle meshes, choosing a higher vertex resolution and thus a lower normal resolution can e.g.\ be advantageous in milling simulations where there is some freedom to discretize input normals.
    }
    \label{fig:method:sizes}
\end{figure}

Given a budget of up to $b$ bits per operation, we can determine the input precision tradeoffs for normals and vertex positions of our method.
Let the biggest absolute coordinate value of any input vertex be $v^+$ and similarly that of any input normal be $n^+$.
Similarly, $d^+$ is the bound for the plane distance coefficient.
The largest intermediate result is $x^T s$ and must not exceed $\pm 2^{b-1}$ (ignoring $-2^{b-1} - 1$ for simplicity).
$x^T s$ is the same as $|p~q~r~s|$ where $p$, $q$, $r$ are the planes that $x$ was created from.
\citet{hadamard93} showed that a $4 \times 4$ determinant with matrix entries $\in [-1,1]$ cannot exceed \num{16}.
$|p~q~r~s|$ has three columns bounded by $n^+$ and one by $d^+$.
Thus, linearity of the determinant implies that $|p~q~r~s|$ cannot exceed $16 \cdot {n^+}^3 \cdot d^+$.
For our planes, $d = -v^T n$ for some input vertex $v$, so we can conservatively estimate $d^+$ as $3 \cdot v^+ \cdot n^+$.
In summary, given $b$ bit integer operations, we can guarantee overflow-free exact predicates if the following is satisfied:
\begin{linenomath*}\begin{equation}
    {n^+}^4 \cdot v^+ \leq \frac{1}{48} \cdot 2^{b - 1}
\end{equation}\end{linenomath*}
Choosing $n^+$ and $v^+$ independently is beneficial for applications that construct planes directly, but when importing a triangle mesh, normals are computed from vertex positions via $e_{12} \times e_{13}$ where $e_{ij} = v_j - v_i$ are the triangle edge vectors.
A (very) conservative estimate of $e^+$ would be $2 \cdot v^+$ and thus $n^+ = 2 \cdot {e^+}^2 = 8 {v^+}^2$ due to the cross product.
Thus, if normals are computed from vertex positions, the condition simplifies to
\begin{linenomath*}\begin{equation}
    {v^+}^9 \leq \frac{1}{8^4 \cdot 48} \cdot 2^{b - 1}
\end{equation}\end{linenomath*}
or equivalently
\begin{linenomath*}\begin{equation}
    v^+ \leq 0.239 \cdot 2^{b / 9} \approx 2^{b/9-2}.
\end{equation}\end{linenomath*}
Therefore, our $ b = 256$ bit arithmetics can exactly import all integer meshes with positions smaller than $8.73 \cdot 10^7$, or slightly above \num{26} bit.
For comparison, single precision floating point has \num{23} bit mantissa and most ASCII meshes we tested do not exceed 7 significant decimal digits.
Figure~\ref{fig:method:sizes} shows different tradeoffs for varying bit widths $b$.

For a simulation with a bounding volume of \SI{1}{\cubic\metre}, we get a resolution of \SI{0.22}{\milli\meter} for \num{128} bit, \SI{1.58}{\micro\meter} for \num{192} bit, and \SI{11.5}{\nano\meter} for \num{256} bit, assuming we want to guarantee that any input normal can be exactly represented.
In Section~\ref{sec:eval:math} we evaluate the performance of operations described in this section and compare \SI{128}{\bit}, \SI{192}{\bit}, \SI{256}{\bit}, and \texttt{gmp} versions.

% -----------------------------------------------------------------------------------------------------
\subsection{Mesh Cutting}
\label{sec:method:cut}

\begin{figure*}
    \centering
    \subcaptionbox{edge descent}{\includegraphics[width=0.3\linewidth]{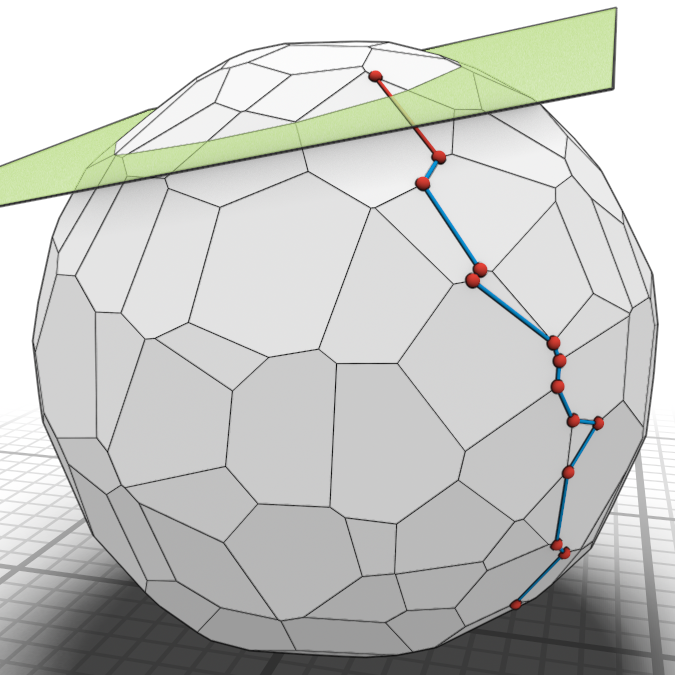}}~~
    \subcaptionbox{marching}{\includegraphics[width=0.3\linewidth]{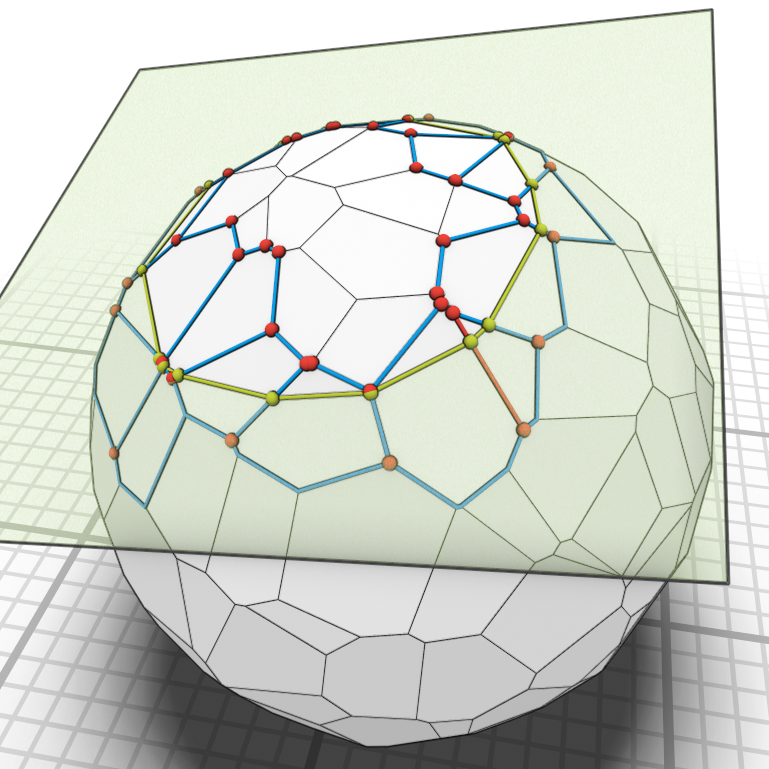}}~~
    \subcaptionbox{result}{\includegraphics[width=0.3\linewidth]{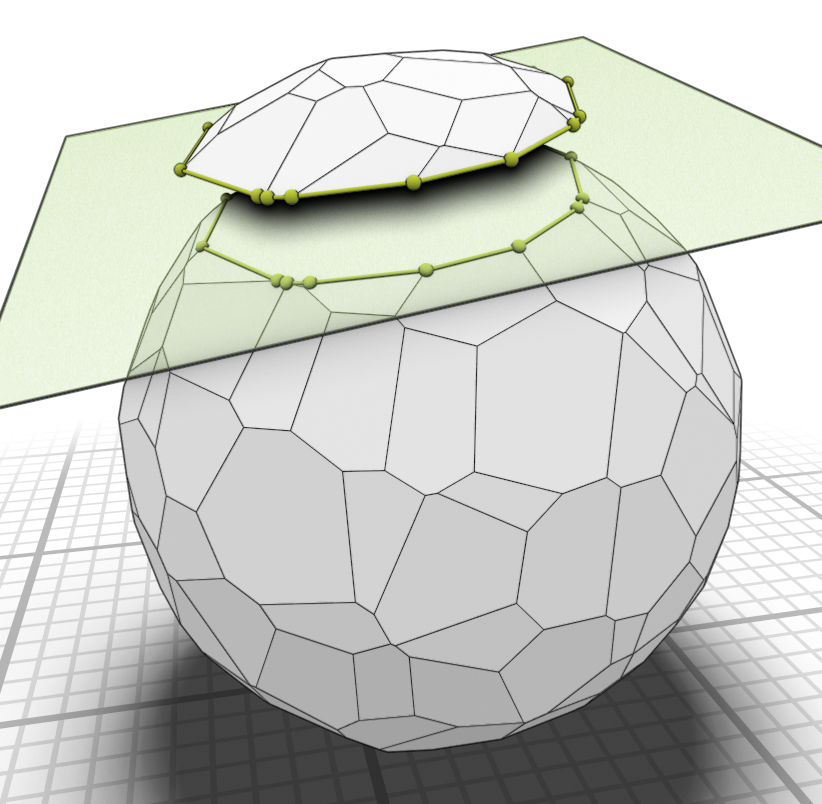}}
    \caption{
        (a) In the \emph{edge descent} step, our mesh cutting algorithm first finds an edge intersecting the cut plane (red edge) by following vertices with shorter distance-to-plane (blue edges).
        Only vertices along the way are tested (red).
        (b) In the \emph{marching} step, we cut each affected face and insert cut edges (green) until reaching the starting point again.
        Again, only the red vertices need to be classified.
        (c) The result is the mesh cut in-place into lower and upper part (explosion view).
    }
    \label{fig:cut}
\end{figure*}

\begin{figure}
    \centering
    \fontsize{7pt}{11pt}\selectfont
    \def\svgwidth{\columnwidth}
    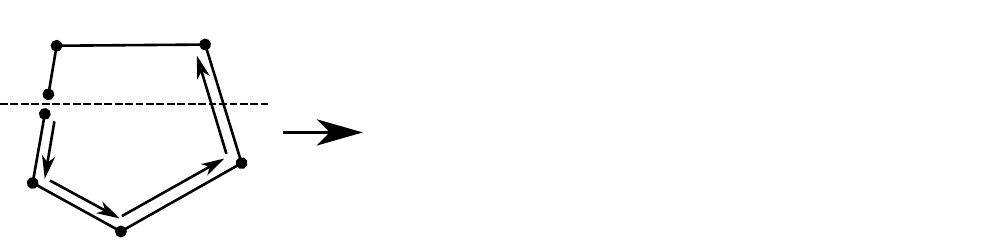
    \caption{
        During the marching there are only two distinct cases when splitting a single face: 
        $a)$ The face is traversed until an edge with vertices on opposite sides of the cutting plane are found. 
        The edge is then split and we are now in case $b)$ 
        A vertex lies exactly on the cutting plane.
        The vertex is duplicated and two new edges on the cutting plane are inserted.
    }
    \label{fig:cut:marching}
\end{figure}

The core of our method is the following operation: cut a convex mesh into two parts along a given plane.
With our numerical foundation, all results are exact and no epsilon tolerance is required.
By modifying the mesh in-place and trying to minimize how many geometric primitives need to be processed we obtain a highly efficient cutting procedure.

This operation serves as the basis for boundary extraction (Section~\ref{sec:method:extract}), BSP Booleans (Section~\ref{sec:method:Booleans}), and BSP simplification (Section~\ref{sec:method:simpl}).
Previous methods use cutting for these as well but suffered from stability issues and overall quadratic scaling \citep{naylor1990merging,lysenko2008improved,Bernstein09}. 

Figure~\ref{fig:cut} illustrates our algorithm.
Given a starting halfedge and the cutting plane, we traverse from vertex to vertex, always choosing the neighbor vertex with smallest distance to the plane, a process we call \emph{edge descent}.
We stop when reaching a vertex on the opposite side of the plane.
If no such vertex can be found, the edge descent terminates in a global minimum (all submeshes are convex) and we know that the plane does not properly intersect the mesh.
In our case, a \emph{proper} intersection means that both resulting meshes from the cut have non-zero volume.
For example, a non-proper intersection would be a cut plane that is coplanar with a mesh face.
In case of a proper intersection, we always find an edge that intersects with or a vertex that lies on the cutting plane.
From there we follow the faces-to-cut around until we reach the initial edge again.
Along the way, we add the new faces and edges produced by the cut and locally apply the necessary changes to the mesh topology.
We call this step \emph{marching}.
The decision which edge or vertex lies on the cutting plane uses the classification function from the previous section.
With exact computations, the meshes stay perfectly convex after cutting.
During marching, only two different cases can occur: we find the next edge intersecting the plane or we find a vertex lying on the plane (see Figure~\ref{fig:cut:marching}).
If an edge lies within the cutting plane, no neighboring face needs to be split.

At the end, the single convex mesh is split into two parts, one belonging to the positive half-space defined by the plane and the other to the negative half-space.
Two new planar faces are inserted.
In case one of the parts has zero volume, the plane does not have a proper intersection and the mesh is not changed.
The main reason why this algorithm is fast is that most of the time only a fraction of the mesh is traversed and the modification is performed in-place.
Our experiments in Figure~\ref{fig:eval:convex} demonstrate superior results compared with the naive, classify-all-vertices method.
Empirically, we observe that the per-cut costs grow sub-linear with the mesh size.

It is important to note that the resulting mesh has neither integer nor floating-point coordinates.
Instead, all vertices are represented as the homogeneous 4D integer coordinates introduced in Section~\ref{sec:method:int}, especially Equation~\ref{eq:homocoords}.
This makes the cutting operation exact and fully robust, allowing us to use it as a basis for the boundary mesh extraction and even the BSP booleans.

There is a certain similarity with the GJK algorithm \cite{GJKAlgo} for collision detection between two convex objects.
While our plane-mesh cutting traverses from edge to edge, choosing those that reduce the distance to the cutting plane the most, the GJK algorithm traverses from simplex to simplex, choosing the simplex closest to the origin.
On a meta level, both algorithms exploit that, on a convex polyhedron, we can minimize certain distance functions by greedily follow edges or sub-simplices without running into local minima.

% -----------------------------------------------------------------------------------------------------
\subsection{Boundary Extraction}
\label{sec:method:extract}

\begin{figure}
    \centering
    
    \subcaptionbox{}{\includegraphics[width=0.47\linewidth]{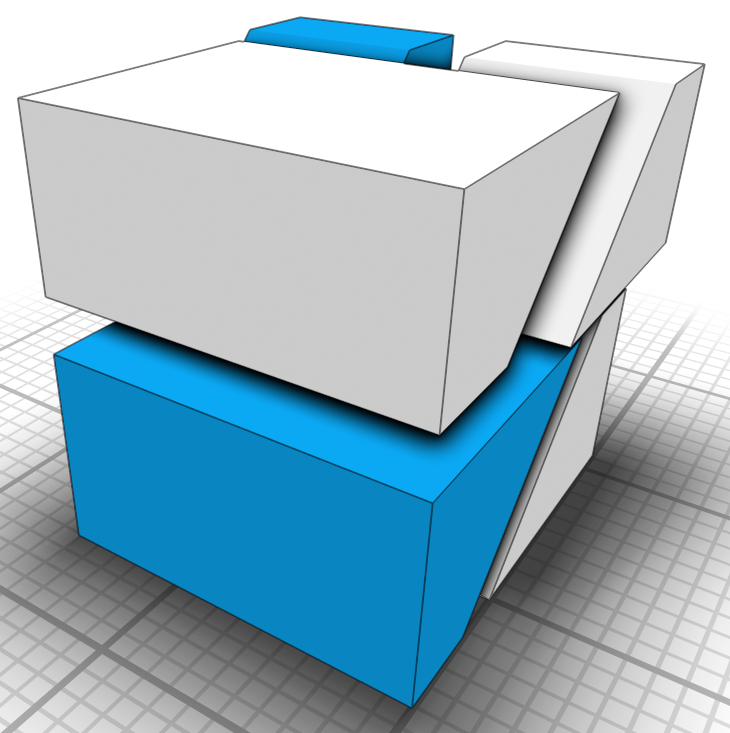}}~
    \subcaptionbox{}{\includegraphics[width=0.47\linewidth]{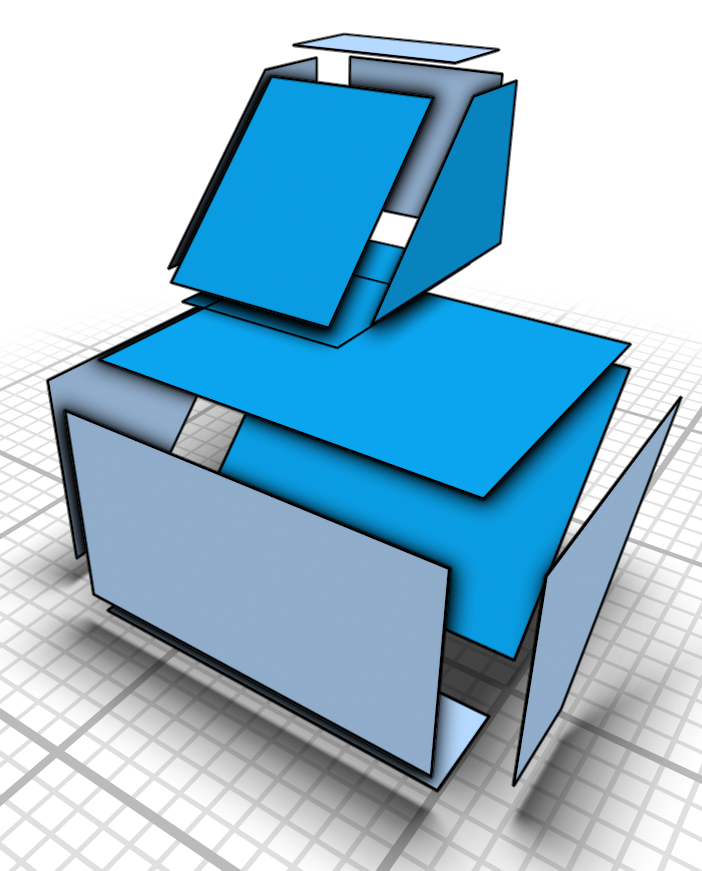}}

    \subcaptionbox{}{\includegraphics[width=0.47\linewidth]{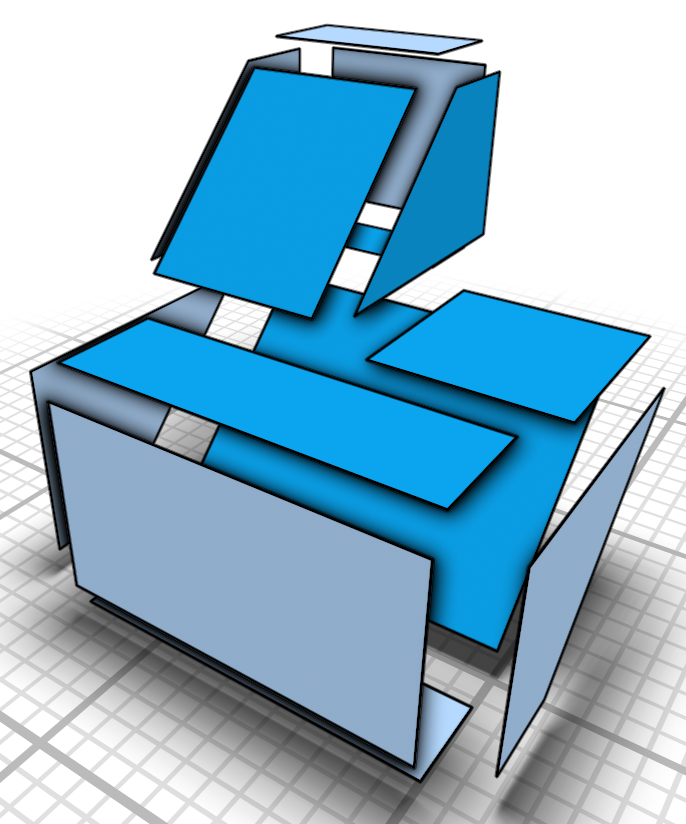}}~
    \subcaptionbox{}{\includegraphics[width=0.47\linewidth]{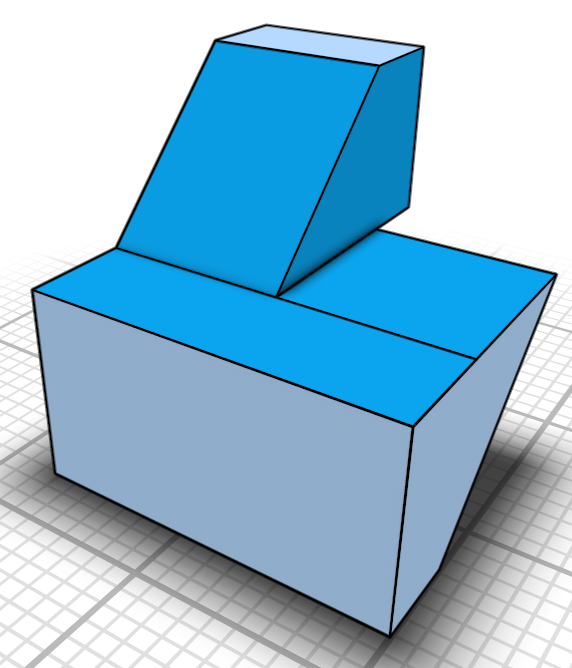}}

    \caption{
        Exact mesh boundary extraction.
        Given an input BSP (a), we use our mesh cutting to extract convex polyhedra for each BSP leaf, discarding \textbf{out}-cells (b, explosion view).
        The \textbf{in}-cells can contain overlapping interior surfaces which are removed by a 2D face-plane clipping (c, explosion view).
        The resulting mesh (d) exactly describes the surface represented by the BSP (light blue are octree cell boundaries, darker blue result from BSP planes).
    }
    \label{fig:method:boundary}
\end{figure}

\begin{figure}
    \centering
    \includegraphics[width=\linewidth]{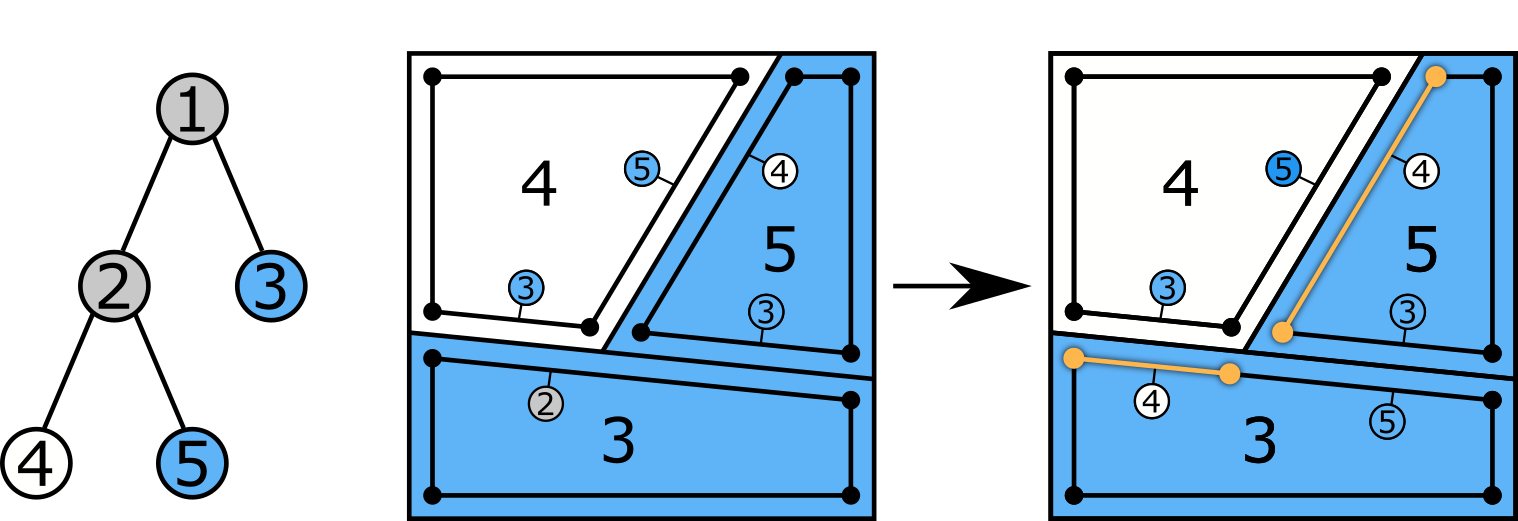}
    \caption{
        Every time a polyhedron is cut during boundary extraction the two new faces inside the cutting plane are assigned the BSP node on the opposite side of the cutting plane (left and center).
        To extract the exact boundary, faces not pointing to leaf nodes are cut again according to their opposite sub-BSP (right).
        Boundary faces (orange) belong to BSP nodes marked as \textbf{in} with opposite nodes \textbf{out}.
    }
    \label{fig:method:boundary-2d}
\end{figure}

Based on mesh-plane cutting, we can formulate an efficient boundary extraction algorithm.
The task is, given a BSP tree and a bounding box, to compute a mesh of the represented surface inside the bounding box.
Our algorithm works in two steps.

First, we compute convex meshes for each BSP leaf node labelled \textbf{in}.
This is done recursively by starting with a mesh for the bounding box and applying our cutting operation for each inner node of the BSP.
During cutting we annotate each face with the index of the \emph{opposite} BSP node.
All \textbf{out} leaves are discarded.

While these convex meshes can be used for rendering, they contain many interior (double) faces.
Thus, in a second step, we remove all parts of the faces that are not \textbf{in-out} transitions.
For each face we know the opposite BSP subtree and cut the face recursively against this tree.
This procedure is shown in Figure~\ref{fig:method:boundary-2d}.
This is a simple convex face against plane cutting, equivalent to polygon clipping.
When the leaf nodes are reached we can classify the face pieces as \textbf{in-out} or \textbf{in-in}, discarding the latter.
Figure~\ref{fig:method:boundary} shows the result of this step.

Note that only clipping faces against the opposite BSP will produce T-junctions that might produce problems with rendering and further processing.
However, during clipping we can again store the opposite BSP and afterwards subdivide polygon edges according to this opposite BSP, introducing valence-2 vertices.
Take for example the upwards pointing face in the middle of Figure~\ref{fig:method:boundary} (c).
After the polygon clipping, this is a rectangle, resulting in a non-conforming mesh with a T-junction.
After introducing the valence-2 vertices, it topologically becomes a pentagon.
Thus, in our 3D use case, the procedure in Figure~\ref{fig:method:boundary-2d} is actually applied two times:
First, to clip the boundary polygons to get rid of inner faces.
Second, to subdivide boundary edges to prevent one-sided T-junctions.

% -----------------------------------------------------------------------------------------------------
\subsection{Booleans on BSPs}
\label{sec:method:Booleans}

Given two BSPs representing geometry, \citet{naylor1990merging} describe how to \emph{merge} these.
The \emph{merge} procedure takes two BSPs and a \textit{merge function} $f_m(s_a, s_b) \rightarrow s$, where $s_a, s_b, s \in \{\textbf{in}, \textbf{out}\}$ are leaf node labels.
The choice of $f_m$ determines which Boolean operation is performed.
For example, the \emph{union} is defined by the function that only returns \textbf{out} when both $s_a$ and $s_b$ are \textbf{out}.

We formulate the BSP \emph{merge} based on our efficient mesh cutting.
Starting with a cube representing the bounding volume and root nodes $A$ and $B$ of two BSPs, we define the following \emph{merge} operation that recurses on $A$:
\begin{enumerate}
    \item Terminate if either $A$ or $B$ is a leaf node (based on the merge function, either \textbf{in}, \textbf{out}, $A$, $B$, or leaf-inverted versions of $A$ or $B$ are inserted at the current node)
    \item Cut $A$'s plane against the current mesh
        \begin{enumerate}
            \item if $A$'s plane produces a proper intersection with the current mesh, recursively call $\text{merge}(A.\text{left}, B)$ and $\text{merge}(A.\text{right}, B)$
            \item otherwise recursively call $\text{merge}(A', B)$ where $A'$ is the child of $A$ that contains $B$
        \end{enumerate}
\end{enumerate}
This algorithm produces a BSP that corresponds to the chosen Boolean operation applied to the two input BSPs as well as a polygonal mesh.
The mesh consists of a set of convex polyhedra that can each be mapped to one BSP node of the output BSP.
Note that they may not always correspond to a leaf node:
The \emph{merge} procedure never recurses $B$ but instead may copy $B$ or its leaf-inverse directly into the output BSP.
In that case the corresponding polyhedron belongs to $B$ which may be an inner node.
In the resulting BSP some nodes that belong to $B$ may not produce any volume because the planes of $B$ are not guaranteed to intersect with its corresponding polyhedron.
In our implementation we circumvent this by also fully cutting the corresponding polyhedron according to $B$.
This fits nicely with our pipeline since each Boolean operation is followed by a \emph{Redundancy Removal} step (see Section~\ref{sec:method:simpl}) where the cut mesh is required anyways.

% -----------------------------------------------------------------------------------------------------
\subsection{BSP Import}
\label{sec:method:bsp-import}

For converting a mesh into BSP representation we apply a commonly used algorithm \citep{naylor1990merging} that iteratively builds a BSP from the given input faces.
At first, the BSP is initialized as a single leaf node.
During each iteration, an input face is chosen and inserted into the BSP.
Traversal starts at the root node. 
Traversed leaf nodes are replaced by inner nodes with the plane equal to the supporting plane of the inserted face and traversal stops.
The plane's normal direction determines which children of the new node are marked \textbf{in} or \textbf{out}.
When an inner node is encountered there are three different cases: 
Either the face intersects with the node's plane.
It is then split along the plane resulting in two faces, each traversing its corresponding subtree.
If the face lies completely on one side of the node's plane, it traverses only that subtree.
Finally, if the face is coplanar to the node's plane, the face is discarded.
This step is important for the merging of split faces in the redundancy removal step (Section~\ref{sec:method:simpl}).

To avoid stability issues caused by floating point inaccuracies, the face-splitting algorithm must be performed in an exact manner.
Therefore we use the same plane-based vertex representation as presented in Section~\ref{sec:method:int}.
\begin{wrapfigure}{r}{0.3\linewidth}
        \includegraphics[trim=40 20 0 20, width=\linewidth]{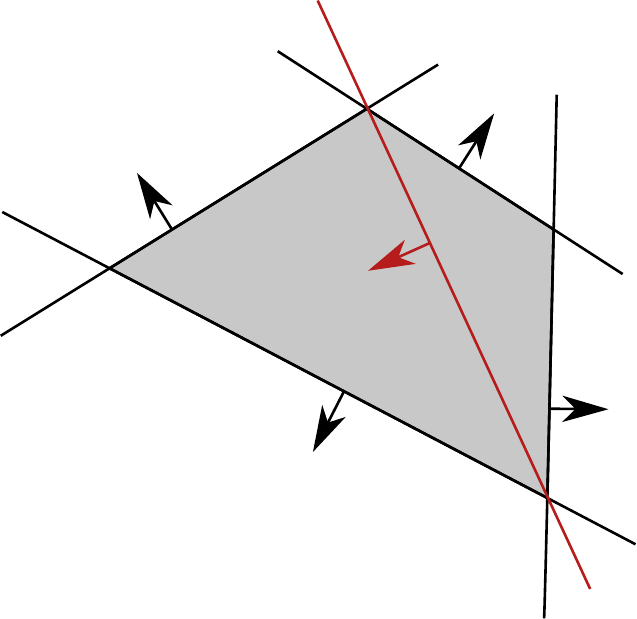}
\end{wrapfigure}
Even if the input faces consist only of triangles, intermediate faces are generally arbitrary convex polygons with corners defined by plane intersections. 
Re-triangulation is not possible because it requires construction of new planes (red in the inset) from points that are not original input vertices and thus may lie on fractional coordinates.
Such new planes might then require more precision than we have available and therefore cannot be constructed in the general case.

A crucial step for creating well balanced BSPs is the choice of the next face that is inserted into the BSP.
Heuristics that aim to improve the \textit{balanced-ness} of the BSP are often expensive and scale poorly \cite{lysenko2008improved}, as they often require global knowledge.
As our BSPs are naturally limited by their enclosing octree and their number of inner nodes is bounded, we found that choosing the faces in a random order is sufficient in our case.

% -----------------------------------------------------------------------------------------------------
\subsection{BSP Redundancy Removal}
\label{sec:method:simpl}

The result of CSG operations on BSP trees often produces suboptimal BSPs that contain redundant or unnecessary nodes.
Figure~\ref{fig:bsp:workflow} shows a simple example.
Although we are unaware of methods that produce optimal trees, a common way to improve the BSP structure is to extract the corresponding mesh and rebuild a new BSP from scratch.
This quickly becomes infeasible in an inexact setting as rounding errors accumulate and prevent the rebuilt BSP from representing the exact same volume as the original BSP.
However, our boundary extraction from Section~\ref{sec:method:extract} is fast and exact, making the simplification-through-rebuilding approach effective.
In practice, we perform the simplification after each BSP merge.
This means little additional computation is required to perform the boundary extraction:
The merge procedure already produces a convex cell for each leaf node.
What remains is the removal of interior overlapping surfaces (cf. Section~\ref{sec:method:extract}).

Other exact methods, like \cite{Bernstein09,Campen10,Zhou16}, could apply this approach as well.
However, this incurs a significant performance penalty when used without our fast mesh cutting from Section~\ref{sec:method:cut}.

% -----------------------------------------------------------------------------------------------------
\subsection{Octree-Embedded BSPs}
\label{sec:method:octree}
\begin{figure}
    \centering
    \includegraphics[width=\linewidth]{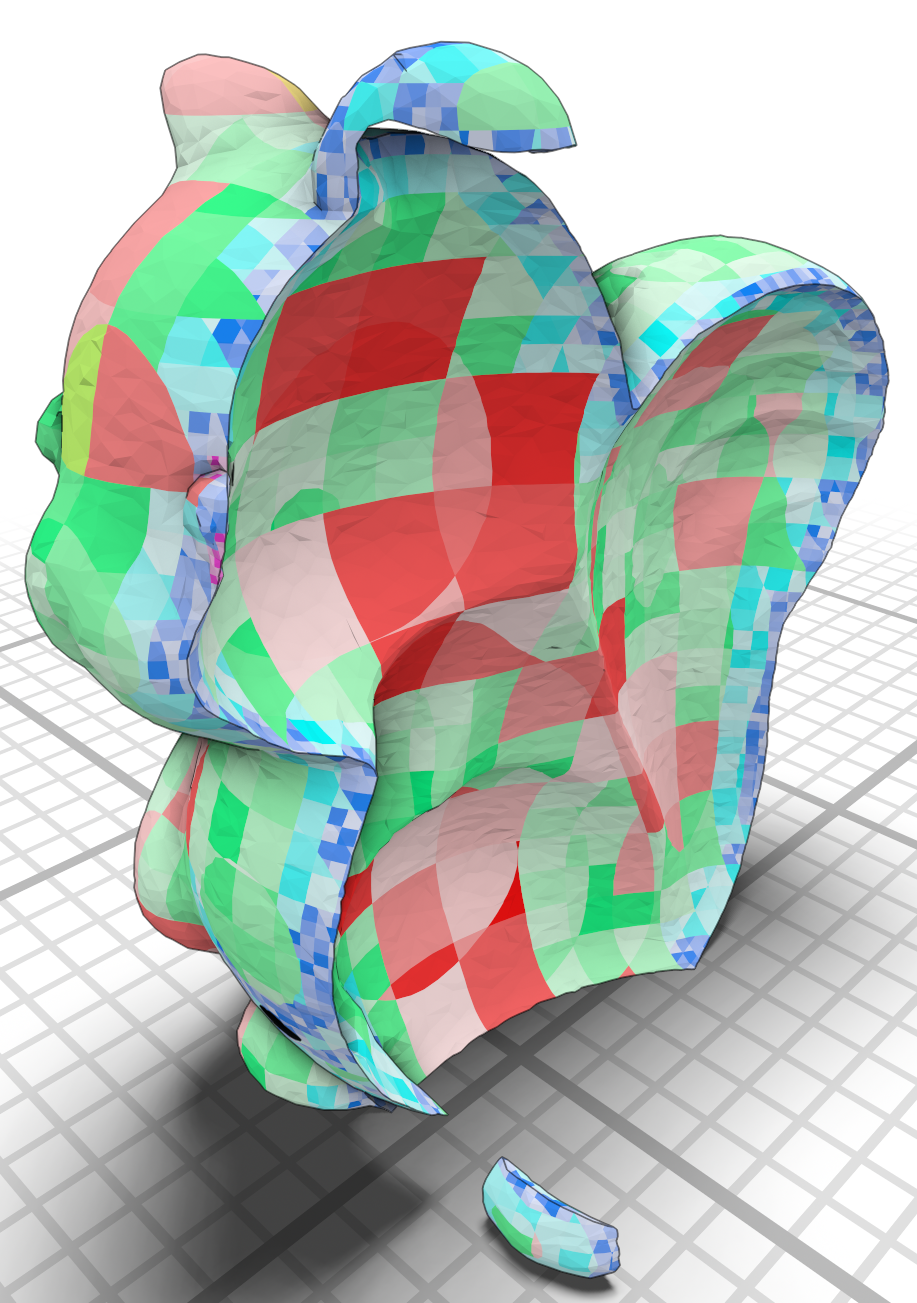}
    \caption{
        Each leaf of the octree contains a BSP. 
        The BSP size is the subdivision criterion for each octree cell.
        The surface is color-coded by octree cell.
    }
    \label{fig:method:octree}
\end{figure}
\begin{figure}
    \centering
    \includegraphics[width=\linewidth]{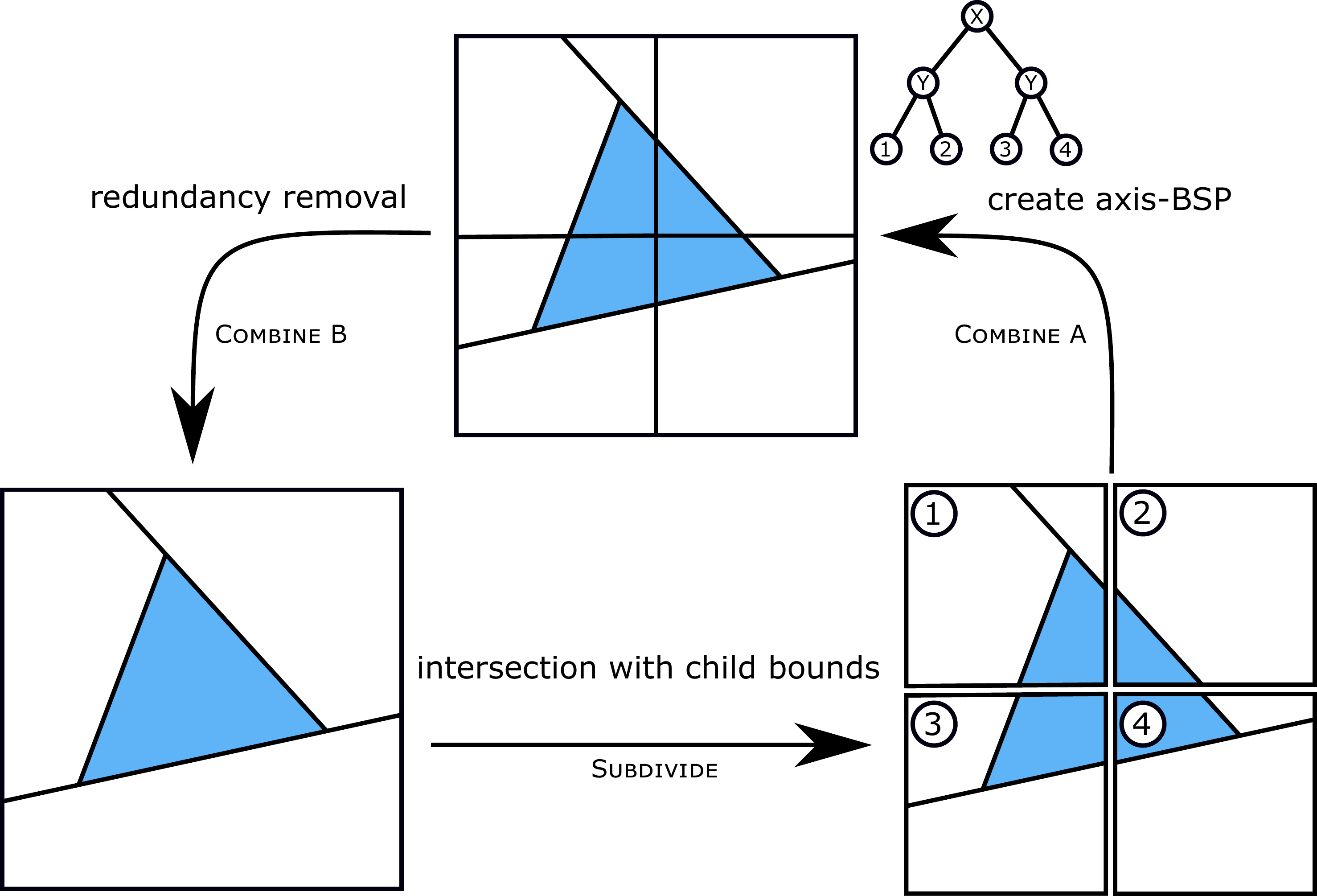}
    \caption{
        Subdividing an octree cell is equivalent to intersecting the cells BSP with the bounding box of each of its child cells.
        For combining the children of a cell, their BSPs are inserted into the leaf nodes of a BSP that contains the axis-aligned planes separating the children (\textsc{Combine A}).
        This is followed by a step (\textsc{Combine B}) of redundancy removal (\cf Section~\ref{sec:method:simpl}) which eliminates the planes from the axis-BSP.
    }
    \label{fig:method:merge-and-split}
\end{figure}
BSPs offer a well defined method to perform CSG operations.
However, these methods scale superlinearly with the size of the BSP (see Figure~\ref{fig:eval:merge}).
To circumvent this, we embed BSPs with a limited number of nodes into a global octree.
An example of this is shown in Figure~\ref{fig:method:octree}.
Changes to the geometry may cause a BSP to exceed the maximal number of nodes.
In that case its corresponding octree cell is subdivided into eight equally sized child cells.
The smaller child bounding boxes paired with a redundancy removal step cause a decrease in BSP sizes per octree cell, with the exception of a few pathological cases.
For example, consider a valence 250 vertex without coplanar adjacent faces that lies at a fractional coordinate.
The cell containing this vertex cannot be subdivided to contain less than 250 BSP nodes.
In that case, some higher complexity nodes have to be accepted.

It may be necessary to perform multiple splits until the BSP size has dropped under the given threshold.
Since the complexity of the merge procedure can result in quadratic runtime and output size, smaller BSP-trees perform better.
However, reducing the maximal BSP size causes an increase in octree refinement.
Experimentally, we found a good compromise for the BSP threshold between \num{100} to \num{200} (\cf Section~\ref{sec:eval:octree}).
To assure the octree complexity stays low, octree leaf nodes that have a combined complexity small enough to fit into a single octree cell are merged together.

As octrees are a special form of BSP trees, cell subdivision and merge can be implemented efficiently using our BSP algorithms.
A 2 dimensional version of the process is shown in Figure~\ref{fig:method:merge-and-split}.
Subdividing an octree cell is equivalent to computing the intersection of its BSP with the boundary of each of its children.
The opposite -- combining eight leaf cells into one -- is technically the union of the eight child BSPs.
However, this requires the inclusion of their boundaries, as BSPs are in general unbounded.
The outer boundary is already implicitly given by the parent octree cell.
To represent the inner boundaries that separate the eight children from each other, we build an \textit{axis-BSP}.
The first three of its levels each contain planes that separate the child cells along one axis.
The eight leaf nodes are then replaced with the BSP from their corresponding octree cell.
This technically concludes the combing step.
However, the additional planes introduced by the \textit{axis-BSP} are rarely part of the solid boundary and the combined child BSPs often share duplicate planes.
Therefore the combine step is always followed by a redundancy removal (\cf Section~\ref{sec:method:simpl}).

% -----------------------------------------------------------------------------------------------------
\subsection{Octree Import}
\label{sec:method:octree-import}

We have already described how to convert a mesh into a BSP.
After the input mesh has been rounded once to integer coordinates (according to Section~\ref{sec:method:int}) further operations make use of exact calculations.

Converting a mesh into the octree could be achieved by creating a single BSP in the octree root and then subdividing until the threshold for each octree cell is reached.
However, the subdivision scales poorly with the large BSP size.
Instead, we first create a triangle octree.
The splitting criterion of an octree cell is first based on the number of triangles.
The result is an octree with triangles in some leaf cells. 
Next, triangles are clipped to their octree node boundary and are then used to create a BSP for the given cell.
The clipping is necessary to ensure the correct creation of BSPs.
After creating the BSP, the octree cell complexity bound is enforced by another round of subdivisions, which is now based on the actual BSP sizes.
At this point, all octree cells that contain parts of the input geometry store a valid BSP.
To correctly classify the remaining octree cells we perform a flood-fill (on octree cells) to propagate the correct \textbf{in} and \textbf{out} labels.

Note that the triangle clipping must be exact and result in polygons.
The newly created vertices on the clipping plane may not necessary lie on integer coordinates.
Instead, they use the same 4D homogeneous representation that we describe in Section~\ref{sec:method:int}.
It is also not possible to re-triangulate the clipped polygons, as that would require constructing new planes which in general need more bits than the input planes.

%% file: figures/bsp_workflow.pdf_tex
%% Accompanies image file 'bsp_workflow.pdf' (pdf, eps, ps)
\begingroup%
  \makeatletter%
  \providecommand\color[2][]{%
    \errmessage{(Inkscape) Color is used for the text in Inkscape, but the package 'color.sty' is not loaded}%
    \renewcommand\color[2][]{}%
  }%
  \providecommand\transparent[1]{%
    \errmessage{(Inkscape) Transparency is used (non-zero) for the text in Inkscape, but the package 'transparent.sty' is not loaded}%
    \renewcommand\transparent[1]{}%
  }%
  \providecommand\rotatebox[2]{#2}%
  \newcommand*\fsize{\dimexpr\f@size pt\relax}%
  \newcommand*\lineheight[1]{\fontsize{\fsize}{#1\fsize}\selectfont}%
  \ifx\svgwidth\undefined%
    \setlength{\unitlength}{525.75bp}%
    \ifx\svgscale\undefined%
      \relax%
    \else%
      \setlength{\unitlength}{\unitlength * \real{\svgscale}}%
    \fi%
  \else%
    \setlength{\unitlength}{\svgwidth}%
  \fi%
  \global\let\svgwidth\undefined%
  \global\let\svgscale\undefined%
  \makeatother%
  \begin{picture}(1,0.38751978)%
    \lineheight{1}%
    \setlength\tabcolsep{0pt}%
    \put(0,0){\includegraphics[width=\unitlength,page=1]{bsp_workflow.pdf}}%
    \put(0.07203994,0.00021593){\makebox(0,0)[t]{\lineheight{1.25}\smash{\begin{tabular}[t]{c}Octree Cell\end{tabular}}}}%
    \put(0,0){\includegraphics[width=\unitlength,page=2]{bsp_workflow.pdf}}%
    \put(0.07203994,0.21419596){\makebox(0,0)[t]{\lineheight{1.25}\smash{\begin{tabular}[t]{c}Other BSP\end{tabular}}}}%
    \put(0,0){\includegraphics[width=\unitlength,page=3]{bsp_workflow.pdf}}%
    \put(0.64265335,0.10720594){\makebox(0,0)[t]{\lineheight{1.25}\smash{\begin{tabular}[t]{c}Mesh\end{tabular}}}}%
    \put(0,0){\includegraphics[width=\unitlength,page=4]{bsp_workflow.pdf}}%
    \put(0.92796006,0.10720594){\makebox(0,0)[t]{\lineheight{1.25}\smash{\begin{tabular}[t]{c}Final BSP\end{tabular}}}}%
    \put(0,0){\includegraphics[width=\unitlength,page=5]{bsp_workflow.pdf}}%
    \put(0.5,0.17995915){\makebox(0,0)[t]{\lineheight{1.25}\smash{\begin{tabular}[t]{c}Extract\end{tabular}}}}%
    \put(0,0){\includegraphics[width=\unitlength,page=6]{bsp_workflow.pdf}}%
    \put(0.7853067,0.17995915){\makebox(0,0)[t]{\lineheight{1.25}\smash{\begin{tabular}[t]{c}Import\end{tabular}}}}%
    \put(0.2169159,0.32933068){\color[rgb]{0,0,0}\makebox(0,0)[t]{\lineheight{0.25}\smash{\begin{tabular}[t]{c}Boolean\end{tabular}}}}%
    \put(0.3575382,0.10720594){\color[rgb]{0,0,0}\makebox(0,0)[t]{\lineheight{1.25}\smash{\begin{tabular}[t]{c}Suboptimal\\BSP\end{tabular}}}}%
  \end{picture}%
\endgroup%

%% file: figures/size-tradeoff.pgf
%% Creator: Matplotlib, PGF backend
%%
%% To include the figure in your LaTeX document, write
%%   \input{<filename>.pgf}
%%
%% Make sure the required packages are loaded in your preamble
%%   \usepackage{pgf}
%%
%% Figures using additional raster images can only be included by \input if
%% they are in the same directory as the main LaTeX file. For loading figures
%% from other directories you can use the `import` package
%% and then include the figures with
%%   \import{<path to file>}{<filename>.pgf}
%%
%% Matplotlib used the following preamble
%%
\begingroup%
\makeatletter%
\begin{pgfpicture}%
\pgfpathrectangle{\pgfpointorigin}{\pgfqpoint{5.000000in}{3.000000in}}%
\pgfusepath{use as bounding box, clip}%
\begin{pgfscope}%
\pgfsetbuttcap%
\pgfsetmiterjoin%
\definecolor{currentfill}{rgb}{1.000000,1.000000,1.000000}%
\pgfsetfillcolor{currentfill}%
\pgfsetlinewidth{0.000000pt}%
\definecolor{currentstroke}{rgb}{1.000000,1.000000,1.000000}%
\pgfsetstrokecolor{currentstroke}%
\pgfsetdash{}{0pt}%
\pgfpathmoveto{\pgfqpoint{0.000000in}{0.000000in}}%
\pgfpathlineto{\pgfqpoint{5.000000in}{0.000000in}}%
\pgfpathlineto{\pgfqpoint{5.000000in}{3.000000in}}%
\pgfpathlineto{\pgfqpoint{0.000000in}{3.000000in}}%
\pgfpathclose%
\pgfusepath{fill}%
\end{pgfscope}%
\begin{pgfscope}%
\pgfsetbuttcap%
\pgfsetmiterjoin%
\definecolor{currentfill}{rgb}{1.000000,1.000000,1.000000}%
\pgfsetfillcolor{currentfill}%
\pgfsetlinewidth{0.000000pt}%
\definecolor{currentstroke}{rgb}{0.000000,0.000000,0.000000}%
\pgfsetstrokecolor{currentstroke}%
\pgfsetstrokeopacity{0.000000}%
\pgfsetdash{}{0pt}%
\pgfpathmoveto{\pgfqpoint{0.700000in}{0.720000in}}%
\pgfpathlineto{\pgfqpoint{5.000000in}{0.720000in}}%
\pgfpathlineto{\pgfqpoint{5.000000in}{2.850000in}}%
\pgfpathlineto{\pgfqpoint{0.700000in}{2.850000in}}%
\pgfpathclose%
\pgfusepath{fill}%
\end{pgfscope}%
\begin{pgfscope}%
\pgfpathrectangle{\pgfqpoint{0.700000in}{0.720000in}}{\pgfqpoint{4.300000in}{2.130000in}}%
\pgfusepath{clip}%
\pgfsetroundcap%
\pgfsetroundjoin%
\pgfsetlinewidth{1.003750pt}%
\definecolor{currentstroke}{rgb}{0.800000,0.800000,0.800000}%
\pgfsetstrokecolor{currentstroke}%
\pgfsetdash{}{0pt}%
\pgfpathmoveto{\pgfqpoint{0.700000in}{0.720000in}}%
\pgfpathlineto{\pgfqpoint{0.700000in}{2.850000in}}%
\pgfusepath{stroke}%
\end{pgfscope}%
\begin{pgfscope}%
\definecolor{textcolor}{rgb}{0.150000,0.150000,0.150000}%
\pgfsetstrokecolor{textcolor}%
\pgfsetfillcolor{textcolor}%
\pgftext[x=0.700000in,y=0.588056in,,top]{\color{textcolor}\sffamily\fontsize{11.000000}{13.200000}\selectfont \(\displaystyle 10^{1}\)}%
\end{pgfscope}%
\begin{pgfscope}%
\pgfpathrectangle{\pgfqpoint{0.700000in}{0.720000in}}{\pgfqpoint{4.300000in}{2.130000in}}%
\pgfusepath{clip}%
\pgfsetroundcap%
\pgfsetroundjoin%
\pgfsetlinewidth{1.003750pt}%
\definecolor{currentstroke}{rgb}{0.800000,0.800000,0.800000}%
\pgfsetstrokecolor{currentstroke}%
\pgfsetdash{}{0pt}%
\pgfpathmoveto{\pgfqpoint{1.624629in}{0.720000in}}%
\pgfpathlineto{\pgfqpoint{1.624629in}{2.850000in}}%
\pgfusepath{stroke}%
\end{pgfscope}%
\begin{pgfscope}%
\definecolor{textcolor}{rgb}{0.150000,0.150000,0.150000}%
\pgfsetstrokecolor{textcolor}%
\pgfsetfillcolor{textcolor}%
\pgftext[x=1.624629in,y=0.588056in,,top]{\color{textcolor}\sffamily\fontsize{11.000000}{13.200000}\selectfont \(\displaystyle 10^{3}\)}%
\end{pgfscope}%
\begin{pgfscope}%
\pgfpathrectangle{\pgfqpoint{0.700000in}{0.720000in}}{\pgfqpoint{4.300000in}{2.130000in}}%
\pgfusepath{clip}%
\pgfsetroundcap%
\pgfsetroundjoin%
\pgfsetlinewidth{1.003750pt}%
\definecolor{currentstroke}{rgb}{0.800000,0.800000,0.800000}%
\pgfsetstrokecolor{currentstroke}%
\pgfsetdash{}{0pt}%
\pgfpathmoveto{\pgfqpoint{2.549258in}{0.720000in}}%
\pgfpathlineto{\pgfqpoint{2.549258in}{2.850000in}}%
\pgfusepath{stroke}%
\end{pgfscope}%
\begin{pgfscope}%
\definecolor{textcolor}{rgb}{0.150000,0.150000,0.150000}%
\pgfsetstrokecolor{textcolor}%
\pgfsetfillcolor{textcolor}%
\pgftext[x=2.549258in,y=0.588056in,,top]{\color{textcolor}\sffamily\fontsize{11.000000}{13.200000}\selectfont \(\displaystyle 10^{5}\)}%
\end{pgfscope}%
\begin{pgfscope}%
\pgfpathrectangle{\pgfqpoint{0.700000in}{0.720000in}}{\pgfqpoint{4.300000in}{2.130000in}}%
\pgfusepath{clip}%
\pgfsetroundcap%
\pgfsetroundjoin%
\pgfsetlinewidth{1.003750pt}%
\definecolor{currentstroke}{rgb}{0.800000,0.800000,0.800000}%
\pgfsetstrokecolor{currentstroke}%
\pgfsetdash{}{0pt}%
\pgfpathmoveto{\pgfqpoint{3.473886in}{0.720000in}}%
\pgfpathlineto{\pgfqpoint{3.473886in}{2.850000in}}%
\pgfusepath{stroke}%
\end{pgfscope}%
\begin{pgfscope}%
\definecolor{textcolor}{rgb}{0.150000,0.150000,0.150000}%
\pgfsetstrokecolor{textcolor}%
\pgfsetfillcolor{textcolor}%
\pgftext[x=3.473886in,y=0.588056in,,top]{\color{textcolor}\sffamily\fontsize{11.000000}{13.200000}\selectfont \(\displaystyle 10^{7}\)}%
\end{pgfscope}%
\begin{pgfscope}%
\pgfpathrectangle{\pgfqpoint{0.700000in}{0.720000in}}{\pgfqpoint{4.300000in}{2.130000in}}%
\pgfusepath{clip}%
\pgfsetroundcap%
\pgfsetroundjoin%
\pgfsetlinewidth{1.003750pt}%
\definecolor{currentstroke}{rgb}{0.800000,0.800000,0.800000}%
\pgfsetstrokecolor{currentstroke}%
\pgfsetdash{}{0pt}%
\pgfpathmoveto{\pgfqpoint{4.398515in}{0.720000in}}%
\pgfpathlineto{\pgfqpoint{4.398515in}{2.850000in}}%
\pgfusepath{stroke}%
\end{pgfscope}%
\begin{pgfscope}%
\definecolor{textcolor}{rgb}{0.150000,0.150000,0.150000}%
\pgfsetstrokecolor{textcolor}%
\pgfsetfillcolor{textcolor}%
\pgftext[x=4.398515in,y=0.588056in,,top]{\color{textcolor}\sffamily\fontsize{11.000000}{13.200000}\selectfont \(\displaystyle 10^{9}\)}%
\end{pgfscope}%
\begin{pgfscope}%
\definecolor{textcolor}{rgb}{0.150000,0.150000,0.150000}%
\pgfsetstrokecolor{textcolor}%
\pgfsetfillcolor{textcolor}%
\pgftext[x=2.850000in,y=0.397315in,,top]{\color{textcolor}\sffamily\fontsize{12.000000}{14.400000}\selectfont maximal vertex coordinate}%
\end{pgfscope}%
\begin{pgfscope}%
\pgfpathrectangle{\pgfqpoint{0.700000in}{0.720000in}}{\pgfqpoint{4.300000in}{2.130000in}}%
\pgfusepath{clip}%
\pgfsetroundcap%
\pgfsetroundjoin%
\pgfsetlinewidth{1.003750pt}%
\definecolor{currentstroke}{rgb}{0.800000,0.800000,0.800000}%
\pgfsetstrokecolor{currentstroke}%
\pgfsetdash{}{0pt}%
\pgfpathmoveto{\pgfqpoint{0.700000in}{0.720000in}}%
\pgfpathlineto{\pgfqpoint{5.000000in}{0.720000in}}%
\pgfusepath{stroke}%
\end{pgfscope}%
\begin{pgfscope}%
\definecolor{textcolor}{rgb}{0.150000,0.150000,0.150000}%
\pgfsetstrokecolor{textcolor}%
\pgfsetfillcolor{textcolor}%
\pgftext[x=0.349999in,y=0.667193in,left,base]{\color{textcolor}\sffamily\fontsize{11.000000}{13.200000}\selectfont \(\displaystyle 10^{1}\)}%
\end{pgfscope}%
\begin{pgfscope}%
\pgfpathrectangle{\pgfqpoint{0.700000in}{0.720000in}}{\pgfqpoint{4.300000in}{2.130000in}}%
\pgfusepath{clip}%
\pgfsetroundcap%
\pgfsetroundjoin%
\pgfsetlinewidth{1.003750pt}%
\definecolor{currentstroke}{rgb}{0.800000,0.800000,0.800000}%
\pgfsetstrokecolor{currentstroke}%
\pgfsetdash{}{0pt}%
\pgfpathmoveto{\pgfqpoint{0.700000in}{1.193333in}}%
\pgfpathlineto{\pgfqpoint{5.000000in}{1.193333in}}%
\pgfusepath{stroke}%
\end{pgfscope}%
\begin{pgfscope}%
\definecolor{textcolor}{rgb}{0.150000,0.150000,0.150000}%
\pgfsetstrokecolor{textcolor}%
\pgfsetfillcolor{textcolor}%
\pgftext[x=0.349999in,y=1.140527in,left,base]{\color{textcolor}\sffamily\fontsize{11.000000}{13.200000}\selectfont \(\displaystyle 10^{5}\)}%
\end{pgfscope}%
\begin{pgfscope}%
\pgfpathrectangle{\pgfqpoint{0.700000in}{0.720000in}}{\pgfqpoint{4.300000in}{2.130000in}}%
\pgfusepath{clip}%
\pgfsetroundcap%
\pgfsetroundjoin%
\pgfsetlinewidth{1.003750pt}%
\definecolor{currentstroke}{rgb}{0.800000,0.800000,0.800000}%
\pgfsetstrokecolor{currentstroke}%
\pgfsetdash{}{0pt}%
\pgfpathmoveto{\pgfqpoint{0.700000in}{1.666667in}}%
\pgfpathlineto{\pgfqpoint{5.000000in}{1.666667in}}%
\pgfusepath{stroke}%
\end{pgfscope}%
\begin{pgfscope}%
\definecolor{textcolor}{rgb}{0.150000,0.150000,0.150000}%
\pgfsetstrokecolor{textcolor}%
\pgfsetfillcolor{textcolor}%
\pgftext[x=0.349999in,y=1.613860in,left,base]{\color{textcolor}\sffamily\fontsize{11.000000}{13.200000}\selectfont \(\displaystyle 10^{9}\)}%
\end{pgfscope}%
\begin{pgfscope}%
\pgfpathrectangle{\pgfqpoint{0.700000in}{0.720000in}}{\pgfqpoint{4.300000in}{2.130000in}}%
\pgfusepath{clip}%
\pgfsetroundcap%
\pgfsetroundjoin%
\pgfsetlinewidth{1.003750pt}%
\definecolor{currentstroke}{rgb}{0.800000,0.800000,0.800000}%
\pgfsetstrokecolor{currentstroke}%
\pgfsetdash{}{0pt}%
\pgfpathmoveto{\pgfqpoint{0.700000in}{2.140000in}}%
\pgfpathlineto{\pgfqpoint{5.000000in}{2.140000in}}%
\pgfusepath{stroke}%
\end{pgfscope}%
\begin{pgfscope}%
\definecolor{textcolor}{rgb}{0.150000,0.150000,0.150000}%
\pgfsetstrokecolor{textcolor}%
\pgfsetfillcolor{textcolor}%
\pgftext[x=0.290970in,y=2.087193in,left,base]{\color{textcolor}\sffamily\fontsize{11.000000}{13.200000}\selectfont \(\displaystyle 10^{13}\)}%
\end{pgfscope}%
\begin{pgfscope}%
\pgfpathrectangle{\pgfqpoint{0.700000in}{0.720000in}}{\pgfqpoint{4.300000in}{2.130000in}}%
\pgfusepath{clip}%
\pgfsetroundcap%
\pgfsetroundjoin%
\pgfsetlinewidth{1.003750pt}%
\definecolor{currentstroke}{rgb}{0.800000,0.800000,0.800000}%
\pgfsetstrokecolor{currentstroke}%
\pgfsetdash{}{0pt}%
\pgfpathmoveto{\pgfqpoint{0.700000in}{2.613333in}}%
\pgfpathlineto{\pgfqpoint{5.000000in}{2.613333in}}%
\pgfusepath{stroke}%
\end{pgfscope}%
\begin{pgfscope}%
\definecolor{textcolor}{rgb}{0.150000,0.150000,0.150000}%
\pgfsetstrokecolor{textcolor}%
\pgfsetfillcolor{textcolor}%
\pgftext[x=0.290970in,y=2.560527in,left,base]{\color{textcolor}\sffamily\fontsize{11.000000}{13.200000}\selectfont \(\displaystyle 10^{17}\)}%
\end{pgfscope}%
\begin{pgfscope}%
\definecolor{textcolor}{rgb}{0.150000,0.150000,0.150000}%
\pgfsetstrokecolor{textcolor}%
\pgfsetfillcolor{textcolor}%
\pgftext[x=0.235415in,y=1.785000in,,bottom,rotate=90.000000]{\color{textcolor}\sffamily\fontsize{12.000000}{14.400000}\selectfont maximal normal coordinate}%
\end{pgfscope}%
\begin{pgfscope}%
\pgfpathrectangle{\pgfqpoint{0.700000in}{0.720000in}}{\pgfqpoint{4.300000in}{2.130000in}}%
\pgfusepath{clip}%
\pgfsetbuttcap%
\pgfsetroundjoin%
\definecolor{currentfill}{rgb}{0.298039,0.686275,0.313725}%
\pgfsetfillcolor{currentfill}%
\pgfsetlinewidth{1.003750pt}%
\definecolor{currentstroke}{rgb}{0.298039,0.686275,0.313725}%
\pgfsetstrokecolor{currentstroke}%
\pgfsetdash{}{0pt}%
\pgfpathmoveto{\pgfqpoint{3.908938in}{2.528977in}}%
\pgfpathlineto{\pgfqpoint{3.967863in}{2.587903in}}%
\pgfpathlineto{\pgfqpoint{3.908938in}{2.646828in}}%
\pgfpathlineto{\pgfqpoint{3.850012in}{2.587903in}}%
\pgfpathclose%
\pgfusepath{stroke,fill}%
\end{pgfscope}%
\begin{pgfscope}%
\pgfpathrectangle{\pgfqpoint{0.700000in}{0.720000in}}{\pgfqpoint{4.300000in}{2.130000in}}%
\pgfusepath{clip}%
\pgfsetbuttcap%
\pgfsetroundjoin%
\definecolor{currentfill}{rgb}{0.011765,0.662745,0.956863}%
\pgfsetfillcolor{currentfill}%
\pgfsetlinewidth{1.003750pt}%
\definecolor{currentstroke}{rgb}{0.011765,0.662745,0.956863}%
\pgfsetstrokecolor{currentstroke}%
\pgfsetdash{}{0pt}%
\pgfpathmoveto{\pgfqpoint{2.919281in}{2.022355in}}%
\pgfpathlineto{\pgfqpoint{2.978206in}{2.081281in}}%
\pgfpathlineto{\pgfqpoint{2.919281in}{2.140206in}}%
\pgfpathlineto{\pgfqpoint{2.860355in}{2.081281in}}%
\pgfpathclose%
\pgfusepath{stroke,fill}%
\end{pgfscope}%
\begin{pgfscope}%
\pgfpathrectangle{\pgfqpoint{0.700000in}{0.720000in}}{\pgfqpoint{4.300000in}{2.130000in}}%
\pgfusepath{clip}%
\pgfsetbuttcap%
\pgfsetroundjoin%
\definecolor{currentfill}{rgb}{0.247059,0.317647,0.709804}%
\pgfsetfillcolor{currentfill}%
\pgfsetlinewidth{1.003750pt}%
\definecolor{currentstroke}{rgb}{0.247059,0.317647,0.709804}%
\pgfsetstrokecolor{currentstroke}%
\pgfsetdash{}{0pt}%
\pgfpathmoveto{\pgfqpoint{1.929624in}{1.515733in}}%
\pgfpathlineto{\pgfqpoint{1.988549in}{1.574658in}}%
\pgfpathlineto{\pgfqpoint{1.929624in}{1.633584in}}%
\pgfpathlineto{\pgfqpoint{1.870698in}{1.574658in}}%
\pgfpathclose%
\pgfusepath{stroke,fill}%
\end{pgfscope}%
\begin{pgfscope}%
\pgfpathrectangle{\pgfqpoint{0.700000in}{0.720000in}}{\pgfqpoint{4.300000in}{2.130000in}}%
\pgfusepath{clip}%
\pgfsetbuttcap%
\pgfsetroundjoin%
\definecolor{currentfill}{rgb}{0.611765,0.152941,0.690196}%
\pgfsetfillcolor{currentfill}%
\pgfsetlinewidth{1.003750pt}%
\definecolor{currentstroke}{rgb}{0.611765,0.152941,0.690196}%
\pgfsetstrokecolor{currentstroke}%
\pgfsetdash{}{0pt}%
\pgfpathmoveto{\pgfqpoint{0.939967in}{1.009110in}}%
\pgfpathlineto{\pgfqpoint{0.998892in}{1.068036in}}%
\pgfpathlineto{\pgfqpoint{0.939967in}{1.126961in}}%
\pgfpathlineto{\pgfqpoint{0.881041in}{1.068036in}}%
\pgfpathclose%
\pgfusepath{stroke,fill}%
\end{pgfscope}%
\begin{pgfscope}%
\pgfpathrectangle{\pgfqpoint{0.700000in}{0.720000in}}{\pgfqpoint{4.300000in}{2.130000in}}%
\pgfusepath{clip}%
\pgfsetroundcap%
\pgfsetroundjoin%
\pgfsetlinewidth{1.505625pt}%
\definecolor{currentstroke}{rgb}{0.298039,0.686275,0.313725}%
\pgfsetstrokecolor{currentstroke}%
\pgfsetdash{}{0pt}%
\pgfpathmoveto{\pgfqpoint{0.700000in}{2.793242in}}%
\pgfpathlineto{\pgfqpoint{0.742029in}{2.790552in}}%
\pgfpathlineto{\pgfqpoint{0.784057in}{2.787863in}}%
\pgfpathlineto{\pgfqpoint{0.826086in}{2.785173in}}%
\pgfpathlineto{\pgfqpoint{0.868114in}{2.782484in}}%
\pgfpathlineto{\pgfqpoint{0.910143in}{2.779795in}}%
\pgfpathlineto{\pgfqpoint{0.952171in}{2.777105in}}%
\pgfpathlineto{\pgfqpoint{0.994200in}{2.774416in}}%
\pgfpathlineto{\pgfqpoint{1.036229in}{2.771726in}}%
\pgfpathlineto{\pgfqpoint{1.078257in}{2.769037in}}%
\pgfpathlineto{\pgfqpoint{1.120286in}{2.766348in}}%
\pgfpathlineto{\pgfqpoint{1.162314in}{2.763658in}}%
\pgfpathlineto{\pgfqpoint{1.204343in}{2.760969in}}%
\pgfpathlineto{\pgfqpoint{1.246372in}{2.758280in}}%
\pgfpathlineto{\pgfqpoint{1.288400in}{2.755590in}}%
\pgfpathlineto{\pgfqpoint{1.330429in}{2.752901in}}%
\pgfpathlineto{\pgfqpoint{1.372457in}{2.750211in}}%
\pgfpathlineto{\pgfqpoint{1.414486in}{2.747522in}}%
\pgfpathlineto{\pgfqpoint{1.456514in}{2.744833in}}%
\pgfpathlineto{\pgfqpoint{1.498543in}{2.742143in}}%
\pgfpathlineto{\pgfqpoint{1.540572in}{2.739454in}}%
\pgfpathlineto{\pgfqpoint{1.582600in}{2.736764in}}%
\pgfpathlineto{\pgfqpoint{1.624629in}{2.734075in}}%
\pgfpathlineto{\pgfqpoint{1.666657in}{2.731386in}}%
\pgfpathlineto{\pgfqpoint{1.708686in}{2.728696in}}%
\pgfpathlineto{\pgfqpoint{1.750715in}{2.726007in}}%
\pgfpathlineto{\pgfqpoint{1.792743in}{2.723317in}}%
\pgfpathlineto{\pgfqpoint{1.834772in}{2.720628in}}%
\pgfpathlineto{\pgfqpoint{1.876800in}{2.717939in}}%
\pgfpathlineto{\pgfqpoint{1.918829in}{2.715249in}}%
\pgfpathlineto{\pgfqpoint{1.960857in}{2.712560in}}%
\pgfpathlineto{\pgfqpoint{2.002886in}{2.709870in}}%
\pgfpathlineto{\pgfqpoint{2.044915in}{2.707181in}}%
\pgfpathlineto{\pgfqpoint{2.086943in}{2.704492in}}%
\pgfpathlineto{\pgfqpoint{2.128972in}{2.701802in}}%
\pgfpathlineto{\pgfqpoint{2.171000in}{2.699113in}}%
\pgfpathlineto{\pgfqpoint{2.213029in}{2.696423in}}%
\pgfpathlineto{\pgfqpoint{2.255057in}{2.693734in}}%
\pgfpathlineto{\pgfqpoint{2.297086in}{2.691045in}}%
\pgfpathlineto{\pgfqpoint{2.339115in}{2.688355in}}%
\pgfpathlineto{\pgfqpoint{2.381143in}{2.685666in}}%
\pgfpathlineto{\pgfqpoint{2.423172in}{2.682976in}}%
\pgfpathlineto{\pgfqpoint{2.465200in}{2.680287in}}%
\pgfpathlineto{\pgfqpoint{2.507229in}{2.677598in}}%
\pgfpathlineto{\pgfqpoint{2.549258in}{2.674908in}}%
\pgfpathlineto{\pgfqpoint{2.591286in}{2.672219in}}%
\pgfpathlineto{\pgfqpoint{2.633315in}{2.669530in}}%
\pgfpathlineto{\pgfqpoint{2.675343in}{2.666840in}}%
\pgfpathlineto{\pgfqpoint{2.717372in}{2.664151in}}%
\pgfpathlineto{\pgfqpoint{2.759400in}{2.661461in}}%
\pgfpathlineto{\pgfqpoint{2.801429in}{2.658772in}}%
\pgfpathlineto{\pgfqpoint{2.843458in}{2.656083in}}%
\pgfpathlineto{\pgfqpoint{2.885486in}{2.653393in}}%
\pgfpathlineto{\pgfqpoint{2.927515in}{2.650704in}}%
\pgfpathlineto{\pgfqpoint{2.969543in}{2.648014in}}%
\pgfpathlineto{\pgfqpoint{3.011572in}{2.645325in}}%
\pgfpathlineto{\pgfqpoint{3.053601in}{2.642636in}}%
\pgfpathlineto{\pgfqpoint{3.095629in}{2.639946in}}%
\pgfpathlineto{\pgfqpoint{3.137658in}{2.637257in}}%
\pgfpathlineto{\pgfqpoint{3.179686in}{2.634567in}}%
\pgfpathlineto{\pgfqpoint{3.221715in}{2.631878in}}%
\pgfpathlineto{\pgfqpoint{3.263743in}{2.629189in}}%
\pgfpathlineto{\pgfqpoint{3.305772in}{2.626499in}}%
\pgfpathlineto{\pgfqpoint{3.347801in}{2.623810in}}%
\pgfpathlineto{\pgfqpoint{3.389829in}{2.621120in}}%
\pgfpathlineto{\pgfqpoint{3.431858in}{2.618431in}}%
\pgfpathlineto{\pgfqpoint{3.473886in}{2.615742in}}%
\pgfpathlineto{\pgfqpoint{3.515915in}{2.613052in}}%
\pgfpathlineto{\pgfqpoint{3.557943in}{2.610363in}}%
\pgfpathlineto{\pgfqpoint{3.599972in}{2.607673in}}%
\pgfpathlineto{\pgfqpoint{3.642001in}{2.604984in}}%
\pgfpathlineto{\pgfqpoint{3.684029in}{2.602295in}}%
\pgfpathlineto{\pgfqpoint{3.726058in}{2.599605in}}%
\pgfpathlineto{\pgfqpoint{3.768086in}{2.596916in}}%
\pgfpathlineto{\pgfqpoint{3.810115in}{2.594226in}}%
\pgfpathlineto{\pgfqpoint{3.852144in}{2.591537in}}%
\pgfpathlineto{\pgfqpoint{3.894172in}{2.588848in}}%
\pgfpathlineto{\pgfqpoint{3.936201in}{2.586158in}}%
\pgfpathlineto{\pgfqpoint{3.978229in}{2.583469in}}%
\pgfpathlineto{\pgfqpoint{4.020258in}{2.580780in}}%
\pgfpathlineto{\pgfqpoint{4.062286in}{2.578090in}}%
\pgfpathlineto{\pgfqpoint{4.104315in}{2.575401in}}%
\pgfpathlineto{\pgfqpoint{4.146344in}{2.572711in}}%
\pgfpathlineto{\pgfqpoint{4.188372in}{2.570022in}}%
\pgfpathlineto{\pgfqpoint{4.230401in}{2.567333in}}%
\pgfpathlineto{\pgfqpoint{4.272429in}{2.564643in}}%
\pgfpathlineto{\pgfqpoint{4.314458in}{2.561954in}}%
\pgfpathlineto{\pgfqpoint{4.356487in}{2.559264in}}%
\pgfpathlineto{\pgfqpoint{4.398515in}{2.556575in}}%
\pgfpathlineto{\pgfqpoint{4.440544in}{2.553886in}}%
\pgfpathlineto{\pgfqpoint{4.482572in}{2.551196in}}%
\pgfpathlineto{\pgfqpoint{4.524601in}{2.548507in}}%
\pgfpathlineto{\pgfqpoint{4.566629in}{2.545817in}}%
\pgfpathlineto{\pgfqpoint{4.608658in}{2.543128in}}%
\pgfpathlineto{\pgfqpoint{4.650687in}{2.540439in}}%
\pgfpathlineto{\pgfqpoint{4.692715in}{2.537749in}}%
\pgfpathlineto{\pgfqpoint{4.734744in}{2.535060in}}%
\pgfpathlineto{\pgfqpoint{4.776772in}{2.532370in}}%
\pgfpathlineto{\pgfqpoint{4.818801in}{2.529681in}}%
\pgfpathlineto{\pgfqpoint{4.860830in}{2.526992in}}%
\pgfusepath{stroke}%
\end{pgfscope}%
\begin{pgfscope}%
\pgfpathrectangle{\pgfqpoint{0.700000in}{0.720000in}}{\pgfqpoint{4.300000in}{2.130000in}}%
\pgfusepath{clip}%
\pgfsetroundcap%
\pgfsetroundjoin%
\pgfsetlinewidth{1.505625pt}%
\definecolor{currentstroke}{rgb}{0.011765,0.662745,0.956863}%
\pgfsetstrokecolor{currentstroke}%
\pgfsetdash{}{0pt}%
\pgfpathmoveto{\pgfqpoint{0.700000in}{2.223292in}}%
\pgfpathlineto{\pgfqpoint{0.742029in}{2.220602in}}%
\pgfpathlineto{\pgfqpoint{0.784057in}{2.217913in}}%
\pgfpathlineto{\pgfqpoint{0.826086in}{2.215223in}}%
\pgfpathlineto{\pgfqpoint{0.868114in}{2.212534in}}%
\pgfpathlineto{\pgfqpoint{0.910143in}{2.209845in}}%
\pgfpathlineto{\pgfqpoint{0.952171in}{2.207155in}}%
\pgfpathlineto{\pgfqpoint{0.994200in}{2.204466in}}%
\pgfpathlineto{\pgfqpoint{1.036229in}{2.201776in}}%
\pgfpathlineto{\pgfqpoint{1.078257in}{2.199087in}}%
\pgfpathlineto{\pgfqpoint{1.120286in}{2.196398in}}%
\pgfpathlineto{\pgfqpoint{1.162314in}{2.193708in}}%
\pgfpathlineto{\pgfqpoint{1.204343in}{2.191019in}}%
\pgfpathlineto{\pgfqpoint{1.246372in}{2.188329in}}%
\pgfpathlineto{\pgfqpoint{1.288400in}{2.185640in}}%
\pgfpathlineto{\pgfqpoint{1.330429in}{2.182951in}}%
\pgfpathlineto{\pgfqpoint{1.372457in}{2.180261in}}%
\pgfpathlineto{\pgfqpoint{1.414486in}{2.177572in}}%
\pgfpathlineto{\pgfqpoint{1.456514in}{2.174882in}}%
\pgfpathlineto{\pgfqpoint{1.498543in}{2.172193in}}%
\pgfpathlineto{\pgfqpoint{1.540572in}{2.169504in}}%
\pgfpathlineto{\pgfqpoint{1.582600in}{2.166814in}}%
\pgfpathlineto{\pgfqpoint{1.624629in}{2.164125in}}%
\pgfpathlineto{\pgfqpoint{1.666657in}{2.161435in}}%
\pgfpathlineto{\pgfqpoint{1.708686in}{2.158746in}}%
\pgfpathlineto{\pgfqpoint{1.750715in}{2.156057in}}%
\pgfpathlineto{\pgfqpoint{1.792743in}{2.153367in}}%
\pgfpathlineto{\pgfqpoint{1.834772in}{2.150678in}}%
\pgfpathlineto{\pgfqpoint{1.876800in}{2.147988in}}%
\pgfpathlineto{\pgfqpoint{1.918829in}{2.145299in}}%
\pgfpathlineto{\pgfqpoint{1.960857in}{2.142610in}}%
\pgfpathlineto{\pgfqpoint{2.002886in}{2.139920in}}%
\pgfpathlineto{\pgfqpoint{2.044915in}{2.137231in}}%
\pgfpathlineto{\pgfqpoint{2.086943in}{2.134542in}}%
\pgfpathlineto{\pgfqpoint{2.128972in}{2.131852in}}%
\pgfpathlineto{\pgfqpoint{2.171000in}{2.129163in}}%
\pgfpathlineto{\pgfqpoint{2.213029in}{2.126473in}}%
\pgfpathlineto{\pgfqpoint{2.255057in}{2.123784in}}%
\pgfpathlineto{\pgfqpoint{2.297086in}{2.121095in}}%
\pgfpathlineto{\pgfqpoint{2.339115in}{2.118405in}}%
\pgfpathlineto{\pgfqpoint{2.381143in}{2.115716in}}%
\pgfpathlineto{\pgfqpoint{2.423172in}{2.113026in}}%
\pgfpathlineto{\pgfqpoint{2.465200in}{2.110337in}}%
\pgfpathlineto{\pgfqpoint{2.507229in}{2.107648in}}%
\pgfpathlineto{\pgfqpoint{2.549258in}{2.104958in}}%
\pgfpathlineto{\pgfqpoint{2.591286in}{2.102269in}}%
\pgfpathlineto{\pgfqpoint{2.633315in}{2.099579in}}%
\pgfpathlineto{\pgfqpoint{2.675343in}{2.096890in}}%
\pgfpathlineto{\pgfqpoint{2.717372in}{2.094201in}}%
\pgfpathlineto{\pgfqpoint{2.759400in}{2.091511in}}%
\pgfpathlineto{\pgfqpoint{2.801429in}{2.088822in}}%
\pgfpathlineto{\pgfqpoint{2.843458in}{2.086132in}}%
\pgfpathlineto{\pgfqpoint{2.885486in}{2.083443in}}%
\pgfpathlineto{\pgfqpoint{2.927515in}{2.080754in}}%
\pgfpathlineto{\pgfqpoint{2.969543in}{2.078064in}}%
\pgfpathlineto{\pgfqpoint{3.011572in}{2.075375in}}%
\pgfpathlineto{\pgfqpoint{3.053601in}{2.072685in}}%
\pgfpathlineto{\pgfqpoint{3.095629in}{2.069996in}}%
\pgfpathlineto{\pgfqpoint{3.137658in}{2.067307in}}%
\pgfpathlineto{\pgfqpoint{3.179686in}{2.064617in}}%
\pgfpathlineto{\pgfqpoint{3.221715in}{2.061928in}}%
\pgfpathlineto{\pgfqpoint{3.263743in}{2.059238in}}%
\pgfpathlineto{\pgfqpoint{3.305772in}{2.056549in}}%
\pgfpathlineto{\pgfqpoint{3.347801in}{2.053860in}}%
\pgfpathlineto{\pgfqpoint{3.389829in}{2.051170in}}%
\pgfpathlineto{\pgfqpoint{3.431858in}{2.048481in}}%
\pgfpathlineto{\pgfqpoint{3.473886in}{2.045792in}}%
\pgfpathlineto{\pgfqpoint{3.515915in}{2.043102in}}%
\pgfpathlineto{\pgfqpoint{3.557943in}{2.040413in}}%
\pgfpathlineto{\pgfqpoint{3.599972in}{2.037723in}}%
\pgfpathlineto{\pgfqpoint{3.642001in}{2.035034in}}%
\pgfpathlineto{\pgfqpoint{3.684029in}{2.032345in}}%
\pgfpathlineto{\pgfqpoint{3.726058in}{2.029655in}}%
\pgfpathlineto{\pgfqpoint{3.768086in}{2.026966in}}%
\pgfpathlineto{\pgfqpoint{3.810115in}{2.024276in}}%
\pgfpathlineto{\pgfqpoint{3.852144in}{2.021587in}}%
\pgfpathlineto{\pgfqpoint{3.894172in}{2.018898in}}%
\pgfpathlineto{\pgfqpoint{3.936201in}{2.016208in}}%
\pgfpathlineto{\pgfqpoint{3.978229in}{2.013519in}}%
\pgfpathlineto{\pgfqpoint{4.020258in}{2.010829in}}%
\pgfpathlineto{\pgfqpoint{4.062286in}{2.008140in}}%
\pgfpathlineto{\pgfqpoint{4.104315in}{2.005451in}}%
\pgfpathlineto{\pgfqpoint{4.146344in}{2.002761in}}%
\pgfpathlineto{\pgfqpoint{4.188372in}{2.000072in}}%
\pgfpathlineto{\pgfqpoint{4.230401in}{1.997382in}}%
\pgfpathlineto{\pgfqpoint{4.272429in}{1.994693in}}%
\pgfpathlineto{\pgfqpoint{4.314458in}{1.992004in}}%
\pgfpathlineto{\pgfqpoint{4.356487in}{1.989314in}}%
\pgfpathlineto{\pgfqpoint{4.398515in}{1.986625in}}%
\pgfpathlineto{\pgfqpoint{4.440544in}{1.983935in}}%
\pgfpathlineto{\pgfqpoint{4.482572in}{1.981246in}}%
\pgfpathlineto{\pgfqpoint{4.524601in}{1.978557in}}%
\pgfpathlineto{\pgfqpoint{4.566629in}{1.975867in}}%
\pgfpathlineto{\pgfqpoint{4.608658in}{1.973178in}}%
\pgfpathlineto{\pgfqpoint{4.650687in}{1.970488in}}%
\pgfpathlineto{\pgfqpoint{4.692715in}{1.967799in}}%
\pgfpathlineto{\pgfqpoint{4.734744in}{1.965110in}}%
\pgfpathlineto{\pgfqpoint{4.776772in}{1.962420in}}%
\pgfpathlineto{\pgfqpoint{4.818801in}{1.959731in}}%
\pgfpathlineto{\pgfqpoint{4.860830in}{1.957042in}}%
\pgfusepath{stroke}%
\end{pgfscope}%
\begin{pgfscope}%
\pgfpathrectangle{\pgfqpoint{0.700000in}{0.720000in}}{\pgfqpoint{4.300000in}{2.130000in}}%
\pgfusepath{clip}%
\pgfsetroundcap%
\pgfsetroundjoin%
\pgfsetlinewidth{1.505625pt}%
\definecolor{currentstroke}{rgb}{0.247059,0.317647,0.709804}%
\pgfsetstrokecolor{currentstroke}%
\pgfsetdash{}{0pt}%
\pgfpathmoveto{\pgfqpoint{0.700000in}{1.653341in}}%
\pgfpathlineto{\pgfqpoint{0.742029in}{1.650652in}}%
\pgfpathlineto{\pgfqpoint{0.784057in}{1.647963in}}%
\pgfpathlineto{\pgfqpoint{0.826086in}{1.645273in}}%
\pgfpathlineto{\pgfqpoint{0.868114in}{1.642584in}}%
\pgfpathlineto{\pgfqpoint{0.910143in}{1.639894in}}%
\pgfpathlineto{\pgfqpoint{0.952171in}{1.637205in}}%
\pgfpathlineto{\pgfqpoint{0.994200in}{1.634516in}}%
\pgfpathlineto{\pgfqpoint{1.036229in}{1.631826in}}%
\pgfpathlineto{\pgfqpoint{1.078257in}{1.629137in}}%
\pgfpathlineto{\pgfqpoint{1.120286in}{1.626447in}}%
\pgfpathlineto{\pgfqpoint{1.162314in}{1.623758in}}%
\pgfpathlineto{\pgfqpoint{1.204343in}{1.621069in}}%
\pgfpathlineto{\pgfqpoint{1.246372in}{1.618379in}}%
\pgfpathlineto{\pgfqpoint{1.288400in}{1.615690in}}%
\pgfpathlineto{\pgfqpoint{1.330429in}{1.613000in}}%
\pgfpathlineto{\pgfqpoint{1.372457in}{1.610311in}}%
\pgfpathlineto{\pgfqpoint{1.414486in}{1.607622in}}%
\pgfpathlineto{\pgfqpoint{1.456514in}{1.604932in}}%
\pgfpathlineto{\pgfqpoint{1.498543in}{1.602243in}}%
\pgfpathlineto{\pgfqpoint{1.540572in}{1.599554in}}%
\pgfpathlineto{\pgfqpoint{1.582600in}{1.596864in}}%
\pgfpathlineto{\pgfqpoint{1.624629in}{1.594175in}}%
\pgfpathlineto{\pgfqpoint{1.666657in}{1.591485in}}%
\pgfpathlineto{\pgfqpoint{1.708686in}{1.588796in}}%
\pgfpathlineto{\pgfqpoint{1.750715in}{1.586107in}}%
\pgfpathlineto{\pgfqpoint{1.792743in}{1.583417in}}%
\pgfpathlineto{\pgfqpoint{1.834772in}{1.580728in}}%
\pgfpathlineto{\pgfqpoint{1.876800in}{1.578038in}}%
\pgfpathlineto{\pgfqpoint{1.918829in}{1.575349in}}%
\pgfpathlineto{\pgfqpoint{1.960857in}{1.572660in}}%
\pgfpathlineto{\pgfqpoint{2.002886in}{1.569970in}}%
\pgfpathlineto{\pgfqpoint{2.044915in}{1.567281in}}%
\pgfpathlineto{\pgfqpoint{2.086943in}{1.564591in}}%
\pgfpathlineto{\pgfqpoint{2.128972in}{1.561902in}}%
\pgfpathlineto{\pgfqpoint{2.171000in}{1.559213in}}%
\pgfpathlineto{\pgfqpoint{2.213029in}{1.556523in}}%
\pgfpathlineto{\pgfqpoint{2.255057in}{1.553834in}}%
\pgfpathlineto{\pgfqpoint{2.297086in}{1.551144in}}%
\pgfpathlineto{\pgfqpoint{2.339115in}{1.548455in}}%
\pgfpathlineto{\pgfqpoint{2.381143in}{1.545766in}}%
\pgfpathlineto{\pgfqpoint{2.423172in}{1.543076in}}%
\pgfpathlineto{\pgfqpoint{2.465200in}{1.540387in}}%
\pgfpathlineto{\pgfqpoint{2.507229in}{1.537697in}}%
\pgfpathlineto{\pgfqpoint{2.549258in}{1.535008in}}%
\pgfpathlineto{\pgfqpoint{2.591286in}{1.532319in}}%
\pgfpathlineto{\pgfqpoint{2.633315in}{1.529629in}}%
\pgfpathlineto{\pgfqpoint{2.675343in}{1.526940in}}%
\pgfpathlineto{\pgfqpoint{2.717372in}{1.524250in}}%
\pgfpathlineto{\pgfqpoint{2.759400in}{1.521561in}}%
\pgfpathlineto{\pgfqpoint{2.801429in}{1.518872in}}%
\pgfpathlineto{\pgfqpoint{2.843458in}{1.516182in}}%
\pgfpathlineto{\pgfqpoint{2.885486in}{1.513493in}}%
\pgfpathlineto{\pgfqpoint{2.927515in}{1.510804in}}%
\pgfpathlineto{\pgfqpoint{2.969543in}{1.508114in}}%
\pgfpathlineto{\pgfqpoint{3.011572in}{1.505425in}}%
\pgfpathlineto{\pgfqpoint{3.053601in}{1.502735in}}%
\pgfpathlineto{\pgfqpoint{3.095629in}{1.500046in}}%
\pgfpathlineto{\pgfqpoint{3.137658in}{1.497357in}}%
\pgfpathlineto{\pgfqpoint{3.179686in}{1.494667in}}%
\pgfpathlineto{\pgfqpoint{3.221715in}{1.491978in}}%
\pgfpathlineto{\pgfqpoint{3.263743in}{1.489288in}}%
\pgfpathlineto{\pgfqpoint{3.305772in}{1.486599in}}%
\pgfpathlineto{\pgfqpoint{3.347801in}{1.483910in}}%
\pgfpathlineto{\pgfqpoint{3.389829in}{1.481220in}}%
\pgfpathlineto{\pgfqpoint{3.431858in}{1.478531in}}%
\pgfpathlineto{\pgfqpoint{3.473886in}{1.475841in}}%
\pgfpathlineto{\pgfqpoint{3.515915in}{1.473152in}}%
\pgfpathlineto{\pgfqpoint{3.557943in}{1.470463in}}%
\pgfpathlineto{\pgfqpoint{3.599972in}{1.467773in}}%
\pgfpathlineto{\pgfqpoint{3.642001in}{1.465084in}}%
\pgfpathlineto{\pgfqpoint{3.684029in}{1.462394in}}%
\pgfpathlineto{\pgfqpoint{3.726058in}{1.459705in}}%
\pgfpathlineto{\pgfqpoint{3.768086in}{1.457016in}}%
\pgfpathlineto{\pgfqpoint{3.810115in}{1.454326in}}%
\pgfpathlineto{\pgfqpoint{3.852144in}{1.451637in}}%
\pgfpathlineto{\pgfqpoint{3.894172in}{1.448947in}}%
\pgfpathlineto{\pgfqpoint{3.936201in}{1.446258in}}%
\pgfpathlineto{\pgfqpoint{3.978229in}{1.443569in}}%
\pgfpathlineto{\pgfqpoint{4.020258in}{1.440879in}}%
\pgfpathlineto{\pgfqpoint{4.062286in}{1.438190in}}%
\pgfpathlineto{\pgfqpoint{4.104315in}{1.435500in}}%
\pgfpathlineto{\pgfqpoint{4.146344in}{1.432811in}}%
\pgfpathlineto{\pgfqpoint{4.188372in}{1.430122in}}%
\pgfpathlineto{\pgfqpoint{4.230401in}{1.427432in}}%
\pgfpathlineto{\pgfqpoint{4.272429in}{1.424743in}}%
\pgfpathlineto{\pgfqpoint{4.314458in}{1.422054in}}%
\pgfpathlineto{\pgfqpoint{4.356487in}{1.419364in}}%
\pgfpathlineto{\pgfqpoint{4.398515in}{1.416675in}}%
\pgfpathlineto{\pgfqpoint{4.440544in}{1.413985in}}%
\pgfpathlineto{\pgfqpoint{4.482572in}{1.411296in}}%
\pgfpathlineto{\pgfqpoint{4.524601in}{1.408607in}}%
\pgfpathlineto{\pgfqpoint{4.566629in}{1.405917in}}%
\pgfpathlineto{\pgfqpoint{4.608658in}{1.403228in}}%
\pgfpathlineto{\pgfqpoint{4.650687in}{1.400538in}}%
\pgfpathlineto{\pgfqpoint{4.692715in}{1.397849in}}%
\pgfpathlineto{\pgfqpoint{4.734744in}{1.395160in}}%
\pgfpathlineto{\pgfqpoint{4.776772in}{1.392470in}}%
\pgfpathlineto{\pgfqpoint{4.818801in}{1.389781in}}%
\pgfpathlineto{\pgfqpoint{4.860830in}{1.387091in}}%
\pgfusepath{stroke}%
\end{pgfscope}%
\begin{pgfscope}%
\pgfpathrectangle{\pgfqpoint{0.700000in}{0.720000in}}{\pgfqpoint{4.300000in}{2.130000in}}%
\pgfusepath{clip}%
\pgfsetroundcap%
\pgfsetroundjoin%
\pgfsetlinewidth{1.505625pt}%
\definecolor{currentstroke}{rgb}{0.611765,0.152941,0.690196}%
\pgfsetstrokecolor{currentstroke}%
\pgfsetdash{}{0pt}%
\pgfpathmoveto{\pgfqpoint{0.700000in}{1.083391in}}%
\pgfpathlineto{\pgfqpoint{0.742029in}{1.080702in}}%
\pgfpathlineto{\pgfqpoint{0.784057in}{1.078012in}}%
\pgfpathlineto{\pgfqpoint{0.826086in}{1.075323in}}%
\pgfpathlineto{\pgfqpoint{0.868114in}{1.072634in}}%
\pgfpathlineto{\pgfqpoint{0.910143in}{1.069944in}}%
\pgfpathlineto{\pgfqpoint{0.952171in}{1.067255in}}%
\pgfpathlineto{\pgfqpoint{0.994200in}{1.064566in}}%
\pgfpathlineto{\pgfqpoint{1.036229in}{1.061876in}}%
\pgfpathlineto{\pgfqpoint{1.078257in}{1.059187in}}%
\pgfpathlineto{\pgfqpoint{1.120286in}{1.056497in}}%
\pgfpathlineto{\pgfqpoint{1.162314in}{1.053808in}}%
\pgfpathlineto{\pgfqpoint{1.204343in}{1.051119in}}%
\pgfpathlineto{\pgfqpoint{1.246372in}{1.048429in}}%
\pgfpathlineto{\pgfqpoint{1.288400in}{1.045740in}}%
\pgfpathlineto{\pgfqpoint{1.330429in}{1.043050in}}%
\pgfpathlineto{\pgfqpoint{1.372457in}{1.040361in}}%
\pgfpathlineto{\pgfqpoint{1.414486in}{1.037672in}}%
\pgfpathlineto{\pgfqpoint{1.456514in}{1.034982in}}%
\pgfpathlineto{\pgfqpoint{1.498543in}{1.032293in}}%
\pgfpathlineto{\pgfqpoint{1.540572in}{1.029603in}}%
\pgfpathlineto{\pgfqpoint{1.582600in}{1.026914in}}%
\pgfpathlineto{\pgfqpoint{1.624629in}{1.024225in}}%
\pgfpathlineto{\pgfqpoint{1.666657in}{1.021535in}}%
\pgfpathlineto{\pgfqpoint{1.708686in}{1.018846in}}%
\pgfpathlineto{\pgfqpoint{1.750715in}{1.016156in}}%
\pgfpathlineto{\pgfqpoint{1.792743in}{1.013467in}}%
\pgfpathlineto{\pgfqpoint{1.834772in}{1.010778in}}%
\pgfpathlineto{\pgfqpoint{1.876800in}{1.008088in}}%
\pgfpathlineto{\pgfqpoint{1.918829in}{1.005399in}}%
\pgfpathlineto{\pgfqpoint{1.960857in}{1.002709in}}%
\pgfpathlineto{\pgfqpoint{2.002886in}{1.000020in}}%
\pgfpathlineto{\pgfqpoint{2.044915in}{0.997331in}}%
\pgfpathlineto{\pgfqpoint{2.086943in}{0.994641in}}%
\pgfpathlineto{\pgfqpoint{2.128972in}{0.991952in}}%
\pgfpathlineto{\pgfqpoint{2.171000in}{0.989262in}}%
\pgfpathlineto{\pgfqpoint{2.213029in}{0.986573in}}%
\pgfpathlineto{\pgfqpoint{2.255057in}{0.983884in}}%
\pgfpathlineto{\pgfqpoint{2.297086in}{0.981194in}}%
\pgfpathlineto{\pgfqpoint{2.339115in}{0.978505in}}%
\pgfpathlineto{\pgfqpoint{2.381143in}{0.975816in}}%
\pgfpathlineto{\pgfqpoint{2.423172in}{0.973126in}}%
\pgfpathlineto{\pgfqpoint{2.465200in}{0.970437in}}%
\pgfpathlineto{\pgfqpoint{2.507229in}{0.967747in}}%
\pgfpathlineto{\pgfqpoint{2.549258in}{0.965058in}}%
\pgfpathlineto{\pgfqpoint{2.591286in}{0.962369in}}%
\pgfpathlineto{\pgfqpoint{2.633315in}{0.959679in}}%
\pgfpathlineto{\pgfqpoint{2.675343in}{0.956990in}}%
\pgfpathlineto{\pgfqpoint{2.717372in}{0.954300in}}%
\pgfpathlineto{\pgfqpoint{2.759400in}{0.951611in}}%
\pgfpathlineto{\pgfqpoint{2.801429in}{0.948922in}}%
\pgfpathlineto{\pgfqpoint{2.843458in}{0.946232in}}%
\pgfpathlineto{\pgfqpoint{2.885486in}{0.943543in}}%
\pgfpathlineto{\pgfqpoint{2.927515in}{0.940853in}}%
\pgfpathlineto{\pgfqpoint{2.969543in}{0.938164in}}%
\pgfpathlineto{\pgfqpoint{3.011572in}{0.935475in}}%
\pgfpathlineto{\pgfqpoint{3.053601in}{0.932785in}}%
\pgfpathlineto{\pgfqpoint{3.095629in}{0.930096in}}%
\pgfpathlineto{\pgfqpoint{3.137658in}{0.927406in}}%
\pgfpathlineto{\pgfqpoint{3.179686in}{0.924717in}}%
\pgfpathlineto{\pgfqpoint{3.221715in}{0.922028in}}%
\pgfpathlineto{\pgfqpoint{3.263743in}{0.919338in}}%
\pgfpathlineto{\pgfqpoint{3.305772in}{0.916649in}}%
\pgfpathlineto{\pgfqpoint{3.347801in}{0.913959in}}%
\pgfpathlineto{\pgfqpoint{3.389829in}{0.911270in}}%
\pgfpathlineto{\pgfqpoint{3.431858in}{0.908581in}}%
\pgfpathlineto{\pgfqpoint{3.473886in}{0.905891in}}%
\pgfpathlineto{\pgfqpoint{3.515915in}{0.903202in}}%
\pgfpathlineto{\pgfqpoint{3.557943in}{0.900512in}}%
\pgfpathlineto{\pgfqpoint{3.599972in}{0.897823in}}%
\pgfpathlineto{\pgfqpoint{3.642001in}{0.895134in}}%
\pgfpathlineto{\pgfqpoint{3.684029in}{0.892444in}}%
\pgfpathlineto{\pgfqpoint{3.726058in}{0.889755in}}%
\pgfpathlineto{\pgfqpoint{3.768086in}{0.887066in}}%
\pgfpathlineto{\pgfqpoint{3.810115in}{0.884376in}}%
\pgfpathlineto{\pgfqpoint{3.852144in}{0.881687in}}%
\pgfpathlineto{\pgfqpoint{3.894172in}{0.878997in}}%
\pgfpathlineto{\pgfqpoint{3.936201in}{0.876308in}}%
\pgfpathlineto{\pgfqpoint{3.978229in}{0.873619in}}%
\pgfpathlineto{\pgfqpoint{4.020258in}{0.870929in}}%
\pgfpathlineto{\pgfqpoint{4.062286in}{0.868240in}}%
\pgfpathlineto{\pgfqpoint{4.104315in}{0.865550in}}%
\pgfpathlineto{\pgfqpoint{4.146344in}{0.862861in}}%
\pgfpathlineto{\pgfqpoint{4.188372in}{0.860172in}}%
\pgfpathlineto{\pgfqpoint{4.230401in}{0.857482in}}%
\pgfpathlineto{\pgfqpoint{4.272429in}{0.854793in}}%
\pgfpathlineto{\pgfqpoint{4.314458in}{0.852103in}}%
\pgfpathlineto{\pgfqpoint{4.356487in}{0.849414in}}%
\pgfpathlineto{\pgfqpoint{4.398515in}{0.846725in}}%
\pgfpathlineto{\pgfqpoint{4.440544in}{0.844035in}}%
\pgfpathlineto{\pgfqpoint{4.482572in}{0.841346in}}%
\pgfpathlineto{\pgfqpoint{4.524601in}{0.838656in}}%
\pgfpathlineto{\pgfqpoint{4.566629in}{0.835967in}}%
\pgfpathlineto{\pgfqpoint{4.608658in}{0.833278in}}%
\pgfpathlineto{\pgfqpoint{4.650687in}{0.830588in}}%
\pgfpathlineto{\pgfqpoint{4.692715in}{0.827899in}}%
\pgfpathlineto{\pgfqpoint{4.734744in}{0.825209in}}%
\pgfpathlineto{\pgfqpoint{4.776772in}{0.822520in}}%
\pgfpathlineto{\pgfqpoint{4.818801in}{0.819831in}}%
\pgfpathlineto{\pgfqpoint{4.860830in}{0.817141in}}%
\pgfusepath{stroke}%
\end{pgfscope}%
\begin{pgfscope}%
\pgfsetrectcap%
\pgfsetmiterjoin%
\pgfsetlinewidth{1.254687pt}%
\definecolor{currentstroke}{rgb}{0.800000,0.800000,0.800000}%
\pgfsetstrokecolor{currentstroke}%
\pgfsetdash{}{0pt}%
\pgfpathmoveto{\pgfqpoint{0.700000in}{0.720000in}}%
\pgfpathlineto{\pgfqpoint{0.700000in}{2.850000in}}%
\pgfusepath{stroke}%
\end{pgfscope}%
\begin{pgfscope}%
\pgfsetrectcap%
\pgfsetmiterjoin%
\pgfsetlinewidth{1.254687pt}%
\definecolor{currentstroke}{rgb}{0.800000,0.800000,0.800000}%
\pgfsetstrokecolor{currentstroke}%
\pgfsetdash{}{0pt}%
\pgfpathmoveto{\pgfqpoint{5.000000in}{0.720000in}}%
\pgfpathlineto{\pgfqpoint{5.000000in}{2.850000in}}%
\pgfusepath{stroke}%
\end{pgfscope}%
\begin{pgfscope}%
\pgfsetrectcap%
\pgfsetmiterjoin%
\pgfsetlinewidth{1.254687pt}%
\definecolor{currentstroke}{rgb}{0.800000,0.800000,0.800000}%
\pgfsetstrokecolor{currentstroke}%
\pgfsetdash{}{0pt}%
\pgfpathmoveto{\pgfqpoint{0.700000in}{0.720000in}}%
\pgfpathlineto{\pgfqpoint{5.000000in}{0.720000in}}%
\pgfusepath{stroke}%
\end{pgfscope}%
\begin{pgfscope}%
\pgfsetrectcap%
\pgfsetmiterjoin%
\pgfsetlinewidth{1.254687pt}%
\definecolor{currentstroke}{rgb}{0.800000,0.800000,0.800000}%
\pgfsetstrokecolor{currentstroke}%
\pgfsetdash{}{0pt}%
\pgfpathmoveto{\pgfqpoint{0.700000in}{2.850000in}}%
\pgfpathlineto{\pgfqpoint{5.000000in}{2.850000in}}%
\pgfusepath{stroke}%
\end{pgfscope}%
\begin{pgfscope}%
\definecolor{textcolor}{rgb}{0.150000,0.150000,0.150000}%
\pgfsetstrokecolor{textcolor}%
\pgfsetfillcolor{textcolor}%
\pgftext[x=3.992271in,y=2.671236in,left,base]{\color{textcolor}\sffamily\fontsize{12.000000}{14.400000}\selectfont \(\displaystyle 8.73 \cdot 10^7\)}%
\end{pgfscope}%
\begin{pgfscope}%
\definecolor{textcolor}{rgb}{0.150000,0.150000,0.150000}%
\pgfsetstrokecolor{textcolor}%
\pgfsetfillcolor{textcolor}%
\pgftext[x=3.002614in,y=2.164614in,left,base]{\color{textcolor}\sffamily\fontsize{12.000000}{14.400000}\selectfont \(\displaystyle 6.31 \cdot 10^5\)}%
\end{pgfscope}%
\begin{pgfscope}%
\definecolor{textcolor}{rgb}{0.150000,0.150000,0.150000}%
\pgfsetstrokecolor{textcolor}%
\pgfsetfillcolor{textcolor}%
\pgftext[x=2.012957in,y=1.657992in,left,base]{\color{textcolor}\sffamily\fontsize{12.000000}{14.400000}\selectfont \(\displaystyle 4.57 \cdot 10^3\)}%
\end{pgfscope}%
\begin{pgfscope}%
\definecolor{textcolor}{rgb}{0.150000,0.150000,0.150000}%
\pgfsetstrokecolor{textcolor}%
\pgfsetfillcolor{textcolor}%
\pgftext[x=1.023300in,y=1.151369in,left,base]{\color{textcolor}\sffamily\fontsize{12.000000}{14.400000}\selectfont \(\displaystyle 3.30 \cdot 10^1\)}%
\end{pgfscope}%
\begin{pgfscope}%
\pgfsetbuttcap%
\pgfsetmiterjoin%
\definecolor{currentfill}{rgb}{1.000000,1.000000,1.000000}%
\pgfsetfillcolor{currentfill}%
\pgfsetfillopacity{0.800000}%
\pgfsetlinewidth{1.003750pt}%
\definecolor{currentstroke}{rgb}{0.800000,0.800000,0.800000}%
\pgfsetstrokecolor{currentstroke}%
\pgfsetstrokeopacity{0.800000}%
\pgfsetdash{}{0pt}%
\pgfpathmoveto{\pgfqpoint{3.955521in}{0.796389in}}%
\pgfpathlineto{\pgfqpoint{4.893056in}{0.796389in}}%
\pgfpathquadraticcurveto{\pgfqpoint{4.923611in}{0.796389in}}{\pgfqpoint{4.923611in}{0.826944in}}%
\pgfpathlineto{\pgfqpoint{4.923611in}{1.663287in}}%
\pgfpathquadraticcurveto{\pgfqpoint{4.923611in}{1.693843in}}{\pgfqpoint{4.893056in}{1.693843in}}%
\pgfpathlineto{\pgfqpoint{3.955521in}{1.693843in}}%
\pgfpathquadraticcurveto{\pgfqpoint{3.924965in}{1.693843in}}{\pgfqpoint{3.924965in}{1.663287in}}%
\pgfpathlineto{\pgfqpoint{3.924965in}{0.826944in}}%
\pgfpathquadraticcurveto{\pgfqpoint{3.924965in}{0.796389in}}{\pgfqpoint{3.955521in}{0.796389in}}%
\pgfpathclose%
\pgfusepath{stroke,fill}%
\end{pgfscope}%
\begin{pgfscope}%
\pgfsetroundcap%
\pgfsetroundjoin%
\pgfsetlinewidth{1.505625pt}%
\definecolor{currentstroke}{rgb}{0.298039,0.686275,0.313725}%
\pgfsetstrokecolor{currentstroke}%
\pgfsetdash{}{0pt}%
\pgfpathmoveto{\pgfqpoint{3.986076in}{1.579259in}}%
\pgfpathlineto{\pgfqpoint{4.291632in}{1.579259in}}%
\pgfusepath{stroke}%
\end{pgfscope}%
\begin{pgfscope}%
\definecolor{textcolor}{rgb}{0.150000,0.150000,0.150000}%
\pgfsetstrokecolor{textcolor}%
\pgfsetfillcolor{textcolor}%
\pgftext[x=4.413854in,y=1.525787in,left,base]{\color{textcolor}\sffamily\fontsize{11.000000}{13.200000}\selectfont 256 bit}%
\end{pgfscope}%
\begin{pgfscope}%
\pgfsetroundcap%
\pgfsetroundjoin%
\pgfsetlinewidth{1.505625pt}%
\definecolor{currentstroke}{rgb}{0.011765,0.662745,0.956863}%
\pgfsetstrokecolor{currentstroke}%
\pgfsetdash{}{0pt}%
\pgfpathmoveto{\pgfqpoint{3.986076in}{1.366354in}}%
\pgfpathlineto{\pgfqpoint{4.291632in}{1.366354in}}%
\pgfusepath{stroke}%
\end{pgfscope}%
\begin{pgfscope}%
\definecolor{textcolor}{rgb}{0.150000,0.150000,0.150000}%
\pgfsetstrokecolor{textcolor}%
\pgfsetfillcolor{textcolor}%
\pgftext[x=4.413854in,y=1.312882in,left,base]{\color{textcolor}\sffamily\fontsize{11.000000}{13.200000}\selectfont 192 bit}%
\end{pgfscope}%
\begin{pgfscope}%
\pgfsetroundcap%
\pgfsetroundjoin%
\pgfsetlinewidth{1.505625pt}%
\definecolor{currentstroke}{rgb}{0.247059,0.317647,0.709804}%
\pgfsetstrokecolor{currentstroke}%
\pgfsetdash{}{0pt}%
\pgfpathmoveto{\pgfqpoint{3.986076in}{1.153449in}}%
\pgfpathlineto{\pgfqpoint{4.291632in}{1.153449in}}%
\pgfusepath{stroke}%
\end{pgfscope}%
\begin{pgfscope}%
\definecolor{textcolor}{rgb}{0.150000,0.150000,0.150000}%
\pgfsetstrokecolor{textcolor}%
\pgfsetfillcolor{textcolor}%
\pgftext[x=4.413854in,y=1.099977in,left,base]{\color{textcolor}\sffamily\fontsize{11.000000}{13.200000}\selectfont 128 bit}%
\end{pgfscope}%
\begin{pgfscope}%
\pgfsetroundcap%
\pgfsetroundjoin%
\pgfsetlinewidth{1.505625pt}%
\definecolor{currentstroke}{rgb}{0.611765,0.152941,0.690196}%
\pgfsetstrokecolor{currentstroke}%
\pgfsetdash{}{0pt}%
\pgfpathmoveto{\pgfqpoint{3.986076in}{0.940544in}}%
\pgfpathlineto{\pgfqpoint{4.291632in}{0.940544in}}%
\pgfusepath{stroke}%
\end{pgfscope}%
\begin{pgfscope}%
\definecolor{textcolor}{rgb}{0.150000,0.150000,0.150000}%
\pgfsetstrokecolor{textcolor}%
\pgfsetfillcolor{textcolor}%
\pgftext[x=4.413854in,y=0.887072in,left,base]{\color{textcolor}\sffamily\fontsize{11.000000}{13.200000}\selectfont 64 bit}%
\end{pgfscope}%
\end{pgfpicture}%
\makeatother%
\endgroup%

%% file: figures/marching-cases.pdf_tex
%% Accompanies image file 'marching-cases.pdf' (pdf, eps, ps)
\begingroup%
  \makeatletter%
  \providecommand\color[2][]{%
    \errmessage{(Inkscape) Color is used for the text in Inkscape, but the package 'color.sty' is not loaded}%
    \renewcommand\color[2][]{}%
  }%
  \providecommand\transparent[1]{%
    \errmessage{(Inkscape) Transparency is used (non-zero) for the text in Inkscape, but the package 'transparent.sty' is not loaded}%
    \renewcommand\transparent[1]{}%
  }%
  \providecommand\rotatebox[2]{#2}%
  \newcommand*\fsize{\dimexpr\f@size pt\relax}%
  \newcommand*\lineheight[1]{\fontsize{\fsize}{#1\fsize}\selectfont}%
  \ifx\svgwidth\undefined%
    \setlength{\unitlength}{287.04426236bp}%
    \ifx\svgscale\undefined%
      \relax%
    \else%
      \setlength{\unitlength}{\unitlength * \real{\svgscale}}%
    \fi%
  \else%
    \setlength{\unitlength}{\svgwidth}%
  \fi%
  \global\let\svgwidth\undefined%
  \global\let\svgscale\undefined%
  \makeatother%
  \begin{picture}(1,0.24133666)%
    \lineheight{1}%
    \setlength\tabcolsep{0pt}%
    \put(0,0){\includegraphics[width=\unitlength,page=1]{marching-cases.pdf}}%
    \put(-0.0004386,0.21486807){\color[rgb]{0,0,0}\makebox(0,0)[lt]{\lineheight{1.25}\smash{\begin{tabular}[t]{l}a)\end{tabular}}}}%
    \put(0,0){\includegraphics[width=\unitlength,page=2]{marching-cases.pdf}}%
    \put(0.36403598,0.21253532){\color[rgb]{0,0,0}\makebox(0,0)[lt]{\lineheight{1.25}\smash{\begin{tabular}[t]{l}b)\end{tabular}}}}%
    \put(0,0){\includegraphics[width=\unitlength,page=3]{marching-cases.pdf}}%
  \end{picture}%
\endgroup%

%% file: paper-04-evaluation.tex
% !TEX root paper.tex
\section{Evaluation}
\label{sec:eval}

Our test system has a \SI{4.20}{\giga\hertz} Intel Core i7-7700K with \SI{16}{\giga\byte} RAM.
All tests were performed on a single thread.
All algorithms were implemented in C++, compiled with Clang~7 using the optimization level \texttt{-O2}.
Our custom fixed-precision arithmetic was implemented with the help of the \texttt{\_addcarry\_u64} and \texttt{\_mulx\_u64} intrinsics.
For comparing with arbitrary-precision arithmetic we use the \texttt{gmp} \cite{Granlund12} library which uses platform-specific optimized assembly code.

% ==============================================================
\subsection{Mathematical Foundation}
\label{sec:eval:math}

\begin{table*}
    \centering
    \begin{tabular}{l r r r r r r}
        operation                      & 128b & 192b & 256b & \texttt{CGAL} & \texttt{gmp}  \\
        \hline
        {\texttt{plane\_from\_points}}   & 226    & 512    & 761    & 2770 & 9100 \\
        {\texttt{are\_planes\_parallel}} & 5      & 15     & 15     &  910 & 1840 \\
        {\texttt{signed\_distance}}      & 4      & 5      & 11     &  530 & 1360 \\
        {\texttt{to\_double\_position}}  & 5      & 62     & 88     &   31 & 120  \\
        {\texttt{intersect\_3\_planes}}  & 23     & 160    & 402    & 4200 & 6380 \\
        {\texttt{classify\_vertex}}      & 13     & 103    & 142    &  710 & 1540
    \end{tabular}
    \caption{
        Performance of the elementary operations used in our method in CPU cycles.
        We compare our custom 128bit, 192bit, and 256bit arithmetic against arbitrary-precision arithmetic provided by \texttt{gmp} \cite{Granlund12} and a lazy exact implementation from \texttt{CGAL} \cite{cgal}, \texttt{Lazy\_exact\_nt<Quotient<MP\_Float>>}. 
        \texttt{gmp} and \texttt{CGAL} were tested with the same numbers as the 256bit arithmetic.
        Due to their internal allocations, the timings of \texttt{gmp} and \texttt{CGAL} showed non-negligible variance of about 5--10\%.
        }
    \label{tab:math}
\end{table*}

Table~\ref{tab:math} shows the performance of the operations presented in Section~\ref{sec:method:int}.
\texttt{plane\_from\_points} is used when converting a mesh into BSP form and it is expensive because we compute the greatest common divisor (gcd) of the normal components to have them in canonical form.
Note that the gcd version is only needed when the normals might otherwise exceed the precision limits (see Figure~\ref{fig:method:sizes}) which is not the case when the quantization bounds of Section~\ref{sec:method:int} are observed.
The two important functions during the mesh cutting (and thus the Boolean operations) are \texttt{intersect\_3\_planes} and \texttt{classify\_vertex}.
\texttt{intersect\_3\_planes} constructs the exact intersection position of three planes in homogeneous 4D integer coordinates by computing four $3 \times 3$ determinants.
\texttt{classify\_vertex} takes such a 4D position and a plane and computes via dot product if the position is on the positive side, the negative side, or exactly on the plane.

% ==============================================================
\subsection{Mesh Cutting}
\begin{figure}
    \centering
    \subcaptionbox{50 planes}{\includegraphics[width=0.43\linewidth]{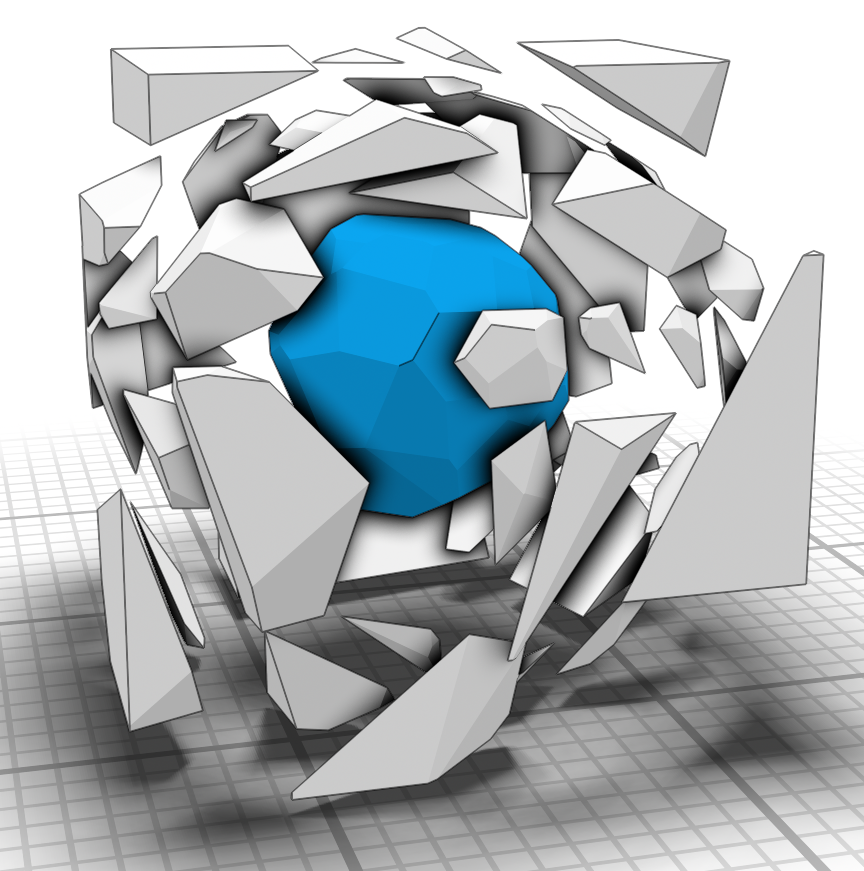}}~~~
    \subcaptionbox{250 planes}{\includegraphics[width=0.43\linewidth]{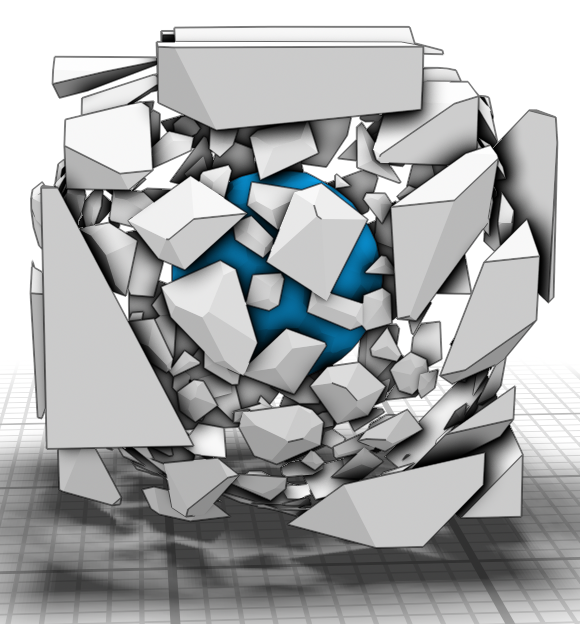}}
    \subcaptionbox{timings for mesh extraction}{\resizebox{\linewidth}{!}{\input{figures/convex-timing-total.pgf}}\vspace{-0.2cm}}

    \subcaptionbox{timings for single mesh-plane cuts}{\resizebox{\linewidth}{!}{\input{figures/convex-timing-depth.pgf}}\vspace{-0.2cm}}
    \subcaptionbox{mesh-plane cut metrics}{\resizebox{\linewidth}{!}{\input{figures/convex-numbers.pgf}}\vspace{-0.2cm}}
    \caption{
        Timings and key metrics for our highly efficient BSP mesh extraction ((c), (d), (e)).
        The tested BSP is the natural worst-case: a linear tree representing a convex object (see (a) and (b) for an explosion view).
        The performance is achieved using algorithmic innovation (in-place mesh cutting using \emph{edge descent} for sublinear scaling) and code-level optimizations (custom-tailored fixed-precision integer arithmetic).
        Note the logarithmic scales for the charts.
		See Table~\ref{tab:math} for the exact \texttt{CGAL} type.
    }
    \label{fig:eval:convex}
\end{figure}
For evaluating the performance of our mesh cutting and extraction, we measured time, number of vertex constructions, and number of vertex classifications.
The results are shown in Figure~\ref{fig:eval:convex}.
Because BSP mesh extraction scales especially poorly with depth, we evaluated on a particularly challenging scenario: a convex object with many faces, corresponding to a purely linear, degenerated BSP tree.
Classical mesh cutting (classify all vertices, add cut geometry) scales quadratically with the number of nodes here because the mesh to cut grows linearly more complex and for each cut, all vertices have to be classified (cf.\ Figure~\ref{fig:eval:convex}~(e)).
Our cutting algorithm with \emph{edge descent} described in Section~\ref{sec:method:cut} usually traverses only a fraction of the mesh, making the method efficient even for deep BSPs.
Note that the BSP-merging algorithms shown in \cite{naylor1990merging, Bernstein09, Campen10} work slightly different but use a subroutine to determine how a newly added plane lies in relation to a BSP node's plane and its subhalfspaces.
During this procedure the new plane is cut with the bounds of the BSP node's region, which scales linearly with the depth of the node and is equivalent to the complexity of the node's cut-mesh in our approach.
This means that \cite{naylor1990merging, Bernstein09, Campen10} scale like the "naive" baseline in Figure~\ref{fig:eval:convex} while our method, at least empirically, scales sub-linear with BSP-depth.

The other significant performance improvement is caused by the custom-tailored fixed-precision integer arithmetic described in Section~\ref{sec:method:int} and consists of two aspects.
Firstly, by storing vertex positions as homogeneous 4D integer coordinates, classification can be performed by a simple dot product.
Vertex construction requires computing four $3 \times 3$ determinants and is more expensive but also needed less often.
Secondly, instead of using arbitrary-precision arithmetic (e.g.\ \texttt{gmp}), we derived the required precision of all operations and intermediary results and use fixed-precision arithmetic instead.

In total, we sped up exact mesh extraction by almost two orders of magnitude, allowing us to compute a few million mesh-plane-intersections per second on a single core, even for challenging BSP trees.
Note that none of these tests exploit the octree, they only test the underlying BSP algorithms.

% ==============================================================
\subsection{BSP Booleans}
\begin{figure}
    \centering
    \subcaptionbox{two random BSPs and their intersection}{\includegraphics[width=\linewidth]{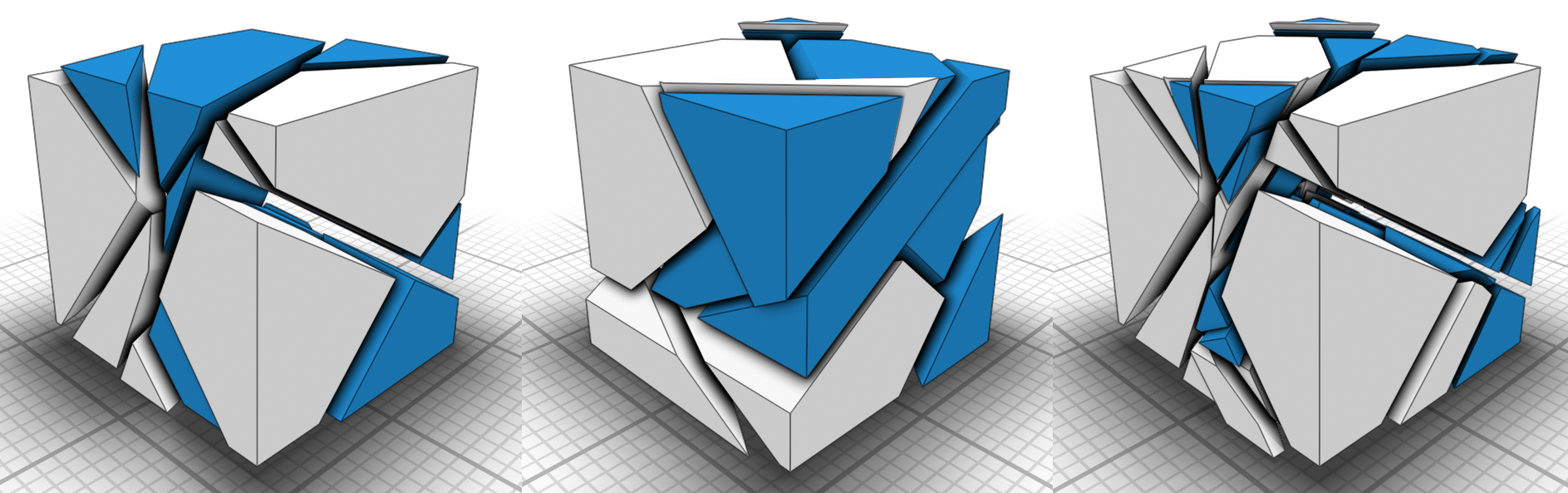}}
    \subcaptionbox{timings for BSP merging (log scale)}{\resizebox{\linewidth}{!}{\input{figures/merge-timing.pgf}}\vspace{-0.2cm}}
    \subcaptionbox{complexity growth}{\resizebox{\linewidth}{!}{\input{figures/merge-growth.pgf}}\vspace{-0.2cm}}
    \caption{
        Timings for merging two random BSPs with 1--200 nodes.
        (a) shows a sample in explosion view (blue is \textbf{in}, white is \textbf{out}).
        Performance mostly depends on the output BSP size.
        See Table~\ref{tab:math} for the exact \texttt{CGAL} type.
    }
    \label{fig:eval:merge}
\end{figure}
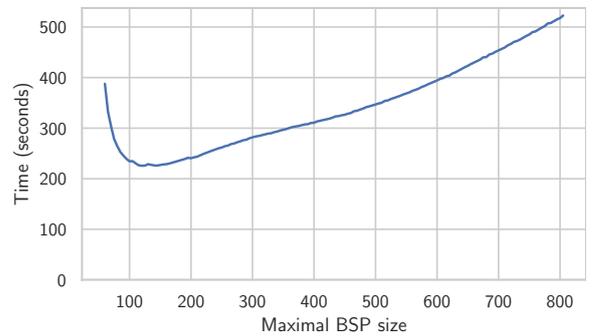
\begin{figure}
    \centering
    \resizebox{\linewidth}{!}{\input{figures/max_bsp_size.pgf}}
    \caption{Time to compute the sweep shown in Figure~\ref{fig:app:sweep-volume} with different maximal BSP sizes.}
    \label{fig:eval:max-bsp-size}
\end{figure}
Figure~\ref{fig:eval:merge} shows the performance of Booleans on random BSPs.
The BSPs are generated by sampling random points in a cube, looking up the leaf node at that position, and splitting it along a random direction.
Leaf nodes are randomly classified as \textbf{in} or \textbf{out}.
We generate two of these BSPs with a size of 1--200 nodes and randomly perform either union, intersection, or difference.
As complete sub-BSPs can be culled during merging, performance is more robustly characterized in terms of output complexity.
For random BSPs, output complexity is roughly quadratic in input complexity.
Real BSPs often only exhibit this quadratic growth close to where the surfaces of the two BSPs meet and otherwise retain their original complexity.
Measured on random BSPs and against output complexity, our method can create around 40--50 million BSP nodes per second per CPU core via CSG.
This again does not yet use the octree and only tests raw BSP boolean performance.

% ==============================================================
\subsection{Memory Consumption}
Our octree-embedded BSPs are a persistent data structure.
Each Octree node stores some metadata and a BSP.
The BSP consists of a flat array of nodes where each node stores index of left and right child and an index into a palette of planes with special indices for leaf nodes.
Planes $(a,b,c,d)$ are either $4+4+4+8 =$ \SI{20}{\byte} or $8+8+8+16=$ \SI{40}{\byte} depending on \SI{128}{\bit} or \SI{256}{\bit} arithmetic (plane coefficients are significantly smaller than the largest intermediate result).
These planes dominate the persistent memory consumption.
The palette can reuse some planes as input triangles might have been split during BSP import, leading to several inner BSP nodes having the same splitting plane.

During mesh extraction and BSP merging, we temporarily build a half-edge mesh with homogeneous 4D integer positions.
In our experiments, this mesh requires around 600--\SI{850}{\byte} per BSP node on average.
In pathological cases it is possible for the mesh to grow quadratically with the BSP size.

When importing a mesh into our data structure we perform polygon clipping where we use a plane palette again.
Each input face corresponds to a plane.
Our polygons reference a supporting plane and a list of edge planes that define edges and vertices.
The clipping process rewrites these references but never creates new planes.
Thus, the total temporary space overhead when importing a mesh amounts to less than \SI{50}{\byte} per input face.

% ==============================================================
\subsection{Octree}
\label{sec:eval:octree}
The octree has only a single parameter: the maximal BSP size. 
To determine which values offer good performance we created the sweep volume shown in Figure~\ref{fig:app:sweep-volume} with varying maximal BSP size.
The results are shown in Figure~\ref{fig:eval:max-bsp-size}.
Good choices for the maximal BSP size can clearly be observed in the range of \num{100} to \num{200} BSP nodes.
% ==============================================================

% ==============================================================
\subsection{Stability and Correctness}
Our core algorithms, BSP mesh extraction and BSP merging, are unconditionally stable and correct.
Any BSP with integer planes that satisfy our mild precision limits (see Section~\ref{sec:method:int}) will work.
The reason is that our mesh cutting can provide strong guarantees:
Given a strictly convex polyhedral input, the cutting always produces two strictly convex polyhedra corresponding to the positive and negative side of the cutting plane (or the unchanged mesh if the cutting plane does not properly intersect the input).
Here, strictly convex means that there are no coplanar faces, i.e.\ each inner dihedral angle is strictly less than $180^\circ$.
Construction and classification of vertex positions are exact due their homogeneous 4D integer representation.
These guarantees also extend to the octree which is, in principle, just a union of BSPs.

The only situations where correctness might get compromised is when importing or exporting floating-point meshes.
When our homogeneous 4D integer coordinates are converted into 3D positions of \texttt{float} or \texttt{double}, rounding must occur.
While we can provide a result that is correctly rounded to the nearest representable number, this could---in theory---still lead to self-intersecting meshes.
In particular, \cite{Milenkovic90} showed that finding a topology-preserving non-intersecting rounding is NP-hard.

Extra care has to be taken when importing meshes.
When the input triangles do not form a closed watertight mesh, the resulting BSP can contain superfluous microgeometry and misclassified cells.
The usual workflow is to scale and round the input vertices to half the normal bits to guarantee that all normals can be represented exactly when computing them via cross product.
Our most permissive setting uses \SI{256}{\bit} arithmetic with \SI{27}{\bit} positions and \SI{55}{\bit} normals which should be sufficient for any practical purposes, though pathological cases can be constructed.
In particular, this guarantees that if a watertight mesh stays self-intersection free under any perturbation with relative magnitude $10^{-8}$ or less then our import finds a BSP corresponding to this mesh within $10^{-8}$ relative Hausdorff distance.

We ran extensive fuzz tests on random BSPs to test stability and correctness.
While we ran out of memory eventually, we were not able to make our algorithm fail.

%% file: figures/convex-timing-total.pgf
%% Creator: Matplotlib, PGF backend
%%
%% To include the figure in your LaTeX document, write
%%   \input{<filename>.pgf}
%%
%% Make sure the required packages are loaded in your preamble
%%   \usepackage{pgf}
%%
%% and, on pdftex
%%   \usepackage[utf8]{inputenc}\DeclareUnicodeCharacter{2212}{-}
%%
%% or, on luatex and xetex
%%   \usepackage{unicode-math}
%%
%% Figures using additional raster images can only be included by \input if
%% they are in the same directory as the main LaTeX file. For loading figures
%% from other directories you can use the `import` package
%% and then include the figures with
%%   \import{<path to file>}{<filename>.pgf}
%%
%% Matplotlib used the following preamble
%%   \usepackage{fontspec}
%%
\begingroup%
\makeatletter%
\begin{pgfpicture}%
\pgfpathrectangle{\pgfpointorigin}{\pgfqpoint{6.000000in}{3.600000in}}%
\pgfusepath{use as bounding box, clip}%
\begin{pgfscope}%
\pgfsetbuttcap%
\pgfsetmiterjoin%
\definecolor{currentfill}{rgb}{1.000000,1.000000,1.000000}%
\pgfsetfillcolor{currentfill}%
\pgfsetlinewidth{0.000000pt}%
\definecolor{currentstroke}{rgb}{1.000000,1.000000,1.000000}%
\pgfsetstrokecolor{currentstroke}%
\pgfsetdash{}{0pt}%
\pgfpathmoveto{\pgfqpoint{0.000000in}{0.000000in}}%
\pgfpathlineto{\pgfqpoint{6.000000in}{0.000000in}}%
\pgfpathlineto{\pgfqpoint{6.000000in}{3.600000in}}%
\pgfpathlineto{\pgfqpoint{0.000000in}{3.600000in}}%
\pgfpathclose%
\pgfusepath{fill}%
\end{pgfscope}%
\begin{pgfscope}%
\pgfsetbuttcap%
\pgfsetmiterjoin%
\definecolor{currentfill}{rgb}{1.000000,1.000000,1.000000}%
\pgfsetfillcolor{currentfill}%
\pgfsetlinewidth{0.000000pt}%
\definecolor{currentstroke}{rgb}{0.000000,0.000000,0.000000}%
\pgfsetstrokecolor{currentstroke}%
\pgfsetstrokeopacity{0.000000}%
\pgfsetdash{}{0pt}%
\pgfpathmoveto{\pgfqpoint{0.900000in}{0.648000in}}%
\pgfpathlineto{\pgfqpoint{6.000000in}{0.648000in}}%
\pgfpathlineto{\pgfqpoint{6.000000in}{3.420000in}}%
\pgfpathlineto{\pgfqpoint{0.900000in}{3.420000in}}%
\pgfpathclose%
\pgfusepath{fill}%
\end{pgfscope}%
\begin{pgfscope}%
\pgfpathrectangle{\pgfqpoint{0.900000in}{0.648000in}}{\pgfqpoint{5.100000in}{2.772000in}}%
\pgfusepath{clip}%
\pgfsetroundcap%
\pgfsetroundjoin%
\pgfsetlinewidth{1.003750pt}%
\definecolor{currentstroke}{rgb}{0.800000,0.800000,0.800000}%
\pgfsetstrokecolor{currentstroke}%
\pgfsetdash{}{0pt}%
\pgfpathmoveto{\pgfqpoint{0.938636in}{0.648000in}}%
\pgfpathlineto{\pgfqpoint{0.938636in}{3.420000in}}%
\pgfusepath{stroke}%
\end{pgfscope}%
\begin{pgfscope}%
\definecolor{textcolor}{rgb}{0.150000,0.150000,0.150000}%
\pgfsetstrokecolor{textcolor}%
\pgfsetfillcolor{textcolor}%
\pgftext[x=0.938636in,y=0.516056in,,top]{\color{textcolor}\sffamily\fontsize{11.000000}{13.200000}\selectfont 0}%
\end{pgfscope}%
\begin{pgfscope}%
\pgfpathrectangle{\pgfqpoint{0.900000in}{0.648000in}}{\pgfqpoint{5.100000in}{2.772000in}}%
\pgfusepath{clip}%
\pgfsetroundcap%
\pgfsetroundjoin%
\pgfsetlinewidth{1.003750pt}%
\definecolor{currentstroke}{rgb}{0.800000,0.800000,0.800000}%
\pgfsetstrokecolor{currentstroke}%
\pgfsetdash{}{0pt}%
\pgfpathmoveto{\pgfqpoint{1.904545in}{0.648000in}}%
\pgfpathlineto{\pgfqpoint{1.904545in}{3.420000in}}%
\pgfusepath{stroke}%
\end{pgfscope}%
\begin{pgfscope}%
\definecolor{textcolor}{rgb}{0.150000,0.150000,0.150000}%
\pgfsetstrokecolor{textcolor}%
\pgfsetfillcolor{textcolor}%
\pgftext[x=1.904545in,y=0.516056in,,top]{\color{textcolor}\sffamily\fontsize{11.000000}{13.200000}\selectfont 50}%
\end{pgfscope}%
\begin{pgfscope}%
\pgfpathrectangle{\pgfqpoint{0.900000in}{0.648000in}}{\pgfqpoint{5.100000in}{2.772000in}}%
\pgfusepath{clip}%
\pgfsetroundcap%
\pgfsetroundjoin%
\pgfsetlinewidth{1.003750pt}%
\definecolor{currentstroke}{rgb}{0.800000,0.800000,0.800000}%
\pgfsetstrokecolor{currentstroke}%
\pgfsetdash{}{0pt}%
\pgfpathmoveto{\pgfqpoint{2.870455in}{0.648000in}}%
\pgfpathlineto{\pgfqpoint{2.870455in}{3.420000in}}%
\pgfusepath{stroke}%
\end{pgfscope}%
\begin{pgfscope}%
\definecolor{textcolor}{rgb}{0.150000,0.150000,0.150000}%
\pgfsetstrokecolor{textcolor}%
\pgfsetfillcolor{textcolor}%
\pgftext[x=2.870455in,y=0.516056in,,top]{\color{textcolor}\sffamily\fontsize{11.000000}{13.200000}\selectfont 100}%
\end{pgfscope}%
\begin{pgfscope}%
\pgfpathrectangle{\pgfqpoint{0.900000in}{0.648000in}}{\pgfqpoint{5.100000in}{2.772000in}}%
\pgfusepath{clip}%
\pgfsetroundcap%
\pgfsetroundjoin%
\pgfsetlinewidth{1.003750pt}%
\definecolor{currentstroke}{rgb}{0.800000,0.800000,0.800000}%
\pgfsetstrokecolor{currentstroke}%
\pgfsetdash{}{0pt}%
\pgfpathmoveto{\pgfqpoint{3.836364in}{0.648000in}}%
\pgfpathlineto{\pgfqpoint{3.836364in}{3.420000in}}%
\pgfusepath{stroke}%
\end{pgfscope}%
\begin{pgfscope}%
\definecolor{textcolor}{rgb}{0.150000,0.150000,0.150000}%
\pgfsetstrokecolor{textcolor}%
\pgfsetfillcolor{textcolor}%
\pgftext[x=3.836364in,y=0.516056in,,top]{\color{textcolor}\sffamily\fontsize{11.000000}{13.200000}\selectfont 150}%
\end{pgfscope}%
\begin{pgfscope}%
\pgfpathrectangle{\pgfqpoint{0.900000in}{0.648000in}}{\pgfqpoint{5.100000in}{2.772000in}}%
\pgfusepath{clip}%
\pgfsetroundcap%
\pgfsetroundjoin%
\pgfsetlinewidth{1.003750pt}%
\definecolor{currentstroke}{rgb}{0.800000,0.800000,0.800000}%
\pgfsetstrokecolor{currentstroke}%
\pgfsetdash{}{0pt}%
\pgfpathmoveto{\pgfqpoint{4.802273in}{0.648000in}}%
\pgfpathlineto{\pgfqpoint{4.802273in}{3.420000in}}%
\pgfusepath{stroke}%
\end{pgfscope}%
\begin{pgfscope}%
\definecolor{textcolor}{rgb}{0.150000,0.150000,0.150000}%
\pgfsetstrokecolor{textcolor}%
\pgfsetfillcolor{textcolor}%
\pgftext[x=4.802273in,y=0.516056in,,top]{\color{textcolor}\sffamily\fontsize{11.000000}{13.200000}\selectfont 200}%
\end{pgfscope}%
\begin{pgfscope}%
\pgfpathrectangle{\pgfqpoint{0.900000in}{0.648000in}}{\pgfqpoint{5.100000in}{2.772000in}}%
\pgfusepath{clip}%
\pgfsetroundcap%
\pgfsetroundjoin%
\pgfsetlinewidth{1.003750pt}%
\definecolor{currentstroke}{rgb}{0.800000,0.800000,0.800000}%
\pgfsetstrokecolor{currentstroke}%
\pgfsetdash{}{0pt}%
\pgfpathmoveto{\pgfqpoint{5.768182in}{0.648000in}}%
\pgfpathlineto{\pgfqpoint{5.768182in}{3.420000in}}%
\pgfusepath{stroke}%
\end{pgfscope}%
\begin{pgfscope}%
\definecolor{textcolor}{rgb}{0.150000,0.150000,0.150000}%
\pgfsetstrokecolor{textcolor}%
\pgfsetfillcolor{textcolor}%
\pgftext[x=5.768182in,y=0.516056in,,top]{\color{textcolor}\sffamily\fontsize{11.000000}{13.200000}\selectfont 250}%
\end{pgfscope}%
\begin{pgfscope}%
\definecolor{textcolor}{rgb}{0.150000,0.150000,0.150000}%
\pgfsetstrokecolor{textcolor}%
\pgfsetfillcolor{textcolor}%
\pgftext[x=3.450000in,y=0.324833in,,top]{\color{textcolor}\sffamily\fontsize{12.000000}{14.400000}\selectfont planes (= bsp depth)}%
\end{pgfscope}%
\begin{pgfscope}%
\pgfpathrectangle{\pgfqpoint{0.900000in}{0.648000in}}{\pgfqpoint{5.100000in}{2.772000in}}%
\pgfusepath{clip}%
\pgfsetroundcap%
\pgfsetroundjoin%
\pgfsetlinewidth{1.003750pt}%
\definecolor{currentstroke}{rgb}{0.800000,0.800000,0.800000}%
\pgfsetstrokecolor{currentstroke}%
\pgfsetdash{}{0pt}%
\pgfpathmoveto{\pgfqpoint{0.900000in}{1.083835in}}%
\pgfpathlineto{\pgfqpoint{6.000000in}{1.083835in}}%
\pgfusepath{stroke}%
\end{pgfscope}%
\begin{pgfscope}%
\definecolor{textcolor}{rgb}{0.150000,0.150000,0.150000}%
\pgfsetstrokecolor{textcolor}%
\pgfsetfillcolor{textcolor}%
\pgftext[x=0.458177in, y=1.030821in, left, base]{\color{textcolor}\sffamily\fontsize{11.000000}{13.200000}\selectfont \(\displaystyle {10^{-2}}\)}%
\end{pgfscope}%
\begin{pgfscope}%
\pgfpathrectangle{\pgfqpoint{0.900000in}{0.648000in}}{\pgfqpoint{5.100000in}{2.772000in}}%
\pgfusepath{clip}%
\pgfsetroundcap%
\pgfsetroundjoin%
\pgfsetlinewidth{1.003750pt}%
\definecolor{currentstroke}{rgb}{0.800000,0.800000,0.800000}%
\pgfsetstrokecolor{currentstroke}%
\pgfsetdash{}{0pt}%
\pgfpathmoveto{\pgfqpoint{0.900000in}{1.800406in}}%
\pgfpathlineto{\pgfqpoint{6.000000in}{1.800406in}}%
\pgfusepath{stroke}%
\end{pgfscope}%
\begin{pgfscope}%
\definecolor{textcolor}{rgb}{0.150000,0.150000,0.150000}%
\pgfsetstrokecolor{textcolor}%
\pgfsetfillcolor{textcolor}%
\pgftext[x=0.458177in, y=1.747392in, left, base]{\color{textcolor}\sffamily\fontsize{11.000000}{13.200000}\selectfont \(\displaystyle {10^{-1}}\)}%
\end{pgfscope}%
\begin{pgfscope}%
\pgfpathrectangle{\pgfqpoint{0.900000in}{0.648000in}}{\pgfqpoint{5.100000in}{2.772000in}}%
\pgfusepath{clip}%
\pgfsetroundcap%
\pgfsetroundjoin%
\pgfsetlinewidth{1.003750pt}%
\definecolor{currentstroke}{rgb}{0.800000,0.800000,0.800000}%
\pgfsetstrokecolor{currentstroke}%
\pgfsetdash{}{0pt}%
\pgfpathmoveto{\pgfqpoint{0.900000in}{2.516978in}}%
\pgfpathlineto{\pgfqpoint{6.000000in}{2.516978in}}%
\pgfusepath{stroke}%
\end{pgfscope}%
\begin{pgfscope}%
\definecolor{textcolor}{rgb}{0.150000,0.150000,0.150000}%
\pgfsetstrokecolor{textcolor}%
\pgfsetfillcolor{textcolor}%
\pgftext[x=0.549999in, y=2.463964in, left, base]{\color{textcolor}\sffamily\fontsize{11.000000}{13.200000}\selectfont \(\displaystyle {10^{0}}\)}%
\end{pgfscope}%
\begin{pgfscope}%
\pgfpathrectangle{\pgfqpoint{0.900000in}{0.648000in}}{\pgfqpoint{5.100000in}{2.772000in}}%
\pgfusepath{clip}%
\pgfsetroundcap%
\pgfsetroundjoin%
\pgfsetlinewidth{1.003750pt}%
\definecolor{currentstroke}{rgb}{0.800000,0.800000,0.800000}%
\pgfsetstrokecolor{currentstroke}%
\pgfsetdash{}{0pt}%
\pgfpathmoveto{\pgfqpoint{0.900000in}{3.233549in}}%
\pgfpathlineto{\pgfqpoint{6.000000in}{3.233549in}}%
\pgfusepath{stroke}%
\end{pgfscope}%
\begin{pgfscope}%
\definecolor{textcolor}{rgb}{0.150000,0.150000,0.150000}%
\pgfsetstrokecolor{textcolor}%
\pgfsetfillcolor{textcolor}%
\pgftext[x=0.549999in, y=3.180535in, left, base]{\color{textcolor}\sffamily\fontsize{11.000000}{13.200000}\selectfont \(\displaystyle {10^{1}}\)}%
\end{pgfscope}%
\begin{pgfscope}%
\definecolor{textcolor}{rgb}{0.150000,0.150000,0.150000}%
\pgfsetstrokecolor{textcolor}%
\pgfsetfillcolor{textcolor}%
\pgftext[x=0.402621in,y=2.034000in,,bottom,rotate=90.000000]{\color{textcolor}\sffamily\fontsize{12.000000}{14.400000}\selectfont time (ms)}%
\end{pgfscope}%
\begin{pgfscope}%
\pgfpathrectangle{\pgfqpoint{0.900000in}{0.648000in}}{\pgfqpoint{5.100000in}{2.772000in}}%
\pgfusepath{clip}%
\pgfsetbuttcap%
\pgfsetroundjoin%
\pgfsetlinewidth{1.505625pt}%
\definecolor{currentstroke}{rgb}{0.545098,0.764706,0.290196}%
\pgfsetstrokecolor{currentstroke}%
\pgfsetdash{{4.500000pt}{4.500000pt}}{0.000000pt}%
\pgfpathmoveto{\pgfqpoint{1.131818in}{0.838571in}}%
\pgfpathlineto{\pgfqpoint{1.228409in}{0.997116in}}%
\pgfpathlineto{\pgfqpoint{1.325000in}{1.120738in}}%
\pgfpathlineto{\pgfqpoint{1.421591in}{1.227994in}}%
\pgfpathlineto{\pgfqpoint{1.518182in}{1.322051in}}%
\pgfpathlineto{\pgfqpoint{1.614773in}{1.396399in}}%
\pgfpathlineto{\pgfqpoint{1.711364in}{1.466421in}}%
\pgfpathlineto{\pgfqpoint{1.807955in}{1.531318in}}%
\pgfpathlineto{\pgfqpoint{1.904545in}{1.588612in}}%
\pgfpathlineto{\pgfqpoint{2.001136in}{1.640511in}}%
\pgfpathlineto{\pgfqpoint{2.097727in}{1.683638in}}%
\pgfpathlineto{\pgfqpoint{2.194318in}{1.726043in}}%
\pgfpathlineto{\pgfqpoint{2.290909in}{1.764832in}}%
\pgfpathlineto{\pgfqpoint{2.387500in}{1.802716in}}%
\pgfpathlineto{\pgfqpoint{2.484091in}{1.836665in}}%
\pgfpathlineto{\pgfqpoint{2.580682in}{1.870036in}}%
\pgfpathlineto{\pgfqpoint{2.677273in}{1.899822in}}%
\pgfpathlineto{\pgfqpoint{2.773864in}{1.928590in}}%
\pgfpathlineto{\pgfqpoint{2.870455in}{1.959323in}}%
\pgfpathlineto{\pgfqpoint{2.967045in}{1.988318in}}%
\pgfpathlineto{\pgfqpoint{3.063636in}{2.014804in}}%
\pgfpathlineto{\pgfqpoint{3.160227in}{2.039456in}}%
\pgfpathlineto{\pgfqpoint{3.256818in}{2.062788in}}%
\pgfpathlineto{\pgfqpoint{3.353409in}{2.085083in}}%
\pgfpathlineto{\pgfqpoint{3.450000in}{2.107082in}}%
\pgfpathlineto{\pgfqpoint{3.546591in}{2.128011in}}%
\pgfpathlineto{\pgfqpoint{3.643182in}{2.148049in}}%
\pgfpathlineto{\pgfqpoint{3.739773in}{2.167298in}}%
\pgfpathlineto{\pgfqpoint{3.836364in}{2.186533in}}%
\pgfpathlineto{\pgfqpoint{3.932955in}{2.205148in}}%
\pgfpathlineto{\pgfqpoint{4.029545in}{2.224037in}}%
\pgfpathlineto{\pgfqpoint{4.126136in}{2.240737in}}%
\pgfpathlineto{\pgfqpoint{4.222727in}{2.258492in}}%
\pgfpathlineto{\pgfqpoint{4.319318in}{2.275192in}}%
\pgfpathlineto{\pgfqpoint{4.415909in}{2.291831in}}%
\pgfpathlineto{\pgfqpoint{4.512500in}{2.307771in}}%
\pgfpathlineto{\pgfqpoint{4.609091in}{2.323279in}}%
\pgfpathlineto{\pgfqpoint{4.705682in}{2.338842in}}%
\pgfpathlineto{\pgfqpoint{4.802273in}{2.354074in}}%
\pgfpathlineto{\pgfqpoint{4.898864in}{2.367922in}}%
\pgfpathlineto{\pgfqpoint{4.995455in}{2.382935in}}%
\pgfpathlineto{\pgfqpoint{5.092045in}{2.396780in}}%
\pgfpathlineto{\pgfqpoint{5.188636in}{2.410993in}}%
\pgfpathlineto{\pgfqpoint{5.285227in}{2.424545in}}%
\pgfpathlineto{\pgfqpoint{5.381818in}{2.437774in}}%
\pgfpathlineto{\pgfqpoint{5.478409in}{2.450704in}}%
\pgfpathlineto{\pgfqpoint{5.575000in}{2.463387in}}%
\pgfpathlineto{\pgfqpoint{5.671591in}{2.475812in}}%
\pgfpathlineto{\pgfqpoint{5.768182in}{2.487574in}}%
\pgfusepath{stroke}%
\end{pgfscope}%
\begin{pgfscope}%
\pgfpathrectangle{\pgfqpoint{0.900000in}{0.648000in}}{\pgfqpoint{5.100000in}{2.772000in}}%
\pgfusepath{clip}%
\pgfsetroundcap%
\pgfsetroundjoin%
\pgfsetlinewidth{1.505625pt}%
\definecolor{currentstroke}{rgb}{0.545098,0.764706,0.290196}%
\pgfsetstrokecolor{currentstroke}%
\pgfsetdash{}{0pt}%
\pgfpathmoveto{\pgfqpoint{1.131818in}{0.774000in}}%
\pgfpathlineto{\pgfqpoint{1.228409in}{0.908957in}}%
\pgfpathlineto{\pgfqpoint{1.325000in}{1.023968in}}%
\pgfpathlineto{\pgfqpoint{1.421591in}{1.098900in}}%
\pgfpathlineto{\pgfqpoint{1.518182in}{1.159441in}}%
\pgfpathlineto{\pgfqpoint{1.614773in}{1.219638in}}%
\pgfpathlineto{\pgfqpoint{1.711364in}{1.265890in}}%
\pgfpathlineto{\pgfqpoint{1.807955in}{1.312736in}}%
\pgfpathlineto{\pgfqpoint{1.904545in}{1.351988in}}%
\pgfpathlineto{\pgfqpoint{2.001136in}{1.388644in}}%
\pgfpathlineto{\pgfqpoint{2.097727in}{1.423101in}}%
\pgfpathlineto{\pgfqpoint{2.194318in}{1.450377in}}%
\pgfpathlineto{\pgfqpoint{2.290909in}{1.475472in}}%
\pgfpathlineto{\pgfqpoint{2.387500in}{1.504433in}}%
\pgfpathlineto{\pgfqpoint{2.484091in}{1.531022in}}%
\pgfpathlineto{\pgfqpoint{2.580682in}{1.551915in}}%
\pgfpathlineto{\pgfqpoint{2.677273in}{1.570628in}}%
\pgfpathlineto{\pgfqpoint{2.773864in}{1.589869in}}%
\pgfpathlineto{\pgfqpoint{2.870455in}{1.609758in}}%
\pgfpathlineto{\pgfqpoint{2.967045in}{1.626416in}}%
\pgfpathlineto{\pgfqpoint{3.063636in}{1.642016in}}%
\pgfpathlineto{\pgfqpoint{3.160227in}{1.658908in}}%
\pgfpathlineto{\pgfqpoint{3.256818in}{1.673132in}}%
\pgfpathlineto{\pgfqpoint{3.353409in}{1.690136in}}%
\pgfpathlineto{\pgfqpoint{3.450000in}{1.707529in}}%
\pgfpathlineto{\pgfqpoint{3.546591in}{1.720870in}}%
\pgfpathlineto{\pgfqpoint{3.643182in}{1.732585in}}%
\pgfpathlineto{\pgfqpoint{3.739773in}{1.747758in}}%
\pgfpathlineto{\pgfqpoint{3.836364in}{1.760153in}}%
\pgfpathlineto{\pgfqpoint{3.932955in}{1.775228in}}%
\pgfpathlineto{\pgfqpoint{4.029545in}{1.787144in}}%
\pgfpathlineto{\pgfqpoint{4.126136in}{1.799908in}}%
\pgfpathlineto{\pgfqpoint{4.222727in}{1.813538in}}%
\pgfpathlineto{\pgfqpoint{4.319318in}{1.824178in}}%
\pgfpathlineto{\pgfqpoint{4.415909in}{1.834776in}}%
\pgfpathlineto{\pgfqpoint{4.512500in}{1.844744in}}%
\pgfpathlineto{\pgfqpoint{4.609091in}{1.862422in}}%
\pgfpathlineto{\pgfqpoint{4.705682in}{1.872840in}}%
\pgfpathlineto{\pgfqpoint{4.802273in}{1.884240in}}%
\pgfpathlineto{\pgfqpoint{4.898864in}{1.893986in}}%
\pgfpathlineto{\pgfqpoint{4.995455in}{1.906059in}}%
\pgfpathlineto{\pgfqpoint{5.092045in}{1.915729in}}%
\pgfpathlineto{\pgfqpoint{5.188636in}{1.924765in}}%
\pgfpathlineto{\pgfqpoint{5.285227in}{1.933557in}}%
\pgfpathlineto{\pgfqpoint{5.381818in}{1.942076in}}%
\pgfpathlineto{\pgfqpoint{5.478409in}{1.950114in}}%
\pgfpathlineto{\pgfqpoint{5.575000in}{1.958354in}}%
\pgfpathlineto{\pgfqpoint{5.671591in}{1.967577in}}%
\pgfpathlineto{\pgfqpoint{5.768182in}{1.975070in}}%
\pgfusepath{stroke}%
\end{pgfscope}%
\begin{pgfscope}%
\pgfpathrectangle{\pgfqpoint{0.900000in}{0.648000in}}{\pgfqpoint{5.100000in}{2.772000in}}%
\pgfusepath{clip}%
\pgfsetbuttcap%
\pgfsetroundjoin%
\pgfsetlinewidth{1.505625pt}%
\definecolor{currentstroke}{rgb}{1.000000,0.756863,0.027451}%
\pgfsetstrokecolor{currentstroke}%
\pgfsetdash{{4.500000pt}{4.500000pt}}{0.000000pt}%
\pgfpathmoveto{\pgfqpoint{1.131818in}{1.057683in}}%
\pgfpathlineto{\pgfqpoint{1.228409in}{1.226049in}}%
\pgfpathlineto{\pgfqpoint{1.325000in}{1.378790in}}%
\pgfpathlineto{\pgfqpoint{1.421591in}{1.460958in}}%
\pgfpathlineto{\pgfqpoint{1.518182in}{1.554052in}}%
\pgfpathlineto{\pgfqpoint{1.614773in}{1.630144in}}%
\pgfpathlineto{\pgfqpoint{1.711364in}{1.699899in}}%
\pgfpathlineto{\pgfqpoint{1.807955in}{1.762138in}}%
\pgfpathlineto{\pgfqpoint{1.904545in}{1.818264in}}%
\pgfpathlineto{\pgfqpoint{2.001136in}{1.870185in}}%
\pgfpathlineto{\pgfqpoint{2.097727in}{1.916413in}}%
\pgfpathlineto{\pgfqpoint{2.194318in}{1.958635in}}%
\pgfpathlineto{\pgfqpoint{2.290909in}{1.999491in}}%
\pgfpathlineto{\pgfqpoint{2.387500in}{2.038035in}}%
\pgfpathlineto{\pgfqpoint{2.484091in}{2.073680in}}%
\pgfpathlineto{\pgfqpoint{2.580682in}{2.106568in}}%
\pgfpathlineto{\pgfqpoint{2.677273in}{2.139002in}}%
\pgfpathlineto{\pgfqpoint{2.773864in}{2.169076in}}%
\pgfpathlineto{\pgfqpoint{2.870455in}{2.198032in}}%
\pgfpathlineto{\pgfqpoint{2.967045in}{2.226149in}}%
\pgfpathlineto{\pgfqpoint{3.063636in}{2.252450in}}%
\pgfpathlineto{\pgfqpoint{3.160227in}{2.278021in}}%
\pgfpathlineto{\pgfqpoint{3.256818in}{2.303398in}}%
\pgfpathlineto{\pgfqpoint{3.353409in}{2.328369in}}%
\pgfpathlineto{\pgfqpoint{3.450000in}{2.351977in}}%
\pgfpathlineto{\pgfqpoint{3.546591in}{2.373854in}}%
\pgfpathlineto{\pgfqpoint{3.643182in}{2.394891in}}%
\pgfpathlineto{\pgfqpoint{3.739773in}{2.415538in}}%
\pgfpathlineto{\pgfqpoint{3.836364in}{2.435299in}}%
\pgfpathlineto{\pgfqpoint{3.932955in}{2.454708in}}%
\pgfpathlineto{\pgfqpoint{4.029545in}{2.473544in}}%
\pgfpathlineto{\pgfqpoint{4.126136in}{2.491491in}}%
\pgfpathlineto{\pgfqpoint{4.222727in}{2.509212in}}%
\pgfpathlineto{\pgfqpoint{4.319318in}{2.526385in}}%
\pgfpathlineto{\pgfqpoint{4.415909in}{2.543271in}}%
\pgfpathlineto{\pgfqpoint{4.512500in}{2.559607in}}%
\pgfpathlineto{\pgfqpoint{4.609091in}{2.577698in}}%
\pgfpathlineto{\pgfqpoint{4.705682in}{2.593237in}}%
\pgfpathlineto{\pgfqpoint{4.802273in}{2.608410in}}%
\pgfpathlineto{\pgfqpoint{4.898864in}{2.623630in}}%
\pgfpathlineto{\pgfqpoint{4.995455in}{2.638199in}}%
\pgfpathlineto{\pgfqpoint{5.092045in}{2.652531in}}%
\pgfpathlineto{\pgfqpoint{5.188636in}{2.668620in}}%
\pgfpathlineto{\pgfqpoint{5.285227in}{2.680381in}}%
\pgfpathlineto{\pgfqpoint{5.381818in}{2.694096in}}%
\pgfpathlineto{\pgfqpoint{5.478409in}{2.706686in}}%
\pgfpathlineto{\pgfqpoint{5.575000in}{2.719425in}}%
\pgfpathlineto{\pgfqpoint{5.671591in}{2.731596in}}%
\pgfpathlineto{\pgfqpoint{5.768182in}{2.741494in}}%
\pgfusepath{stroke}%
\end{pgfscope}%
\begin{pgfscope}%
\pgfpathrectangle{\pgfqpoint{0.900000in}{0.648000in}}{\pgfqpoint{5.100000in}{2.772000in}}%
\pgfusepath{clip}%
\pgfsetroundcap%
\pgfsetroundjoin%
\pgfsetlinewidth{1.505625pt}%
\definecolor{currentstroke}{rgb}{1.000000,0.756863,0.027451}%
\pgfsetstrokecolor{currentstroke}%
\pgfsetdash{}{0pt}%
\pgfpathmoveto{\pgfqpoint{1.131818in}{1.012754in}}%
\pgfpathlineto{\pgfqpoint{1.228409in}{1.151956in}}%
\pgfpathlineto{\pgfqpoint{1.325000in}{1.271066in}}%
\pgfpathlineto{\pgfqpoint{1.421591in}{1.350168in}}%
\pgfpathlineto{\pgfqpoint{1.518182in}{1.411897in}}%
\pgfpathlineto{\pgfqpoint{1.614773in}{1.466857in}}%
\pgfpathlineto{\pgfqpoint{1.711364in}{1.512652in}}%
\pgfpathlineto{\pgfqpoint{1.807955in}{1.556215in}}%
\pgfpathlineto{\pgfqpoint{1.904545in}{1.594807in}}%
\pgfpathlineto{\pgfqpoint{2.001136in}{1.631823in}}%
\pgfpathlineto{\pgfqpoint{2.097727in}{1.659349in}}%
\pgfpathlineto{\pgfqpoint{2.194318in}{1.683918in}}%
\pgfpathlineto{\pgfqpoint{2.290909in}{1.708199in}}%
\pgfpathlineto{\pgfqpoint{2.387500in}{1.736637in}}%
\pgfpathlineto{\pgfqpoint{2.484091in}{1.760529in}}%
\pgfpathlineto{\pgfqpoint{2.580682in}{1.781352in}}%
\pgfpathlineto{\pgfqpoint{2.677273in}{1.803133in}}%
\pgfpathlineto{\pgfqpoint{2.773864in}{1.823494in}}%
\pgfpathlineto{\pgfqpoint{2.870455in}{1.843608in}}%
\pgfpathlineto{\pgfqpoint{2.967045in}{1.861446in}}%
\pgfpathlineto{\pgfqpoint{3.063636in}{1.876921in}}%
\pgfpathlineto{\pgfqpoint{3.160227in}{1.891453in}}%
\pgfpathlineto{\pgfqpoint{3.256818in}{1.907114in}}%
\pgfpathlineto{\pgfqpoint{3.353409in}{1.923748in}}%
\pgfpathlineto{\pgfqpoint{3.450000in}{1.939537in}}%
\pgfpathlineto{\pgfqpoint{3.546591in}{1.951747in}}%
\pgfpathlineto{\pgfqpoint{3.643182in}{1.964423in}}%
\pgfpathlineto{\pgfqpoint{3.739773in}{1.977038in}}%
\pgfpathlineto{\pgfqpoint{3.836364in}{1.988667in}}%
\pgfpathlineto{\pgfqpoint{3.932955in}{2.000796in}}%
\pgfpathlineto{\pgfqpoint{4.029545in}{2.012549in}}%
\pgfpathlineto{\pgfqpoint{4.126136in}{2.023717in}}%
\pgfpathlineto{\pgfqpoint{4.222727in}{2.035200in}}%
\pgfpathlineto{\pgfqpoint{4.319318in}{2.046066in}}%
\pgfpathlineto{\pgfqpoint{4.415909in}{2.056739in}}%
\pgfpathlineto{\pgfqpoint{4.512500in}{2.067581in}}%
\pgfpathlineto{\pgfqpoint{4.609091in}{2.076244in}}%
\pgfpathlineto{\pgfqpoint{4.705682in}{2.084711in}}%
\pgfpathlineto{\pgfqpoint{4.802273in}{2.095364in}}%
\pgfpathlineto{\pgfqpoint{4.898864in}{2.104377in}}%
\pgfpathlineto{\pgfqpoint{4.995455in}{2.114882in}}%
\pgfpathlineto{\pgfqpoint{5.092045in}{2.123840in}}%
\pgfpathlineto{\pgfqpoint{5.188636in}{2.133159in}}%
\pgfpathlineto{\pgfqpoint{5.285227in}{2.141718in}}%
\pgfpathlineto{\pgfqpoint{5.381818in}{2.149968in}}%
\pgfpathlineto{\pgfqpoint{5.478409in}{2.158057in}}%
\pgfpathlineto{\pgfqpoint{5.575000in}{2.166914in}}%
\pgfpathlineto{\pgfqpoint{5.671591in}{2.174718in}}%
\pgfpathlineto{\pgfqpoint{5.768182in}{2.181839in}}%
\pgfusepath{stroke}%
\end{pgfscope}%
\begin{pgfscope}%
\pgfpathrectangle{\pgfqpoint{0.900000in}{0.648000in}}{\pgfqpoint{5.100000in}{2.772000in}}%
\pgfusepath{clip}%
\pgfsetbuttcap%
\pgfsetroundjoin%
\pgfsetlinewidth{1.505625pt}%
\definecolor{currentstroke}{rgb}{0.129412,0.588235,0.952941}%
\pgfsetstrokecolor{currentstroke}%
\pgfsetdash{{4.500000pt}{4.500000pt}}{0.000000pt}%
\pgfpathmoveto{\pgfqpoint{1.131818in}{1.150502in}}%
\pgfpathlineto{\pgfqpoint{1.228409in}{1.308320in}}%
\pgfpathlineto{\pgfqpoint{1.325000in}{1.433552in}}%
\pgfpathlineto{\pgfqpoint{1.421591in}{1.531303in}}%
\pgfpathlineto{\pgfqpoint{1.518182in}{1.613491in}}%
\pgfpathlineto{\pgfqpoint{1.614773in}{1.690212in}}%
\pgfpathlineto{\pgfqpoint{1.711364in}{1.757007in}}%
\pgfpathlineto{\pgfqpoint{1.807955in}{1.818836in}}%
\pgfpathlineto{\pgfqpoint{1.904545in}{1.873059in}}%
\pgfpathlineto{\pgfqpoint{2.001136in}{1.923270in}}%
\pgfpathlineto{\pgfqpoint{2.097727in}{1.967501in}}%
\pgfpathlineto{\pgfqpoint{2.194318in}{2.008906in}}%
\pgfpathlineto{\pgfqpoint{2.290909in}{2.048113in}}%
\pgfpathlineto{\pgfqpoint{2.387500in}{2.085653in}}%
\pgfpathlineto{\pgfqpoint{2.484091in}{2.121551in}}%
\pgfpathlineto{\pgfqpoint{2.580682in}{2.153388in}}%
\pgfpathlineto{\pgfqpoint{2.677273in}{2.184350in}}%
\pgfpathlineto{\pgfqpoint{2.773864in}{2.214222in}}%
\pgfpathlineto{\pgfqpoint{2.870455in}{2.242759in}}%
\pgfpathlineto{\pgfqpoint{2.967045in}{2.269350in}}%
\pgfpathlineto{\pgfqpoint{3.063636in}{2.295584in}}%
\pgfpathlineto{\pgfqpoint{3.160227in}{2.320399in}}%
\pgfpathlineto{\pgfqpoint{3.256818in}{2.345366in}}%
\pgfpathlineto{\pgfqpoint{3.353409in}{2.368777in}}%
\pgfpathlineto{\pgfqpoint{3.450000in}{2.391407in}}%
\pgfpathlineto{\pgfqpoint{3.546591in}{2.413543in}}%
\pgfpathlineto{\pgfqpoint{3.643182in}{2.434175in}}%
\pgfpathlineto{\pgfqpoint{3.739773in}{2.454466in}}%
\pgfpathlineto{\pgfqpoint{3.836364in}{2.473895in}}%
\pgfpathlineto{\pgfqpoint{3.932955in}{2.492851in}}%
\pgfpathlineto{\pgfqpoint{4.029545in}{2.511045in}}%
\pgfpathlineto{\pgfqpoint{4.126136in}{2.529177in}}%
\pgfpathlineto{\pgfqpoint{4.222727in}{2.546658in}}%
\pgfpathlineto{\pgfqpoint{4.319318in}{2.563892in}}%
\pgfpathlineto{\pgfqpoint{4.415909in}{2.580889in}}%
\pgfpathlineto{\pgfqpoint{4.512500in}{2.597071in}}%
\pgfpathlineto{\pgfqpoint{4.609091in}{2.612679in}}%
\pgfpathlineto{\pgfqpoint{4.705682in}{2.627917in}}%
\pgfpathlineto{\pgfqpoint{4.802273in}{2.643984in}}%
\pgfpathlineto{\pgfqpoint{4.898864in}{2.657544in}}%
\pgfpathlineto{\pgfqpoint{4.995455in}{2.671940in}}%
\pgfpathlineto{\pgfqpoint{5.092045in}{2.685816in}}%
\pgfpathlineto{\pgfqpoint{5.188636in}{2.701519in}}%
\pgfpathlineto{\pgfqpoint{5.285227in}{2.715707in}}%
\pgfpathlineto{\pgfqpoint{5.381818in}{2.727781in}}%
\pgfpathlineto{\pgfqpoint{5.478409in}{2.740629in}}%
\pgfpathlineto{\pgfqpoint{5.575000in}{2.753350in}}%
\pgfpathlineto{\pgfqpoint{5.671591in}{2.765477in}}%
\pgfpathlineto{\pgfqpoint{5.768182in}{2.777954in}}%
\pgfusepath{stroke}%
\end{pgfscope}%
\begin{pgfscope}%
\pgfpathrectangle{\pgfqpoint{0.900000in}{0.648000in}}{\pgfqpoint{5.100000in}{2.772000in}}%
\pgfusepath{clip}%
\pgfsetroundcap%
\pgfsetroundjoin%
\pgfsetlinewidth{1.505625pt}%
\definecolor{currentstroke}{rgb}{0.129412,0.588235,0.952941}%
\pgfsetstrokecolor{currentstroke}%
\pgfsetdash{}{0pt}%
\pgfpathmoveto{\pgfqpoint{1.131818in}{1.113976in}}%
\pgfpathlineto{\pgfqpoint{1.228409in}{1.247058in}}%
\pgfpathlineto{\pgfqpoint{1.325000in}{1.362954in}}%
\pgfpathlineto{\pgfqpoint{1.421591in}{1.440228in}}%
\pgfpathlineto{\pgfqpoint{1.518182in}{1.498955in}}%
\pgfpathlineto{\pgfqpoint{1.614773in}{1.552737in}}%
\pgfpathlineto{\pgfqpoint{1.711364in}{1.597727in}}%
\pgfpathlineto{\pgfqpoint{1.807955in}{1.643869in}}%
\pgfpathlineto{\pgfqpoint{1.904545in}{1.682412in}}%
\pgfpathlineto{\pgfqpoint{2.001136in}{1.718381in}}%
\pgfpathlineto{\pgfqpoint{2.097727in}{1.745876in}}%
\pgfpathlineto{\pgfqpoint{2.194318in}{1.769758in}}%
\pgfpathlineto{\pgfqpoint{2.290909in}{1.793210in}}%
\pgfpathlineto{\pgfqpoint{2.387500in}{1.820355in}}%
\pgfpathlineto{\pgfqpoint{2.484091in}{1.844155in}}%
\pgfpathlineto{\pgfqpoint{2.580682in}{1.863844in}}%
\pgfpathlineto{\pgfqpoint{2.677273in}{1.885324in}}%
\pgfpathlineto{\pgfqpoint{2.773864in}{1.904037in}}%
\pgfpathlineto{\pgfqpoint{2.870455in}{1.923921in}}%
\pgfpathlineto{\pgfqpoint{2.967045in}{1.940782in}}%
\pgfpathlineto{\pgfqpoint{3.063636in}{1.955977in}}%
\pgfpathlineto{\pgfqpoint{3.160227in}{1.970427in}}%
\pgfpathlineto{\pgfqpoint{3.256818in}{1.984976in}}%
\pgfpathlineto{\pgfqpoint{3.353409in}{2.001436in}}%
\pgfpathlineto{\pgfqpoint{3.450000in}{2.016669in}}%
\pgfpathlineto{\pgfqpoint{3.546591in}{2.028119in}}%
\pgfpathlineto{\pgfqpoint{3.643182in}{2.040596in}}%
\pgfpathlineto{\pgfqpoint{3.739773in}{2.052473in}}%
\pgfpathlineto{\pgfqpoint{3.836364in}{2.063779in}}%
\pgfpathlineto{\pgfqpoint{3.932955in}{2.075991in}}%
\pgfpathlineto{\pgfqpoint{4.029545in}{2.086999in}}%
\pgfpathlineto{\pgfqpoint{4.126136in}{2.098080in}}%
\pgfpathlineto{\pgfqpoint{4.222727in}{2.109559in}}%
\pgfpathlineto{\pgfqpoint{4.319318in}{2.119621in}}%
\pgfpathlineto{\pgfqpoint{4.415909in}{2.129798in}}%
\pgfpathlineto{\pgfqpoint{4.512500in}{2.139746in}}%
\pgfpathlineto{\pgfqpoint{4.609091in}{2.149587in}}%
\pgfpathlineto{\pgfqpoint{4.705682in}{2.158346in}}%
\pgfpathlineto{\pgfqpoint{4.802273in}{2.168809in}}%
\pgfpathlineto{\pgfqpoint{4.898864in}{2.177166in}}%
\pgfpathlineto{\pgfqpoint{4.995455in}{2.186763in}}%
\pgfpathlineto{\pgfqpoint{5.092045in}{2.195589in}}%
\pgfpathlineto{\pgfqpoint{5.188636in}{2.203944in}}%
\pgfpathlineto{\pgfqpoint{5.285227in}{2.213000in}}%
\pgfpathlineto{\pgfqpoint{5.381818in}{2.220926in}}%
\pgfpathlineto{\pgfqpoint{5.478409in}{2.228810in}}%
\pgfpathlineto{\pgfqpoint{5.575000in}{2.237400in}}%
\pgfpathlineto{\pgfqpoint{5.671591in}{2.245008in}}%
\pgfpathlineto{\pgfqpoint{5.768182in}{2.252090in}}%
\pgfusepath{stroke}%
\end{pgfscope}%
\begin{pgfscope}%
\pgfpathrectangle{\pgfqpoint{0.900000in}{0.648000in}}{\pgfqpoint{5.100000in}{2.772000in}}%
\pgfusepath{clip}%
\pgfsetbuttcap%
\pgfsetroundjoin%
\pgfsetlinewidth{1.505625pt}%
\definecolor{currentstroke}{rgb}{1.000000,0.341176,0.133333}%
\pgfsetstrokecolor{currentstroke}%
\pgfsetdash{{4.500000pt}{4.500000pt}}{0.000000pt}%
\pgfpathmoveto{\pgfqpoint{1.131818in}{1.740913in}}%
\pgfpathlineto{\pgfqpoint{1.228409in}{1.892363in}}%
\pgfpathlineto{\pgfqpoint{1.325000in}{2.022307in}}%
\pgfpathlineto{\pgfqpoint{1.421591in}{2.116817in}}%
\pgfpathlineto{\pgfqpoint{1.518182in}{2.199788in}}%
\pgfpathlineto{\pgfqpoint{1.614773in}{2.266751in}}%
\pgfpathlineto{\pgfqpoint{1.711364in}{2.327176in}}%
\pgfpathlineto{\pgfqpoint{1.807955in}{2.386640in}}%
\pgfpathlineto{\pgfqpoint{1.904545in}{2.437662in}}%
\pgfpathlineto{\pgfqpoint{2.001136in}{2.483419in}}%
\pgfpathlineto{\pgfqpoint{2.097727in}{2.525808in}}%
\pgfpathlineto{\pgfqpoint{2.194318in}{2.562522in}}%
\pgfpathlineto{\pgfqpoint{2.290909in}{2.598360in}}%
\pgfpathlineto{\pgfqpoint{2.387500in}{2.634304in}}%
\pgfpathlineto{\pgfqpoint{2.484091in}{2.668624in}}%
\pgfpathlineto{\pgfqpoint{2.580682in}{2.698731in}}%
\pgfpathlineto{\pgfqpoint{2.677273in}{2.729070in}}%
\pgfpathlineto{\pgfqpoint{2.773864in}{2.756604in}}%
\pgfpathlineto{\pgfqpoint{2.870455in}{2.782828in}}%
\pgfpathlineto{\pgfqpoint{2.967045in}{2.809098in}}%
\pgfpathlineto{\pgfqpoint{3.063636in}{2.833307in}}%
\pgfpathlineto{\pgfqpoint{3.160227in}{2.857531in}}%
\pgfpathlineto{\pgfqpoint{3.256818in}{2.881051in}}%
\pgfpathlineto{\pgfqpoint{3.353409in}{2.903224in}}%
\pgfpathlineto{\pgfqpoint{3.450000in}{2.925006in}}%
\pgfpathlineto{\pgfqpoint{3.546591in}{2.945716in}}%
\pgfpathlineto{\pgfqpoint{3.643182in}{2.965790in}}%
\pgfpathlineto{\pgfqpoint{3.739773in}{2.984272in}}%
\pgfpathlineto{\pgfqpoint{3.836364in}{3.003375in}}%
\pgfpathlineto{\pgfqpoint{3.932955in}{3.021426in}}%
\pgfpathlineto{\pgfqpoint{4.029545in}{3.038795in}}%
\pgfpathlineto{\pgfqpoint{4.126136in}{3.055553in}}%
\pgfpathlineto{\pgfqpoint{4.222727in}{3.072829in}}%
\pgfpathlineto{\pgfqpoint{4.319318in}{3.089999in}}%
\pgfpathlineto{\pgfqpoint{4.415909in}{3.105657in}}%
\pgfpathlineto{\pgfqpoint{4.512500in}{3.121540in}}%
\pgfpathlineto{\pgfqpoint{4.609091in}{3.136630in}}%
\pgfpathlineto{\pgfqpoint{4.705682in}{3.150789in}}%
\pgfpathlineto{\pgfqpoint{4.802273in}{3.165048in}}%
\pgfpathlineto{\pgfqpoint{4.898864in}{3.179743in}}%
\pgfpathlineto{\pgfqpoint{4.995455in}{3.194011in}}%
\pgfpathlineto{\pgfqpoint{5.092045in}{3.207778in}}%
\pgfpathlineto{\pgfqpoint{5.188636in}{3.220382in}}%
\pgfpathlineto{\pgfqpoint{5.285227in}{3.233189in}}%
\pgfpathlineto{\pgfqpoint{5.381818in}{3.245437in}}%
\pgfpathlineto{\pgfqpoint{5.478409in}{3.258388in}}%
\pgfpathlineto{\pgfqpoint{5.575000in}{3.271029in}}%
\pgfpathlineto{\pgfqpoint{5.671591in}{3.282334in}}%
\pgfpathlineto{\pgfqpoint{5.768182in}{3.294000in}}%
\pgfusepath{stroke}%
\end{pgfscope}%
\begin{pgfscope}%
\pgfpathrectangle{\pgfqpoint{0.900000in}{0.648000in}}{\pgfqpoint{5.100000in}{2.772000in}}%
\pgfusepath{clip}%
\pgfsetroundcap%
\pgfsetroundjoin%
\pgfsetlinewidth{1.505625pt}%
\definecolor{currentstroke}{rgb}{1.000000,0.341176,0.133333}%
\pgfsetstrokecolor{currentstroke}%
\pgfsetdash{}{0pt}%
\pgfpathmoveto{\pgfqpoint{1.131818in}{1.718559in}}%
\pgfpathlineto{\pgfqpoint{1.228409in}{1.859030in}}%
\pgfpathlineto{\pgfqpoint{1.325000in}{1.975120in}}%
\pgfpathlineto{\pgfqpoint{1.421591in}{2.054048in}}%
\pgfpathlineto{\pgfqpoint{1.518182in}{2.114708in}}%
\pgfpathlineto{\pgfqpoint{1.614773in}{2.169859in}}%
\pgfpathlineto{\pgfqpoint{1.711364in}{2.215479in}}%
\pgfpathlineto{\pgfqpoint{1.807955in}{2.260335in}}%
\pgfpathlineto{\pgfqpoint{1.904545in}{2.299002in}}%
\pgfpathlineto{\pgfqpoint{2.001136in}{2.335672in}}%
\pgfpathlineto{\pgfqpoint{2.097727in}{2.361954in}}%
\pgfpathlineto{\pgfqpoint{2.194318in}{2.387850in}}%
\pgfpathlineto{\pgfqpoint{2.290909in}{2.411896in}}%
\pgfpathlineto{\pgfqpoint{2.387500in}{2.438328in}}%
\pgfpathlineto{\pgfqpoint{2.484091in}{2.461736in}}%
\pgfpathlineto{\pgfqpoint{2.580682in}{2.480620in}}%
\pgfpathlineto{\pgfqpoint{2.677273in}{2.498249in}}%
\pgfpathlineto{\pgfqpoint{2.773864in}{2.517304in}}%
\pgfpathlineto{\pgfqpoint{2.870455in}{2.535469in}}%
\pgfpathlineto{\pgfqpoint{2.967045in}{2.552068in}}%
\pgfpathlineto{\pgfqpoint{3.063636in}{2.566064in}}%
\pgfpathlineto{\pgfqpoint{3.160227in}{2.581340in}}%
\pgfpathlineto{\pgfqpoint{3.256818in}{2.594362in}}%
\pgfpathlineto{\pgfqpoint{3.353409in}{2.609288in}}%
\pgfpathlineto{\pgfqpoint{3.450000in}{2.626418in}}%
\pgfpathlineto{\pgfqpoint{3.546591in}{2.635153in}}%
\pgfpathlineto{\pgfqpoint{3.643182in}{2.646697in}}%
\pgfpathlineto{\pgfqpoint{3.739773in}{2.660795in}}%
\pgfpathlineto{\pgfqpoint{3.836364in}{2.670505in}}%
\pgfpathlineto{\pgfqpoint{3.932955in}{2.681660in}}%
\pgfpathlineto{\pgfqpoint{4.029545in}{2.691494in}}%
\pgfpathlineto{\pgfqpoint{4.126136in}{2.701681in}}%
\pgfpathlineto{\pgfqpoint{4.222727in}{2.712179in}}%
\pgfpathlineto{\pgfqpoint{4.319318in}{2.723157in}}%
\pgfpathlineto{\pgfqpoint{4.415909in}{2.733932in}}%
\pgfpathlineto{\pgfqpoint{4.512500in}{2.743149in}}%
\pgfpathlineto{\pgfqpoint{4.609091in}{2.753965in}}%
\pgfpathlineto{\pgfqpoint{4.705682in}{2.759474in}}%
\pgfpathlineto{\pgfqpoint{4.802273in}{2.769556in}}%
\pgfpathlineto{\pgfqpoint{4.898864in}{2.780763in}}%
\pgfpathlineto{\pgfqpoint{4.995455in}{2.789917in}}%
\pgfpathlineto{\pgfqpoint{5.092045in}{2.795202in}}%
\pgfpathlineto{\pgfqpoint{5.188636in}{2.806777in}}%
\pgfpathlineto{\pgfqpoint{5.285227in}{2.816298in}}%
\pgfpathlineto{\pgfqpoint{5.381818in}{2.820977in}}%
\pgfpathlineto{\pgfqpoint{5.478409in}{2.828746in}}%
\pgfpathlineto{\pgfqpoint{5.575000in}{2.834096in}}%
\pgfpathlineto{\pgfqpoint{5.671591in}{2.840069in}}%
\pgfpathlineto{\pgfqpoint{5.768182in}{2.846782in}}%
\pgfusepath{stroke}%
\end{pgfscope}%
\begin{pgfscope}%
\pgfpathrectangle{\pgfqpoint{0.900000in}{0.648000in}}{\pgfqpoint{5.100000in}{2.772000in}}%
\pgfusepath{clip}%
\pgfsetbuttcap%
\pgfsetroundjoin%
\pgfsetlinewidth{1.505625pt}%
\definecolor{currentstroke}{rgb}{0.486275,0.301961,1.000000}%
\pgfsetstrokecolor{currentstroke}%
\pgfsetdash{{4.500000pt}{4.500000pt}}{0.000000pt}%
\pgfpathmoveto{\pgfqpoint{1.131818in}{1.723096in}}%
\pgfpathlineto{\pgfqpoint{1.228409in}{1.874181in}}%
\pgfpathlineto{\pgfqpoint{1.325000in}{2.003329in}}%
\pgfpathlineto{\pgfqpoint{1.421591in}{2.095251in}}%
\pgfpathlineto{\pgfqpoint{1.518182in}{2.171452in}}%
\pgfpathlineto{\pgfqpoint{1.614773in}{2.238825in}}%
\pgfpathlineto{\pgfqpoint{1.711364in}{2.299167in}}%
\pgfpathlineto{\pgfqpoint{1.807955in}{2.354472in}}%
\pgfpathlineto{\pgfqpoint{1.904545in}{2.402870in}}%
\pgfpathlineto{\pgfqpoint{2.001136in}{2.447670in}}%
\pgfpathlineto{\pgfqpoint{2.097727in}{2.489233in}}%
\pgfpathlineto{\pgfqpoint{2.194318in}{2.526819in}}%
\pgfpathlineto{\pgfqpoint{2.290909in}{2.563121in}}%
\pgfpathlineto{\pgfqpoint{2.387500in}{2.599021in}}%
\pgfpathlineto{\pgfqpoint{2.484091in}{2.632356in}}%
\pgfpathlineto{\pgfqpoint{2.580682in}{2.661809in}}%
\pgfpathlineto{\pgfqpoint{2.677273in}{2.691446in}}%
\pgfpathlineto{\pgfqpoint{2.773864in}{2.720419in}}%
\pgfpathlineto{\pgfqpoint{2.870455in}{2.746928in}}%
\pgfpathlineto{\pgfqpoint{2.967045in}{2.773395in}}%
\pgfpathlineto{\pgfqpoint{3.063636in}{2.798067in}}%
\pgfpathlineto{\pgfqpoint{3.160227in}{2.823048in}}%
\pgfpathlineto{\pgfqpoint{3.256818in}{2.847103in}}%
\pgfpathlineto{\pgfqpoint{3.353409in}{2.868981in}}%
\pgfpathlineto{\pgfqpoint{3.450000in}{2.889981in}}%
\pgfpathlineto{\pgfqpoint{3.546591in}{2.910742in}}%
\pgfpathlineto{\pgfqpoint{3.643182in}{2.933803in}}%
\pgfpathlineto{\pgfqpoint{3.739773in}{2.953367in}}%
\pgfpathlineto{\pgfqpoint{3.836364in}{2.974198in}}%
\pgfpathlineto{\pgfqpoint{3.932955in}{2.992764in}}%
\pgfpathlineto{\pgfqpoint{4.029545in}{3.009450in}}%
\pgfpathlineto{\pgfqpoint{4.126136in}{3.027292in}}%
\pgfpathlineto{\pgfqpoint{4.222727in}{3.044831in}}%
\pgfpathlineto{\pgfqpoint{4.319318in}{3.061614in}}%
\pgfpathlineto{\pgfqpoint{4.415909in}{3.076890in}}%
\pgfpathlineto{\pgfqpoint{4.512500in}{3.091209in}}%
\pgfpathlineto{\pgfqpoint{4.609091in}{3.105516in}}%
\pgfpathlineto{\pgfqpoint{4.705682in}{3.120666in}}%
\pgfpathlineto{\pgfqpoint{4.802273in}{3.135880in}}%
\pgfpathlineto{\pgfqpoint{4.898864in}{3.154876in}}%
\pgfpathlineto{\pgfqpoint{4.995455in}{3.167143in}}%
\pgfpathlineto{\pgfqpoint{5.092045in}{3.179806in}}%
\pgfpathlineto{\pgfqpoint{5.188636in}{3.193150in}}%
\pgfpathlineto{\pgfqpoint{5.285227in}{3.206658in}}%
\pgfpathlineto{\pgfqpoint{5.381818in}{3.220517in}}%
\pgfpathlineto{\pgfqpoint{5.478409in}{3.232550in}}%
\pgfpathlineto{\pgfqpoint{5.575000in}{3.245736in}}%
\pgfpathlineto{\pgfqpoint{5.671591in}{3.258477in}}%
\pgfpathlineto{\pgfqpoint{5.768182in}{3.270477in}}%
\pgfusepath{stroke}%
\end{pgfscope}%
\begin{pgfscope}%
\pgfpathrectangle{\pgfqpoint{0.900000in}{0.648000in}}{\pgfqpoint{5.100000in}{2.772000in}}%
\pgfusepath{clip}%
\pgfsetroundcap%
\pgfsetroundjoin%
\pgfsetlinewidth{1.505625pt}%
\definecolor{currentstroke}{rgb}{0.486275,0.301961,1.000000}%
\pgfsetstrokecolor{currentstroke}%
\pgfsetdash{}{0pt}%
\pgfpathmoveto{\pgfqpoint{1.131818in}{1.699779in}}%
\pgfpathlineto{\pgfqpoint{1.228409in}{1.837279in}}%
\pgfpathlineto{\pgfqpoint{1.325000in}{1.954634in}}%
\pgfpathlineto{\pgfqpoint{1.421591in}{2.030739in}}%
\pgfpathlineto{\pgfqpoint{1.518182in}{2.089552in}}%
\pgfpathlineto{\pgfqpoint{1.614773in}{2.140772in}}%
\pgfpathlineto{\pgfqpoint{1.711364in}{2.185904in}}%
\pgfpathlineto{\pgfqpoint{1.807955in}{2.228929in}}%
\pgfpathlineto{\pgfqpoint{1.904545in}{2.262613in}}%
\pgfpathlineto{\pgfqpoint{2.001136in}{2.296681in}}%
\pgfpathlineto{\pgfqpoint{2.097727in}{2.325039in}}%
\pgfpathlineto{\pgfqpoint{2.194318in}{2.349793in}}%
\pgfpathlineto{\pgfqpoint{2.290909in}{2.373536in}}%
\pgfpathlineto{\pgfqpoint{2.387500in}{2.400029in}}%
\pgfpathlineto{\pgfqpoint{2.484091in}{2.423538in}}%
\pgfpathlineto{\pgfqpoint{2.580682in}{2.443281in}}%
\pgfpathlineto{\pgfqpoint{2.677273in}{2.462200in}}%
\pgfpathlineto{\pgfqpoint{2.773864in}{2.479172in}}%
\pgfpathlineto{\pgfqpoint{2.870455in}{2.496174in}}%
\pgfpathlineto{\pgfqpoint{2.967045in}{2.513450in}}%
\pgfpathlineto{\pgfqpoint{3.063636in}{2.527888in}}%
\pgfpathlineto{\pgfqpoint{3.160227in}{2.544318in}}%
\pgfpathlineto{\pgfqpoint{3.256818in}{2.556523in}}%
\pgfpathlineto{\pgfqpoint{3.353409in}{2.571023in}}%
\pgfpathlineto{\pgfqpoint{3.450000in}{2.586179in}}%
\pgfpathlineto{\pgfqpoint{3.546591in}{2.597948in}}%
\pgfpathlineto{\pgfqpoint{3.643182in}{2.610143in}}%
\pgfpathlineto{\pgfqpoint{3.739773in}{2.620745in}}%
\pgfpathlineto{\pgfqpoint{3.836364in}{2.631461in}}%
\pgfpathlineto{\pgfqpoint{3.932955in}{2.644786in}}%
\pgfpathlineto{\pgfqpoint{4.029545in}{2.653877in}}%
\pgfpathlineto{\pgfqpoint{4.126136in}{2.665853in}}%
\pgfpathlineto{\pgfqpoint{4.222727in}{2.678157in}}%
\pgfpathlineto{\pgfqpoint{4.319318in}{2.687461in}}%
\pgfpathlineto{\pgfqpoint{4.415909in}{2.695202in}}%
\pgfpathlineto{\pgfqpoint{4.512500in}{2.705607in}}%
\pgfpathlineto{\pgfqpoint{4.609091in}{2.713746in}}%
\pgfpathlineto{\pgfqpoint{4.705682in}{2.722365in}}%
\pgfpathlineto{\pgfqpoint{4.802273in}{2.732279in}}%
\pgfpathlineto{\pgfqpoint{4.898864in}{2.743361in}}%
\pgfpathlineto{\pgfqpoint{4.995455in}{2.749581in}}%
\pgfpathlineto{\pgfqpoint{5.092045in}{2.762693in}}%
\pgfpathlineto{\pgfqpoint{5.188636in}{2.765715in}}%
\pgfpathlineto{\pgfqpoint{5.285227in}{2.774464in}}%
\pgfpathlineto{\pgfqpoint{5.381818in}{2.781707in}}%
\pgfpathlineto{\pgfqpoint{5.478409in}{2.789615in}}%
\pgfpathlineto{\pgfqpoint{5.575000in}{2.796203in}}%
\pgfpathlineto{\pgfqpoint{5.671591in}{2.805888in}}%
\pgfpathlineto{\pgfqpoint{5.768182in}{2.813045in}}%
\pgfusepath{stroke}%
\end{pgfscope}%
\begin{pgfscope}%
\pgfsetrectcap%
\pgfsetmiterjoin%
\pgfsetlinewidth{1.254687pt}%
\definecolor{currentstroke}{rgb}{0.800000,0.800000,0.800000}%
\pgfsetstrokecolor{currentstroke}%
\pgfsetdash{}{0pt}%
\pgfpathmoveto{\pgfqpoint{0.900000in}{0.648000in}}%
\pgfpathlineto{\pgfqpoint{0.900000in}{3.420000in}}%
\pgfusepath{stroke}%
\end{pgfscope}%
\begin{pgfscope}%
\pgfsetrectcap%
\pgfsetmiterjoin%
\pgfsetlinewidth{1.254687pt}%
\definecolor{currentstroke}{rgb}{0.800000,0.800000,0.800000}%
\pgfsetstrokecolor{currentstroke}%
\pgfsetdash{}{0pt}%
\pgfpathmoveto{\pgfqpoint{6.000000in}{0.648000in}}%
\pgfpathlineto{\pgfqpoint{6.000000in}{3.420000in}}%
\pgfusepath{stroke}%
\end{pgfscope}%
\begin{pgfscope}%
\pgfsetrectcap%
\pgfsetmiterjoin%
\pgfsetlinewidth{1.254687pt}%
\definecolor{currentstroke}{rgb}{0.800000,0.800000,0.800000}%
\pgfsetstrokecolor{currentstroke}%
\pgfsetdash{}{0pt}%
\pgfpathmoveto{\pgfqpoint{0.900000in}{0.648000in}}%
\pgfpathlineto{\pgfqpoint{6.000000in}{0.648000in}}%
\pgfusepath{stroke}%
\end{pgfscope}%
\begin{pgfscope}%
\pgfsetrectcap%
\pgfsetmiterjoin%
\pgfsetlinewidth{1.254687pt}%
\definecolor{currentstroke}{rgb}{0.800000,0.800000,0.800000}%
\pgfsetstrokecolor{currentstroke}%
\pgfsetdash{}{0pt}%
\pgfpathmoveto{\pgfqpoint{0.900000in}{3.420000in}}%
\pgfpathlineto{\pgfqpoint{6.000000in}{3.420000in}}%
\pgfusepath{stroke}%
\end{pgfscope}%
\begin{pgfscope}%
\pgfsetbuttcap%
\pgfsetmiterjoin%
\definecolor{currentfill}{rgb}{1.000000,1.000000,1.000000}%
\pgfsetfillcolor{currentfill}%
\pgfsetfillopacity{0.800000}%
\pgfsetlinewidth{1.003750pt}%
\definecolor{currentstroke}{rgb}{0.800000,0.800000,0.800000}%
\pgfsetstrokecolor{currentstroke}%
\pgfsetstrokeopacity{0.800000}%
\pgfsetdash{}{0pt}%
\pgfpathmoveto{\pgfqpoint{4.963861in}{0.724389in}}%
\pgfpathlineto{\pgfqpoint{5.893056in}{0.724389in}}%
\pgfpathquadraticcurveto{\pgfqpoint{5.923611in}{0.724389in}}{\pgfqpoint{5.923611in}{0.754944in}}%
\pgfpathlineto{\pgfqpoint{5.923611in}{1.806208in}}%
\pgfpathquadraticcurveto{\pgfqpoint{5.923611in}{1.836763in}}{\pgfqpoint{5.893056in}{1.836763in}}%
\pgfpathlineto{\pgfqpoint{4.963861in}{1.836763in}}%
\pgfpathquadraticcurveto{\pgfqpoint{4.933306in}{1.836763in}}{\pgfqpoint{4.933306in}{1.806208in}}%
\pgfpathlineto{\pgfqpoint{4.933306in}{0.754944in}}%
\pgfpathquadraticcurveto{\pgfqpoint{4.933306in}{0.724389in}}{\pgfqpoint{4.963861in}{0.724389in}}%
\pgfpathclose%
\pgfusepath{stroke,fill}%
\end{pgfscope}%
\begin{pgfscope}%
\pgfsetroundcap%
\pgfsetroundjoin%
\pgfsetlinewidth{1.505625pt}%
\definecolor{currentstroke}{rgb}{0.545098,0.764706,0.290196}%
\pgfsetstrokecolor{currentstroke}%
\pgfsetdash{}{0pt}%
\pgfpathmoveto{\pgfqpoint{4.994417in}{1.722180in}}%
\pgfpathlineto{\pgfqpoint{5.299972in}{1.722180in}}%
\pgfusepath{stroke}%
\end{pgfscope}%
\begin{pgfscope}%
\definecolor{textcolor}{rgb}{0.150000,0.150000,0.150000}%
\pgfsetstrokecolor{textcolor}%
\pgfsetfillcolor{textcolor}%
\pgftext[x=5.422194in,y=1.668708in,left,base]{\color{textcolor}\sffamily\fontsize{11.000000}{13.200000}\selectfont 128 bit}%
\end{pgfscope}%
\begin{pgfscope}%
\pgfsetroundcap%
\pgfsetroundjoin%
\pgfsetlinewidth{1.505625pt}%
\definecolor{currentstroke}{rgb}{1.000000,0.756863,0.027451}%
\pgfsetstrokecolor{currentstroke}%
\pgfsetdash{}{0pt}%
\pgfpathmoveto{\pgfqpoint{4.994417in}{1.509208in}}%
\pgfpathlineto{\pgfqpoint{5.299972in}{1.509208in}}%
\pgfusepath{stroke}%
\end{pgfscope}%
\begin{pgfscope}%
\definecolor{textcolor}{rgb}{0.150000,0.150000,0.150000}%
\pgfsetstrokecolor{textcolor}%
\pgfsetfillcolor{textcolor}%
\pgftext[x=5.422194in,y=1.455736in,left,base]{\color{textcolor}\sffamily\fontsize{11.000000}{13.200000}\selectfont 192 bit}%
\end{pgfscope}%
\begin{pgfscope}%
\pgfsetroundcap%
\pgfsetroundjoin%
\pgfsetlinewidth{1.505625pt}%
\definecolor{currentstroke}{rgb}{0.129412,0.588235,0.952941}%
\pgfsetstrokecolor{currentstroke}%
\pgfsetdash{}{0pt}%
\pgfpathmoveto{\pgfqpoint{4.994417in}{1.296236in}}%
\pgfpathlineto{\pgfqpoint{5.299972in}{1.296236in}}%
\pgfusepath{stroke}%
\end{pgfscope}%
\begin{pgfscope}%
\definecolor{textcolor}{rgb}{0.150000,0.150000,0.150000}%
\pgfsetstrokecolor{textcolor}%
\pgfsetfillcolor{textcolor}%
\pgftext[x=5.422194in,y=1.242763in,left,base]{\color{textcolor}\sffamily\fontsize{11.000000}{13.200000}\selectfont 256 bit}%
\end{pgfscope}%
\begin{pgfscope}%
\pgfsetroundcap%
\pgfsetroundjoin%
\pgfsetlinewidth{1.505625pt}%
\definecolor{currentstroke}{rgb}{1.000000,0.341176,0.133333}%
\pgfsetstrokecolor{currentstroke}%
\pgfsetdash{}{0pt}%
\pgfpathmoveto{\pgfqpoint{4.994417in}{1.083264in}}%
\pgfpathlineto{\pgfqpoint{5.299972in}{1.083264in}}%
\pgfusepath{stroke}%
\end{pgfscope}%
\begin{pgfscope}%
\definecolor{textcolor}{rgb}{0.150000,0.150000,0.150000}%
\pgfsetstrokecolor{textcolor}%
\pgfsetfillcolor{textcolor}%
\pgftext[x=5.422194in,y=1.029791in,left,base]{\color{textcolor}\sffamily\fontsize{11.000000}{13.200000}\selectfont gmp}%
\end{pgfscope}%
\begin{pgfscope}%
\pgfsetroundcap%
\pgfsetroundjoin%
\pgfsetlinewidth{1.505625pt}%
\definecolor{currentstroke}{rgb}{0.486275,0.301961,1.000000}%
\pgfsetstrokecolor{currentstroke}%
\pgfsetdash{}{0pt}%
\pgfpathmoveto{\pgfqpoint{4.994417in}{0.868611in}}%
\pgfpathlineto{\pgfqpoint{5.299972in}{0.868611in}}%
\pgfusepath{stroke}%
\end{pgfscope}%
\begin{pgfscope}%
\definecolor{textcolor}{rgb}{0.150000,0.150000,0.150000}%
\pgfsetstrokecolor{textcolor}%
\pgfsetfillcolor{textcolor}%
\pgftext[x=5.422194in,y=0.815139in,left,base]{\color{textcolor}\sffamily\fontsize{11.000000}{13.200000}\selectfont CGAL}%
\end{pgfscope}%
\begin{pgfscope}%
\pgfsetbuttcap%
\pgfsetmiterjoin%
\definecolor{currentfill}{rgb}{1.000000,1.000000,1.000000}%
\pgfsetfillcolor{currentfill}%
\pgfsetfillopacity{0.800000}%
\pgfsetlinewidth{1.003750pt}%
\definecolor{currentstroke}{rgb}{0.800000,0.800000,0.800000}%
\pgfsetstrokecolor{currentstroke}%
\pgfsetstrokeopacity{0.800000}%
\pgfsetdash{}{0pt}%
\pgfpathmoveto{\pgfqpoint{1.006944in}{2.870153in}}%
\pgfpathlineto{\pgfqpoint{2.294250in}{2.870153in}}%
\pgfpathquadraticcurveto{\pgfqpoint{2.324806in}{2.870153in}}{\pgfqpoint{2.324806in}{2.900709in}}%
\pgfpathlineto{\pgfqpoint{2.324806in}{3.313056in}}%
\pgfpathquadraticcurveto{\pgfqpoint{2.324806in}{3.343611in}}{\pgfqpoint{2.294250in}{3.343611in}}%
\pgfpathlineto{\pgfqpoint{1.006944in}{3.343611in}}%
\pgfpathquadraticcurveto{\pgfqpoint{0.976389in}{3.343611in}}{\pgfqpoint{0.976389in}{3.313056in}}%
\pgfpathlineto{\pgfqpoint{0.976389in}{2.900709in}}%
\pgfpathquadraticcurveto{\pgfqpoint{0.976389in}{2.870153in}}{\pgfqpoint{1.006944in}{2.870153in}}%
\pgfpathclose%
\pgfusepath{stroke,fill}%
\end{pgfscope}%
\begin{pgfscope}%
\pgfsetbuttcap%
\pgfsetroundjoin%
\pgfsetlinewidth{1.505625pt}%
\definecolor{currentstroke}{rgb}{0.100000,0.100000,0.100000}%
\pgfsetstrokecolor{currentstroke}%
\pgfsetdash{{4.500000pt}{4.500000pt}}{0.000000pt}%
\pgfpathmoveto{\pgfqpoint{1.037500in}{3.229028in}}%
\pgfpathlineto{\pgfqpoint{1.343056in}{3.229028in}}%
\pgfusepath{stroke}%
\end{pgfscope}%
\begin{pgfscope}%
\definecolor{textcolor}{rgb}{0.150000,0.150000,0.150000}%
\pgfsetstrokecolor{textcolor}%
\pgfsetfillcolor{textcolor}%
\pgftext[x=1.465278in,y=3.175556in,left,base]{\color{textcolor}\sffamily\fontsize{11.000000}{13.200000}\selectfont naive}%
\end{pgfscope}%
\begin{pgfscope}%
\pgfsetroundcap%
\pgfsetroundjoin%
\pgfsetlinewidth{1.505625pt}%
\definecolor{currentstroke}{rgb}{0.100000,0.100000,0.100000}%
\pgfsetstrokecolor{currentstroke}%
\pgfsetdash{}{0pt}%
\pgfpathmoveto{\pgfqpoint{1.037500in}{3.016056in}}%
\pgfpathlineto{\pgfqpoint{1.343056in}{3.016056in}}%
\pgfusepath{stroke}%
\end{pgfscope}%
\begin{pgfscope}%
\definecolor{textcolor}{rgb}{0.150000,0.150000,0.150000}%
\pgfsetstrokecolor{textcolor}%
\pgfsetfillcolor{textcolor}%
\pgftext[x=1.465278in,y=2.962583in,left,base]{\color{textcolor}\sffamily\fontsize{11.000000}{13.200000}\selectfont edge descent}%
\end{pgfscope}%
\end{pgfpicture}%
\makeatother%
\endgroup%

%% file: figures/convex-timing-depth.pgf
%% Creator: Matplotlib, PGF backend
%%
%% To include the figure in your LaTeX document, write
%%   \input{<filename>.pgf}
%%
%% Make sure the required packages are loaded in your preamble
%%   \usepackage{pgf}
%%
%% and, on pdftex
%%   \usepackage[utf8]{inputenc}\DeclareUnicodeCharacter{2212}{-}
%%
%% or, on luatex and xetex
%%   \usepackage{unicode-math}
%%
%% Figures using additional raster images can only be included by \input if
%% they are in the same directory as the main LaTeX file. For loading figures
%% from other directories you can use the `import` package
%% and then include the figures with
%%   \import{<path to file>}{<filename>.pgf}
%%
%% Matplotlib used the following preamble
%%   \usepackage{fontspec}
%%
\begingroup%
\makeatletter%
\begin{pgfpicture}%
\pgfpathrectangle{\pgfpointorigin}{\pgfqpoint{6.000000in}{3.600000in}}%
\pgfusepath{use as bounding box, clip}%
\begin{pgfscope}%
\pgfsetbuttcap%
\pgfsetmiterjoin%
\definecolor{currentfill}{rgb}{1.000000,1.000000,1.000000}%
\pgfsetfillcolor{currentfill}%
\pgfsetlinewidth{0.000000pt}%
\definecolor{currentstroke}{rgb}{1.000000,1.000000,1.000000}%
\pgfsetstrokecolor{currentstroke}%
\pgfsetdash{}{0pt}%
\pgfpathmoveto{\pgfqpoint{0.000000in}{0.000000in}}%
\pgfpathlineto{\pgfqpoint{6.000000in}{0.000000in}}%
\pgfpathlineto{\pgfqpoint{6.000000in}{3.600000in}}%
\pgfpathlineto{\pgfqpoint{0.000000in}{3.600000in}}%
\pgfpathclose%
\pgfusepath{fill}%
\end{pgfscope}%
\begin{pgfscope}%
\pgfsetbuttcap%
\pgfsetmiterjoin%
\definecolor{currentfill}{rgb}{1.000000,1.000000,1.000000}%
\pgfsetfillcolor{currentfill}%
\pgfsetlinewidth{0.000000pt}%
\definecolor{currentstroke}{rgb}{0.000000,0.000000,0.000000}%
\pgfsetstrokecolor{currentstroke}%
\pgfsetstrokeopacity{0.000000}%
\pgfsetdash{}{0pt}%
\pgfpathmoveto{\pgfqpoint{0.900000in}{0.648000in}}%
\pgfpathlineto{\pgfqpoint{6.000000in}{0.648000in}}%
\pgfpathlineto{\pgfqpoint{6.000000in}{3.420000in}}%
\pgfpathlineto{\pgfqpoint{0.900000in}{3.420000in}}%
\pgfpathclose%
\pgfusepath{fill}%
\end{pgfscope}%
\begin{pgfscope}%
\pgfpathrectangle{\pgfqpoint{0.900000in}{0.648000in}}{\pgfqpoint{5.100000in}{2.772000in}}%
\pgfusepath{clip}%
\pgfsetroundcap%
\pgfsetroundjoin%
\pgfsetlinewidth{1.003750pt}%
\definecolor{currentstroke}{rgb}{0.800000,0.800000,0.800000}%
\pgfsetstrokecolor{currentstroke}%
\pgfsetdash{}{0pt}%
\pgfpathmoveto{\pgfqpoint{0.938636in}{0.648000in}}%
\pgfpathlineto{\pgfqpoint{0.938636in}{3.420000in}}%
\pgfusepath{stroke}%
\end{pgfscope}%
\begin{pgfscope}%
\definecolor{textcolor}{rgb}{0.150000,0.150000,0.150000}%
\pgfsetstrokecolor{textcolor}%
\pgfsetfillcolor{textcolor}%
\pgftext[x=0.938636in,y=0.516056in,,top]{\color{textcolor}\sffamily\fontsize{11.000000}{13.200000}\selectfont 0}%
\end{pgfscope}%
\begin{pgfscope}%
\pgfpathrectangle{\pgfqpoint{0.900000in}{0.648000in}}{\pgfqpoint{5.100000in}{2.772000in}}%
\pgfusepath{clip}%
\pgfsetroundcap%
\pgfsetroundjoin%
\pgfsetlinewidth{1.003750pt}%
\definecolor{currentstroke}{rgb}{0.800000,0.800000,0.800000}%
\pgfsetstrokecolor{currentstroke}%
\pgfsetdash{}{0pt}%
\pgfpathmoveto{\pgfqpoint{1.904545in}{0.648000in}}%
\pgfpathlineto{\pgfqpoint{1.904545in}{3.420000in}}%
\pgfusepath{stroke}%
\end{pgfscope}%
\begin{pgfscope}%
\definecolor{textcolor}{rgb}{0.150000,0.150000,0.150000}%
\pgfsetstrokecolor{textcolor}%
\pgfsetfillcolor{textcolor}%
\pgftext[x=1.904545in,y=0.516056in,,top]{\color{textcolor}\sffamily\fontsize{11.000000}{13.200000}\selectfont 50}%
\end{pgfscope}%
\begin{pgfscope}%
\pgfpathrectangle{\pgfqpoint{0.900000in}{0.648000in}}{\pgfqpoint{5.100000in}{2.772000in}}%
\pgfusepath{clip}%
\pgfsetroundcap%
\pgfsetroundjoin%
\pgfsetlinewidth{1.003750pt}%
\definecolor{currentstroke}{rgb}{0.800000,0.800000,0.800000}%
\pgfsetstrokecolor{currentstroke}%
\pgfsetdash{}{0pt}%
\pgfpathmoveto{\pgfqpoint{2.870455in}{0.648000in}}%
\pgfpathlineto{\pgfqpoint{2.870455in}{3.420000in}}%
\pgfusepath{stroke}%
\end{pgfscope}%
\begin{pgfscope}%
\definecolor{textcolor}{rgb}{0.150000,0.150000,0.150000}%
\pgfsetstrokecolor{textcolor}%
\pgfsetfillcolor{textcolor}%
\pgftext[x=2.870455in,y=0.516056in,,top]{\color{textcolor}\sffamily\fontsize{11.000000}{13.200000}\selectfont 100}%
\end{pgfscope}%
\begin{pgfscope}%
\pgfpathrectangle{\pgfqpoint{0.900000in}{0.648000in}}{\pgfqpoint{5.100000in}{2.772000in}}%
\pgfusepath{clip}%
\pgfsetroundcap%
\pgfsetroundjoin%
\pgfsetlinewidth{1.003750pt}%
\definecolor{currentstroke}{rgb}{0.800000,0.800000,0.800000}%
\pgfsetstrokecolor{currentstroke}%
\pgfsetdash{}{0pt}%
\pgfpathmoveto{\pgfqpoint{3.836364in}{0.648000in}}%
\pgfpathlineto{\pgfqpoint{3.836364in}{3.420000in}}%
\pgfusepath{stroke}%
\end{pgfscope}%
\begin{pgfscope}%
\definecolor{textcolor}{rgb}{0.150000,0.150000,0.150000}%
\pgfsetstrokecolor{textcolor}%
\pgfsetfillcolor{textcolor}%
\pgftext[x=3.836364in,y=0.516056in,,top]{\color{textcolor}\sffamily\fontsize{11.000000}{13.200000}\selectfont 150}%
\end{pgfscope}%
\begin{pgfscope}%
\pgfpathrectangle{\pgfqpoint{0.900000in}{0.648000in}}{\pgfqpoint{5.100000in}{2.772000in}}%
\pgfusepath{clip}%
\pgfsetroundcap%
\pgfsetroundjoin%
\pgfsetlinewidth{1.003750pt}%
\definecolor{currentstroke}{rgb}{0.800000,0.800000,0.800000}%
\pgfsetstrokecolor{currentstroke}%
\pgfsetdash{}{0pt}%
\pgfpathmoveto{\pgfqpoint{4.802273in}{0.648000in}}%
\pgfpathlineto{\pgfqpoint{4.802273in}{3.420000in}}%
\pgfusepath{stroke}%
\end{pgfscope}%
\begin{pgfscope}%
\definecolor{textcolor}{rgb}{0.150000,0.150000,0.150000}%
\pgfsetstrokecolor{textcolor}%
\pgfsetfillcolor{textcolor}%
\pgftext[x=4.802273in,y=0.516056in,,top]{\color{textcolor}\sffamily\fontsize{11.000000}{13.200000}\selectfont 200}%
\end{pgfscope}%
\begin{pgfscope}%
\pgfpathrectangle{\pgfqpoint{0.900000in}{0.648000in}}{\pgfqpoint{5.100000in}{2.772000in}}%
\pgfusepath{clip}%
\pgfsetroundcap%
\pgfsetroundjoin%
\pgfsetlinewidth{1.003750pt}%
\definecolor{currentstroke}{rgb}{0.800000,0.800000,0.800000}%
\pgfsetstrokecolor{currentstroke}%
\pgfsetdash{}{0pt}%
\pgfpathmoveto{\pgfqpoint{5.768182in}{0.648000in}}%
\pgfpathlineto{\pgfqpoint{5.768182in}{3.420000in}}%
\pgfusepath{stroke}%
\end{pgfscope}%
\begin{pgfscope}%
\definecolor{textcolor}{rgb}{0.150000,0.150000,0.150000}%
\pgfsetstrokecolor{textcolor}%
\pgfsetfillcolor{textcolor}%
\pgftext[x=5.768182in,y=0.516056in,,top]{\color{textcolor}\sffamily\fontsize{11.000000}{13.200000}\selectfont 250}%
\end{pgfscope}%
\begin{pgfscope}%
\definecolor{textcolor}{rgb}{0.150000,0.150000,0.150000}%
\pgfsetstrokecolor{textcolor}%
\pgfsetfillcolor{textcolor}%
\pgftext[x=3.450000in,y=0.324833in,,top]{\color{textcolor}\sffamily\fontsize{12.000000}{14.400000}\selectfont depth}%
\end{pgfscope}%
\begin{pgfscope}%
\pgfpathrectangle{\pgfqpoint{0.900000in}{0.648000in}}{\pgfqpoint{5.100000in}{2.772000in}}%
\pgfusepath{clip}%
\pgfsetroundcap%
\pgfsetroundjoin%
\pgfsetlinewidth{1.003750pt}%
\definecolor{currentstroke}{rgb}{0.800000,0.800000,0.800000}%
\pgfsetstrokecolor{currentstroke}%
\pgfsetdash{}{0pt}%
\pgfpathmoveto{\pgfqpoint{0.900000in}{0.648000in}}%
\pgfpathlineto{\pgfqpoint{6.000000in}{0.648000in}}%
\pgfusepath{stroke}%
\end{pgfscope}%
\begin{pgfscope}%
\definecolor{textcolor}{rgb}{0.150000,0.150000,0.150000}%
\pgfsetstrokecolor{textcolor}%
\pgfsetfillcolor{textcolor}%
\pgftext[x=0.458177in, y=0.594986in, left, base]{\color{textcolor}\sffamily\fontsize{11.000000}{13.200000}\selectfont \(\displaystyle {10^{-4}}\)}%
\end{pgfscope}%
\begin{pgfscope}%
\pgfpathrectangle{\pgfqpoint{0.900000in}{0.648000in}}{\pgfqpoint{5.100000in}{2.772000in}}%
\pgfusepath{clip}%
\pgfsetroundcap%
\pgfsetroundjoin%
\pgfsetlinewidth{1.003750pt}%
\definecolor{currentstroke}{rgb}{0.800000,0.800000,0.800000}%
\pgfsetstrokecolor{currentstroke}%
\pgfsetdash{}{0pt}%
\pgfpathmoveto{\pgfqpoint{0.900000in}{1.572000in}}%
\pgfpathlineto{\pgfqpoint{6.000000in}{1.572000in}}%
\pgfusepath{stroke}%
\end{pgfscope}%
\begin{pgfscope}%
\definecolor{textcolor}{rgb}{0.150000,0.150000,0.150000}%
\pgfsetstrokecolor{textcolor}%
\pgfsetfillcolor{textcolor}%
\pgftext[x=0.458177in, y=1.518986in, left, base]{\color{textcolor}\sffamily\fontsize{11.000000}{13.200000}\selectfont \(\displaystyle {10^{-3}}\)}%
\end{pgfscope}%
\begin{pgfscope}%
\pgfpathrectangle{\pgfqpoint{0.900000in}{0.648000in}}{\pgfqpoint{5.100000in}{2.772000in}}%
\pgfusepath{clip}%
\pgfsetroundcap%
\pgfsetroundjoin%
\pgfsetlinewidth{1.003750pt}%
\definecolor{currentstroke}{rgb}{0.800000,0.800000,0.800000}%
\pgfsetstrokecolor{currentstroke}%
\pgfsetdash{}{0pt}%
\pgfpathmoveto{\pgfqpoint{0.900000in}{2.496000in}}%
\pgfpathlineto{\pgfqpoint{6.000000in}{2.496000in}}%
\pgfusepath{stroke}%
\end{pgfscope}%
\begin{pgfscope}%
\definecolor{textcolor}{rgb}{0.150000,0.150000,0.150000}%
\pgfsetstrokecolor{textcolor}%
\pgfsetfillcolor{textcolor}%
\pgftext[x=0.458177in, y=2.442986in, left, base]{\color{textcolor}\sffamily\fontsize{11.000000}{13.200000}\selectfont \(\displaystyle {10^{-2}}\)}%
\end{pgfscope}%
\begin{pgfscope}%
\pgfpathrectangle{\pgfqpoint{0.900000in}{0.648000in}}{\pgfqpoint{5.100000in}{2.772000in}}%
\pgfusepath{clip}%
\pgfsetroundcap%
\pgfsetroundjoin%
\pgfsetlinewidth{1.003750pt}%
\definecolor{currentstroke}{rgb}{0.800000,0.800000,0.800000}%
\pgfsetstrokecolor{currentstroke}%
\pgfsetdash{}{0pt}%
\pgfpathmoveto{\pgfqpoint{0.900000in}{3.420000in}}%
\pgfpathlineto{\pgfqpoint{6.000000in}{3.420000in}}%
\pgfusepath{stroke}%
\end{pgfscope}%
\begin{pgfscope}%
\definecolor{textcolor}{rgb}{0.150000,0.150000,0.150000}%
\pgfsetstrokecolor{textcolor}%
\pgfsetfillcolor{textcolor}%
\pgftext[x=0.458177in, y=3.366986in, left, base]{\color{textcolor}\sffamily\fontsize{11.000000}{13.200000}\selectfont \(\displaystyle {10^{-1}}\)}%
\end{pgfscope}%
\begin{pgfscope}%
\definecolor{textcolor}{rgb}{0.150000,0.150000,0.150000}%
\pgfsetstrokecolor{textcolor}%
\pgfsetfillcolor{textcolor}%
\pgftext[x=0.402621in,y=2.034000in,,bottom,rotate=90.000000]{\color{textcolor}\sffamily\fontsize{12.000000}{14.400000}\selectfont ms/plane}%
\end{pgfscope}%
\begin{pgfscope}%
\pgfpathrectangle{\pgfqpoint{0.900000in}{0.648000in}}{\pgfqpoint{5.100000in}{2.772000in}}%
\pgfusepath{clip}%
\pgfsetbuttcap%
\pgfsetroundjoin%
\pgfsetlinewidth{1.505625pt}%
\definecolor{currentstroke}{rgb}{0.545098,0.764706,0.290196}%
\pgfsetstrokecolor{currentstroke}%
\pgfsetdash{{4.500000pt}{4.500000pt}}{0.000000pt}%
\pgfpathmoveto{\pgfqpoint{1.131818in}{1.255738in}}%
\pgfpathlineto{\pgfqpoint{1.228409in}{1.297470in}}%
\pgfpathlineto{\pgfqpoint{1.325000in}{1.341434in}}%
\pgfpathlineto{\pgfqpoint{1.421591in}{1.390192in}}%
\pgfpathlineto{\pgfqpoint{1.518182in}{1.438313in}}%
\pgfpathlineto{\pgfqpoint{1.614773in}{1.472325in}}%
\pgfpathlineto{\pgfqpoint{1.711364in}{1.509031in}}%
\pgfpathlineto{\pgfqpoint{1.807955in}{1.545449in}}%
\pgfpathlineto{\pgfqpoint{1.904545in}{1.577048in}}%
\pgfpathlineto{\pgfqpoint{2.001136in}{1.605724in}}%
\pgfpathlineto{\pgfqpoint{2.097727in}{1.626418in}}%
\pgfpathlineto{\pgfqpoint{2.194318in}{1.648978in}}%
\pgfpathlineto{\pgfqpoint{2.290909in}{1.669258in}}%
\pgfpathlineto{\pgfqpoint{2.387500in}{1.690422in}}%
\pgfpathlineto{\pgfqpoint{2.484091in}{1.708299in}}%
\pgfpathlineto{\pgfqpoint{2.580682in}{1.727002in}}%
\pgfpathlineto{\pgfqpoint{2.677273in}{1.742474in}}%
\pgfpathlineto{\pgfqpoint{2.773864in}{1.757873in}}%
\pgfpathlineto{\pgfqpoint{2.870455in}{1.776919in}}%
\pgfpathlineto{\pgfqpoint{2.967045in}{1.794728in}}%
\pgfpathlineto{\pgfqpoint{3.063636in}{1.810214in}}%
\pgfpathlineto{\pgfqpoint{3.160227in}{1.824164in}}%
\pgfpathlineto{\pgfqpoint{3.256818in}{1.837171in}}%
\pgfpathlineto{\pgfqpoint{3.353409in}{1.849538in}}%
\pgfpathlineto{\pgfqpoint{3.450000in}{1.862166in}}%
\pgfpathlineto{\pgfqpoint{3.546591in}{1.874009in}}%
\pgfpathlineto{\pgfqpoint{3.643182in}{1.885254in}}%
\pgfpathlineto{\pgfqpoint{3.739773in}{1.895993in}}%
\pgfpathlineto{\pgfqpoint{3.836364in}{1.907193in}}%
\pgfpathlineto{\pgfqpoint{3.932955in}{1.918038in}}%
\pgfpathlineto{\pgfqpoint{4.029545in}{1.929654in}}%
\pgfpathlineto{\pgfqpoint{4.126136in}{1.938840in}}%
\pgfpathlineto{\pgfqpoint{4.222727in}{1.949755in}}%
\pgfpathlineto{\pgfqpoint{4.319318in}{1.959657in}}%
\pgfpathlineto{\pgfqpoint{4.415909in}{1.969807in}}%
\pgfpathlineto{\pgfqpoint{4.512500in}{1.979367in}}%
\pgfpathlineto{\pgfqpoint{4.609091in}{1.988663in}}%
\pgfpathlineto{\pgfqpoint{4.705682in}{1.998307in}}%
\pgfpathlineto{\pgfqpoint{4.802273in}{2.007788in}}%
\pgfpathlineto{\pgfqpoint{4.898864in}{2.015736in}}%
\pgfpathlineto{\pgfqpoint{4.995455in}{2.025425in}}%
\pgfpathlineto{\pgfqpoint{5.092045in}{2.033835in}}%
\pgfpathlineto{\pgfqpoint{5.188636in}{2.042938in}}%
\pgfpathlineto{\pgfqpoint{5.285227in}{2.051394in}}%
\pgfpathlineto{\pgfqpoint{5.381818in}{2.059633in}}%
\pgfpathlineto{\pgfqpoint{5.478409in}{2.067675in}}%
\pgfpathlineto{\pgfqpoint{5.575000in}{2.075581in}}%
\pgfpathlineto{\pgfqpoint{5.671591in}{2.083329in}}%
\pgfpathlineto{\pgfqpoint{5.768182in}{2.090389in}}%
\pgfusepath{stroke}%
\end{pgfscope}%
\begin{pgfscope}%
\pgfpathrectangle{\pgfqpoint{0.900000in}{0.648000in}}{\pgfqpoint{5.100000in}{2.772000in}}%
\pgfusepath{clip}%
\pgfsetroundcap%
\pgfsetroundjoin%
\pgfsetlinewidth{1.505625pt}%
\definecolor{currentstroke}{rgb}{0.545098,0.764706,0.290196}%
\pgfsetstrokecolor{currentstroke}%
\pgfsetdash{}{0pt}%
\pgfpathmoveto{\pgfqpoint{1.131818in}{1.172476in}}%
\pgfpathlineto{\pgfqpoint{1.228409in}{1.183790in}}%
\pgfpathlineto{\pgfqpoint{1.325000in}{1.216652in}}%
\pgfpathlineto{\pgfqpoint{1.421591in}{1.223729in}}%
\pgfpathlineto{\pgfqpoint{1.518182in}{1.228631in}}%
\pgfpathlineto{\pgfqpoint{1.614773in}{1.244396in}}%
\pgfpathlineto{\pgfqpoint{1.711364in}{1.250452in}}%
\pgfpathlineto{\pgfqpoint{1.807955in}{1.263593in}}%
\pgfpathlineto{\pgfqpoint{1.904545in}{1.271928in}}%
\pgfpathlineto{\pgfqpoint{2.001136in}{1.280948in}}%
\pgfpathlineto{\pgfqpoint{2.097727in}{1.290463in}}%
\pgfpathlineto{\pgfqpoint{2.194318in}{1.293515in}}%
\pgfpathlineto{\pgfqpoint{2.290909in}{1.296135in}}%
\pgfpathlineto{\pgfqpoint{2.387500in}{1.305793in}}%
\pgfpathlineto{\pgfqpoint{2.484091in}{1.314181in}}%
\pgfpathlineto{\pgfqpoint{2.580682in}{1.316794in}}%
\pgfpathlineto{\pgfqpoint{2.677273in}{1.317987in}}%
\pgfpathlineto{\pgfqpoint{2.773864in}{1.321102in}}%
\pgfpathlineto{\pgfqpoint{2.870455in}{1.326164in}}%
\pgfpathlineto{\pgfqpoint{2.967045in}{1.328066in}}%
\pgfpathlineto{\pgfqpoint{3.063636in}{1.329513in}}%
\pgfpathlineto{\pgfqpoint{3.160227in}{1.333457in}}%
\pgfpathlineto{\pgfqpoint{3.256818in}{1.334719in}}%
\pgfpathlineto{\pgfqpoint{3.353409in}{1.340265in}}%
\pgfpathlineto{\pgfqpoint{3.450000in}{1.346953in}}%
\pgfpathlineto{\pgfqpoint{3.546591in}{1.349012in}}%
\pgfpathlineto{\pgfqpoint{3.643182in}{1.349524in}}%
\pgfpathlineto{\pgfqpoint{3.739773in}{1.355008in}}%
\pgfpathlineto{\pgfqpoint{3.836364in}{1.357387in}}%
\pgfpathlineto{\pgfqpoint{3.932955in}{1.363667in}}%
\pgfpathlineto{\pgfqpoint{4.029545in}{1.366292in}}%
\pgfpathlineto{\pgfqpoint{4.126136in}{1.370402in}}%
\pgfpathlineto{\pgfqpoint{4.222727in}{1.375998in}}%
\pgfpathlineto{\pgfqpoint{4.319318in}{1.378086in}}%
\pgfpathlineto{\pgfqpoint{4.415909in}{1.380447in}}%
\pgfpathlineto{\pgfqpoint{4.512500in}{1.382305in}}%
\pgfpathlineto{\pgfqpoint{4.609091in}{1.394399in}}%
\pgfpathlineto{\pgfqpoint{4.705682in}{1.397409in}}%
\pgfpathlineto{\pgfqpoint{4.802273in}{1.401950in}}%
\pgfpathlineto{\pgfqpoint{4.898864in}{1.404609in}}%
\pgfpathlineto{\pgfqpoint{4.995455in}{1.410505in}}%
\pgfpathlineto{\pgfqpoint{5.092045in}{1.413532in}}%
\pgfpathlineto{\pgfqpoint{5.188636in}{1.415959in}}%
\pgfpathlineto{\pgfqpoint{5.285227in}{1.418278in}}%
\pgfpathlineto{\pgfqpoint{5.381818in}{1.420443in}}%
\pgfpathlineto{\pgfqpoint{5.478409in}{1.422177in}}%
\pgfpathlineto{\pgfqpoint{5.575000in}{1.424355in}}%
\pgfpathlineto{\pgfqpoint{5.671591in}{1.427973in}}%
\pgfpathlineto{\pgfqpoint{5.768182in}{1.429528in}}%
\pgfusepath{stroke}%
\end{pgfscope}%
\begin{pgfscope}%
\pgfpathrectangle{\pgfqpoint{0.900000in}{0.648000in}}{\pgfqpoint{5.100000in}{2.772000in}}%
\pgfusepath{clip}%
\pgfsetbuttcap%
\pgfsetroundjoin%
\pgfsetlinewidth{1.505625pt}%
\definecolor{currentstroke}{rgb}{1.000000,0.756863,0.027451}%
\pgfsetstrokecolor{currentstroke}%
\pgfsetdash{{4.500000pt}{4.500000pt}}{0.000000pt}%
\pgfpathmoveto{\pgfqpoint{1.131818in}{1.538278in}}%
\pgfpathlineto{\pgfqpoint{1.228409in}{1.592673in}}%
\pgfpathlineto{\pgfqpoint{1.325000in}{1.674185in}}%
\pgfpathlineto{\pgfqpoint{1.421591in}{1.690593in}}%
\pgfpathlineto{\pgfqpoint{1.518182in}{1.737472in}}%
\pgfpathlineto{\pgfqpoint{1.614773in}{1.773733in}}%
\pgfpathlineto{\pgfqpoint{1.711364in}{1.810095in}}%
\pgfpathlineto{\pgfqpoint{1.807955in}{1.843085in}}%
\pgfpathlineto{\pgfqpoint{1.904545in}{1.873178in}}%
\pgfpathlineto{\pgfqpoint{2.001136in}{1.901883in}}%
\pgfpathlineto{\pgfqpoint{2.097727in}{1.926576in}}%
\pgfpathlineto{\pgfqpoint{2.194318in}{1.948900in}}%
\pgfpathlineto{\pgfqpoint{2.290909in}{1.971844in}}%
\pgfpathlineto{\pgfqpoint{2.387500in}{1.993860in}}%
\pgfpathlineto{\pgfqpoint{2.484091in}{2.013924in}}%
\pgfpathlineto{\pgfqpoint{2.580682in}{2.032004in}}%
\pgfpathlineto{\pgfqpoint{2.677273in}{2.050891in}}%
\pgfpathlineto{\pgfqpoint{2.773864in}{2.067974in}}%
\pgfpathlineto{\pgfqpoint{2.870455in}{2.084728in}}%
\pgfpathlineto{\pgfqpoint{2.967045in}{2.101405in}}%
\pgfpathlineto{\pgfqpoint{3.063636in}{2.116652in}}%
\pgfpathlineto{\pgfqpoint{3.160227in}{2.131786in}}%
\pgfpathlineto{\pgfqpoint{3.256818in}{2.147431in}}%
\pgfpathlineto{\pgfqpoint{3.353409in}{2.163250in}}%
\pgfpathlineto{\pgfqpoint{3.450000in}{2.177952in}}%
\pgfpathlineto{\pgfqpoint{3.546591in}{2.191018in}}%
\pgfpathlineto{\pgfqpoint{3.643182in}{2.203550in}}%
\pgfpathlineto{\pgfqpoint{3.739773in}{2.216092in}}%
\pgfpathlineto{\pgfqpoint{3.836364in}{2.227969in}}%
\pgfpathlineto{\pgfqpoint{3.932955in}{2.239838in}}%
\pgfpathlineto{\pgfqpoint{4.029545in}{2.251387in}}%
\pgfpathlineto{\pgfqpoint{4.126136in}{2.262180in}}%
\pgfpathlineto{\pgfqpoint{4.222727in}{2.273052in}}%
\pgfpathlineto{\pgfqpoint{4.319318in}{2.283563in}}%
\pgfpathlineto{\pgfqpoint{4.415909in}{2.294032in}}%
\pgfpathlineto{\pgfqpoint{4.512500in}{2.304103in}}%
\pgfpathlineto{\pgfqpoint{4.609091in}{2.316729in}}%
\pgfpathlineto{\pgfqpoint{4.705682in}{2.326342in}}%
\pgfpathlineto{\pgfqpoint{4.802273in}{2.335748in}}%
\pgfpathlineto{\pgfqpoint{4.898864in}{2.345465in}}%
\pgfpathlineto{\pgfqpoint{4.995455in}{2.354581in}}%
\pgfpathlineto{\pgfqpoint{5.092045in}{2.363620in}}%
\pgfpathlineto{\pgfqpoint{5.188636in}{2.375141in}}%
\pgfpathlineto{\pgfqpoint{5.285227in}{2.381288in}}%
\pgfpathlineto{\pgfqpoint{5.381818in}{2.390153in}}%
\pgfpathlineto{\pgfqpoint{5.478409in}{2.397757in}}%
\pgfpathlineto{\pgfqpoint{5.575000in}{2.405736in}}%
\pgfpathlineto{\pgfqpoint{5.671591in}{2.413155in}}%
\pgfpathlineto{\pgfqpoint{5.768182in}{2.417812in}}%
\pgfusepath{stroke}%
\end{pgfscope}%
\begin{pgfscope}%
\pgfpathrectangle{\pgfqpoint{0.900000in}{0.648000in}}{\pgfqpoint{5.100000in}{2.772000in}}%
\pgfusepath{clip}%
\pgfsetroundcap%
\pgfsetroundjoin%
\pgfsetlinewidth{1.505625pt}%
\definecolor{currentstroke}{rgb}{1.000000,0.756863,0.027451}%
\pgfsetstrokecolor{currentstroke}%
\pgfsetdash{}{0pt}%
\pgfpathmoveto{\pgfqpoint{1.131818in}{1.480343in}}%
\pgfpathlineto{\pgfqpoint{1.228409in}{1.497131in}}%
\pgfpathlineto{\pgfqpoint{1.325000in}{1.535277in}}%
\pgfpathlineto{\pgfqpoint{1.421591in}{1.547733in}}%
\pgfpathlineto{\pgfqpoint{1.518182in}{1.554167in}}%
\pgfpathlineto{\pgfqpoint{1.614773in}{1.563179in}}%
\pgfpathlineto{\pgfqpoint{1.711364in}{1.568645in}}%
\pgfpathlineto{\pgfqpoint{1.807955in}{1.577553in}}%
\pgfpathlineto{\pgfqpoint{1.904545in}{1.585036in}}%
\pgfpathlineto{\pgfqpoint{2.001136in}{1.594521in}}%
\pgfpathlineto{\pgfqpoint{2.097727in}{1.595099in}}%
\pgfpathlineto{\pgfqpoint{2.194318in}{1.594660in}}%
\pgfpathlineto{\pgfqpoint{2.290909in}{1.596231in}}%
\pgfpathlineto{\pgfqpoint{2.387500in}{1.605215in}}%
\pgfpathlineto{\pgfqpoint{2.484091in}{1.610124in}}%
\pgfpathlineto{\pgfqpoint{2.580682in}{1.612647in}}%
\pgfpathlineto{\pgfqpoint{2.677273in}{1.617796in}}%
\pgfpathlineto{\pgfqpoint{2.773864in}{1.622354in}}%
\pgfpathlineto{\pgfqpoint{2.870455in}{1.627708in}}%
\pgfpathlineto{\pgfqpoint{2.967045in}{1.631131in}}%
\pgfpathlineto{\pgfqpoint{3.063636in}{1.632417in}}%
\pgfpathlineto{\pgfqpoint{3.160227in}{1.633318in}}%
\pgfpathlineto{\pgfqpoint{3.256818in}{1.636433in}}%
\pgfpathlineto{\pgfqpoint{3.353409in}{1.641500in}}%
\pgfpathlineto{\pgfqpoint{3.450000in}{1.646122in}}%
\pgfpathlineto{\pgfqpoint{3.546591in}{1.646721in}}%
\pgfpathlineto{\pgfqpoint{3.643182in}{1.648472in}}%
\pgfpathlineto{\pgfqpoint{3.739773in}{1.650658in}}%
\pgfpathlineto{\pgfqpoint{3.836364in}{1.652049in}}%
\pgfpathlineto{\pgfqpoint{3.932955in}{1.654531in}}%
\pgfpathlineto{\pgfqpoint{4.029545in}{1.656946in}}%
\pgfpathlineto{\pgfqpoint{4.126136in}{1.658998in}}%
\pgfpathlineto{\pgfqpoint{4.222727in}{1.661826in}}%
\pgfpathlineto{\pgfqpoint{4.319318in}{1.664205in}}%
\pgfpathlineto{\pgfqpoint{4.415909in}{1.666663in}}%
\pgfpathlineto{\pgfqpoint{4.512500in}{1.669648in}}%
\pgfpathlineto{\pgfqpoint{4.609091in}{1.670117in}}%
\pgfpathlineto{\pgfqpoint{4.705682in}{1.670612in}}%
\pgfpathlineto{\pgfqpoint{4.802273in}{1.674188in}}%
\pgfpathlineto{\pgfqpoint{4.898864in}{1.675902in}}%
\pgfpathlineto{\pgfqpoint{4.995455in}{1.679778in}}%
\pgfpathlineto{\pgfqpoint{5.092045in}{1.681886in}}%
\pgfpathlineto{\pgfqpoint{5.188636in}{1.684677in}}%
\pgfpathlineto{\pgfqpoint{5.285227in}{1.686696in}}%
\pgfpathlineto{\pgfqpoint{5.381818in}{1.688514in}}%
\pgfpathlineto{\pgfqpoint{5.478409in}{1.690314in}}%
\pgfpathlineto{\pgfqpoint{5.575000in}{1.693287in}}%
\pgfpathlineto{\pgfqpoint{5.671591in}{1.695075in}}%
\pgfpathlineto{\pgfqpoint{5.768182in}{1.696150in}}%
\pgfusepath{stroke}%
\end{pgfscope}%
\begin{pgfscope}%
\pgfpathrectangle{\pgfqpoint{0.900000in}{0.648000in}}{\pgfqpoint{5.100000in}{2.772000in}}%
\pgfusepath{clip}%
\pgfsetbuttcap%
\pgfsetroundjoin%
\pgfsetlinewidth{1.505625pt}%
\definecolor{currentstroke}{rgb}{0.129412,0.588235,0.952941}%
\pgfsetstrokecolor{currentstroke}%
\pgfsetdash{{4.500000pt}{4.500000pt}}{0.000000pt}%
\pgfpathmoveto{\pgfqpoint{1.131818in}{1.657965in}}%
\pgfpathlineto{\pgfqpoint{1.228409in}{1.698759in}}%
\pgfpathlineto{\pgfqpoint{1.325000in}{1.744800in}}%
\pgfpathlineto{\pgfqpoint{1.421591in}{1.781302in}}%
\pgfpathlineto{\pgfqpoint{1.518182in}{1.814117in}}%
\pgfpathlineto{\pgfqpoint{1.614773in}{1.851188in}}%
\pgfpathlineto{\pgfqpoint{1.711364in}{1.883734in}}%
\pgfpathlineto{\pgfqpoint{1.807955in}{1.916196in}}%
\pgfpathlineto{\pgfqpoint{1.904545in}{1.943836in}}%
\pgfpathlineto{\pgfqpoint{2.001136in}{1.970334in}}%
\pgfpathlineto{\pgfqpoint{2.097727in}{1.992453in}}%
\pgfpathlineto{\pgfqpoint{2.194318in}{2.013723in}}%
\pgfpathlineto{\pgfqpoint{2.290909in}{2.034541in}}%
\pgfpathlineto{\pgfqpoint{2.387500in}{2.055262in}}%
\pgfpathlineto{\pgfqpoint{2.484091in}{2.075652in}}%
\pgfpathlineto{\pgfqpoint{2.580682in}{2.092378in}}%
\pgfpathlineto{\pgfqpoint{2.677273in}{2.109365in}}%
\pgfpathlineto{\pgfqpoint{2.773864in}{2.126188in}}%
\pgfpathlineto{\pgfqpoint{2.870455in}{2.142402in}}%
\pgfpathlineto{\pgfqpoint{2.967045in}{2.157112in}}%
\pgfpathlineto{\pgfqpoint{3.063636in}{2.172272in}}%
\pgfpathlineto{\pgfqpoint{3.160227in}{2.186432in}}%
\pgfpathlineto{\pgfqpoint{3.256818in}{2.201548in}}%
\pgfpathlineto{\pgfqpoint{3.353409in}{2.215354in}}%
\pgfpathlineto{\pgfqpoint{3.450000in}{2.228796in}}%
\pgfpathlineto{\pgfqpoint{3.546591in}{2.242196in}}%
\pgfpathlineto{\pgfqpoint{3.643182in}{2.254206in}}%
\pgfpathlineto{\pgfqpoint{3.739773in}{2.266289in}}%
\pgfpathlineto{\pgfqpoint{3.836364in}{2.277738in}}%
\pgfpathlineto{\pgfqpoint{3.932955in}{2.289023in}}%
\pgfpathlineto{\pgfqpoint{4.029545in}{2.299743in}}%
\pgfpathlineto{\pgfqpoint{4.126136in}{2.310776in}}%
\pgfpathlineto{\pgfqpoint{4.222727in}{2.321338in}}%
\pgfpathlineto{\pgfqpoint{4.319318in}{2.331928in}}%
\pgfpathlineto{\pgfqpoint{4.415909in}{2.342541in}}%
\pgfpathlineto{\pgfqpoint{4.512500in}{2.352412in}}%
\pgfpathlineto{\pgfqpoint{4.609091in}{2.361836in}}%
\pgfpathlineto{\pgfqpoint{4.705682in}{2.371061in}}%
\pgfpathlineto{\pgfqpoint{4.802273in}{2.381620in}}%
\pgfpathlineto{\pgfqpoint{4.898864in}{2.389196in}}%
\pgfpathlineto{\pgfqpoint{4.995455in}{2.398090in}}%
\pgfpathlineto{\pgfqpoint{5.092045in}{2.406539in}}%
\pgfpathlineto{\pgfqpoint{5.188636in}{2.417563in}}%
\pgfpathlineto{\pgfqpoint{5.285227in}{2.426840in}}%
\pgfpathlineto{\pgfqpoint{5.381818in}{2.433589in}}%
\pgfpathlineto{\pgfqpoint{5.478409in}{2.441527in}}%
\pgfpathlineto{\pgfqpoint{5.575000in}{2.449481in}}%
\pgfpathlineto{\pgfqpoint{5.671591in}{2.456844in}}%
\pgfpathlineto{\pgfqpoint{5.768182in}{2.464826in}}%
\pgfusepath{stroke}%
\end{pgfscope}%
\begin{pgfscope}%
\pgfpathrectangle{\pgfqpoint{0.900000in}{0.648000in}}{\pgfqpoint{5.100000in}{2.772000in}}%
\pgfusepath{clip}%
\pgfsetroundcap%
\pgfsetroundjoin%
\pgfsetlinewidth{1.505625pt}%
\definecolor{currentstroke}{rgb}{0.129412,0.588235,0.952941}%
\pgfsetstrokecolor{currentstroke}%
\pgfsetdash{}{0pt}%
\pgfpathmoveto{\pgfqpoint{1.131818in}{1.610867in}}%
\pgfpathlineto{\pgfqpoint{1.228409in}{1.619764in}}%
\pgfpathlineto{\pgfqpoint{1.325000in}{1.653765in}}%
\pgfpathlineto{\pgfqpoint{1.421591in}{1.663862in}}%
\pgfpathlineto{\pgfqpoint{1.518182in}{1.666427in}}%
\pgfpathlineto{\pgfqpoint{1.614773in}{1.673918in}}%
\pgfpathlineto{\pgfqpoint{1.711364in}{1.678347in}}%
\pgfpathlineto{\pgfqpoint{1.807955in}{1.690581in}}%
\pgfpathlineto{\pgfqpoint{1.904545in}{1.698002in}}%
\pgfpathlineto{\pgfqpoint{2.001136in}{1.706135in}}%
\pgfpathlineto{\pgfqpoint{2.097727in}{1.706673in}}%
\pgfpathlineto{\pgfqpoint{2.194318in}{1.705348in}}%
\pgfpathlineto{\pgfqpoint{2.290909in}{1.705849in}}%
\pgfpathlineto{\pgfqpoint{2.387500in}{1.713166in}}%
\pgfpathlineto{\pgfqpoint{2.484091in}{1.717958in}}%
\pgfpathlineto{\pgfqpoint{2.580682in}{1.719018in}}%
\pgfpathlineto{\pgfqpoint{2.677273in}{1.723780in}}%
\pgfpathlineto{\pgfqpoint{2.773864in}{1.726213in}}%
\pgfpathlineto{\pgfqpoint{2.870455in}{1.731269in}}%
\pgfpathlineto{\pgfqpoint{2.967045in}{1.733432in}}%
\pgfpathlineto{\pgfqpoint{3.063636in}{1.734358in}}%
\pgfpathlineto{\pgfqpoint{3.160227in}{1.735152in}}%
\pgfpathlineto{\pgfqpoint{3.256818in}{1.736834in}}%
\pgfpathlineto{\pgfqpoint{3.353409in}{1.741678in}}%
\pgfpathlineto{\pgfqpoint{3.450000in}{1.745582in}}%
\pgfpathlineto{\pgfqpoint{3.546591in}{1.745202in}}%
\pgfpathlineto{\pgfqpoint{3.643182in}{1.746696in}}%
\pgfpathlineto{\pgfqpoint{3.739773in}{1.747929in}}%
\pgfpathlineto{\pgfqpoint{3.836364in}{1.748904in}}%
\pgfpathlineto{\pgfqpoint{3.932955in}{1.751493in}}%
\pgfpathlineto{\pgfqpoint{4.029545in}{1.752947in}}%
\pgfpathlineto{\pgfqpoint{4.126136in}{1.754887in}}%
\pgfpathlineto{\pgfqpoint{4.222727in}{1.757709in}}%
\pgfpathlineto{\pgfqpoint{4.319318in}{1.759052in}}%
\pgfpathlineto{\pgfqpoint{4.415909in}{1.760870in}}%
\pgfpathlineto{\pgfqpoint{4.512500in}{1.762703in}}%
\pgfpathlineto{\pgfqpoint{4.609091in}{1.764691in}}%
\pgfpathlineto{\pgfqpoint{4.705682in}{1.765561in}}%
\pgfpathlineto{\pgfqpoint{4.802273in}{1.768894in}}%
\pgfpathlineto{\pgfqpoint{4.898864in}{1.769761in}}%
\pgfpathlineto{\pgfqpoint{4.995455in}{1.772466in}}%
\pgfpathlineto{\pgfqpoint{5.092045in}{1.774405in}}%
\pgfpathlineto{\pgfqpoint{5.188636in}{1.775953in}}%
\pgfpathlineto{\pgfqpoint{5.285227in}{1.778613in}}%
\pgfpathlineto{\pgfqpoint{5.381818in}{1.780013in}}%
\pgfpathlineto{\pgfqpoint{5.478409in}{1.781549in}}%
\pgfpathlineto{\pgfqpoint{5.575000in}{1.784177in}}%
\pgfpathlineto{\pgfqpoint{5.671591in}{1.785713in}}%
\pgfpathlineto{\pgfqpoint{5.768182in}{1.786738in}}%
\pgfusepath{stroke}%
\end{pgfscope}%
\begin{pgfscope}%
\pgfpathrectangle{\pgfqpoint{0.900000in}{0.648000in}}{\pgfqpoint{5.100000in}{2.772000in}}%
\pgfusepath{clip}%
\pgfsetbuttcap%
\pgfsetroundjoin%
\pgfsetlinewidth{1.505625pt}%
\definecolor{currentstroke}{rgb}{1.000000,0.341176,0.133333}%
\pgfsetstrokecolor{currentstroke}%
\pgfsetdash{{4.500000pt}{4.500000pt}}{0.000000pt}%
\pgfpathmoveto{\pgfqpoint{1.131818in}{2.419285in}}%
\pgfpathlineto{\pgfqpoint{1.228409in}{2.451867in}}%
\pgfpathlineto{\pgfqpoint{1.325000in}{2.503984in}}%
\pgfpathlineto{\pgfqpoint{1.421591in}{2.536306in}}%
\pgfpathlineto{\pgfqpoint{1.518182in}{2.570132in}}%
\pgfpathlineto{\pgfqpoint{1.614773in}{2.594621in}}%
\pgfpathlineto{\pgfqpoint{1.711364in}{2.618953in}}%
\pgfpathlineto{\pgfqpoint{1.807955in}{2.648365in}}%
\pgfpathlineto{\pgfqpoint{1.904545in}{2.671876in}}%
\pgfpathlineto{\pgfqpoint{2.001136in}{2.692632in}}%
\pgfpathlineto{\pgfqpoint{2.097727in}{2.712374in}}%
\pgfpathlineto{\pgfqpoint{2.194318in}{2.727596in}}%
\pgfpathlineto{\pgfqpoint{2.290909in}{2.744070in}}%
\pgfpathlineto{\pgfqpoint{2.387500in}{2.762733in}}%
\pgfpathlineto{\pgfqpoint{2.484091in}{2.781089in}}%
\pgfpathlineto{\pgfqpoint{2.580682in}{2.795583in}}%
\pgfpathlineto{\pgfqpoint{2.677273in}{2.811768in}}%
\pgfpathlineto{\pgfqpoint{2.773864in}{2.825576in}}%
\pgfpathlineto{\pgfqpoint{2.870455in}{2.838807in}}%
\pgfpathlineto{\pgfqpoint{2.967045in}{2.853103in}}%
\pgfpathlineto{\pgfqpoint{3.063636in}{2.865652in}}%
\pgfpathlineto{\pgfqpoint{3.160227in}{2.879050in}}%
\pgfpathlineto{\pgfqpoint{3.256818in}{2.892300in}}%
\pgfpathlineto{\pgfqpoint{3.353409in}{2.904510in}}%
\pgfpathlineto{\pgfqpoint{3.450000in}{2.916858in}}%
\pgfpathlineto{\pgfqpoint{3.546591in}{2.928418in}}%
\pgfpathlineto{\pgfqpoint{3.643182in}{2.939710in}}%
\pgfpathlineto{\pgfqpoint{3.739773in}{2.949460in}}%
\pgfpathlineto{\pgfqpoint{3.836364in}{2.960488in}}%
\pgfpathlineto{\pgfqpoint{3.932955in}{2.970607in}}%
\pgfpathlineto{\pgfqpoint{4.029545in}{2.980263in}}%
\pgfpathlineto{\pgfqpoint{4.126136in}{2.989523in}}%
\pgfpathlineto{\pgfqpoint{4.222727in}{2.999821in}}%
\pgfpathlineto{\pgfqpoint{4.319318in}{3.010329in}}%
\pgfpathlineto{\pgfqpoint{4.415909in}{3.019215in}}%
\pgfpathlineto{\pgfqpoint{4.512500in}{3.028701in}}%
\pgfpathlineto{\pgfqpoint{4.609091in}{3.037458in}}%
\pgfpathlineto{\pgfqpoint{4.705682in}{3.045291in}}%
\pgfpathlineto{\pgfqpoint{4.802273in}{3.053518in}}%
\pgfpathlineto{\pgfqpoint{4.898864in}{3.062558in}}%
\pgfpathlineto{\pgfqpoint{4.995455in}{3.071286in}}%
\pgfpathlineto{\pgfqpoint{5.092045in}{3.079597in}}%
\pgfpathlineto{\pgfqpoint{5.188636in}{3.086623in}}%
\pgfpathlineto{\pgfqpoint{5.285227in}{3.094119in}}%
\pgfpathlineto{\pgfqpoint{5.381818in}{3.101093in}}%
\pgfpathlineto{\pgfqpoint{5.478409in}{3.109163in}}%
\pgfpathlineto{\pgfqpoint{5.575000in}{3.117015in}}%
\pgfpathlineto{\pgfqpoint{5.671591in}{3.123318in}}%
\pgfpathlineto{\pgfqpoint{5.768182in}{3.130254in}}%
\pgfusepath{stroke}%
\end{pgfscope}%
\begin{pgfscope}%
\pgfpathrectangle{\pgfqpoint{0.900000in}{0.648000in}}{\pgfqpoint{5.100000in}{2.772000in}}%
\pgfusepath{clip}%
\pgfsetroundcap%
\pgfsetroundjoin%
\pgfsetlinewidth{1.505625pt}%
\definecolor{currentstroke}{rgb}{1.000000,0.341176,0.133333}%
\pgfsetstrokecolor{currentstroke}%
\pgfsetdash{}{0pt}%
\pgfpathmoveto{\pgfqpoint{1.131818in}{2.390460in}}%
\pgfpathlineto{\pgfqpoint{1.228409in}{2.408886in}}%
\pgfpathlineto{\pgfqpoint{1.325000in}{2.443137in}}%
\pgfpathlineto{\pgfqpoint{1.421591in}{2.455368in}}%
\pgfpathlineto{\pgfqpoint{1.518182in}{2.460423in}}%
\pgfpathlineto{\pgfqpoint{1.614773in}{2.469681in}}%
\pgfpathlineto{\pgfqpoint{1.711364in}{2.474922in}}%
\pgfpathlineto{\pgfqpoint{1.807955in}{2.485498in}}%
\pgfpathlineto{\pgfqpoint{1.904545in}{2.493078in}}%
\pgfpathlineto{\pgfqpoint{2.001136in}{2.502116in}}%
\pgfpathlineto{\pgfqpoint{2.097727in}{2.501089in}}%
\pgfpathlineto{\pgfqpoint{2.194318in}{2.502362in}}%
\pgfpathlineto{\pgfqpoint{2.290909in}{2.503629in}}%
\pgfpathlineto{\pgfqpoint{2.387500in}{2.510027in}}%
\pgfpathlineto{\pgfqpoint{2.484091in}{2.514313in}}%
\pgfpathlineto{\pgfqpoint{2.580682in}{2.514335in}}%
\pgfpathlineto{\pgfqpoint{2.677273in}{2.514130in}}%
\pgfpathlineto{\pgfqpoint{2.773864in}{2.517005in}}%
\pgfpathlineto{\pgfqpoint{2.870455in}{2.519844in}}%
\pgfpathlineto{\pgfqpoint{2.967045in}{2.521669in}}%
\pgfpathlineto{\pgfqpoint{3.063636in}{2.521048in}}%
\pgfpathlineto{\pgfqpoint{3.160227in}{2.522909in}}%
\pgfpathlineto{\pgfqpoint{3.256818in}{2.522622in}}%
\pgfpathlineto{\pgfqpoint{3.353409in}{2.525487in}}%
\pgfpathlineto{\pgfqpoint{3.450000in}{2.531838in}}%
\pgfpathlineto{\pgfqpoint{3.546591in}{2.527956in}}%
\pgfpathlineto{\pgfqpoint{3.643182in}{2.528248in}}%
\pgfpathlineto{\pgfqpoint{3.739773in}{2.532345in}}%
\pgfpathlineto{\pgfqpoint{3.836364in}{2.531261in}}%
\pgfpathlineto{\pgfqpoint{3.932955in}{2.532487in}}%
\pgfpathlineto{\pgfqpoint{4.029545in}{2.532428in}}%
\pgfpathlineto{\pgfqpoint{4.126136in}{2.533215in}}%
\pgfpathlineto{\pgfqpoint{4.222727in}{2.534772in}}%
\pgfpathlineto{\pgfqpoint{4.319318in}{2.537295in}}%
\pgfpathlineto{\pgfqpoint{4.415909in}{2.539886in}}%
\pgfpathlineto{\pgfqpoint{4.512500in}{2.540776in}}%
\pgfpathlineto{\pgfqpoint{4.609091in}{2.544021in}}%
\pgfpathlineto{\pgfqpoint{4.705682in}{2.540701in}}%
\pgfpathlineto{\pgfqpoint{4.802273in}{2.543541in}}%
\pgfpathlineto{\pgfqpoint{4.898864in}{2.548084in}}%
\pgfpathlineto{\pgfqpoint{4.995455in}{2.550218in}}%
\pgfpathlineto{\pgfqpoint{5.092045in}{2.547590in}}%
\pgfpathlineto{\pgfqpoint{5.188636in}{2.553290in}}%
\pgfpathlineto{\pgfqpoint{5.285227in}{2.556550in}}%
\pgfpathlineto{\pgfqpoint{5.381818in}{2.553762in}}%
\pgfpathlineto{\pgfqpoint{5.478409in}{2.555151in}}%
\pgfpathlineto{\pgfqpoint{5.575000in}{2.553601in}}%
\pgfpathlineto{\pgfqpoint{5.671591in}{2.553028in}}%
\pgfpathlineto{\pgfqpoint{5.768182in}{2.553578in}}%
\pgfusepath{stroke}%
\end{pgfscope}%
\begin{pgfscope}%
\pgfpathrectangle{\pgfqpoint{0.900000in}{0.648000in}}{\pgfqpoint{5.100000in}{2.772000in}}%
\pgfusepath{clip}%
\pgfsetbuttcap%
\pgfsetroundjoin%
\pgfsetlinewidth{1.505625pt}%
\definecolor{currentstroke}{rgb}{0.486275,0.301961,1.000000}%
\pgfsetstrokecolor{currentstroke}%
\pgfsetdash{{4.500000pt}{4.500000pt}}{0.000000pt}%
\pgfpathmoveto{\pgfqpoint{1.131818in}{2.396311in}}%
\pgfpathlineto{\pgfqpoint{1.228409in}{2.428422in}}%
\pgfpathlineto{\pgfqpoint{1.325000in}{2.479512in}}%
\pgfpathlineto{\pgfqpoint{1.421591in}{2.508498in}}%
\pgfpathlineto{\pgfqpoint{1.518182in}{2.533593in}}%
\pgfpathlineto{\pgfqpoint{1.614773in}{2.558611in}}%
\pgfpathlineto{\pgfqpoint{1.711364in}{2.582835in}}%
\pgfpathlineto{\pgfqpoint{1.807955in}{2.606885in}}%
\pgfpathlineto{\pgfqpoint{1.904545in}{2.627014in}}%
\pgfpathlineto{\pgfqpoint{2.001136in}{2.646535in}}%
\pgfpathlineto{\pgfqpoint{2.097727in}{2.665212in}}%
\pgfpathlineto{\pgfqpoint{2.194318in}{2.681559in}}%
\pgfpathlineto{\pgfqpoint{2.290909in}{2.698630in}}%
\pgfpathlineto{\pgfqpoint{2.387500in}{2.717236in}}%
\pgfpathlineto{\pgfqpoint{2.484091in}{2.734322in}}%
\pgfpathlineto{\pgfqpoint{2.580682in}{2.747974in}}%
\pgfpathlineto{\pgfqpoint{2.677273in}{2.763253in}}%
\pgfpathlineto{\pgfqpoint{2.773864in}{2.778916in}}%
\pgfpathlineto{\pgfqpoint{2.870455in}{2.792515in}}%
\pgfpathlineto{\pgfqpoint{2.967045in}{2.807065in}}%
\pgfpathlineto{\pgfqpoint{3.063636in}{2.820210in}}%
\pgfpathlineto{\pgfqpoint{3.160227in}{2.834585in}}%
\pgfpathlineto{\pgfqpoint{3.256818in}{2.848524in}}%
\pgfpathlineto{\pgfqpoint{3.353409in}{2.860354in}}%
\pgfpathlineto{\pgfqpoint{3.450000in}{2.871695in}}%
\pgfpathlineto{\pgfqpoint{3.546591in}{2.883321in}}%
\pgfpathlineto{\pgfqpoint{3.643182in}{2.898463in}}%
\pgfpathlineto{\pgfqpoint{3.739773in}{2.909609in}}%
\pgfpathlineto{\pgfqpoint{3.836364in}{2.922865in}}%
\pgfpathlineto{\pgfqpoint{3.932955in}{2.933647in}}%
\pgfpathlineto{\pgfqpoint{4.029545in}{2.942423in}}%
\pgfpathlineto{\pgfqpoint{4.126136in}{2.953082in}}%
\pgfpathlineto{\pgfqpoint{4.222727in}{2.963719in}}%
\pgfpathlineto{\pgfqpoint{4.319318in}{2.973727in}}%
\pgfpathlineto{\pgfqpoint{4.415909in}{2.982120in}}%
\pgfpathlineto{\pgfqpoint{4.512500in}{2.989590in}}%
\pgfpathlineto{\pgfqpoint{4.609091in}{2.997336in}}%
\pgfpathlineto{\pgfqpoint{4.705682in}{3.006449in}}%
\pgfpathlineto{\pgfqpoint{4.802273in}{3.015907in}}%
\pgfpathlineto{\pgfqpoint{4.898864in}{3.030493in}}%
\pgfpathlineto{\pgfqpoint{4.995455in}{3.036640in}}%
\pgfpathlineto{\pgfqpoint{5.092045in}{3.043527in}}%
\pgfpathlineto{\pgfqpoint{5.188636in}{3.051509in}}%
\pgfpathlineto{\pgfqpoint{5.285227in}{3.059909in}}%
\pgfpathlineto{\pgfqpoint{5.381818in}{3.068959in}}%
\pgfpathlineto{\pgfqpoint{5.478409in}{3.075845in}}%
\pgfpathlineto{\pgfqpoint{5.575000in}{3.084400in}}%
\pgfpathlineto{\pgfqpoint{5.671591in}{3.092555in}}%
\pgfpathlineto{\pgfqpoint{5.768182in}{3.099921in}}%
\pgfusepath{stroke}%
\end{pgfscope}%
\begin{pgfscope}%
\pgfpathrectangle{\pgfqpoint{0.900000in}{0.648000in}}{\pgfqpoint{5.100000in}{2.772000in}}%
\pgfusepath{clip}%
\pgfsetroundcap%
\pgfsetroundjoin%
\pgfsetlinewidth{1.505625pt}%
\definecolor{currentstroke}{rgb}{0.486275,0.301961,1.000000}%
\pgfsetstrokecolor{currentstroke}%
\pgfsetdash{}{0pt}%
\pgfpathmoveto{\pgfqpoint{1.131818in}{2.366243in}}%
\pgfpathlineto{\pgfqpoint{1.228409in}{2.380838in}}%
\pgfpathlineto{\pgfqpoint{1.325000in}{2.416721in}}%
\pgfpathlineto{\pgfqpoint{1.421591in}{2.425311in}}%
\pgfpathlineto{\pgfqpoint{1.518182in}{2.427986in}}%
\pgfpathlineto{\pgfqpoint{1.614773in}{2.432174in}}%
\pgfpathlineto{\pgfqpoint{1.711364in}{2.436786in}}%
\pgfpathlineto{\pgfqpoint{1.807955in}{2.445001in}}%
\pgfpathlineto{\pgfqpoint{1.904545in}{2.446156in}}%
\pgfpathlineto{\pgfqpoint{2.001136in}{2.451838in}}%
\pgfpathlineto{\pgfqpoint{2.097727in}{2.453488in}}%
\pgfpathlineto{\pgfqpoint{2.194318in}{2.453288in}}%
\pgfpathlineto{\pgfqpoint{2.290909in}{2.454166in}}%
\pgfpathlineto{\pgfqpoint{2.387500in}{2.460641in}}%
\pgfpathlineto{\pgfqpoint{2.484091in}{2.465058in}}%
\pgfpathlineto{\pgfqpoint{2.580682in}{2.466187in}}%
\pgfpathlineto{\pgfqpoint{2.677273in}{2.467646in}}%
\pgfpathlineto{\pgfqpoint{2.773864in}{2.467835in}}%
\pgfpathlineto{\pgfqpoint{2.870455in}{2.469175in}}%
\pgfpathlineto{\pgfqpoint{2.967045in}{2.471872in}}%
\pgfpathlineto{\pgfqpoint{3.063636in}{2.471822in}}%
\pgfpathlineto{\pgfqpoint{3.160227in}{2.475170in}}%
\pgfpathlineto{\pgfqpoint{3.256818in}{2.473829in}}%
\pgfpathlineto{\pgfqpoint{3.353409in}{2.476146in}}%
\pgfpathlineto{\pgfqpoint{3.450000in}{2.479950in}}%
\pgfpathlineto{\pgfqpoint{3.546591in}{2.479981in}}%
\pgfpathlineto{\pgfqpoint{3.643182in}{2.481112in}}%
\pgfpathlineto{\pgfqpoint{3.739773in}{2.480701in}}%
\pgfpathlineto{\pgfqpoint{3.836364in}{2.480915in}}%
\pgfpathlineto{\pgfqpoint{3.932955in}{2.484939in}}%
\pgfpathlineto{\pgfqpoint{4.029545in}{2.483922in}}%
\pgfpathlineto{\pgfqpoint{4.126136in}{2.487016in}}%
\pgfpathlineto{\pgfqpoint{4.222727in}{2.490902in}}%
\pgfpathlineto{\pgfqpoint{4.319318in}{2.491267in}}%
\pgfpathlineto{\pgfqpoint{4.415909in}{2.489944in}}%
\pgfpathlineto{\pgfqpoint{4.512500in}{2.492365in}}%
\pgfpathlineto{\pgfqpoint{4.609091in}{2.492159in}}%
\pgfpathlineto{\pgfqpoint{4.705682in}{2.492849in}}%
\pgfpathlineto{\pgfqpoint{4.802273in}{2.495474in}}%
\pgfpathlineto{\pgfqpoint{4.898864in}{2.499855in}}%
\pgfpathlineto{\pgfqpoint{4.995455in}{2.498205in}}%
\pgfpathlineto{\pgfqpoint{5.092045in}{2.505670in}}%
\pgfpathlineto{\pgfqpoint{5.188636in}{2.500341in}}%
\pgfpathlineto{\pgfqpoint{5.285227in}{2.502605in}}%
\pgfpathlineto{\pgfqpoint{5.381818in}{2.503125in}}%
\pgfpathlineto{\pgfqpoint{5.478409in}{2.504693in}}%
\pgfpathlineto{\pgfqpoint{5.575000in}{2.504739in}}%
\pgfpathlineto{\pgfqpoint{5.671591in}{2.508953in}}%
\pgfpathlineto{\pgfqpoint{5.768182in}{2.510075in}}%
\pgfusepath{stroke}%
\end{pgfscope}%
\begin{pgfscope}%
\pgfsetrectcap%
\pgfsetmiterjoin%
\pgfsetlinewidth{1.254687pt}%
\definecolor{currentstroke}{rgb}{0.800000,0.800000,0.800000}%
\pgfsetstrokecolor{currentstroke}%
\pgfsetdash{}{0pt}%
\pgfpathmoveto{\pgfqpoint{0.900000in}{0.648000in}}%
\pgfpathlineto{\pgfqpoint{0.900000in}{3.420000in}}%
\pgfusepath{stroke}%
\end{pgfscope}%
\begin{pgfscope}%
\pgfsetrectcap%
\pgfsetmiterjoin%
\pgfsetlinewidth{1.254687pt}%
\definecolor{currentstroke}{rgb}{0.800000,0.800000,0.800000}%
\pgfsetstrokecolor{currentstroke}%
\pgfsetdash{}{0pt}%
\pgfpathmoveto{\pgfqpoint{6.000000in}{0.648000in}}%
\pgfpathlineto{\pgfqpoint{6.000000in}{3.420000in}}%
\pgfusepath{stroke}%
\end{pgfscope}%
\begin{pgfscope}%
\pgfsetrectcap%
\pgfsetmiterjoin%
\pgfsetlinewidth{1.254687pt}%
\definecolor{currentstroke}{rgb}{0.800000,0.800000,0.800000}%
\pgfsetstrokecolor{currentstroke}%
\pgfsetdash{}{0pt}%
\pgfpathmoveto{\pgfqpoint{0.900000in}{0.648000in}}%
\pgfpathlineto{\pgfqpoint{6.000000in}{0.648000in}}%
\pgfusepath{stroke}%
\end{pgfscope}%
\begin{pgfscope}%
\pgfsetrectcap%
\pgfsetmiterjoin%
\pgfsetlinewidth{1.254687pt}%
\definecolor{currentstroke}{rgb}{0.800000,0.800000,0.800000}%
\pgfsetstrokecolor{currentstroke}%
\pgfsetdash{}{0pt}%
\pgfpathmoveto{\pgfqpoint{0.900000in}{3.420000in}}%
\pgfpathlineto{\pgfqpoint{6.000000in}{3.420000in}}%
\pgfusepath{stroke}%
\end{pgfscope}%
\begin{pgfscope}%
\pgfsetbuttcap%
\pgfsetmiterjoin%
\definecolor{currentfill}{rgb}{1.000000,1.000000,1.000000}%
\pgfsetfillcolor{currentfill}%
\pgfsetfillopacity{0.800000}%
\pgfsetlinewidth{1.003750pt}%
\definecolor{currentstroke}{rgb}{0.800000,0.800000,0.800000}%
\pgfsetstrokecolor{currentstroke}%
\pgfsetstrokeopacity{0.800000}%
\pgfsetdash{}{0pt}%
\pgfpathmoveto{\pgfqpoint{1.006944in}{2.231237in}}%
\pgfpathlineto{\pgfqpoint{1.936139in}{2.231237in}}%
\pgfpathquadraticcurveto{\pgfqpoint{1.966694in}{2.231237in}}{\pgfqpoint{1.966694in}{2.261792in}}%
\pgfpathlineto{\pgfqpoint{1.966694in}{3.313056in}}%
\pgfpathquadraticcurveto{\pgfqpoint{1.966694in}{3.343611in}}{\pgfqpoint{1.936139in}{3.343611in}}%
\pgfpathlineto{\pgfqpoint{1.006944in}{3.343611in}}%
\pgfpathquadraticcurveto{\pgfqpoint{0.976389in}{3.343611in}}{\pgfqpoint{0.976389in}{3.313056in}}%
\pgfpathlineto{\pgfqpoint{0.976389in}{2.261792in}}%
\pgfpathquadraticcurveto{\pgfqpoint{0.976389in}{2.231237in}}{\pgfqpoint{1.006944in}{2.231237in}}%
\pgfpathclose%
\pgfusepath{stroke,fill}%
\end{pgfscope}%
\begin{pgfscope}%
\pgfsetroundcap%
\pgfsetroundjoin%
\pgfsetlinewidth{1.505625pt}%
\definecolor{currentstroke}{rgb}{0.545098,0.764706,0.290196}%
\pgfsetstrokecolor{currentstroke}%
\pgfsetdash{}{0pt}%
\pgfpathmoveto{\pgfqpoint{1.037500in}{3.229028in}}%
\pgfpathlineto{\pgfqpoint{1.343056in}{3.229028in}}%
\pgfusepath{stroke}%
\end{pgfscope}%
\begin{pgfscope}%
\definecolor{textcolor}{rgb}{0.150000,0.150000,0.150000}%
\pgfsetstrokecolor{textcolor}%
\pgfsetfillcolor{textcolor}%
\pgftext[x=1.465278in,y=3.175556in,left,base]{\color{textcolor}\sffamily\fontsize{11.000000}{13.200000}\selectfont 128 bit}%
\end{pgfscope}%
\begin{pgfscope}%
\pgfsetroundcap%
\pgfsetroundjoin%
\pgfsetlinewidth{1.505625pt}%
\definecolor{currentstroke}{rgb}{1.000000,0.756863,0.027451}%
\pgfsetstrokecolor{currentstroke}%
\pgfsetdash{}{0pt}%
\pgfpathmoveto{\pgfqpoint{1.037500in}{3.016056in}}%
\pgfpathlineto{\pgfqpoint{1.343056in}{3.016056in}}%
\pgfusepath{stroke}%
\end{pgfscope}%
\begin{pgfscope}%
\definecolor{textcolor}{rgb}{0.150000,0.150000,0.150000}%
\pgfsetstrokecolor{textcolor}%
\pgfsetfillcolor{textcolor}%
\pgftext[x=1.465278in,y=2.962583in,left,base]{\color{textcolor}\sffamily\fontsize{11.000000}{13.200000}\selectfont 192 bit}%
\end{pgfscope}%
\begin{pgfscope}%
\pgfsetroundcap%
\pgfsetroundjoin%
\pgfsetlinewidth{1.505625pt}%
\definecolor{currentstroke}{rgb}{0.129412,0.588235,0.952941}%
\pgfsetstrokecolor{currentstroke}%
\pgfsetdash{}{0pt}%
\pgfpathmoveto{\pgfqpoint{1.037500in}{2.803084in}}%
\pgfpathlineto{\pgfqpoint{1.343056in}{2.803084in}}%
\pgfusepath{stroke}%
\end{pgfscope}%
\begin{pgfscope}%
\definecolor{textcolor}{rgb}{0.150000,0.150000,0.150000}%
\pgfsetstrokecolor{textcolor}%
\pgfsetfillcolor{textcolor}%
\pgftext[x=1.465278in,y=2.749611in,left,base]{\color{textcolor}\sffamily\fontsize{11.000000}{13.200000}\selectfont 256 bit}%
\end{pgfscope}%
\begin{pgfscope}%
\pgfsetroundcap%
\pgfsetroundjoin%
\pgfsetlinewidth{1.505625pt}%
\definecolor{currentstroke}{rgb}{1.000000,0.341176,0.133333}%
\pgfsetstrokecolor{currentstroke}%
\pgfsetdash{}{0pt}%
\pgfpathmoveto{\pgfqpoint{1.037500in}{2.590112in}}%
\pgfpathlineto{\pgfqpoint{1.343056in}{2.590112in}}%
\pgfusepath{stroke}%
\end{pgfscope}%
\begin{pgfscope}%
\definecolor{textcolor}{rgb}{0.150000,0.150000,0.150000}%
\pgfsetstrokecolor{textcolor}%
\pgfsetfillcolor{textcolor}%
\pgftext[x=1.465278in,y=2.536639in,left,base]{\color{textcolor}\sffamily\fontsize{11.000000}{13.200000}\selectfont gmp}%
\end{pgfscope}%
\begin{pgfscope}%
\pgfsetroundcap%
\pgfsetroundjoin%
\pgfsetlinewidth{1.505625pt}%
\definecolor{currentstroke}{rgb}{0.486275,0.301961,1.000000}%
\pgfsetstrokecolor{currentstroke}%
\pgfsetdash{}{0pt}%
\pgfpathmoveto{\pgfqpoint{1.037500in}{2.375459in}}%
\pgfpathlineto{\pgfqpoint{1.343056in}{2.375459in}}%
\pgfusepath{stroke}%
\end{pgfscope}%
\begin{pgfscope}%
\definecolor{textcolor}{rgb}{0.150000,0.150000,0.150000}%
\pgfsetstrokecolor{textcolor}%
\pgfsetfillcolor{textcolor}%
\pgftext[x=1.465278in,y=2.321987in,left,base]{\color{textcolor}\sffamily\fontsize{11.000000}{13.200000}\selectfont CGAL}%
\end{pgfscope}%
\begin{pgfscope}%
\pgfsetbuttcap%
\pgfsetmiterjoin%
\definecolor{currentfill}{rgb}{1.000000,1.000000,1.000000}%
\pgfsetfillcolor{currentfill}%
\pgfsetfillopacity{0.800000}%
\pgfsetlinewidth{1.003750pt}%
\definecolor{currentstroke}{rgb}{0.800000,0.800000,0.800000}%
\pgfsetstrokecolor{currentstroke}%
\pgfsetstrokeopacity{0.800000}%
\pgfsetdash{}{0pt}%
\pgfpathmoveto{\pgfqpoint{4.605750in}{0.724389in}}%
\pgfpathlineto{\pgfqpoint{5.893056in}{0.724389in}}%
\pgfpathquadraticcurveto{\pgfqpoint{5.923611in}{0.724389in}}{\pgfqpoint{5.923611in}{0.754944in}}%
\pgfpathlineto{\pgfqpoint{5.923611in}{1.167291in}}%
\pgfpathquadraticcurveto{\pgfqpoint{5.923611in}{1.197847in}}{\pgfqpoint{5.893056in}{1.197847in}}%
\pgfpathlineto{\pgfqpoint{4.605750in}{1.197847in}}%
\pgfpathquadraticcurveto{\pgfqpoint{4.575194in}{1.197847in}}{\pgfqpoint{4.575194in}{1.167291in}}%
\pgfpathlineto{\pgfqpoint{4.575194in}{0.754944in}}%
\pgfpathquadraticcurveto{\pgfqpoint{4.575194in}{0.724389in}}{\pgfqpoint{4.605750in}{0.724389in}}%
\pgfpathclose%
\pgfusepath{stroke,fill}%
\end{pgfscope}%
\begin{pgfscope}%
\pgfsetbuttcap%
\pgfsetroundjoin%
\pgfsetlinewidth{1.505625pt}%
\definecolor{currentstroke}{rgb}{0.100000,0.100000,0.100000}%
\pgfsetstrokecolor{currentstroke}%
\pgfsetdash{{4.500000pt}{4.500000pt}}{0.000000pt}%
\pgfpathmoveto{\pgfqpoint{4.636306in}{1.083264in}}%
\pgfpathlineto{\pgfqpoint{4.941861in}{1.083264in}}%
\pgfusepath{stroke}%
\end{pgfscope}%
\begin{pgfscope}%
\definecolor{textcolor}{rgb}{0.150000,0.150000,0.150000}%
\pgfsetstrokecolor{textcolor}%
\pgfsetfillcolor{textcolor}%
\pgftext[x=5.064083in,y=1.029791in,left,base]{\color{textcolor}\sffamily\fontsize{11.000000}{13.200000}\selectfont naive}%
\end{pgfscope}%
\begin{pgfscope}%
\pgfsetroundcap%
\pgfsetroundjoin%
\pgfsetlinewidth{1.505625pt}%
\definecolor{currentstroke}{rgb}{0.100000,0.100000,0.100000}%
\pgfsetstrokecolor{currentstroke}%
\pgfsetdash{}{0pt}%
\pgfpathmoveto{\pgfqpoint{4.636306in}{0.870292in}}%
\pgfpathlineto{\pgfqpoint{4.941861in}{0.870292in}}%
\pgfusepath{stroke}%
\end{pgfscope}%
\begin{pgfscope}%
\definecolor{textcolor}{rgb}{0.150000,0.150000,0.150000}%
\pgfsetstrokecolor{textcolor}%
\pgfsetfillcolor{textcolor}%
\pgftext[x=5.064083in,y=0.816819in,left,base]{\color{textcolor}\sffamily\fontsize{11.000000}{13.200000}\selectfont edge descent}%
\end{pgfscope}%
\end{pgfpicture}%
\makeatother%
\endgroup%

%% file: figures/convex-numbers.pgf
%% Creator: Matplotlib, PGF backend
%%
%% To include the figure in your LaTeX document, write
%%   \input{<filename>.pgf}
%%
%% Make sure the required packages are loaded in your preamble
%%   \usepackage{pgf}
%%
%% and, on pdftex
%%   \usepackage[utf8]{inputenc}\DeclareUnicodeCharacter{2212}{-}
%%
%% or, on luatex and xetex
%%   \usepackage{unicode-math}
%%
%% Figures using additional raster images can only be included by \input if
%% they are in the same directory as the main LaTeX file. For loading figures
%% from other directories you can use the `import` package
%% and then include the figures with
%%   \import{<path to file>}{<filename>.pgf}
%%
%% Matplotlib used the following preamble
%%   \usepackage{fontspec}
%%
\begingroup%
\makeatletter%
\begin{pgfpicture}%
\pgfpathrectangle{\pgfpointorigin}{\pgfqpoint{5.000000in}{2.500000in}}%
\pgfusepath{use as bounding box, clip}%
\begin{pgfscope}%
\pgfsetbuttcap%
\pgfsetmiterjoin%
\definecolor{currentfill}{rgb}{1.000000,1.000000,1.000000}%
\pgfsetfillcolor{currentfill}%
\pgfsetlinewidth{0.000000pt}%
\definecolor{currentstroke}{rgb}{1.000000,1.000000,1.000000}%
\pgfsetstrokecolor{currentstroke}%
\pgfsetdash{}{0pt}%
\pgfpathmoveto{\pgfqpoint{0.000000in}{0.000000in}}%
\pgfpathlineto{\pgfqpoint{5.000000in}{0.000000in}}%
\pgfpathlineto{\pgfqpoint{5.000000in}{2.500000in}}%
\pgfpathlineto{\pgfqpoint{0.000000in}{2.500000in}}%
\pgfpathclose%
\pgfusepath{fill}%
\end{pgfscope}%
\begin{pgfscope}%
\pgfsetbuttcap%
\pgfsetmiterjoin%
\definecolor{currentfill}{rgb}{1.000000,1.000000,1.000000}%
\pgfsetfillcolor{currentfill}%
\pgfsetlinewidth{0.000000pt}%
\definecolor{currentstroke}{rgb}{0.000000,0.000000,0.000000}%
\pgfsetstrokecolor{currentstroke}%
\pgfsetstrokeopacity{0.000000}%
\pgfsetdash{}{0pt}%
\pgfpathmoveto{\pgfqpoint{0.750000in}{0.525000in}}%
\pgfpathlineto{\pgfqpoint{5.000000in}{0.525000in}}%
\pgfpathlineto{\pgfqpoint{5.000000in}{2.375000in}}%
\pgfpathlineto{\pgfqpoint{0.750000in}{2.375000in}}%
\pgfpathclose%
\pgfusepath{fill}%
\end{pgfscope}%
\begin{pgfscope}%
\pgfpathrectangle{\pgfqpoint{0.750000in}{0.525000in}}{\pgfqpoint{4.250000in}{1.850000in}}%
\pgfusepath{clip}%
\pgfsetroundcap%
\pgfsetroundjoin%
\pgfsetlinewidth{1.003750pt}%
\definecolor{currentstroke}{rgb}{0.800000,0.800000,0.800000}%
\pgfsetstrokecolor{currentstroke}%
\pgfsetdash{}{0pt}%
\pgfpathmoveto{\pgfqpoint{0.912023in}{0.525000in}}%
\pgfpathlineto{\pgfqpoint{0.912023in}{2.375000in}}%
\pgfusepath{stroke}%
\end{pgfscope}%
\begin{pgfscope}%
\definecolor{textcolor}{rgb}{0.150000,0.150000,0.150000}%
\pgfsetstrokecolor{textcolor}%
\pgfsetfillcolor{textcolor}%
\pgftext[x=0.912023in,y=0.393056in,,top]{\color{textcolor}\sffamily\fontsize{11.000000}{13.200000}\selectfont 0}%
\end{pgfscope}%
\begin{pgfscope}%
\pgfpathrectangle{\pgfqpoint{0.750000in}{0.525000in}}{\pgfqpoint{4.250000in}{1.850000in}}%
\pgfusepath{clip}%
\pgfsetroundcap%
\pgfsetroundjoin%
\pgfsetlinewidth{1.003750pt}%
\definecolor{currentstroke}{rgb}{0.800000,0.800000,0.800000}%
\pgfsetstrokecolor{currentstroke}%
\pgfsetdash{}{0pt}%
\pgfpathmoveto{\pgfqpoint{1.690982in}{0.525000in}}%
\pgfpathlineto{\pgfqpoint{1.690982in}{2.375000in}}%
\pgfusepath{stroke}%
\end{pgfscope}%
\begin{pgfscope}%
\definecolor{textcolor}{rgb}{0.150000,0.150000,0.150000}%
\pgfsetstrokecolor{textcolor}%
\pgfsetfillcolor{textcolor}%
\pgftext[x=1.690982in,y=0.393056in,,top]{\color{textcolor}\sffamily\fontsize{11.000000}{13.200000}\selectfont 50}%
\end{pgfscope}%
\begin{pgfscope}%
\pgfpathrectangle{\pgfqpoint{0.750000in}{0.525000in}}{\pgfqpoint{4.250000in}{1.850000in}}%
\pgfusepath{clip}%
\pgfsetroundcap%
\pgfsetroundjoin%
\pgfsetlinewidth{1.003750pt}%
\definecolor{currentstroke}{rgb}{0.800000,0.800000,0.800000}%
\pgfsetstrokecolor{currentstroke}%
\pgfsetdash{}{0pt}%
\pgfpathmoveto{\pgfqpoint{2.469941in}{0.525000in}}%
\pgfpathlineto{\pgfqpoint{2.469941in}{2.375000in}}%
\pgfusepath{stroke}%
\end{pgfscope}%
\begin{pgfscope}%
\definecolor{textcolor}{rgb}{0.150000,0.150000,0.150000}%
\pgfsetstrokecolor{textcolor}%
\pgfsetfillcolor{textcolor}%
\pgftext[x=2.469941in,y=0.393056in,,top]{\color{textcolor}\sffamily\fontsize{11.000000}{13.200000}\selectfont 100}%
\end{pgfscope}%
\begin{pgfscope}%
\pgfpathrectangle{\pgfqpoint{0.750000in}{0.525000in}}{\pgfqpoint{4.250000in}{1.850000in}}%
\pgfusepath{clip}%
\pgfsetroundcap%
\pgfsetroundjoin%
\pgfsetlinewidth{1.003750pt}%
\definecolor{currentstroke}{rgb}{0.800000,0.800000,0.800000}%
\pgfsetstrokecolor{currentstroke}%
\pgfsetdash{}{0pt}%
\pgfpathmoveto{\pgfqpoint{3.248900in}{0.525000in}}%
\pgfpathlineto{\pgfqpoint{3.248900in}{2.375000in}}%
\pgfusepath{stroke}%
\end{pgfscope}%
\begin{pgfscope}%
\definecolor{textcolor}{rgb}{0.150000,0.150000,0.150000}%
\pgfsetstrokecolor{textcolor}%
\pgfsetfillcolor{textcolor}%
\pgftext[x=3.248900in,y=0.393056in,,top]{\color{textcolor}\sffamily\fontsize{11.000000}{13.200000}\selectfont 150}%
\end{pgfscope}%
\begin{pgfscope}%
\pgfpathrectangle{\pgfqpoint{0.750000in}{0.525000in}}{\pgfqpoint{4.250000in}{1.850000in}}%
\pgfusepath{clip}%
\pgfsetroundcap%
\pgfsetroundjoin%
\pgfsetlinewidth{1.003750pt}%
\definecolor{currentstroke}{rgb}{0.800000,0.800000,0.800000}%
\pgfsetstrokecolor{currentstroke}%
\pgfsetdash{}{0pt}%
\pgfpathmoveto{\pgfqpoint{4.027859in}{0.525000in}}%
\pgfpathlineto{\pgfqpoint{4.027859in}{2.375000in}}%
\pgfusepath{stroke}%
\end{pgfscope}%
\begin{pgfscope}%
\definecolor{textcolor}{rgb}{0.150000,0.150000,0.150000}%
\pgfsetstrokecolor{textcolor}%
\pgfsetfillcolor{textcolor}%
\pgftext[x=4.027859in,y=0.393056in,,top]{\color{textcolor}\sffamily\fontsize{11.000000}{13.200000}\selectfont 200}%
\end{pgfscope}%
\begin{pgfscope}%
\pgfpathrectangle{\pgfqpoint{0.750000in}{0.525000in}}{\pgfqpoint{4.250000in}{1.850000in}}%
\pgfusepath{clip}%
\pgfsetroundcap%
\pgfsetroundjoin%
\pgfsetlinewidth{1.003750pt}%
\definecolor{currentstroke}{rgb}{0.800000,0.800000,0.800000}%
\pgfsetstrokecolor{currentstroke}%
\pgfsetdash{}{0pt}%
\pgfpathmoveto{\pgfqpoint{4.806818in}{0.525000in}}%
\pgfpathlineto{\pgfqpoint{4.806818in}{2.375000in}}%
\pgfusepath{stroke}%
\end{pgfscope}%
\begin{pgfscope}%
\definecolor{textcolor}{rgb}{0.150000,0.150000,0.150000}%
\pgfsetstrokecolor{textcolor}%
\pgfsetfillcolor{textcolor}%
\pgftext[x=4.806818in,y=0.393056in,,top]{\color{textcolor}\sffamily\fontsize{11.000000}{13.200000}\selectfont 250}%
\end{pgfscope}%
\begin{pgfscope}%
\definecolor{textcolor}{rgb}{0.150000,0.150000,0.150000}%
\pgfsetstrokecolor{textcolor}%
\pgfsetfillcolor{textcolor}%
\pgftext[x=2.875000in,y=0.201833in,,top]{\color{textcolor}\sffamily\fontsize{12.000000}{14.400000}\selectfont depth}%
\end{pgfscope}%
\begin{pgfscope}%
\pgfpathrectangle{\pgfqpoint{0.750000in}{0.525000in}}{\pgfqpoint{4.250000in}{1.850000in}}%
\pgfusepath{clip}%
\pgfsetroundcap%
\pgfsetroundjoin%
\pgfsetlinewidth{1.003750pt}%
\definecolor{currentstroke}{rgb}{0.800000,0.800000,0.800000}%
\pgfsetstrokecolor{currentstroke}%
\pgfsetdash{}{0pt}%
\pgfpathmoveto{\pgfqpoint{0.750000in}{0.598815in}}%
\pgfpathlineto{\pgfqpoint{5.000000in}{0.598815in}}%
\pgfusepath{stroke}%
\end{pgfscope}%
\begin{pgfscope}%
\definecolor{textcolor}{rgb}{0.150000,0.150000,0.150000}%
\pgfsetstrokecolor{textcolor}%
\pgfsetfillcolor{textcolor}%
\pgftext[x=0.543194in, y=0.545801in, left, base]{\color{textcolor}\sffamily\fontsize{11.000000}{13.200000}\selectfont 0}%
\end{pgfscope}%
\begin{pgfscope}%
\pgfpathrectangle{\pgfqpoint{0.750000in}{0.525000in}}{\pgfqpoint{4.250000in}{1.850000in}}%
\pgfusepath{clip}%
\pgfsetroundcap%
\pgfsetroundjoin%
\pgfsetlinewidth{1.003750pt}%
\definecolor{currentstroke}{rgb}{0.800000,0.800000,0.800000}%
\pgfsetstrokecolor{currentstroke}%
\pgfsetdash{}{0pt}%
\pgfpathmoveto{\pgfqpoint{0.750000in}{0.941344in}}%
\pgfpathlineto{\pgfqpoint{5.000000in}{0.941344in}}%
\pgfusepath{stroke}%
\end{pgfscope}%
\begin{pgfscope}%
\definecolor{textcolor}{rgb}{0.150000,0.150000,0.150000}%
\pgfsetstrokecolor{textcolor}%
\pgfsetfillcolor{textcolor}%
\pgftext[x=0.393472in, y=0.888330in, left, base]{\color{textcolor}\sffamily\fontsize{11.000000}{13.200000}\selectfont 100}%
\end{pgfscope}%
\begin{pgfscope}%
\pgfpathrectangle{\pgfqpoint{0.750000in}{0.525000in}}{\pgfqpoint{4.250000in}{1.850000in}}%
\pgfusepath{clip}%
\pgfsetroundcap%
\pgfsetroundjoin%
\pgfsetlinewidth{1.003750pt}%
\definecolor{currentstroke}{rgb}{0.800000,0.800000,0.800000}%
\pgfsetstrokecolor{currentstroke}%
\pgfsetdash{}{0pt}%
\pgfpathmoveto{\pgfqpoint{0.750000in}{1.283873in}}%
\pgfpathlineto{\pgfqpoint{5.000000in}{1.283873in}}%
\pgfusepath{stroke}%
\end{pgfscope}%
\begin{pgfscope}%
\definecolor{textcolor}{rgb}{0.150000,0.150000,0.150000}%
\pgfsetstrokecolor{textcolor}%
\pgfsetfillcolor{textcolor}%
\pgftext[x=0.393472in, y=1.230859in, left, base]{\color{textcolor}\sffamily\fontsize{11.000000}{13.200000}\selectfont 200}%
\end{pgfscope}%
\begin{pgfscope}%
\pgfpathrectangle{\pgfqpoint{0.750000in}{0.525000in}}{\pgfqpoint{4.250000in}{1.850000in}}%
\pgfusepath{clip}%
\pgfsetroundcap%
\pgfsetroundjoin%
\pgfsetlinewidth{1.003750pt}%
\definecolor{currentstroke}{rgb}{0.800000,0.800000,0.800000}%
\pgfsetstrokecolor{currentstroke}%
\pgfsetdash{}{0pt}%
\pgfpathmoveto{\pgfqpoint{0.750000in}{1.626403in}}%
\pgfpathlineto{\pgfqpoint{5.000000in}{1.626403in}}%
\pgfusepath{stroke}%
\end{pgfscope}%
\begin{pgfscope}%
\definecolor{textcolor}{rgb}{0.150000,0.150000,0.150000}%
\pgfsetstrokecolor{textcolor}%
\pgfsetfillcolor{textcolor}%
\pgftext[x=0.393472in, y=1.573389in, left, base]{\color{textcolor}\sffamily\fontsize{11.000000}{13.200000}\selectfont 300}%
\end{pgfscope}%
\begin{pgfscope}%
\pgfpathrectangle{\pgfqpoint{0.750000in}{0.525000in}}{\pgfqpoint{4.250000in}{1.850000in}}%
\pgfusepath{clip}%
\pgfsetroundcap%
\pgfsetroundjoin%
\pgfsetlinewidth{1.003750pt}%
\definecolor{currentstroke}{rgb}{0.800000,0.800000,0.800000}%
\pgfsetstrokecolor{currentstroke}%
\pgfsetdash{}{0pt}%
\pgfpathmoveto{\pgfqpoint{0.750000in}{1.968932in}}%
\pgfpathlineto{\pgfqpoint{5.000000in}{1.968932in}}%
\pgfusepath{stroke}%
\end{pgfscope}%
\begin{pgfscope}%
\definecolor{textcolor}{rgb}{0.150000,0.150000,0.150000}%
\pgfsetstrokecolor{textcolor}%
\pgfsetfillcolor{textcolor}%
\pgftext[x=0.393472in, y=1.915918in, left, base]{\color{textcolor}\sffamily\fontsize{11.000000}{13.200000}\selectfont 400}%
\end{pgfscope}%
\begin{pgfscope}%
\pgfpathrectangle{\pgfqpoint{0.750000in}{0.525000in}}{\pgfqpoint{4.250000in}{1.850000in}}%
\pgfusepath{clip}%
\pgfsetroundcap%
\pgfsetroundjoin%
\pgfsetlinewidth{1.003750pt}%
\definecolor{currentstroke}{rgb}{0.800000,0.800000,0.800000}%
\pgfsetstrokecolor{currentstroke}%
\pgfsetdash{}{0pt}%
\pgfpathmoveto{\pgfqpoint{0.750000in}{2.311461in}}%
\pgfpathlineto{\pgfqpoint{5.000000in}{2.311461in}}%
\pgfusepath{stroke}%
\end{pgfscope}%
\begin{pgfscope}%
\definecolor{textcolor}{rgb}{0.150000,0.150000,0.150000}%
\pgfsetstrokecolor{textcolor}%
\pgfsetfillcolor{textcolor}%
\pgftext[x=0.393472in, y=2.258447in, left, base]{\color{textcolor}\sffamily\fontsize{11.000000}{13.200000}\selectfont 500}%
\end{pgfscope}%
\begin{pgfscope}%
\definecolor{textcolor}{rgb}{0.150000,0.150000,0.150000}%
\pgfsetstrokecolor{textcolor}%
\pgfsetfillcolor{textcolor}%
\pgftext[x=0.337917in,y=1.450000in,,bottom,rotate=90.000000]{\color{textcolor}\sffamily\fontsize{12.000000}{14.400000}\selectfont per cut}%
\end{pgfscope}%
\begin{pgfscope}%
\pgfpathrectangle{\pgfqpoint{0.750000in}{0.525000in}}{\pgfqpoint{4.250000in}{1.850000in}}%
\pgfusepath{clip}%
\pgfsetroundcap%
\pgfsetroundjoin%
\pgfsetlinewidth{1.505625pt}%
\definecolor{currentstroke}{rgb}{0.129412,0.588235,0.952941}%
\pgfsetstrokecolor{currentstroke}%
\pgfsetdash{}{0pt}%
\pgfpathmoveto{\pgfqpoint{0.943182in}{0.612516in}}%
\pgfpathlineto{\pgfqpoint{0.958761in}{0.619367in}}%
\pgfpathlineto{\pgfqpoint{0.974340in}{0.612516in}}%
\pgfpathlineto{\pgfqpoint{0.989919in}{0.619367in}}%
\pgfpathlineto{\pgfqpoint{1.021078in}{0.612516in}}%
\pgfpathlineto{\pgfqpoint{1.036657in}{0.612516in}}%
\pgfpathlineto{\pgfqpoint{1.067815in}{0.619367in}}%
\pgfpathlineto{\pgfqpoint{1.083394in}{0.615941in}}%
\pgfpathlineto{\pgfqpoint{1.098974in}{0.615941in}}%
\pgfpathlineto{\pgfqpoint{1.114553in}{0.619367in}}%
\pgfpathlineto{\pgfqpoint{1.130132in}{0.615941in}}%
\pgfpathlineto{\pgfqpoint{1.145711in}{0.619367in}}%
\pgfpathlineto{\pgfqpoint{1.161290in}{0.619367in}}%
\pgfpathlineto{\pgfqpoint{1.176870in}{0.615941in}}%
\pgfpathlineto{\pgfqpoint{1.192449in}{0.622792in}}%
\pgfpathlineto{\pgfqpoint{1.208028in}{0.622792in}}%
\pgfpathlineto{\pgfqpoint{1.223607in}{0.626217in}}%
\pgfpathlineto{\pgfqpoint{1.239186in}{0.622792in}}%
\pgfpathlineto{\pgfqpoint{1.254765in}{0.622792in}}%
\pgfpathlineto{\pgfqpoint{1.270345in}{0.615941in}}%
\pgfpathlineto{\pgfqpoint{1.285924in}{0.612516in}}%
\pgfpathlineto{\pgfqpoint{1.301503in}{0.615941in}}%
\pgfpathlineto{\pgfqpoint{1.348240in}{0.615941in}}%
\pgfpathlineto{\pgfqpoint{1.363820in}{0.619367in}}%
\pgfpathlineto{\pgfqpoint{1.379399in}{0.619367in}}%
\pgfpathlineto{\pgfqpoint{1.394978in}{0.615941in}}%
\pgfpathlineto{\pgfqpoint{1.410557in}{0.619367in}}%
\pgfpathlineto{\pgfqpoint{1.426136in}{0.615941in}}%
\pgfpathlineto{\pgfqpoint{1.441716in}{0.622792in}}%
\pgfpathlineto{\pgfqpoint{1.457295in}{0.615941in}}%
\pgfpathlineto{\pgfqpoint{1.488453in}{0.615941in}}%
\pgfpathlineto{\pgfqpoint{1.504032in}{0.626217in}}%
\pgfpathlineto{\pgfqpoint{1.519611in}{0.615941in}}%
\pgfpathlineto{\pgfqpoint{1.535191in}{0.612516in}}%
\pgfpathlineto{\pgfqpoint{1.550770in}{0.615941in}}%
\pgfpathlineto{\pgfqpoint{1.566349in}{0.622792in}}%
\pgfpathlineto{\pgfqpoint{1.581928in}{0.615941in}}%
\pgfpathlineto{\pgfqpoint{1.597507in}{0.626217in}}%
\pgfpathlineto{\pgfqpoint{1.613087in}{0.619367in}}%
\pgfpathlineto{\pgfqpoint{1.628666in}{0.619367in}}%
\pgfpathlineto{\pgfqpoint{1.644245in}{0.622792in}}%
\pgfpathlineto{\pgfqpoint{1.659824in}{0.619367in}}%
\pgfpathlineto{\pgfqpoint{1.675403in}{0.619367in}}%
\pgfpathlineto{\pgfqpoint{1.690982in}{0.612516in}}%
\pgfpathlineto{\pgfqpoint{1.706562in}{0.619367in}}%
\pgfpathlineto{\pgfqpoint{1.722141in}{0.622792in}}%
\pgfpathlineto{\pgfqpoint{1.737720in}{0.622792in}}%
\pgfpathlineto{\pgfqpoint{1.784457in}{0.612516in}}%
\pgfpathlineto{\pgfqpoint{1.800037in}{0.619367in}}%
\pgfpathlineto{\pgfqpoint{1.815616in}{0.615941in}}%
\pgfpathlineto{\pgfqpoint{1.831195in}{0.619367in}}%
\pgfpathlineto{\pgfqpoint{1.862353in}{0.619367in}}%
\pgfpathlineto{\pgfqpoint{1.877933in}{0.612516in}}%
\pgfpathlineto{\pgfqpoint{1.893512in}{0.619367in}}%
\pgfpathlineto{\pgfqpoint{1.909091in}{0.615941in}}%
\pgfpathlineto{\pgfqpoint{1.955828in}{0.615941in}}%
\pgfpathlineto{\pgfqpoint{1.971408in}{0.622792in}}%
\pgfpathlineto{\pgfqpoint{1.986987in}{0.615941in}}%
\pgfpathlineto{\pgfqpoint{2.002566in}{0.615941in}}%
\pgfpathlineto{\pgfqpoint{2.018145in}{0.622792in}}%
\pgfpathlineto{\pgfqpoint{2.033724in}{0.615941in}}%
\pgfpathlineto{\pgfqpoint{2.049304in}{0.629643in}}%
\pgfpathlineto{\pgfqpoint{2.064883in}{0.615941in}}%
\pgfpathlineto{\pgfqpoint{2.080462in}{0.622792in}}%
\pgfpathlineto{\pgfqpoint{2.096041in}{0.615941in}}%
\pgfpathlineto{\pgfqpoint{2.111620in}{0.615941in}}%
\pgfpathlineto{\pgfqpoint{2.142779in}{0.622792in}}%
\pgfpathlineto{\pgfqpoint{2.158358in}{0.622792in}}%
\pgfpathlineto{\pgfqpoint{2.173937in}{0.612516in}}%
\pgfpathlineto{\pgfqpoint{2.189516in}{0.615941in}}%
\pgfpathlineto{\pgfqpoint{2.205095in}{0.615941in}}%
\pgfpathlineto{\pgfqpoint{2.220674in}{0.619367in}}%
\pgfpathlineto{\pgfqpoint{2.236254in}{0.619367in}}%
\pgfpathlineto{\pgfqpoint{2.251833in}{0.612516in}}%
\pgfpathlineto{\pgfqpoint{2.267412in}{0.619367in}}%
\pgfpathlineto{\pgfqpoint{2.282991in}{0.619367in}}%
\pgfpathlineto{\pgfqpoint{2.298570in}{0.622792in}}%
\pgfpathlineto{\pgfqpoint{2.314150in}{0.612516in}}%
\pgfpathlineto{\pgfqpoint{2.329729in}{0.615941in}}%
\pgfpathlineto{\pgfqpoint{2.345308in}{0.615941in}}%
\pgfpathlineto{\pgfqpoint{2.360887in}{0.619367in}}%
\pgfpathlineto{\pgfqpoint{2.376466in}{0.612516in}}%
\pgfpathlineto{\pgfqpoint{2.392045in}{0.619367in}}%
\pgfpathlineto{\pgfqpoint{2.407625in}{0.619367in}}%
\pgfpathlineto{\pgfqpoint{2.423204in}{0.615941in}}%
\pgfpathlineto{\pgfqpoint{2.438783in}{0.615941in}}%
\pgfpathlineto{\pgfqpoint{2.454362in}{0.612516in}}%
\pgfpathlineto{\pgfqpoint{2.469941in}{0.622792in}}%
\pgfpathlineto{\pgfqpoint{2.485521in}{0.629643in}}%
\pgfpathlineto{\pgfqpoint{2.501100in}{0.622792in}}%
\pgfpathlineto{\pgfqpoint{2.516679in}{0.619367in}}%
\pgfpathlineto{\pgfqpoint{2.532258in}{0.612516in}}%
\pgfpathlineto{\pgfqpoint{2.547837in}{0.612516in}}%
\pgfpathlineto{\pgfqpoint{2.563416in}{0.615941in}}%
\pgfpathlineto{\pgfqpoint{2.578996in}{0.615941in}}%
\pgfpathlineto{\pgfqpoint{2.594575in}{0.612516in}}%
\pgfpathlineto{\pgfqpoint{2.610154in}{0.622792in}}%
\pgfpathlineto{\pgfqpoint{2.625733in}{0.619367in}}%
\pgfpathlineto{\pgfqpoint{2.641312in}{0.619367in}}%
\pgfpathlineto{\pgfqpoint{2.656891in}{0.615941in}}%
\pgfpathlineto{\pgfqpoint{2.672471in}{0.622792in}}%
\pgfpathlineto{\pgfqpoint{2.688050in}{0.615941in}}%
\pgfpathlineto{\pgfqpoint{2.703629in}{0.622792in}}%
\pgfpathlineto{\pgfqpoint{2.719208in}{0.619367in}}%
\pgfpathlineto{\pgfqpoint{2.734787in}{0.619367in}}%
\pgfpathlineto{\pgfqpoint{2.750367in}{0.612516in}}%
\pgfpathlineto{\pgfqpoint{2.765946in}{0.619367in}}%
\pgfpathlineto{\pgfqpoint{2.781525in}{0.615941in}}%
\pgfpathlineto{\pgfqpoint{2.797104in}{0.615941in}}%
\pgfpathlineto{\pgfqpoint{2.812683in}{0.626217in}}%
\pgfpathlineto{\pgfqpoint{2.828262in}{0.615941in}}%
\pgfpathlineto{\pgfqpoint{2.875000in}{0.615941in}}%
\pgfpathlineto{\pgfqpoint{2.890579in}{0.622792in}}%
\pgfpathlineto{\pgfqpoint{2.906158in}{0.619367in}}%
\pgfpathlineto{\pgfqpoint{2.921738in}{0.626217in}}%
\pgfpathlineto{\pgfqpoint{2.937317in}{0.615941in}}%
\pgfpathlineto{\pgfqpoint{2.968475in}{0.622792in}}%
\pgfpathlineto{\pgfqpoint{2.984054in}{0.612516in}}%
\pgfpathlineto{\pgfqpoint{2.999633in}{0.622792in}}%
\pgfpathlineto{\pgfqpoint{3.015213in}{0.612516in}}%
\pgfpathlineto{\pgfqpoint{3.030792in}{0.622792in}}%
\pgfpathlineto{\pgfqpoint{3.046371in}{0.612516in}}%
\pgfpathlineto{\pgfqpoint{3.061950in}{0.615941in}}%
\pgfpathlineto{\pgfqpoint{3.077529in}{0.615941in}}%
\pgfpathlineto{\pgfqpoint{3.108688in}{0.609091in}}%
\pgfpathlineto{\pgfqpoint{3.124267in}{0.622792in}}%
\pgfpathlineto{\pgfqpoint{3.139846in}{0.619367in}}%
\pgfpathlineto{\pgfqpoint{3.186584in}{0.619367in}}%
\pgfpathlineto{\pgfqpoint{3.217742in}{0.612516in}}%
\pgfpathlineto{\pgfqpoint{3.233321in}{0.615941in}}%
\pgfpathlineto{\pgfqpoint{3.248900in}{0.615941in}}%
\pgfpathlineto{\pgfqpoint{3.280059in}{0.629643in}}%
\pgfpathlineto{\pgfqpoint{3.295638in}{0.622792in}}%
\pgfpathlineto{\pgfqpoint{3.311217in}{0.629643in}}%
\pgfpathlineto{\pgfqpoint{3.326796in}{0.612516in}}%
\pgfpathlineto{\pgfqpoint{3.342375in}{0.615941in}}%
\pgfpathlineto{\pgfqpoint{3.357955in}{0.612516in}}%
\pgfpathlineto{\pgfqpoint{3.373534in}{0.615941in}}%
\pgfpathlineto{\pgfqpoint{3.389113in}{0.612516in}}%
\pgfpathlineto{\pgfqpoint{3.404692in}{0.615941in}}%
\pgfpathlineto{\pgfqpoint{3.420271in}{0.612516in}}%
\pgfpathlineto{\pgfqpoint{3.435850in}{0.619367in}}%
\pgfpathlineto{\pgfqpoint{3.451430in}{0.622792in}}%
\pgfpathlineto{\pgfqpoint{3.467009in}{0.615941in}}%
\pgfpathlineto{\pgfqpoint{3.482588in}{0.615941in}}%
\pgfpathlineto{\pgfqpoint{3.498167in}{0.622792in}}%
\pgfpathlineto{\pgfqpoint{3.513746in}{0.626217in}}%
\pgfpathlineto{\pgfqpoint{3.529326in}{0.619367in}}%
\pgfpathlineto{\pgfqpoint{3.544905in}{0.626217in}}%
\pgfpathlineto{\pgfqpoint{3.560484in}{0.622792in}}%
\pgfpathlineto{\pgfqpoint{3.576063in}{0.612516in}}%
\pgfpathlineto{\pgfqpoint{3.591642in}{0.622792in}}%
\pgfpathlineto{\pgfqpoint{3.607221in}{0.615941in}}%
\pgfpathlineto{\pgfqpoint{3.622801in}{0.612516in}}%
\pgfpathlineto{\pgfqpoint{3.638380in}{0.619367in}}%
\pgfpathlineto{\pgfqpoint{3.669538in}{0.619367in}}%
\pgfpathlineto{\pgfqpoint{3.685117in}{0.612516in}}%
\pgfpathlineto{\pgfqpoint{3.700696in}{0.615941in}}%
\pgfpathlineto{\pgfqpoint{3.716276in}{0.612516in}}%
\pgfpathlineto{\pgfqpoint{3.747434in}{0.619367in}}%
\pgfpathlineto{\pgfqpoint{3.825330in}{0.619367in}}%
\pgfpathlineto{\pgfqpoint{3.840909in}{0.615941in}}%
\pgfpathlineto{\pgfqpoint{3.856488in}{0.619367in}}%
\pgfpathlineto{\pgfqpoint{3.872067in}{0.609091in}}%
\pgfpathlineto{\pgfqpoint{3.887647in}{0.615941in}}%
\pgfpathlineto{\pgfqpoint{3.903226in}{0.612516in}}%
\pgfpathlineto{\pgfqpoint{3.918805in}{0.626217in}}%
\pgfpathlineto{\pgfqpoint{3.934384in}{0.615941in}}%
\pgfpathlineto{\pgfqpoint{3.949963in}{0.612516in}}%
\pgfpathlineto{\pgfqpoint{3.965543in}{0.619367in}}%
\pgfpathlineto{\pgfqpoint{3.981122in}{0.612516in}}%
\pgfpathlineto{\pgfqpoint{3.996701in}{0.619367in}}%
\pgfpathlineto{\pgfqpoint{4.012280in}{0.629643in}}%
\pgfpathlineto{\pgfqpoint{4.027859in}{0.622792in}}%
\pgfpathlineto{\pgfqpoint{4.043438in}{0.612516in}}%
\pgfpathlineto{\pgfqpoint{4.059018in}{0.619367in}}%
\pgfpathlineto{\pgfqpoint{4.074597in}{0.615941in}}%
\pgfpathlineto{\pgfqpoint{4.090176in}{0.615941in}}%
\pgfpathlineto{\pgfqpoint{4.136913in}{0.626217in}}%
\pgfpathlineto{\pgfqpoint{4.152493in}{0.622792in}}%
\pgfpathlineto{\pgfqpoint{4.168072in}{0.609091in}}%
\pgfpathlineto{\pgfqpoint{4.183651in}{0.629643in}}%
\pgfpathlineto{\pgfqpoint{4.199230in}{0.622792in}}%
\pgfpathlineto{\pgfqpoint{4.214809in}{0.622792in}}%
\pgfpathlineto{\pgfqpoint{4.230389in}{0.615941in}}%
\pgfpathlineto{\pgfqpoint{4.245968in}{0.615941in}}%
\pgfpathlineto{\pgfqpoint{4.261547in}{0.622792in}}%
\pgfpathlineto{\pgfqpoint{4.277126in}{0.619367in}}%
\pgfpathlineto{\pgfqpoint{4.292705in}{0.612516in}}%
\pgfpathlineto{\pgfqpoint{4.308284in}{0.619367in}}%
\pgfpathlineto{\pgfqpoint{4.323864in}{0.619367in}}%
\pgfpathlineto{\pgfqpoint{4.339443in}{0.612516in}}%
\pgfpathlineto{\pgfqpoint{4.355022in}{0.619367in}}%
\pgfpathlineto{\pgfqpoint{4.370601in}{0.622792in}}%
\pgfpathlineto{\pgfqpoint{4.386180in}{0.615941in}}%
\pgfpathlineto{\pgfqpoint{4.401760in}{0.619367in}}%
\pgfpathlineto{\pgfqpoint{4.417339in}{0.626217in}}%
\pgfpathlineto{\pgfqpoint{4.432918in}{0.626217in}}%
\pgfpathlineto{\pgfqpoint{4.448497in}{0.615941in}}%
\pgfpathlineto{\pgfqpoint{4.464076in}{0.615941in}}%
\pgfpathlineto{\pgfqpoint{4.479655in}{0.619367in}}%
\pgfpathlineto{\pgfqpoint{4.495235in}{0.615941in}}%
\pgfpathlineto{\pgfqpoint{4.510814in}{0.615941in}}%
\pgfpathlineto{\pgfqpoint{4.526393in}{0.619367in}}%
\pgfpathlineto{\pgfqpoint{4.541972in}{0.619367in}}%
\pgfpathlineto{\pgfqpoint{4.557551in}{0.615941in}}%
\pgfpathlineto{\pgfqpoint{4.573130in}{0.619367in}}%
\pgfpathlineto{\pgfqpoint{4.588710in}{0.612516in}}%
\pgfpathlineto{\pgfqpoint{4.604289in}{0.612516in}}%
\pgfpathlineto{\pgfqpoint{4.619868in}{0.619367in}}%
\pgfpathlineto{\pgfqpoint{4.635447in}{0.619367in}}%
\pgfpathlineto{\pgfqpoint{4.651026in}{0.609091in}}%
\pgfpathlineto{\pgfqpoint{4.666606in}{0.619367in}}%
\pgfpathlineto{\pgfqpoint{4.682185in}{0.626217in}}%
\pgfpathlineto{\pgfqpoint{4.697764in}{0.622792in}}%
\pgfpathlineto{\pgfqpoint{4.713343in}{0.615941in}}%
\pgfpathlineto{\pgfqpoint{4.728922in}{0.612516in}}%
\pgfpathlineto{\pgfqpoint{4.760081in}{0.626217in}}%
\pgfpathlineto{\pgfqpoint{4.775660in}{0.619367in}}%
\pgfpathlineto{\pgfqpoint{4.791239in}{0.619367in}}%
\pgfpathlineto{\pgfqpoint{4.806818in}{0.622792in}}%
\pgfpathlineto{\pgfqpoint{4.806818in}{0.622792in}}%
\pgfusepath{stroke}%
\end{pgfscope}%
\begin{pgfscope}%
\pgfpathrectangle{\pgfqpoint{0.750000in}{0.525000in}}{\pgfqpoint{4.250000in}{1.850000in}}%
\pgfusepath{clip}%
\pgfsetroundcap%
\pgfsetroundjoin%
\pgfsetlinewidth{1.505625pt}%
\definecolor{currentstroke}{rgb}{1.000000,0.341176,0.133333}%
\pgfsetstrokecolor{currentstroke}%
\pgfsetdash{}{0pt}%
\pgfpathmoveto{\pgfqpoint{0.943182in}{0.633068in}}%
\pgfpathlineto{\pgfqpoint{0.989919in}{0.653620in}}%
\pgfpathlineto{\pgfqpoint{1.005499in}{0.653620in}}%
\pgfpathlineto{\pgfqpoint{1.036657in}{0.667321in}}%
\pgfpathlineto{\pgfqpoint{1.052236in}{0.667321in}}%
\pgfpathlineto{\pgfqpoint{1.504032in}{0.865988in}}%
\pgfpathlineto{\pgfqpoint{1.519611in}{0.865988in}}%
\pgfpathlineto{\pgfqpoint{1.644245in}{0.920792in}}%
\pgfpathlineto{\pgfqpoint{1.659824in}{0.920792in}}%
\pgfpathlineto{\pgfqpoint{1.722141in}{0.948195in}}%
\pgfpathlineto{\pgfqpoint{1.737720in}{0.948195in}}%
\pgfpathlineto{\pgfqpoint{2.049304in}{1.085206in}}%
\pgfpathlineto{\pgfqpoint{2.064883in}{1.085206in}}%
\pgfpathlineto{\pgfqpoint{4.806818in}{2.290909in}}%
\pgfpathlineto{\pgfqpoint{4.806818in}{2.290909in}}%
\pgfusepath{stroke}%
\end{pgfscope}%
\begin{pgfscope}%
\pgfpathrectangle{\pgfqpoint{0.750000in}{0.525000in}}{\pgfqpoint{4.250000in}{1.850000in}}%
\pgfusepath{clip}%
\pgfsetroundcap%
\pgfsetroundjoin%
\pgfsetlinewidth{1.505625pt}%
\definecolor{currentstroke}{rgb}{0.545098,0.764706,0.290196}%
\pgfsetstrokecolor{currentstroke}%
\pgfsetdash{}{0pt}%
\pgfpathmoveto{\pgfqpoint{0.943182in}{0.629643in}}%
\pgfpathlineto{\pgfqpoint{0.958761in}{0.639919in}}%
\pgfpathlineto{\pgfqpoint{0.989919in}{0.646769in}}%
\pgfpathlineto{\pgfqpoint{1.005499in}{0.643344in}}%
\pgfpathlineto{\pgfqpoint{1.021078in}{0.650194in}}%
\pgfpathlineto{\pgfqpoint{1.036657in}{0.646769in}}%
\pgfpathlineto{\pgfqpoint{1.052236in}{0.646769in}}%
\pgfpathlineto{\pgfqpoint{1.067815in}{0.650194in}}%
\pgfpathlineto{\pgfqpoint{1.083394in}{0.646769in}}%
\pgfpathlineto{\pgfqpoint{1.098974in}{0.663896in}}%
\pgfpathlineto{\pgfqpoint{1.114553in}{0.660470in}}%
\pgfpathlineto{\pgfqpoint{1.130132in}{0.663896in}}%
\pgfpathlineto{\pgfqpoint{1.145711in}{0.663896in}}%
\pgfpathlineto{\pgfqpoint{1.161290in}{0.667321in}}%
\pgfpathlineto{\pgfqpoint{1.176870in}{0.663896in}}%
\pgfpathlineto{\pgfqpoint{1.192449in}{0.681022in}}%
\pgfpathlineto{\pgfqpoint{1.208028in}{0.670746in}}%
\pgfpathlineto{\pgfqpoint{1.223607in}{0.677597in}}%
\pgfpathlineto{\pgfqpoint{1.239186in}{0.687873in}}%
\pgfpathlineto{\pgfqpoint{1.254765in}{0.677597in}}%
\pgfpathlineto{\pgfqpoint{1.270345in}{0.663896in}}%
\pgfpathlineto{\pgfqpoint{1.285924in}{0.667321in}}%
\pgfpathlineto{\pgfqpoint{1.301503in}{0.684447in}}%
\pgfpathlineto{\pgfqpoint{1.317082in}{0.677597in}}%
\pgfpathlineto{\pgfqpoint{1.332661in}{0.653620in}}%
\pgfpathlineto{\pgfqpoint{1.348240in}{0.670746in}}%
\pgfpathlineto{\pgfqpoint{1.363820in}{0.684447in}}%
\pgfpathlineto{\pgfqpoint{1.379399in}{0.667321in}}%
\pgfpathlineto{\pgfqpoint{1.394978in}{0.670746in}}%
\pgfpathlineto{\pgfqpoint{1.410557in}{0.691298in}}%
\pgfpathlineto{\pgfqpoint{1.426136in}{0.674171in}}%
\pgfpathlineto{\pgfqpoint{1.441716in}{0.674171in}}%
\pgfpathlineto{\pgfqpoint{1.457295in}{0.677597in}}%
\pgfpathlineto{\pgfqpoint{1.472874in}{0.667321in}}%
\pgfpathlineto{\pgfqpoint{1.488453in}{0.687873in}}%
\pgfpathlineto{\pgfqpoint{1.504032in}{0.701574in}}%
\pgfpathlineto{\pgfqpoint{1.519611in}{0.670746in}}%
\pgfpathlineto{\pgfqpoint{1.535191in}{0.681022in}}%
\pgfpathlineto{\pgfqpoint{1.550770in}{0.670746in}}%
\pgfpathlineto{\pgfqpoint{1.566349in}{0.691298in}}%
\pgfpathlineto{\pgfqpoint{1.581928in}{0.670746in}}%
\pgfpathlineto{\pgfqpoint{1.597507in}{0.698148in}}%
\pgfpathlineto{\pgfqpoint{1.613087in}{0.677597in}}%
\pgfpathlineto{\pgfqpoint{1.644245in}{0.684447in}}%
\pgfpathlineto{\pgfqpoint{1.659824in}{0.677597in}}%
\pgfpathlineto{\pgfqpoint{1.675403in}{0.677597in}}%
\pgfpathlineto{\pgfqpoint{1.690982in}{0.681022in}}%
\pgfpathlineto{\pgfqpoint{1.706562in}{0.681022in}}%
\pgfpathlineto{\pgfqpoint{1.722141in}{0.694723in}}%
\pgfpathlineto{\pgfqpoint{1.737720in}{0.704999in}}%
\pgfpathlineto{\pgfqpoint{1.753299in}{0.677597in}}%
\pgfpathlineto{\pgfqpoint{1.800037in}{0.687873in}}%
\pgfpathlineto{\pgfqpoint{1.815616in}{0.694723in}}%
\pgfpathlineto{\pgfqpoint{1.846774in}{0.681022in}}%
\pgfpathlineto{\pgfqpoint{1.862353in}{0.691298in}}%
\pgfpathlineto{\pgfqpoint{1.877933in}{0.657045in}}%
\pgfpathlineto{\pgfqpoint{1.893512in}{0.681022in}}%
\pgfpathlineto{\pgfqpoint{1.909091in}{0.677597in}}%
\pgfpathlineto{\pgfqpoint{1.924670in}{0.687873in}}%
\pgfpathlineto{\pgfqpoint{1.940249in}{0.681022in}}%
\pgfpathlineto{\pgfqpoint{1.955828in}{0.677597in}}%
\pgfpathlineto{\pgfqpoint{1.971408in}{0.701574in}}%
\pgfpathlineto{\pgfqpoint{1.986987in}{0.677597in}}%
\pgfpathlineto{\pgfqpoint{2.002566in}{0.667321in}}%
\pgfpathlineto{\pgfqpoint{2.018145in}{0.715275in}}%
\pgfpathlineto{\pgfqpoint{2.033724in}{0.698148in}}%
\pgfpathlineto{\pgfqpoint{2.049304in}{0.711850in}}%
\pgfpathlineto{\pgfqpoint{2.064883in}{0.677597in}}%
\pgfpathlineto{\pgfqpoint{2.080462in}{0.711850in}}%
\pgfpathlineto{\pgfqpoint{2.096041in}{0.691298in}}%
\pgfpathlineto{\pgfqpoint{2.111620in}{0.687873in}}%
\pgfpathlineto{\pgfqpoint{2.127199in}{0.701574in}}%
\pgfpathlineto{\pgfqpoint{2.142779in}{0.728976in}}%
\pgfpathlineto{\pgfqpoint{2.158358in}{0.684447in}}%
\pgfpathlineto{\pgfqpoint{2.173937in}{0.670746in}}%
\pgfpathlineto{\pgfqpoint{2.205095in}{0.711850in}}%
\pgfpathlineto{\pgfqpoint{2.220674in}{0.677597in}}%
\pgfpathlineto{\pgfqpoint{2.236254in}{0.708424in}}%
\pgfpathlineto{\pgfqpoint{2.251833in}{0.667321in}}%
\pgfpathlineto{\pgfqpoint{2.267412in}{0.728976in}}%
\pgfpathlineto{\pgfqpoint{2.282991in}{0.681022in}}%
\pgfpathlineto{\pgfqpoint{2.298570in}{0.722126in}}%
\pgfpathlineto{\pgfqpoint{2.314150in}{0.694723in}}%
\pgfpathlineto{\pgfqpoint{2.329729in}{0.684447in}}%
\pgfpathlineto{\pgfqpoint{2.360887in}{0.725551in}}%
\pgfpathlineto{\pgfqpoint{2.376466in}{0.670746in}}%
\pgfpathlineto{\pgfqpoint{2.392045in}{0.718700in}}%
\pgfpathlineto{\pgfqpoint{2.407625in}{0.694723in}}%
\pgfpathlineto{\pgfqpoint{2.423204in}{0.684447in}}%
\pgfpathlineto{\pgfqpoint{2.438783in}{0.687873in}}%
\pgfpathlineto{\pgfqpoint{2.454362in}{0.701574in}}%
\pgfpathlineto{\pgfqpoint{2.469941in}{0.746103in}}%
\pgfpathlineto{\pgfqpoint{2.485521in}{0.698148in}}%
\pgfpathlineto{\pgfqpoint{2.501100in}{0.704999in}}%
\pgfpathlineto{\pgfqpoint{2.516679in}{0.701574in}}%
\pgfpathlineto{\pgfqpoint{2.532258in}{0.670746in}}%
\pgfpathlineto{\pgfqpoint{2.547837in}{0.701574in}}%
\pgfpathlineto{\pgfqpoint{2.563416in}{0.691298in}}%
\pgfpathlineto{\pgfqpoint{2.578996in}{0.684447in}}%
\pgfpathlineto{\pgfqpoint{2.594575in}{0.687873in}}%
\pgfpathlineto{\pgfqpoint{2.610154in}{0.701574in}}%
\pgfpathlineto{\pgfqpoint{2.625733in}{0.674171in}}%
\pgfpathlineto{\pgfqpoint{2.641312in}{0.694723in}}%
\pgfpathlineto{\pgfqpoint{2.656891in}{0.674171in}}%
\pgfpathlineto{\pgfqpoint{2.672471in}{0.715275in}}%
\pgfpathlineto{\pgfqpoint{2.688050in}{0.691298in}}%
\pgfpathlineto{\pgfqpoint{2.703629in}{0.704999in}}%
\pgfpathlineto{\pgfqpoint{2.719208in}{0.681022in}}%
\pgfpathlineto{\pgfqpoint{2.734787in}{0.715275in}}%
\pgfpathlineto{\pgfqpoint{2.750367in}{0.677597in}}%
\pgfpathlineto{\pgfqpoint{2.765946in}{0.698148in}}%
\pgfpathlineto{\pgfqpoint{2.781525in}{0.677597in}}%
\pgfpathlineto{\pgfqpoint{2.797104in}{0.687873in}}%
\pgfpathlineto{\pgfqpoint{2.812683in}{0.732401in}}%
\pgfpathlineto{\pgfqpoint{2.828262in}{0.708424in}}%
\pgfpathlineto{\pgfqpoint{2.843842in}{0.718700in}}%
\pgfpathlineto{\pgfqpoint{2.859421in}{0.694723in}}%
\pgfpathlineto{\pgfqpoint{2.875000in}{0.694723in}}%
\pgfpathlineto{\pgfqpoint{2.890579in}{0.704999in}}%
\pgfpathlineto{\pgfqpoint{2.906158in}{0.698148in}}%
\pgfpathlineto{\pgfqpoint{2.921738in}{0.746103in}}%
\pgfpathlineto{\pgfqpoint{2.937317in}{0.718700in}}%
\pgfpathlineto{\pgfqpoint{2.952896in}{0.711850in}}%
\pgfpathlineto{\pgfqpoint{2.968475in}{0.687873in}}%
\pgfpathlineto{\pgfqpoint{2.984054in}{0.660470in}}%
\pgfpathlineto{\pgfqpoint{2.999633in}{0.722126in}}%
\pgfpathlineto{\pgfqpoint{3.015213in}{0.681022in}}%
\pgfpathlineto{\pgfqpoint{3.030792in}{0.681022in}}%
\pgfpathlineto{\pgfqpoint{3.093109in}{0.708424in}}%
\pgfpathlineto{\pgfqpoint{3.108688in}{0.677597in}}%
\pgfpathlineto{\pgfqpoint{3.124267in}{0.728976in}}%
\pgfpathlineto{\pgfqpoint{3.139846in}{0.715275in}}%
\pgfpathlineto{\pgfqpoint{3.155425in}{0.691298in}}%
\pgfpathlineto{\pgfqpoint{3.171004in}{0.704999in}}%
\pgfpathlineto{\pgfqpoint{3.186584in}{0.681022in}}%
\pgfpathlineto{\pgfqpoint{3.202163in}{0.704999in}}%
\pgfpathlineto{\pgfqpoint{3.217742in}{0.701574in}}%
\pgfpathlineto{\pgfqpoint{3.233321in}{0.711850in}}%
\pgfpathlineto{\pgfqpoint{3.248900in}{0.694723in}}%
\pgfpathlineto{\pgfqpoint{3.264479in}{0.711850in}}%
\pgfpathlineto{\pgfqpoint{3.280059in}{0.752953in}}%
\pgfpathlineto{\pgfqpoint{3.295638in}{0.687873in}}%
\pgfpathlineto{\pgfqpoint{3.311217in}{0.752953in}}%
\pgfpathlineto{\pgfqpoint{3.326796in}{0.711850in}}%
\pgfpathlineto{\pgfqpoint{3.342375in}{0.718700in}}%
\pgfpathlineto{\pgfqpoint{3.373534in}{0.677597in}}%
\pgfpathlineto{\pgfqpoint{3.389113in}{0.684447in}}%
\pgfpathlineto{\pgfqpoint{3.404692in}{0.722126in}}%
\pgfpathlineto{\pgfqpoint{3.420271in}{0.715275in}}%
\pgfpathlineto{\pgfqpoint{3.435850in}{0.718700in}}%
\pgfpathlineto{\pgfqpoint{3.467009in}{0.704999in}}%
\pgfpathlineto{\pgfqpoint{3.482588in}{0.704999in}}%
\pgfpathlineto{\pgfqpoint{3.513746in}{0.739252in}}%
\pgfpathlineto{\pgfqpoint{3.529326in}{0.687873in}}%
\pgfpathlineto{\pgfqpoint{3.544905in}{0.735827in}}%
\pgfpathlineto{\pgfqpoint{3.560484in}{0.722126in}}%
\pgfpathlineto{\pgfqpoint{3.576063in}{0.677597in}}%
\pgfpathlineto{\pgfqpoint{3.591642in}{0.725551in}}%
\pgfpathlineto{\pgfqpoint{3.607221in}{0.691298in}}%
\pgfpathlineto{\pgfqpoint{3.622801in}{0.708424in}}%
\pgfpathlineto{\pgfqpoint{3.638380in}{0.701574in}}%
\pgfpathlineto{\pgfqpoint{3.653959in}{0.674171in}}%
\pgfpathlineto{\pgfqpoint{3.669538in}{0.728976in}}%
\pgfpathlineto{\pgfqpoint{3.685117in}{0.704999in}}%
\pgfpathlineto{\pgfqpoint{3.700696in}{0.698148in}}%
\pgfpathlineto{\pgfqpoint{3.716276in}{0.725551in}}%
\pgfpathlineto{\pgfqpoint{3.731855in}{0.687873in}}%
\pgfpathlineto{\pgfqpoint{3.747434in}{0.732401in}}%
\pgfpathlineto{\pgfqpoint{3.763013in}{0.715275in}}%
\pgfpathlineto{\pgfqpoint{3.778592in}{0.718700in}}%
\pgfpathlineto{\pgfqpoint{3.794172in}{0.691298in}}%
\pgfpathlineto{\pgfqpoint{3.809751in}{0.701574in}}%
\pgfpathlineto{\pgfqpoint{3.825330in}{0.718700in}}%
\pgfpathlineto{\pgfqpoint{3.840909in}{0.704999in}}%
\pgfpathlineto{\pgfqpoint{3.856488in}{0.718700in}}%
\pgfpathlineto{\pgfqpoint{3.872067in}{0.674171in}}%
\pgfpathlineto{\pgfqpoint{3.887647in}{0.708424in}}%
\pgfpathlineto{\pgfqpoint{3.903226in}{0.691298in}}%
\pgfpathlineto{\pgfqpoint{3.918805in}{0.704999in}}%
\pgfpathlineto{\pgfqpoint{3.934384in}{0.715275in}}%
\pgfpathlineto{\pgfqpoint{3.949963in}{0.704999in}}%
\pgfpathlineto{\pgfqpoint{3.965543in}{0.739252in}}%
\pgfpathlineto{\pgfqpoint{3.981122in}{0.711850in}}%
\pgfpathlineto{\pgfqpoint{3.996701in}{0.735827in}}%
\pgfpathlineto{\pgfqpoint{4.012280in}{0.766654in}}%
\pgfpathlineto{\pgfqpoint{4.027859in}{0.715275in}}%
\pgfpathlineto{\pgfqpoint{4.043438in}{0.687873in}}%
\pgfpathlineto{\pgfqpoint{4.059018in}{0.708424in}}%
\pgfpathlineto{\pgfqpoint{4.074597in}{0.718700in}}%
\pgfpathlineto{\pgfqpoint{4.090176in}{0.711850in}}%
\pgfpathlineto{\pgfqpoint{4.105755in}{0.694723in}}%
\pgfpathlineto{\pgfqpoint{4.121334in}{0.735827in}}%
\pgfpathlineto{\pgfqpoint{4.136913in}{0.718700in}}%
\pgfpathlineto{\pgfqpoint{4.152493in}{0.752953in}}%
\pgfpathlineto{\pgfqpoint{4.168072in}{0.701574in}}%
\pgfpathlineto{\pgfqpoint{4.183651in}{0.725551in}}%
\pgfpathlineto{\pgfqpoint{4.199230in}{0.739252in}}%
\pgfpathlineto{\pgfqpoint{4.214809in}{0.677597in}}%
\pgfpathlineto{\pgfqpoint{4.230389in}{0.698148in}}%
\pgfpathlineto{\pgfqpoint{4.245968in}{0.694723in}}%
\pgfpathlineto{\pgfqpoint{4.261547in}{0.773505in}}%
\pgfpathlineto{\pgfqpoint{4.277126in}{0.701574in}}%
\pgfpathlineto{\pgfqpoint{4.308284in}{0.728976in}}%
\pgfpathlineto{\pgfqpoint{4.323864in}{0.722126in}}%
\pgfpathlineto{\pgfqpoint{4.339443in}{0.708424in}}%
\pgfpathlineto{\pgfqpoint{4.355022in}{0.715275in}}%
\pgfpathlineto{\pgfqpoint{4.370601in}{0.715275in}}%
\pgfpathlineto{\pgfqpoint{4.386180in}{0.728976in}}%
\pgfpathlineto{\pgfqpoint{4.401760in}{0.704999in}}%
\pgfpathlineto{\pgfqpoint{4.417339in}{0.746103in}}%
\pgfpathlineto{\pgfqpoint{4.432918in}{0.722126in}}%
\pgfpathlineto{\pgfqpoint{4.448497in}{0.739252in}}%
\pgfpathlineto{\pgfqpoint{4.464076in}{0.694723in}}%
\pgfpathlineto{\pgfqpoint{4.479655in}{0.722126in}}%
\pgfpathlineto{\pgfqpoint{4.495235in}{0.663896in}}%
\pgfpathlineto{\pgfqpoint{4.510814in}{0.725551in}}%
\pgfpathlineto{\pgfqpoint{4.526393in}{0.725551in}}%
\pgfpathlineto{\pgfqpoint{4.541972in}{0.718700in}}%
\pgfpathlineto{\pgfqpoint{4.588710in}{0.708424in}}%
\pgfpathlineto{\pgfqpoint{4.604289in}{0.708424in}}%
\pgfpathlineto{\pgfqpoint{4.619868in}{0.763229in}}%
\pgfpathlineto{\pgfqpoint{4.635447in}{0.722126in}}%
\pgfpathlineto{\pgfqpoint{4.651026in}{0.722126in}}%
\pgfpathlineto{\pgfqpoint{4.666606in}{0.718700in}}%
\pgfpathlineto{\pgfqpoint{4.682185in}{0.718700in}}%
\pgfpathlineto{\pgfqpoint{4.697764in}{0.701574in}}%
\pgfpathlineto{\pgfqpoint{4.713343in}{0.749528in}}%
\pgfpathlineto{\pgfqpoint{4.728922in}{0.725551in}}%
\pgfpathlineto{\pgfqpoint{4.744501in}{0.725551in}}%
\pgfpathlineto{\pgfqpoint{4.760081in}{0.698148in}}%
\pgfpathlineto{\pgfqpoint{4.775660in}{0.694723in}}%
\pgfpathlineto{\pgfqpoint{4.791239in}{0.746103in}}%
\pgfpathlineto{\pgfqpoint{4.806818in}{0.701574in}}%
\pgfpathlineto{\pgfqpoint{4.806818in}{0.701574in}}%
\pgfusepath{stroke}%
\end{pgfscope}%
\begin{pgfscope}%
\pgfsetrectcap%
\pgfsetmiterjoin%
\pgfsetlinewidth{1.254687pt}%
\definecolor{currentstroke}{rgb}{0.800000,0.800000,0.800000}%
\pgfsetstrokecolor{currentstroke}%
\pgfsetdash{}{0pt}%
\pgfpathmoveto{\pgfqpoint{0.750000in}{0.525000in}}%
\pgfpathlineto{\pgfqpoint{0.750000in}{2.375000in}}%
\pgfusepath{stroke}%
\end{pgfscope}%
\begin{pgfscope}%
\pgfsetrectcap%
\pgfsetmiterjoin%
\pgfsetlinewidth{1.254687pt}%
\definecolor{currentstroke}{rgb}{0.800000,0.800000,0.800000}%
\pgfsetstrokecolor{currentstroke}%
\pgfsetdash{}{0pt}%
\pgfpathmoveto{\pgfqpoint{5.000000in}{0.525000in}}%
\pgfpathlineto{\pgfqpoint{5.000000in}{2.375000in}}%
\pgfusepath{stroke}%
\end{pgfscope}%
\begin{pgfscope}%
\pgfsetrectcap%
\pgfsetmiterjoin%
\pgfsetlinewidth{1.254687pt}%
\definecolor{currentstroke}{rgb}{0.800000,0.800000,0.800000}%
\pgfsetstrokecolor{currentstroke}%
\pgfsetdash{}{0pt}%
\pgfpathmoveto{\pgfqpoint{0.750000in}{0.525000in}}%
\pgfpathlineto{\pgfqpoint{5.000000in}{0.525000in}}%
\pgfusepath{stroke}%
\end{pgfscope}%
\begin{pgfscope}%
\pgfsetrectcap%
\pgfsetmiterjoin%
\pgfsetlinewidth{1.254687pt}%
\definecolor{currentstroke}{rgb}{0.800000,0.800000,0.800000}%
\pgfsetstrokecolor{currentstroke}%
\pgfsetdash{}{0pt}%
\pgfpathmoveto{\pgfqpoint{0.750000in}{2.375000in}}%
\pgfpathlineto{\pgfqpoint{5.000000in}{2.375000in}}%
\pgfusepath{stroke}%
\end{pgfscope}%
\begin{pgfscope}%
\pgfsetbuttcap%
\pgfsetmiterjoin%
\definecolor{currentfill}{rgb}{1.000000,1.000000,1.000000}%
\pgfsetfillcolor{currentfill}%
\pgfsetfillopacity{0.800000}%
\pgfsetlinewidth{1.003750pt}%
\definecolor{currentstroke}{rgb}{0.800000,0.800000,0.800000}%
\pgfsetstrokecolor{currentstroke}%
\pgfsetstrokeopacity{0.800000}%
\pgfsetdash{}{0pt}%
\pgfpathmoveto{\pgfqpoint{0.856944in}{1.581167in}}%
\pgfpathlineto{\pgfqpoint{2.815250in}{1.581167in}}%
\pgfpathquadraticcurveto{\pgfqpoint{2.845806in}{1.581167in}}{\pgfqpoint{2.845806in}{1.611723in}}%
\pgfpathlineto{\pgfqpoint{2.845806in}{2.268056in}}%
\pgfpathquadraticcurveto{\pgfqpoint{2.845806in}{2.298611in}}{\pgfqpoint{2.815250in}{2.298611in}}%
\pgfpathlineto{\pgfqpoint{0.856944in}{2.298611in}}%
\pgfpathquadraticcurveto{\pgfqpoint{0.826389in}{2.298611in}}{\pgfqpoint{0.826389in}{2.268056in}}%
\pgfpathlineto{\pgfqpoint{0.826389in}{1.611723in}}%
\pgfpathquadraticcurveto{\pgfqpoint{0.826389in}{1.581167in}}{\pgfqpoint{0.856944in}{1.581167in}}%
\pgfpathclose%
\pgfusepath{stroke,fill}%
\end{pgfscope}%
\begin{pgfscope}%
\pgfsetroundcap%
\pgfsetroundjoin%
\pgfsetlinewidth{1.505625pt}%
\definecolor{currentstroke}{rgb}{0.129412,0.588235,0.952941}%
\pgfsetstrokecolor{currentstroke}%
\pgfsetdash{}{0pt}%
\pgfpathmoveto{\pgfqpoint{0.887500in}{2.184028in}}%
\pgfpathlineto{\pgfqpoint{1.193056in}{2.184028in}}%
\pgfusepath{stroke}%
\end{pgfscope}%
\begin{pgfscope}%
\definecolor{textcolor}{rgb}{0.150000,0.150000,0.150000}%
\pgfsetstrokecolor{textcolor}%
\pgfsetfillcolor{textcolor}%
\pgftext[x=1.315278in,y=2.130556in,left,base]{\color{textcolor}\sffamily\fontsize{11.000000}{13.200000}\selectfont vertex constructions}%
\end{pgfscope}%
\begin{pgfscope}%
\pgfsetroundcap%
\pgfsetroundjoin%
\pgfsetlinewidth{1.505625pt}%
\definecolor{currentstroke}{rgb}{1.000000,0.341176,0.133333}%
\pgfsetstrokecolor{currentstroke}%
\pgfsetdash{}{0pt}%
\pgfpathmoveto{\pgfqpoint{0.887500in}{1.963417in}}%
\pgfpathlineto{\pgfqpoint{1.193056in}{1.963417in}}%
\pgfusepath{stroke}%
\end{pgfscope}%
\begin{pgfscope}%
\definecolor{textcolor}{rgb}{0.150000,0.150000,0.150000}%
\pgfsetstrokecolor{textcolor}%
\pgfsetfillcolor{textcolor}%
\pgftext[x=1.315278in,y=1.909945in,left,base]{\color{textcolor}\sffamily\fontsize{11.000000}{13.200000}\selectfont classifications (naive)}%
\end{pgfscope}%
\begin{pgfscope}%
\pgfsetroundcap%
\pgfsetroundjoin%
\pgfsetlinewidth{1.505625pt}%
\definecolor{currentstroke}{rgb}{0.545098,0.764706,0.290196}%
\pgfsetstrokecolor{currentstroke}%
\pgfsetdash{}{0pt}%
\pgfpathmoveto{\pgfqpoint{0.887500in}{1.734098in}}%
\pgfpathlineto{\pgfqpoint{1.193056in}{1.734098in}}%
\pgfusepath{stroke}%
\end{pgfscope}%
\begin{pgfscope}%
\definecolor{textcolor}{rgb}{0.150000,0.150000,0.150000}%
\pgfsetstrokecolor{textcolor}%
\pgfsetfillcolor{textcolor}%
\pgftext[x=1.315278in,y=1.680625in,left,base]{\color{textcolor}\sffamily\fontsize{11.000000}{13.200000}\selectfont classifications (descent)}%
\end{pgfscope}%
\end{pgfpicture}%
\makeatother%
\endgroup%

%% file: figures/max_bsp_size.pgf
%% Creator: Matplotlib, PGF backend
%%
%% To include the figure in your LaTeX document, write
%%   \input{<filename>.pgf}
%%
%% Make sure the required packages are loaded in your preamble
%%   \usepackage{pgf}
%%
%% and, on pdftex
%%   \usepackage[utf8]{inputenc}\DeclareUnicodeCharacter{2212}{-}
%%
%% or, on luatex and xetex
%%   \usepackage{unicode-math}
%%
%% Figures using additional raster images can only be included by \input if
%% they are in the same directory as the main LaTeX file. For loading figures
%% from other directories you can use the `import` package
%% and then include the figures with
%%   \import{<path to file>}{<filename>.pgf}
%%
%% Matplotlib used the following preamble
%%   \usepackage{fontspec}
%%
\begingroup%
\makeatletter%
\begin{pgfpicture}%
\pgfpathrectangle{\pgfpointorigin}{\pgfqpoint{5.000000in}{3.000000in}}%
\pgfusepath{use as bounding box, clip}%
\begin{pgfscope}%
\pgfsetbuttcap%
\pgfsetmiterjoin%
\definecolor{currentfill}{rgb}{1.000000,1.000000,1.000000}%
\pgfsetfillcolor{currentfill}%
\pgfsetlinewidth{0.000000pt}%
\definecolor{currentstroke}{rgb}{1.000000,1.000000,1.000000}%
\pgfsetstrokecolor{currentstroke}%
\pgfsetdash{}{0pt}%
\pgfpathmoveto{\pgfqpoint{0.000000in}{0.000000in}}%
\pgfpathlineto{\pgfqpoint{5.000000in}{0.000000in}}%
\pgfpathlineto{\pgfqpoint{5.000000in}{3.000000in}}%
\pgfpathlineto{\pgfqpoint{0.000000in}{3.000000in}}%
\pgfpathclose%
\pgfusepath{fill}%
\end{pgfscope}%
\begin{pgfscope}%
\pgfsetbuttcap%
\pgfsetmiterjoin%
\definecolor{currentfill}{rgb}{1.000000,1.000000,1.000000}%
\pgfsetfillcolor{currentfill}%
\pgfsetlinewidth{0.000000pt}%
\definecolor{currentstroke}{rgb}{0.000000,0.000000,0.000000}%
\pgfsetstrokecolor{currentstroke}%
\pgfsetstrokeopacity{0.000000}%
\pgfsetdash{}{0pt}%
\pgfpathmoveto{\pgfqpoint{0.750000in}{0.540000in}}%
\pgfpathlineto{\pgfqpoint{5.000000in}{0.540000in}}%
\pgfpathlineto{\pgfqpoint{5.000000in}{2.850000in}}%
\pgfpathlineto{\pgfqpoint{0.750000in}{2.850000in}}%
\pgfpathclose%
\pgfusepath{fill}%
\end{pgfscope}%
\begin{pgfscope}%
\pgfpathrectangle{\pgfqpoint{0.750000in}{0.540000in}}{\pgfqpoint{4.250000in}{2.310000in}}%
\pgfusepath{clip}%
\pgfsetroundcap%
\pgfsetroundjoin%
\pgfsetlinewidth{1.003750pt}%
\definecolor{currentstroke}{rgb}{0.800000,0.800000,0.800000}%
\pgfsetstrokecolor{currentstroke}%
\pgfsetdash{}{0pt}%
\pgfpathmoveto{\pgfqpoint{1.150625in}{0.540000in}}%
\pgfpathlineto{\pgfqpoint{1.150625in}{2.850000in}}%
\pgfusepath{stroke}%
\end{pgfscope}%
\begin{pgfscope}%
\definecolor{textcolor}{rgb}{0.150000,0.150000,0.150000}%
\pgfsetstrokecolor{textcolor}%
\pgfsetfillcolor{textcolor}%
\pgftext[x=1.150625in,y=0.408056in,,top]{\color{textcolor}\sffamily\fontsize{11.000000}{13.200000}\selectfont 100}%
\end{pgfscope}%
\begin{pgfscope}%
\pgfpathrectangle{\pgfqpoint{0.750000in}{0.540000in}}{\pgfqpoint{4.250000in}{2.310000in}}%
\pgfusepath{clip}%
\pgfsetroundcap%
\pgfsetroundjoin%
\pgfsetlinewidth{1.003750pt}%
\definecolor{currentstroke}{rgb}{0.800000,0.800000,0.800000}%
\pgfsetstrokecolor{currentstroke}%
\pgfsetdash{}{0pt}%
\pgfpathmoveto{\pgfqpoint{1.669234in}{0.540000in}}%
\pgfpathlineto{\pgfqpoint{1.669234in}{2.850000in}}%
\pgfusepath{stroke}%
\end{pgfscope}%
\begin{pgfscope}%
\definecolor{textcolor}{rgb}{0.150000,0.150000,0.150000}%
\pgfsetstrokecolor{textcolor}%
\pgfsetfillcolor{textcolor}%
\pgftext[x=1.669234in,y=0.408056in,,top]{\color{textcolor}\sffamily\fontsize{11.000000}{13.200000}\selectfont 200}%
\end{pgfscope}%
\begin{pgfscope}%
\pgfpathrectangle{\pgfqpoint{0.750000in}{0.540000in}}{\pgfqpoint{4.250000in}{2.310000in}}%
\pgfusepath{clip}%
\pgfsetroundcap%
\pgfsetroundjoin%
\pgfsetlinewidth{1.003750pt}%
\definecolor{currentstroke}{rgb}{0.800000,0.800000,0.800000}%
\pgfsetstrokecolor{currentstroke}%
\pgfsetdash{}{0pt}%
\pgfpathmoveto{\pgfqpoint{2.187843in}{0.540000in}}%
\pgfpathlineto{\pgfqpoint{2.187843in}{2.850000in}}%
\pgfusepath{stroke}%
\end{pgfscope}%
\begin{pgfscope}%
\definecolor{textcolor}{rgb}{0.150000,0.150000,0.150000}%
\pgfsetstrokecolor{textcolor}%
\pgfsetfillcolor{textcolor}%
\pgftext[x=2.187843in,y=0.408056in,,top]{\color{textcolor}\sffamily\fontsize{11.000000}{13.200000}\selectfont 300}%
\end{pgfscope}%
\begin{pgfscope}%
\pgfpathrectangle{\pgfqpoint{0.750000in}{0.540000in}}{\pgfqpoint{4.250000in}{2.310000in}}%
\pgfusepath{clip}%
\pgfsetroundcap%
\pgfsetroundjoin%
\pgfsetlinewidth{1.003750pt}%
\definecolor{currentstroke}{rgb}{0.800000,0.800000,0.800000}%
\pgfsetstrokecolor{currentstroke}%
\pgfsetdash{}{0pt}%
\pgfpathmoveto{\pgfqpoint{2.706452in}{0.540000in}}%
\pgfpathlineto{\pgfqpoint{2.706452in}{2.850000in}}%
\pgfusepath{stroke}%
\end{pgfscope}%
\begin{pgfscope}%
\definecolor{textcolor}{rgb}{0.150000,0.150000,0.150000}%
\pgfsetstrokecolor{textcolor}%
\pgfsetfillcolor{textcolor}%
\pgftext[x=2.706452in,y=0.408056in,,top]{\color{textcolor}\sffamily\fontsize{11.000000}{13.200000}\selectfont 400}%
\end{pgfscope}%
\begin{pgfscope}%
\pgfpathrectangle{\pgfqpoint{0.750000in}{0.540000in}}{\pgfqpoint{4.250000in}{2.310000in}}%
\pgfusepath{clip}%
\pgfsetroundcap%
\pgfsetroundjoin%
\pgfsetlinewidth{1.003750pt}%
\definecolor{currentstroke}{rgb}{0.800000,0.800000,0.800000}%
\pgfsetstrokecolor{currentstroke}%
\pgfsetdash{}{0pt}%
\pgfpathmoveto{\pgfqpoint{3.225061in}{0.540000in}}%
\pgfpathlineto{\pgfqpoint{3.225061in}{2.850000in}}%
\pgfusepath{stroke}%
\end{pgfscope}%
\begin{pgfscope}%
\definecolor{textcolor}{rgb}{0.150000,0.150000,0.150000}%
\pgfsetstrokecolor{textcolor}%
\pgfsetfillcolor{textcolor}%
\pgftext[x=3.225061in,y=0.408056in,,top]{\color{textcolor}\sffamily\fontsize{11.000000}{13.200000}\selectfont 500}%
\end{pgfscope}%
\begin{pgfscope}%
\pgfpathrectangle{\pgfqpoint{0.750000in}{0.540000in}}{\pgfqpoint{4.250000in}{2.310000in}}%
\pgfusepath{clip}%
\pgfsetroundcap%
\pgfsetroundjoin%
\pgfsetlinewidth{1.003750pt}%
\definecolor{currentstroke}{rgb}{0.800000,0.800000,0.800000}%
\pgfsetstrokecolor{currentstroke}%
\pgfsetdash{}{0pt}%
\pgfpathmoveto{\pgfqpoint{3.743670in}{0.540000in}}%
\pgfpathlineto{\pgfqpoint{3.743670in}{2.850000in}}%
\pgfusepath{stroke}%
\end{pgfscope}%
\begin{pgfscope}%
\definecolor{textcolor}{rgb}{0.150000,0.150000,0.150000}%
\pgfsetstrokecolor{textcolor}%
\pgfsetfillcolor{textcolor}%
\pgftext[x=3.743670in,y=0.408056in,,top]{\color{textcolor}\sffamily\fontsize{11.000000}{13.200000}\selectfont 600}%
\end{pgfscope}%
\begin{pgfscope}%
\pgfpathrectangle{\pgfqpoint{0.750000in}{0.540000in}}{\pgfqpoint{4.250000in}{2.310000in}}%
\pgfusepath{clip}%
\pgfsetroundcap%
\pgfsetroundjoin%
\pgfsetlinewidth{1.003750pt}%
\definecolor{currentstroke}{rgb}{0.800000,0.800000,0.800000}%
\pgfsetstrokecolor{currentstroke}%
\pgfsetdash{}{0pt}%
\pgfpathmoveto{\pgfqpoint{4.262279in}{0.540000in}}%
\pgfpathlineto{\pgfqpoint{4.262279in}{2.850000in}}%
\pgfusepath{stroke}%
\end{pgfscope}%
\begin{pgfscope}%
\definecolor{textcolor}{rgb}{0.150000,0.150000,0.150000}%
\pgfsetstrokecolor{textcolor}%
\pgfsetfillcolor{textcolor}%
\pgftext[x=4.262279in,y=0.408056in,,top]{\color{textcolor}\sffamily\fontsize{11.000000}{13.200000}\selectfont 700}%
\end{pgfscope}%
\begin{pgfscope}%
\pgfpathrectangle{\pgfqpoint{0.750000in}{0.540000in}}{\pgfqpoint{4.250000in}{2.310000in}}%
\pgfusepath{clip}%
\pgfsetroundcap%
\pgfsetroundjoin%
\pgfsetlinewidth{1.003750pt}%
\definecolor{currentstroke}{rgb}{0.800000,0.800000,0.800000}%
\pgfsetstrokecolor{currentstroke}%
\pgfsetdash{}{0pt}%
\pgfpathmoveto{\pgfqpoint{4.780888in}{0.540000in}}%
\pgfpathlineto{\pgfqpoint{4.780888in}{2.850000in}}%
\pgfusepath{stroke}%
\end{pgfscope}%
\begin{pgfscope}%
\definecolor{textcolor}{rgb}{0.150000,0.150000,0.150000}%
\pgfsetstrokecolor{textcolor}%
\pgfsetfillcolor{textcolor}%
\pgftext[x=4.780888in,y=0.408056in,,top]{\color{textcolor}\sffamily\fontsize{11.000000}{13.200000}\selectfont 800}%
\end{pgfscope}%
\begin{pgfscope}%
\definecolor{textcolor}{rgb}{0.150000,0.150000,0.150000}%
\pgfsetstrokecolor{textcolor}%
\pgfsetfillcolor{textcolor}%
\pgftext[x=2.875000in,y=0.216833in,,top]{\color{textcolor}\sffamily\fontsize{12.000000}{14.400000}\selectfont Maximal BSP size}%
\end{pgfscope}%
\begin{pgfscope}%
\pgfpathrectangle{\pgfqpoint{0.750000in}{0.540000in}}{\pgfqpoint{4.250000in}{2.310000in}}%
\pgfusepath{clip}%
\pgfsetroundcap%
\pgfsetroundjoin%
\pgfsetlinewidth{1.003750pt}%
\definecolor{currentstroke}{rgb}{0.800000,0.800000,0.800000}%
\pgfsetstrokecolor{currentstroke}%
\pgfsetdash{}{0pt}%
\pgfpathmoveto{\pgfqpoint{0.750000in}{0.540000in}}%
\pgfpathlineto{\pgfqpoint{5.000000in}{0.540000in}}%
\pgfusepath{stroke}%
\end{pgfscope}%
\begin{pgfscope}%
\definecolor{textcolor}{rgb}{0.150000,0.150000,0.150000}%
\pgfsetstrokecolor{textcolor}%
\pgfsetfillcolor{textcolor}%
\pgftext[x=0.543194in, y=0.486986in, left, base]{\color{textcolor}\sffamily\fontsize{11.000000}{13.200000}\selectfont 0}%
\end{pgfscope}%
\begin{pgfscope}%
\pgfpathrectangle{\pgfqpoint{0.750000in}{0.540000in}}{\pgfqpoint{4.250000in}{2.310000in}}%
\pgfusepath{clip}%
\pgfsetroundcap%
\pgfsetroundjoin%
\pgfsetlinewidth{1.003750pt}%
\definecolor{currentstroke}{rgb}{0.800000,0.800000,0.800000}%
\pgfsetstrokecolor{currentstroke}%
\pgfsetdash{}{0pt}%
\pgfpathmoveto{\pgfqpoint{0.750000in}{0.969720in}}%
\pgfpathlineto{\pgfqpoint{5.000000in}{0.969720in}}%
\pgfusepath{stroke}%
\end{pgfscope}%
\begin{pgfscope}%
\definecolor{textcolor}{rgb}{0.150000,0.150000,0.150000}%
\pgfsetstrokecolor{textcolor}%
\pgfsetfillcolor{textcolor}%
\pgftext[x=0.393472in, y=0.916706in, left, base]{\color{textcolor}\sffamily\fontsize{11.000000}{13.200000}\selectfont 100}%
\end{pgfscope}%
\begin{pgfscope}%
\pgfpathrectangle{\pgfqpoint{0.750000in}{0.540000in}}{\pgfqpoint{4.250000in}{2.310000in}}%
\pgfusepath{clip}%
\pgfsetroundcap%
\pgfsetroundjoin%
\pgfsetlinewidth{1.003750pt}%
\definecolor{currentstroke}{rgb}{0.800000,0.800000,0.800000}%
\pgfsetstrokecolor{currentstroke}%
\pgfsetdash{}{0pt}%
\pgfpathmoveto{\pgfqpoint{0.750000in}{1.399441in}}%
\pgfpathlineto{\pgfqpoint{5.000000in}{1.399441in}}%
\pgfusepath{stroke}%
\end{pgfscope}%
\begin{pgfscope}%
\definecolor{textcolor}{rgb}{0.150000,0.150000,0.150000}%
\pgfsetstrokecolor{textcolor}%
\pgfsetfillcolor{textcolor}%
\pgftext[x=0.393472in, y=1.346427in, left, base]{\color{textcolor}\sffamily\fontsize{11.000000}{13.200000}\selectfont 200}%
\end{pgfscope}%
\begin{pgfscope}%
\pgfpathrectangle{\pgfqpoint{0.750000in}{0.540000in}}{\pgfqpoint{4.250000in}{2.310000in}}%
\pgfusepath{clip}%
\pgfsetroundcap%
\pgfsetroundjoin%
\pgfsetlinewidth{1.003750pt}%
\definecolor{currentstroke}{rgb}{0.800000,0.800000,0.800000}%
\pgfsetstrokecolor{currentstroke}%
\pgfsetdash{}{0pt}%
\pgfpathmoveto{\pgfqpoint{0.750000in}{1.829161in}}%
\pgfpathlineto{\pgfqpoint{5.000000in}{1.829161in}}%
\pgfusepath{stroke}%
\end{pgfscope}%
\begin{pgfscope}%
\definecolor{textcolor}{rgb}{0.150000,0.150000,0.150000}%
\pgfsetstrokecolor{textcolor}%
\pgfsetfillcolor{textcolor}%
\pgftext[x=0.393472in, y=1.776147in, left, base]{\color{textcolor}\sffamily\fontsize{11.000000}{13.200000}\selectfont 300}%
\end{pgfscope}%
\begin{pgfscope}%
\pgfpathrectangle{\pgfqpoint{0.750000in}{0.540000in}}{\pgfqpoint{4.250000in}{2.310000in}}%
\pgfusepath{clip}%
\pgfsetroundcap%
\pgfsetroundjoin%
\pgfsetlinewidth{1.003750pt}%
\definecolor{currentstroke}{rgb}{0.800000,0.800000,0.800000}%
\pgfsetstrokecolor{currentstroke}%
\pgfsetdash{}{0pt}%
\pgfpathmoveto{\pgfqpoint{0.750000in}{2.258881in}}%
\pgfpathlineto{\pgfqpoint{5.000000in}{2.258881in}}%
\pgfusepath{stroke}%
\end{pgfscope}%
\begin{pgfscope}%
\definecolor{textcolor}{rgb}{0.150000,0.150000,0.150000}%
\pgfsetstrokecolor{textcolor}%
\pgfsetfillcolor{textcolor}%
\pgftext[x=0.393472in, y=2.205867in, left, base]{\color{textcolor}\sffamily\fontsize{11.000000}{13.200000}\selectfont 400}%
\end{pgfscope}%
\begin{pgfscope}%
\pgfpathrectangle{\pgfqpoint{0.750000in}{0.540000in}}{\pgfqpoint{4.250000in}{2.310000in}}%
\pgfusepath{clip}%
\pgfsetroundcap%
\pgfsetroundjoin%
\pgfsetlinewidth{1.003750pt}%
\definecolor{currentstroke}{rgb}{0.800000,0.800000,0.800000}%
\pgfsetstrokecolor{currentstroke}%
\pgfsetdash{}{0pt}%
\pgfpathmoveto{\pgfqpoint{0.750000in}{2.688602in}}%
\pgfpathlineto{\pgfqpoint{5.000000in}{2.688602in}}%
\pgfusepath{stroke}%
\end{pgfscope}%
\begin{pgfscope}%
\definecolor{textcolor}{rgb}{0.150000,0.150000,0.150000}%
\pgfsetstrokecolor{textcolor}%
\pgfsetfillcolor{textcolor}%
\pgftext[x=0.393472in, y=2.635588in, left, base]{\color{textcolor}\sffamily\fontsize{11.000000}{13.200000}\selectfont 500}%
\end{pgfscope}%
\begin{pgfscope}%
\definecolor{textcolor}{rgb}{0.150000,0.150000,0.150000}%
\pgfsetstrokecolor{textcolor}%
\pgfsetfillcolor{textcolor}%
\pgftext[x=0.337917in,y=1.695000in,,bottom,rotate=90.000000]{\color{textcolor}\sffamily\fontsize{12.000000}{14.400000}\selectfont Time (seconds)}%
\end{pgfscope}%
\begin{pgfscope}%
\pgfpathrectangle{\pgfqpoint{0.750000in}{0.540000in}}{\pgfqpoint{4.250000in}{2.310000in}}%
\pgfusepath{clip}%
\pgfsetroundcap%
\pgfsetroundjoin%
\pgfsetlinewidth{1.505625pt}%
\definecolor{currentstroke}{rgb}{0.298039,0.447059,0.690196}%
\pgfsetstrokecolor{currentstroke}%
\pgfsetdash{}{0pt}%
\pgfpathmoveto{\pgfqpoint{0.943182in}{2.205596in}}%
\pgfpathlineto{\pgfqpoint{0.969112in}{1.968240in}}%
\pgfpathlineto{\pgfqpoint{0.995043in}{1.845288in}}%
\pgfpathlineto{\pgfqpoint{1.020973in}{1.736359in}}%
\pgfpathlineto{\pgfqpoint{1.046904in}{1.676915in}}%
\pgfpathlineto{\pgfqpoint{1.072834in}{1.627648in}}%
\pgfpathlineto{\pgfqpoint{1.098764in}{1.595836in}}%
\pgfpathlineto{\pgfqpoint{1.124695in}{1.568269in}}%
\pgfpathlineto{\pgfqpoint{1.150625in}{1.546946in}}%
\pgfpathlineto{\pgfqpoint{1.176556in}{1.548609in}}%
\pgfpathlineto{\pgfqpoint{1.202486in}{1.529972in}}%
\pgfpathlineto{\pgfqpoint{1.228417in}{1.513390in}}%
\pgfpathlineto{\pgfqpoint{1.254347in}{1.509831in}}%
\pgfpathlineto{\pgfqpoint{1.280278in}{1.510639in}}%
\pgfpathlineto{\pgfqpoint{1.306208in}{1.522723in}}%
\pgfpathlineto{\pgfqpoint{1.358069in}{1.512706in}}%
\pgfpathlineto{\pgfqpoint{1.383999in}{1.511219in}}%
\pgfpathlineto{\pgfqpoint{1.435860in}{1.520850in}}%
\pgfpathlineto{\pgfqpoint{1.461791in}{1.523028in}}%
\pgfpathlineto{\pgfqpoint{1.487721in}{1.528859in}}%
\pgfpathlineto{\pgfqpoint{1.617373in}{1.566980in}}%
\pgfpathlineto{\pgfqpoint{1.643304in}{1.577173in}}%
\pgfpathlineto{\pgfqpoint{1.669234in}{1.573589in}}%
\pgfpathlineto{\pgfqpoint{1.695165in}{1.581028in}}%
\pgfpathlineto{\pgfqpoint{1.721095in}{1.586584in}}%
\pgfpathlineto{\pgfqpoint{1.772956in}{1.609776in}}%
\pgfpathlineto{\pgfqpoint{1.850747in}{1.639186in}}%
\pgfpathlineto{\pgfqpoint{1.902608in}{1.658111in}}%
\pgfpathlineto{\pgfqpoint{1.928539in}{1.664496in}}%
\pgfpathlineto{\pgfqpoint{1.954469in}{1.675721in}}%
\pgfpathlineto{\pgfqpoint{1.980400in}{1.680890in}}%
\pgfpathlineto{\pgfqpoint{2.006330in}{1.693477in}}%
\pgfpathlineto{\pgfqpoint{2.032261in}{1.698462in}}%
\pgfpathlineto{\pgfqpoint{2.058191in}{1.708805in}}%
\pgfpathlineto{\pgfqpoint{2.084121in}{1.716733in}}%
\pgfpathlineto{\pgfqpoint{2.110052in}{1.726359in}}%
\pgfpathlineto{\pgfqpoint{2.135982in}{1.730910in}}%
\pgfpathlineto{\pgfqpoint{2.161913in}{1.743303in}}%
\pgfpathlineto{\pgfqpoint{2.213774in}{1.758532in}}%
\pgfpathlineto{\pgfqpoint{2.239704in}{1.762455in}}%
\pgfpathlineto{\pgfqpoint{2.317495in}{1.782334in}}%
\pgfpathlineto{\pgfqpoint{2.343426in}{1.785570in}}%
\pgfpathlineto{\pgfqpoint{2.369356in}{1.794392in}}%
\pgfpathlineto{\pgfqpoint{2.395287in}{1.800292in}}%
\pgfpathlineto{\pgfqpoint{2.421217in}{1.808913in}}%
\pgfpathlineto{\pgfqpoint{2.473078in}{1.821916in}}%
\pgfpathlineto{\pgfqpoint{2.524939in}{1.838855in}}%
\pgfpathlineto{\pgfqpoint{2.576800in}{1.847665in}}%
\pgfpathlineto{\pgfqpoint{2.628661in}{1.861141in}}%
\pgfpathlineto{\pgfqpoint{2.654591in}{1.862946in}}%
\pgfpathlineto{\pgfqpoint{2.680522in}{1.872928in}}%
\pgfpathlineto{\pgfqpoint{2.706452in}{1.875210in}}%
\pgfpathlineto{\pgfqpoint{2.732383in}{1.886172in}}%
\pgfpathlineto{\pgfqpoint{2.836104in}{1.909841in}}%
\pgfpathlineto{\pgfqpoint{2.862035in}{1.918023in}}%
\pgfpathlineto{\pgfqpoint{2.887965in}{1.929333in}}%
\pgfpathlineto{\pgfqpoint{2.913896in}{1.931477in}}%
\pgfpathlineto{\pgfqpoint{2.939826in}{1.938619in}}%
\pgfpathlineto{\pgfqpoint{2.965757in}{1.943118in}}%
\pgfpathlineto{\pgfqpoint{2.991687in}{1.951992in}}%
\pgfpathlineto{\pgfqpoint{3.017617in}{1.957905in}}%
\pgfpathlineto{\pgfqpoint{3.043548in}{1.972692in}}%
\pgfpathlineto{\pgfqpoint{3.069478in}{1.977359in}}%
\pgfpathlineto{\pgfqpoint{3.121339in}{1.996318in}}%
\pgfpathlineto{\pgfqpoint{3.147270in}{2.007899in}}%
\pgfpathlineto{\pgfqpoint{3.173200in}{2.014272in}}%
\pgfpathlineto{\pgfqpoint{3.250991in}{2.038400in}}%
\pgfpathlineto{\pgfqpoint{3.276922in}{2.045194in}}%
\pgfpathlineto{\pgfqpoint{3.302852in}{2.060565in}}%
\pgfpathlineto{\pgfqpoint{3.328783in}{2.064355in}}%
\pgfpathlineto{\pgfqpoint{3.354713in}{2.076615in}}%
\pgfpathlineto{\pgfqpoint{3.432505in}{2.103576in}}%
\pgfpathlineto{\pgfqpoint{3.458435in}{2.115075in}}%
\pgfpathlineto{\pgfqpoint{3.510296in}{2.132806in}}%
\pgfpathlineto{\pgfqpoint{3.536226in}{2.145981in}}%
\pgfpathlineto{\pgfqpoint{3.588087in}{2.164162in}}%
\pgfpathlineto{\pgfqpoint{3.614018in}{2.178760in}}%
\pgfpathlineto{\pgfqpoint{3.639948in}{2.188098in}}%
\pgfpathlineto{\pgfqpoint{3.665879in}{2.201170in}}%
\pgfpathlineto{\pgfqpoint{3.743670in}{2.234134in}}%
\pgfpathlineto{\pgfqpoint{3.769600in}{2.247657in}}%
\pgfpathlineto{\pgfqpoint{3.795531in}{2.255315in}}%
\pgfpathlineto{\pgfqpoint{3.821461in}{2.268816in}}%
\pgfpathlineto{\pgfqpoint{3.847392in}{2.275567in}}%
\pgfpathlineto{\pgfqpoint{3.873322in}{2.293585in}}%
\pgfpathlineto{\pgfqpoint{3.899253in}{2.303671in}}%
\pgfpathlineto{\pgfqpoint{3.977044in}{2.346596in}}%
\pgfpathlineto{\pgfqpoint{4.002974in}{2.358469in}}%
\pgfpathlineto{\pgfqpoint{4.028905in}{2.373638in}}%
\pgfpathlineto{\pgfqpoint{4.106696in}{2.411363in}}%
\pgfpathlineto{\pgfqpoint{4.132627in}{2.430653in}}%
\pgfpathlineto{\pgfqpoint{4.158557in}{2.432737in}}%
\pgfpathlineto{\pgfqpoint{4.184487in}{2.455074in}}%
\pgfpathlineto{\pgfqpoint{4.210418in}{2.462500in}}%
\pgfpathlineto{\pgfqpoint{4.236348in}{2.477918in}}%
\pgfpathlineto{\pgfqpoint{4.314140in}{2.514032in}}%
\pgfpathlineto{\pgfqpoint{4.340070in}{2.532643in}}%
\pgfpathlineto{\pgfqpoint{4.366001in}{2.544624in}}%
\pgfpathlineto{\pgfqpoint{4.391931in}{2.562156in}}%
\pgfpathlineto{\pgfqpoint{4.417862in}{2.569470in}}%
\pgfpathlineto{\pgfqpoint{4.443792in}{2.582319in}}%
\pgfpathlineto{\pgfqpoint{4.495653in}{2.612404in}}%
\pgfpathlineto{\pgfqpoint{4.521583in}{2.624865in}}%
\pgfpathlineto{\pgfqpoint{4.547514in}{2.644633in}}%
\pgfpathlineto{\pgfqpoint{4.573444in}{2.651160in}}%
\pgfpathlineto{\pgfqpoint{4.625305in}{2.682276in}}%
\pgfpathlineto{\pgfqpoint{4.651236in}{2.695924in}}%
\pgfpathlineto{\pgfqpoint{4.677166in}{2.718867in}}%
\pgfpathlineto{\pgfqpoint{4.703096in}{2.722928in}}%
\pgfpathlineto{\pgfqpoint{4.754957in}{2.753060in}}%
\pgfpathlineto{\pgfqpoint{4.780888in}{2.764284in}}%
\pgfpathlineto{\pgfqpoint{4.806818in}{2.786182in}}%
\pgfpathlineto{\pgfqpoint{4.806818in}{2.786182in}}%
\pgfusepath{stroke}%
\end{pgfscope}%
\begin{pgfscope}%
\pgfsetrectcap%
\pgfsetmiterjoin%
\pgfsetlinewidth{1.254687pt}%
\definecolor{currentstroke}{rgb}{0.800000,0.800000,0.800000}%
\pgfsetstrokecolor{currentstroke}%
\pgfsetdash{}{0pt}%
\pgfpathmoveto{\pgfqpoint{0.750000in}{0.540000in}}%
\pgfpathlineto{\pgfqpoint{0.750000in}{2.850000in}}%
\pgfusepath{stroke}%
\end{pgfscope}%
\begin{pgfscope}%
\pgfsetrectcap%
\pgfsetmiterjoin%
\pgfsetlinewidth{1.254687pt}%
\definecolor{currentstroke}{rgb}{0.800000,0.800000,0.800000}%
\pgfsetstrokecolor{currentstroke}%
\pgfsetdash{}{0pt}%
\pgfpathmoveto{\pgfqpoint{5.000000in}{0.540000in}}%
\pgfpathlineto{\pgfqpoint{5.000000in}{2.850000in}}%
\pgfusepath{stroke}%
\end{pgfscope}%
\begin{pgfscope}%
\pgfsetrectcap%
\pgfsetmiterjoin%
\pgfsetlinewidth{1.254687pt}%
\definecolor{currentstroke}{rgb}{0.800000,0.800000,0.800000}%
\pgfsetstrokecolor{currentstroke}%
\pgfsetdash{}{0pt}%
\pgfpathmoveto{\pgfqpoint{0.750000in}{0.540000in}}%
\pgfpathlineto{\pgfqpoint{5.000000in}{0.540000in}}%
\pgfusepath{stroke}%
\end{pgfscope}%
\begin{pgfscope}%
\pgfsetrectcap%
\pgfsetmiterjoin%
\pgfsetlinewidth{1.254687pt}%
\definecolor{currentstroke}{rgb}{0.800000,0.800000,0.800000}%
\pgfsetstrokecolor{currentstroke}%
\pgfsetdash{}{0pt}%
\pgfpathmoveto{\pgfqpoint{0.750000in}{2.850000in}}%
\pgfpathlineto{\pgfqpoint{5.000000in}{2.850000in}}%
\pgfusepath{stroke}%
\end{pgfscope}%
\end{pgfpicture}%
\makeatother%
\endgroup%

%% file: paper-05-applications.tex
% !TEX root paper.tex

\section{Applications}
\label{sec:app}
\begin{figure}
    \centerline
    {
        \includegraphics[width=0.3\linewidth]{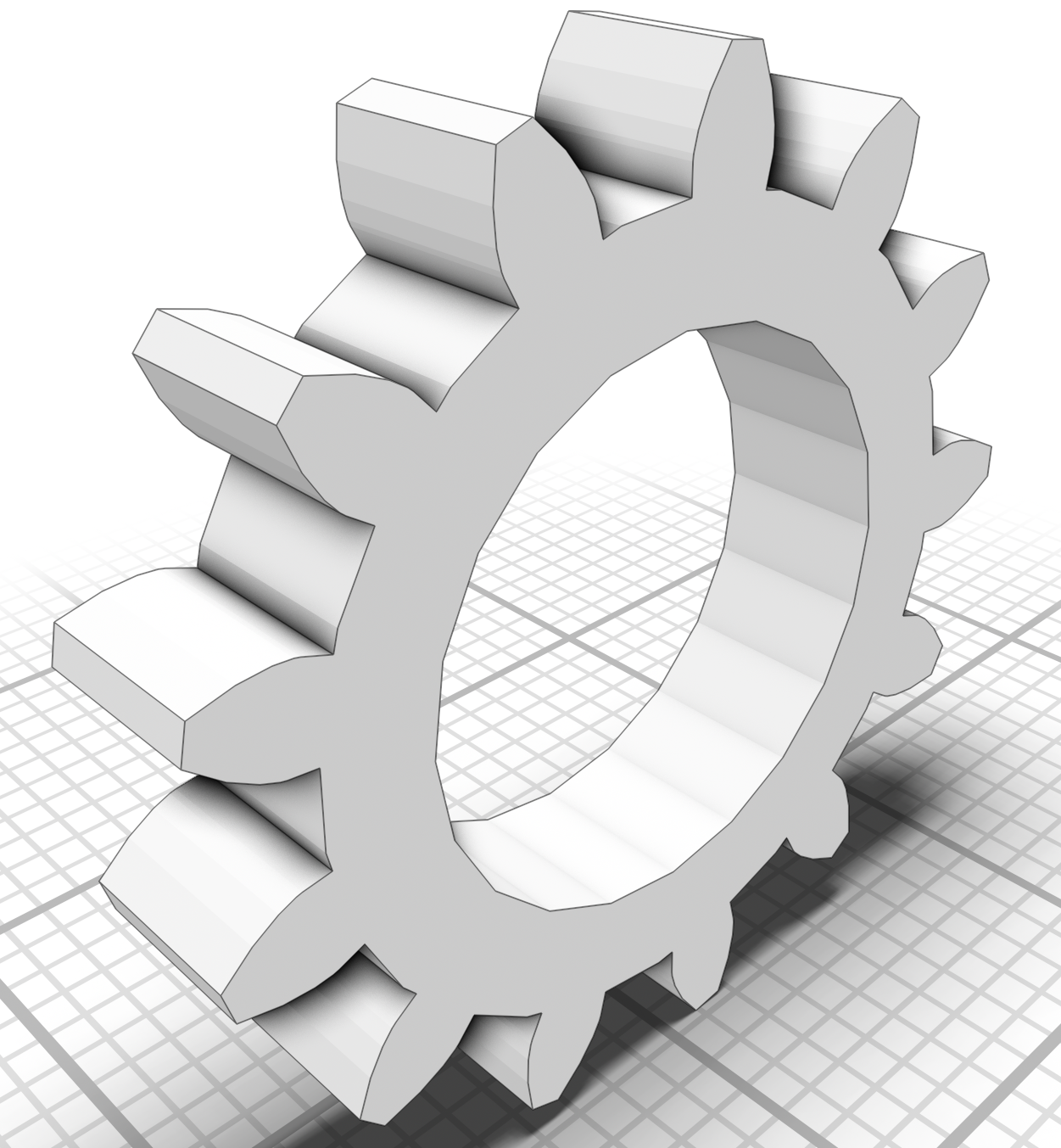}
        \includegraphics[width=0.69\linewidth]{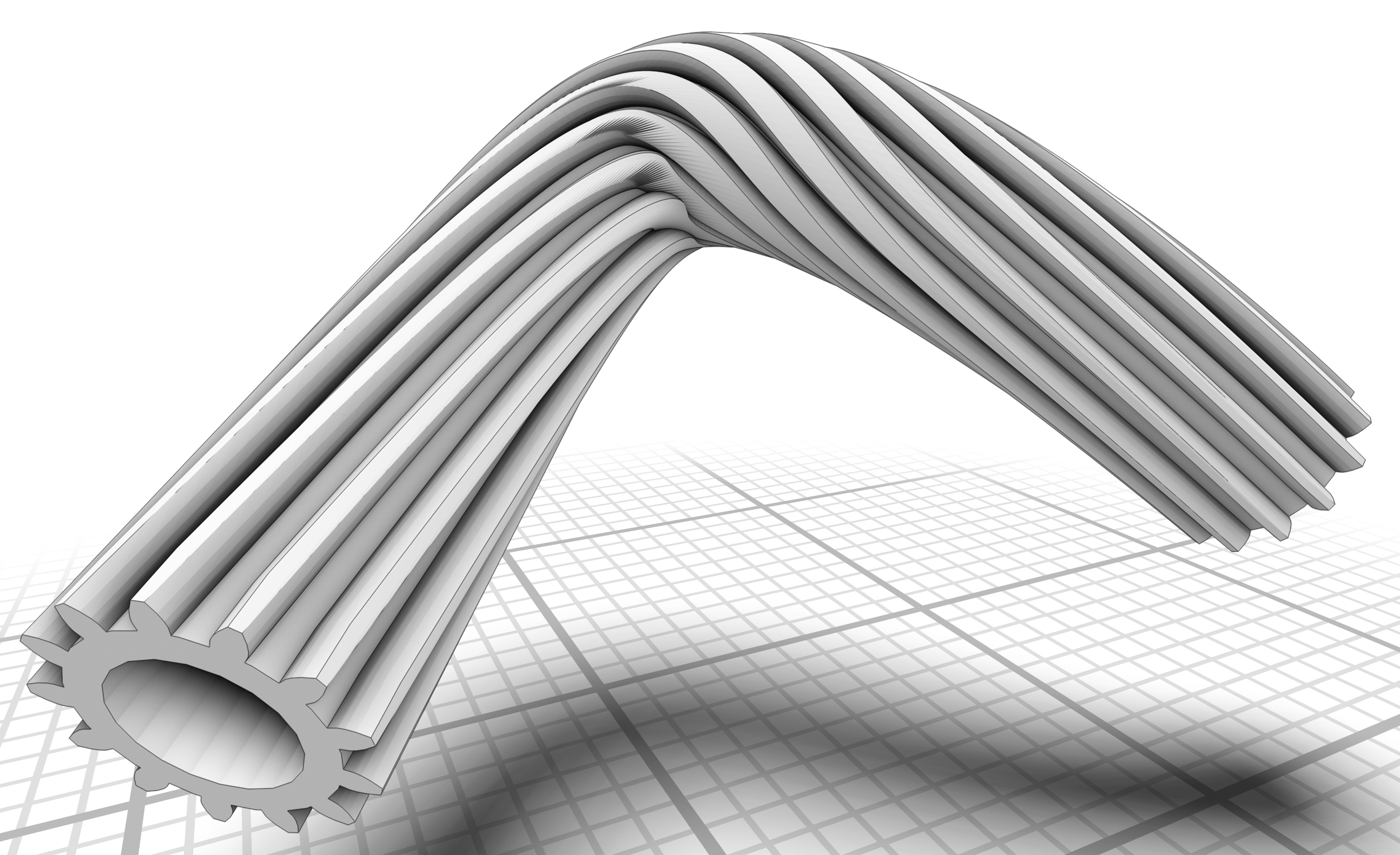}            
    }
    \caption
    {
        Sweep of a gear along a Bezier curve with \num{100} sample positions. 
        This is the result of \num{764501} union operations and took \num{281} seconds to create at a maximal BSP size of \num{300} nodes per octree cell.
    }
    \label{fig:app:sweep-volume}
\end{figure}

\begin{figure}
    \centering
    \subcaptionbox{The drill bit.}{\includegraphics[trim={750px 0 700px 0}, clip, width=0.39\linewidth]{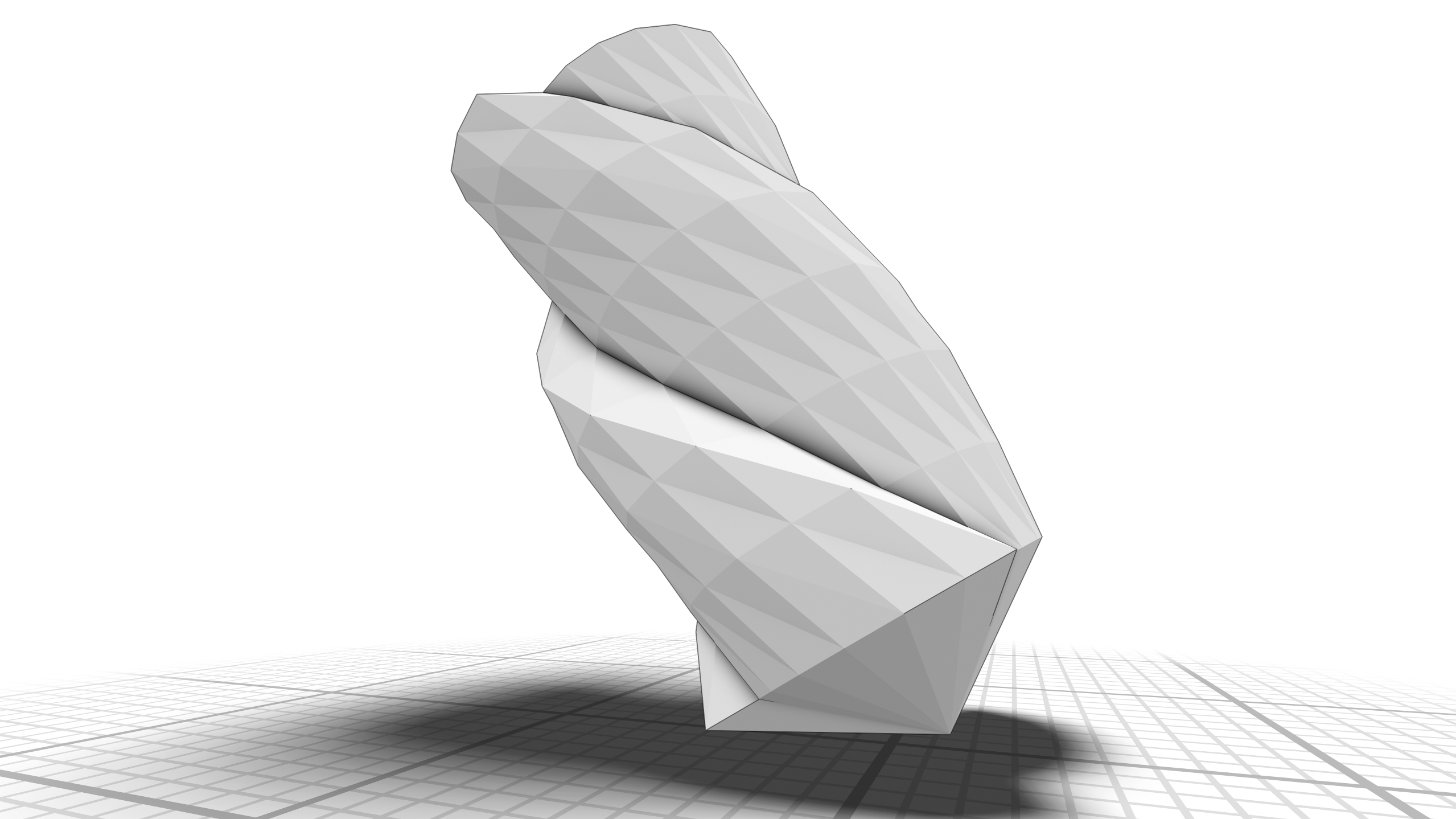}}
    \subcaptionbox{Union of the 90 sweep volumes. Each shade of red is one sweep.}{\includegraphics[trim={550px 0 450px 0}, clip, width=0.59\linewidth]{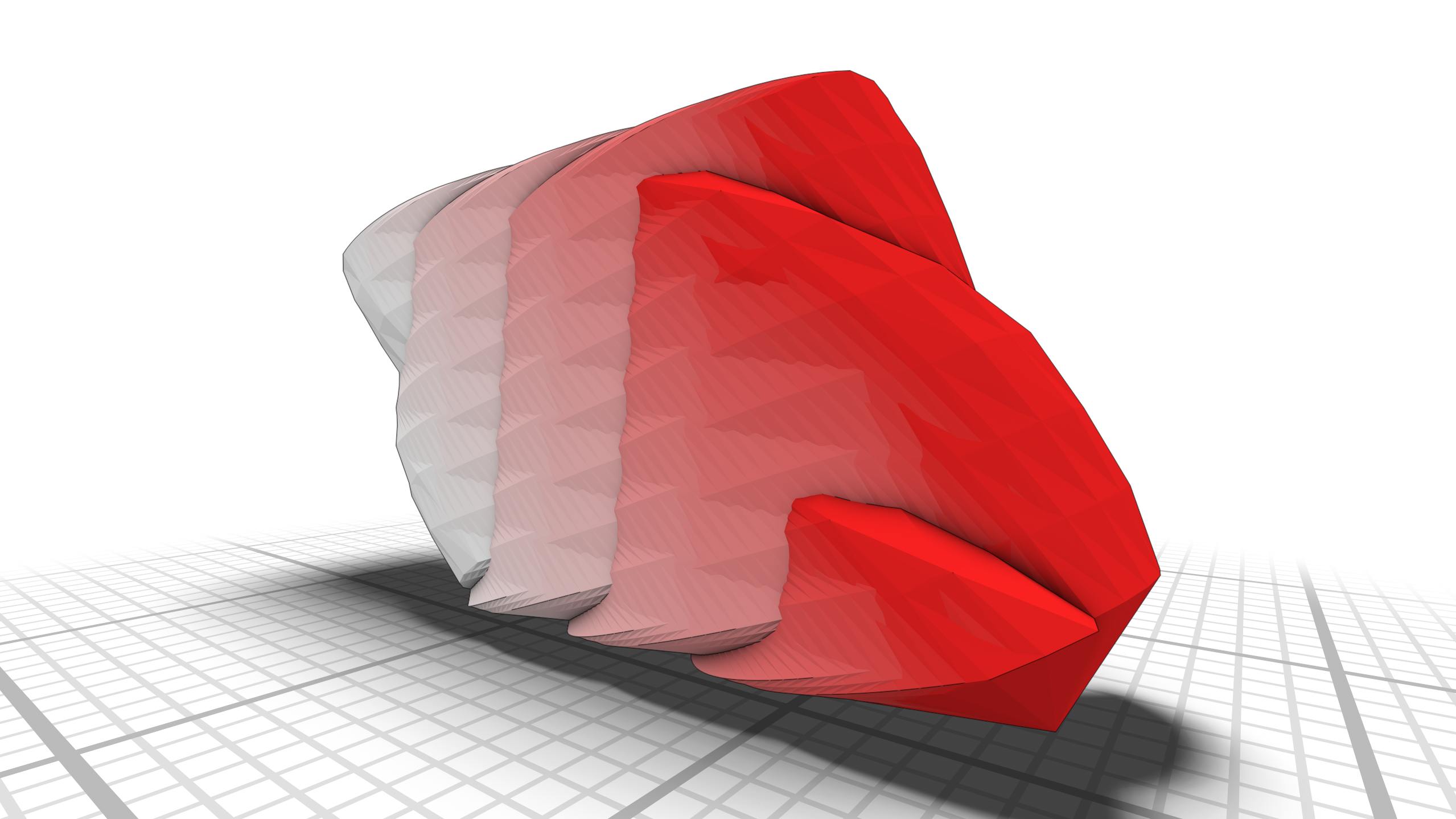}}
    \subcaptionbox{Result of the milling simulation.}{\includegraphics[trim=600 0 600 200, clip, width=0.9\linewidth]{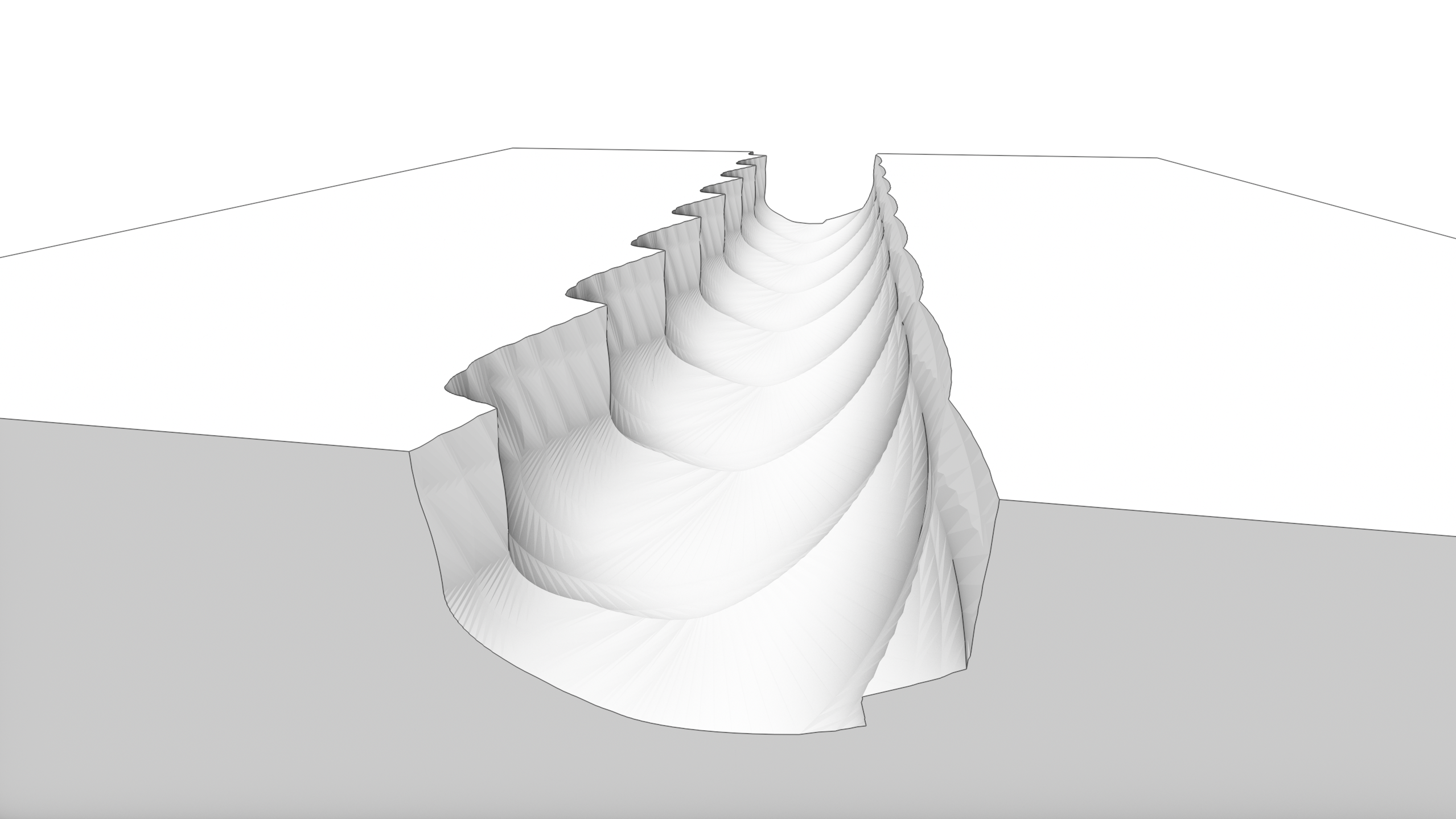}}
    \subcaptionbox{Wireframe of step 71 of our approach (left) and mesh-arrangements \cite{Zhou16} (right).}{\includegraphics[width=\linewidth]{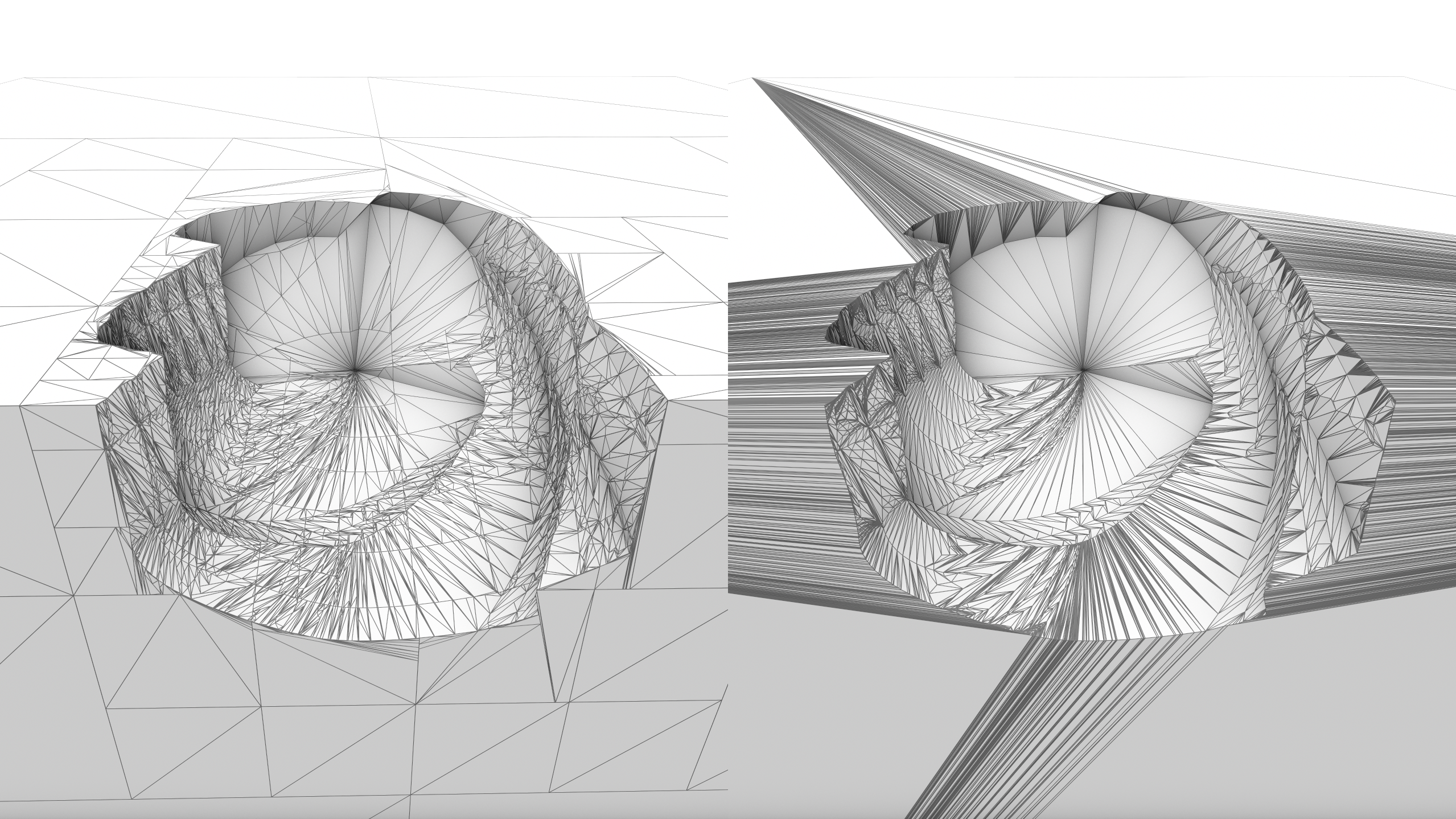}}
    \caption
    {   
        Subtracting the sweep volumes (b) of a rotating drill bit (a). 
        The sweeps were precomputed in \num{4} degree steps, simulation took \num{29} seconds with our approach and performed a total of \num{58472} individual BSP subtractions.
        Note the large amount of poor-quality triangles in the result of the last successful subtraction using mesh-arrangements~\cite{Zhou16} (d).
    }
    \label{fig:app:milling:images}
\end{figure}

\begin{figure}
    \centering
    \subcaptionbox{Per-step timings}{\resizebox{\linewidth}{!}{\input{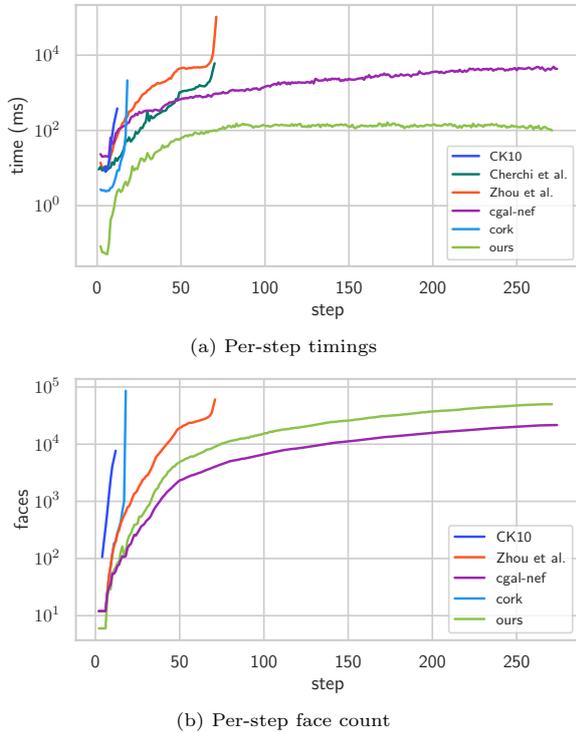}}\vspace{-0.25cm}}
    \subcaptionbox{Per-step face count}{\resizebox{\linewidth}{!}{\input{figures/milling-face-count.pgf}}\vspace{-0.25cm}}
    \caption{
        Milling simulation per-step timings are given in (a), comparing different methods.
        Cork \citep{cork2013} crashes after 17 steps, both mesh-arrangement approaches \cite{Zhou16,CLSA20} after 71, and CK10 \cite{Campen10} after 12.
        Per-step number of faces of the work-piece (b). 
        Our approach produces more faces than cgal-nef \cite{Hachenberger07} due to faces being split across octree cell boundaries.
    }
    \label{fig:app:milling:graphs}
\end{figure}

\begin{figure}
    \centering
    \subcaptionbox{}{
        \includegraphics[width=0.24\linewidth]{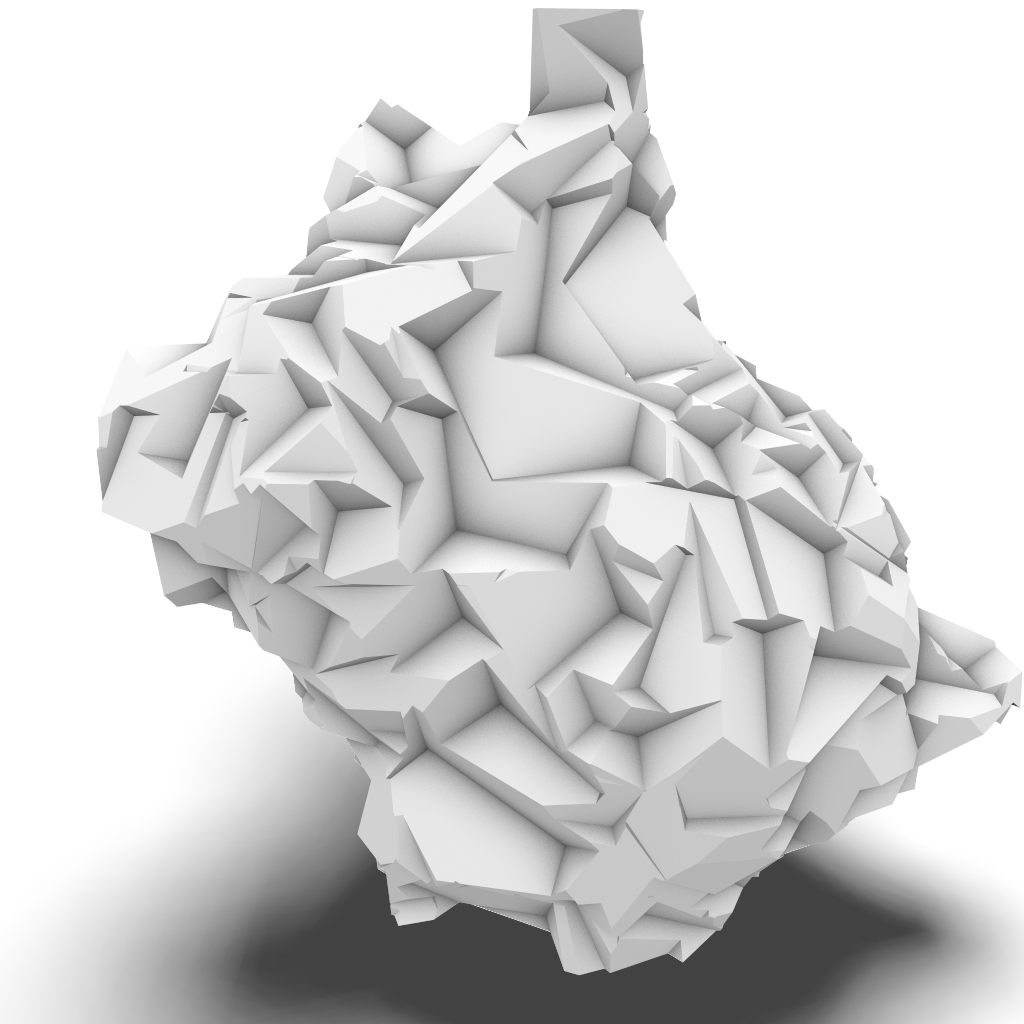}
        \includegraphics[width=0.24\linewidth]{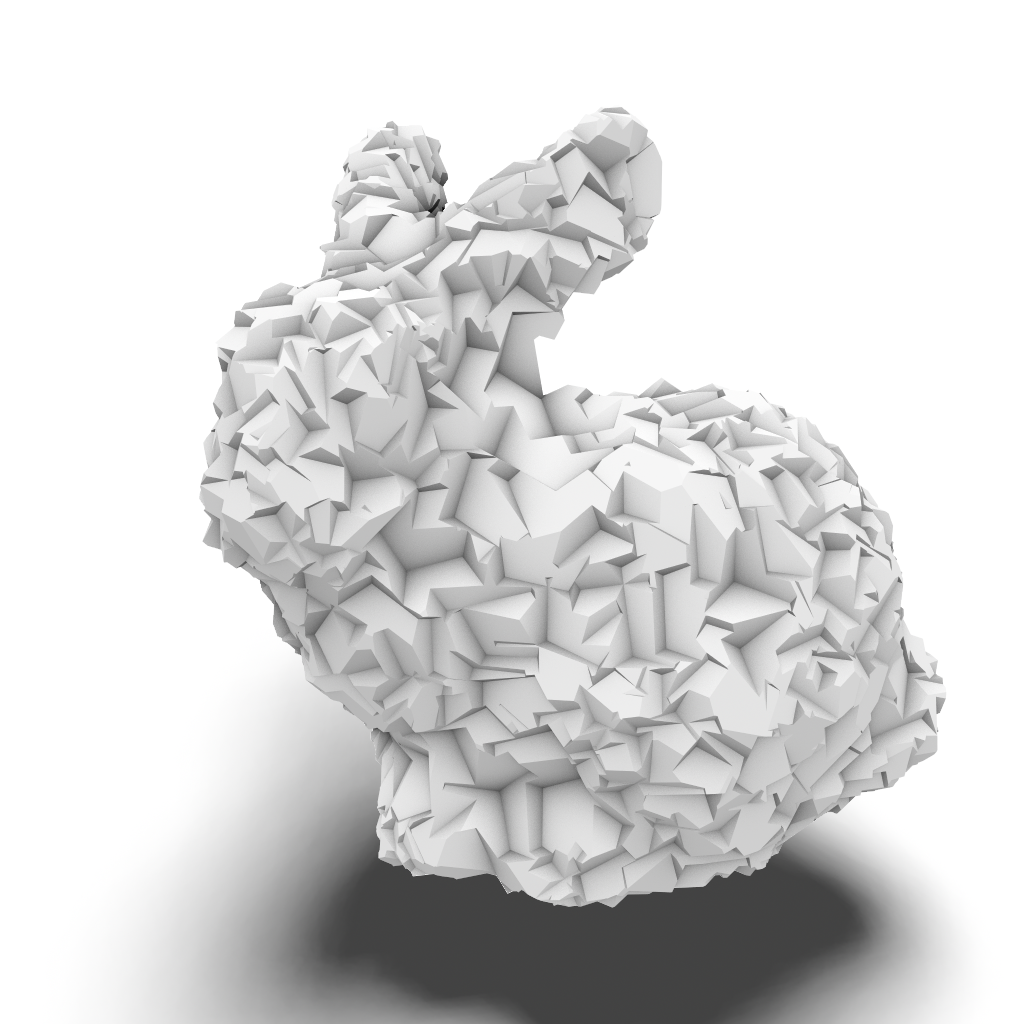}
        \includegraphics[width=0.24\linewidth]{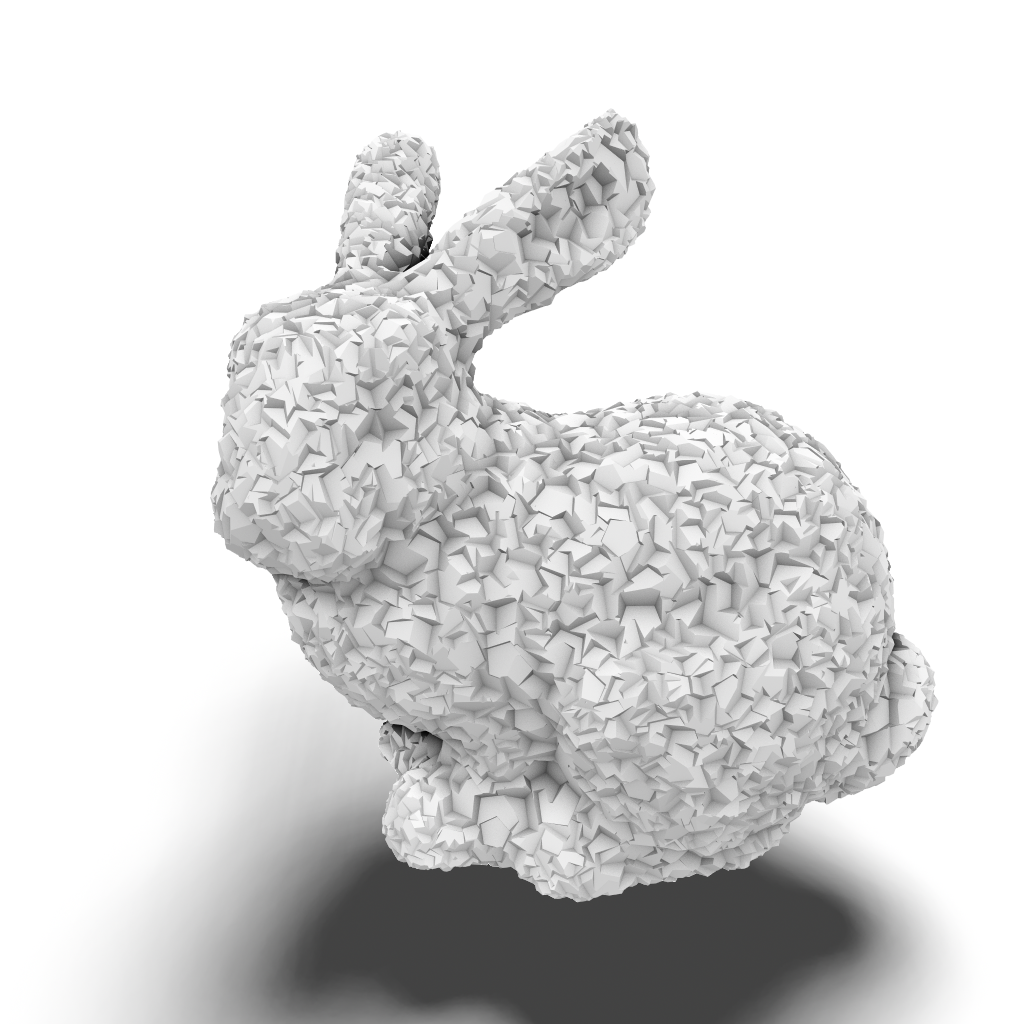}
        \includegraphics[width=0.24\linewidth]{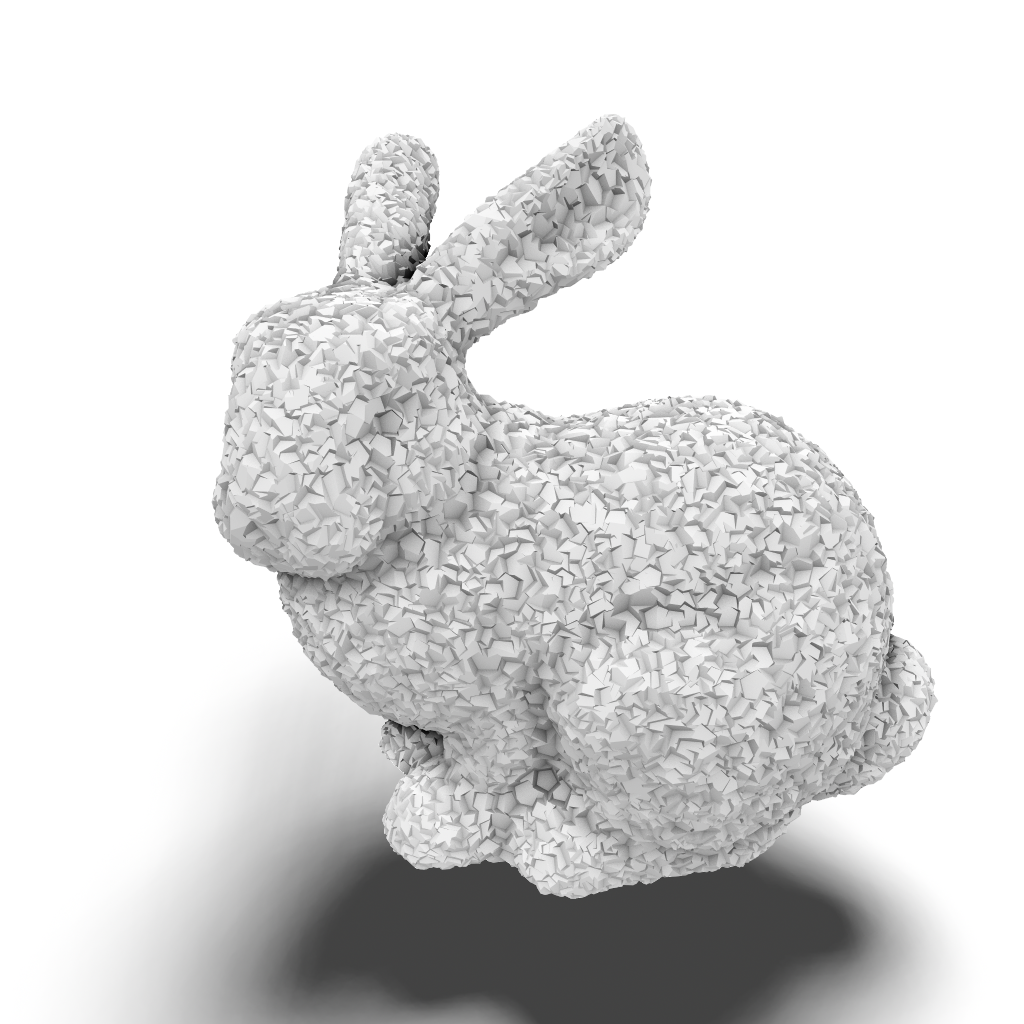}
    }
    \subcaptionbox{}{
        \includegraphics[width=0.24\linewidth]{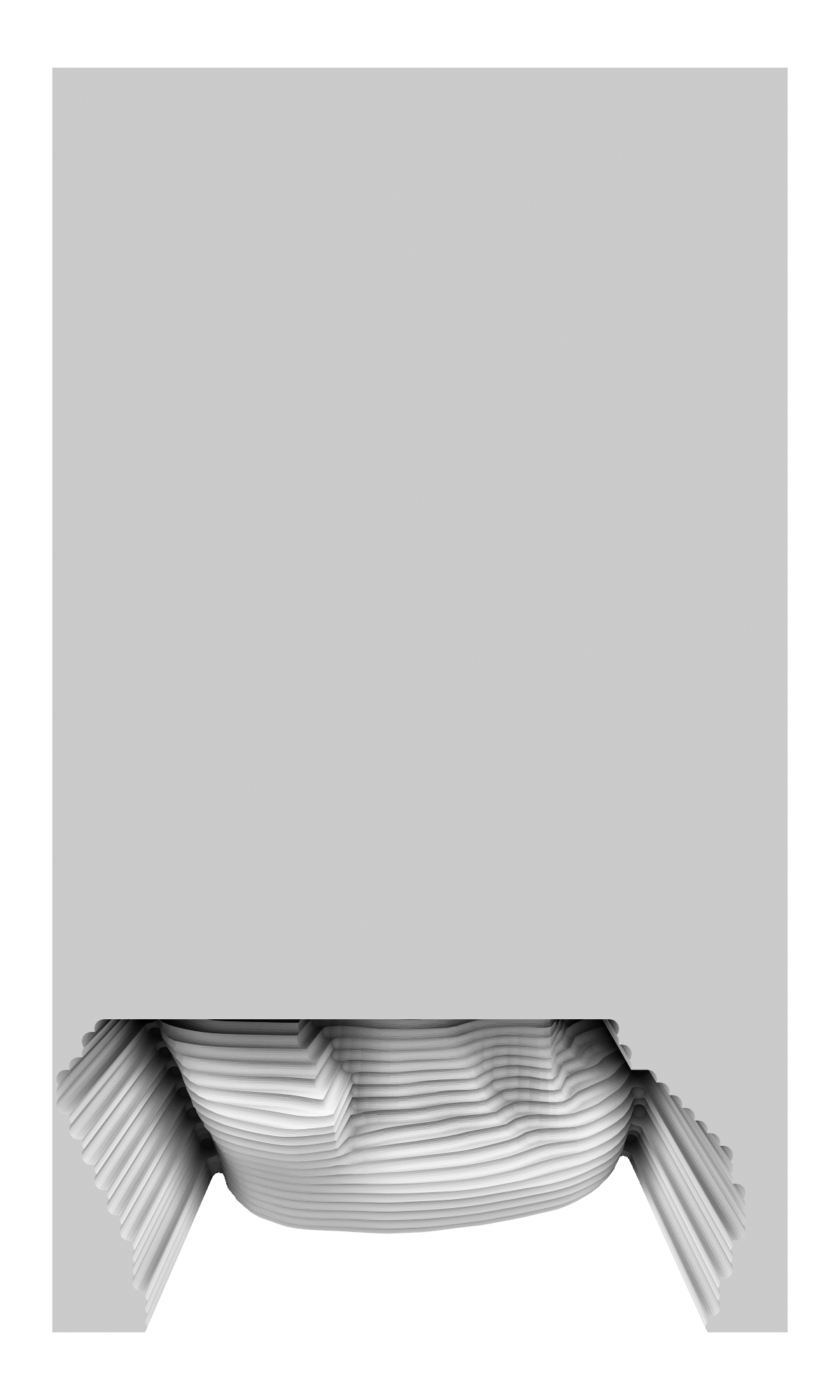}
        \includegraphics[width=0.24\linewidth]{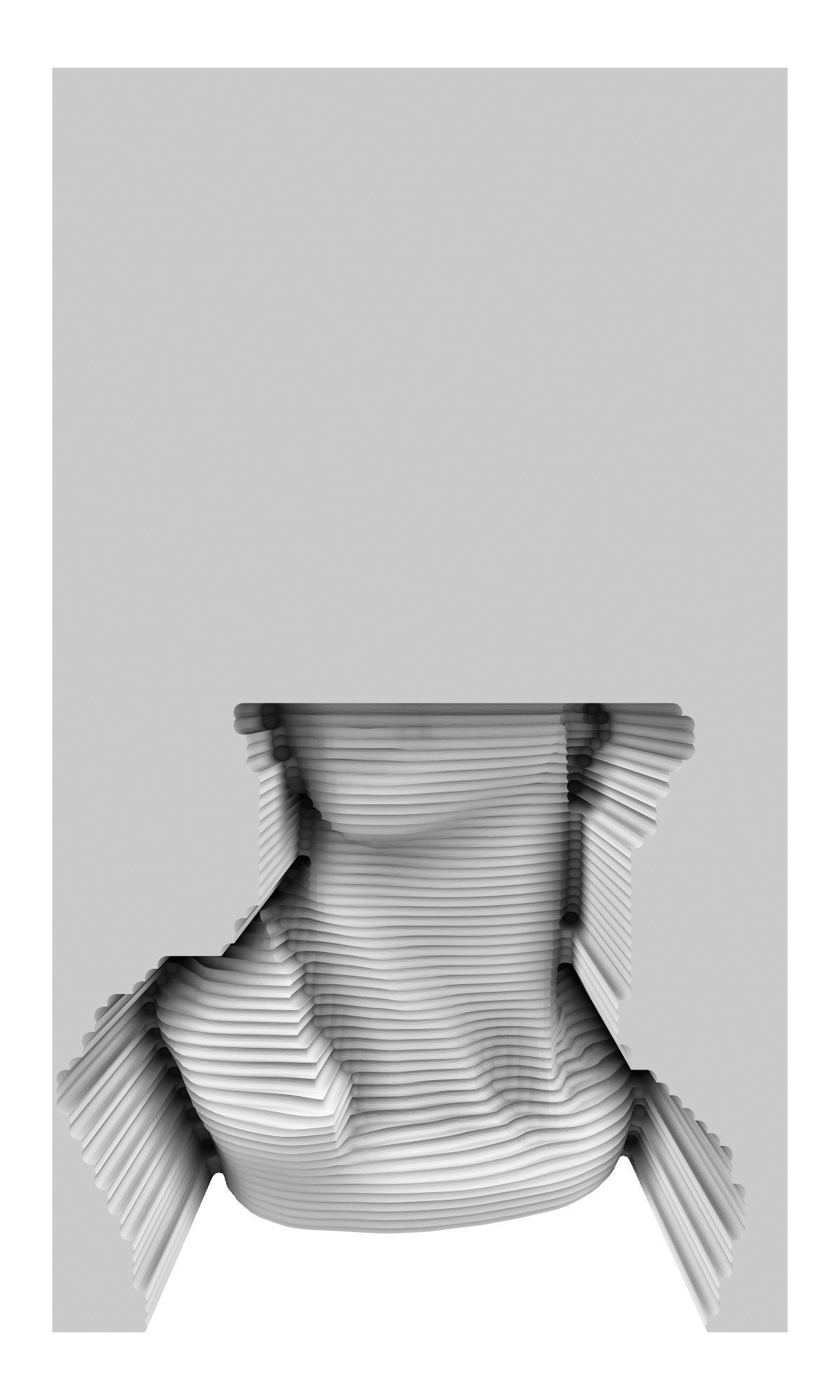}
        \includegraphics[width=0.24\linewidth]{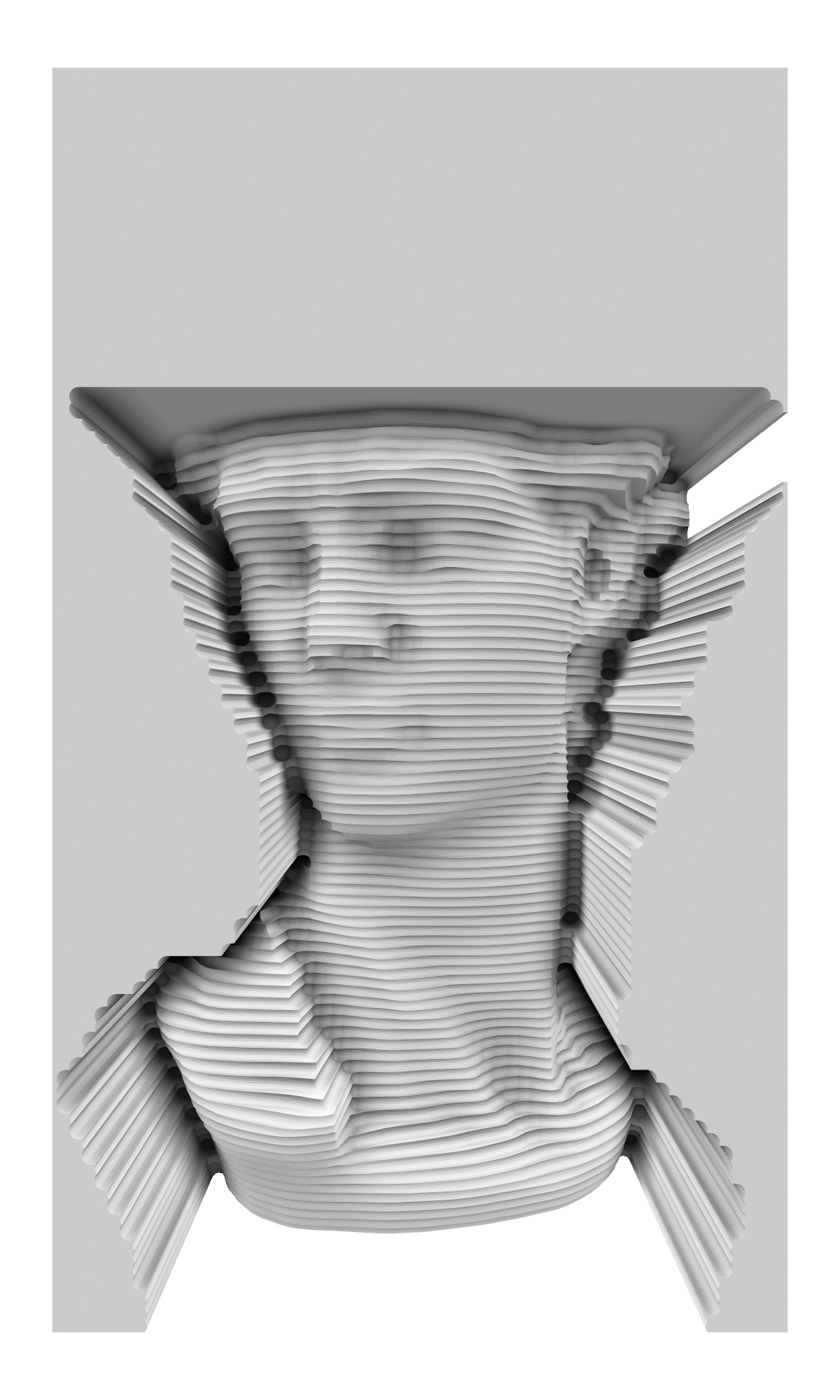}
        \includegraphics[width=0.24\linewidth]{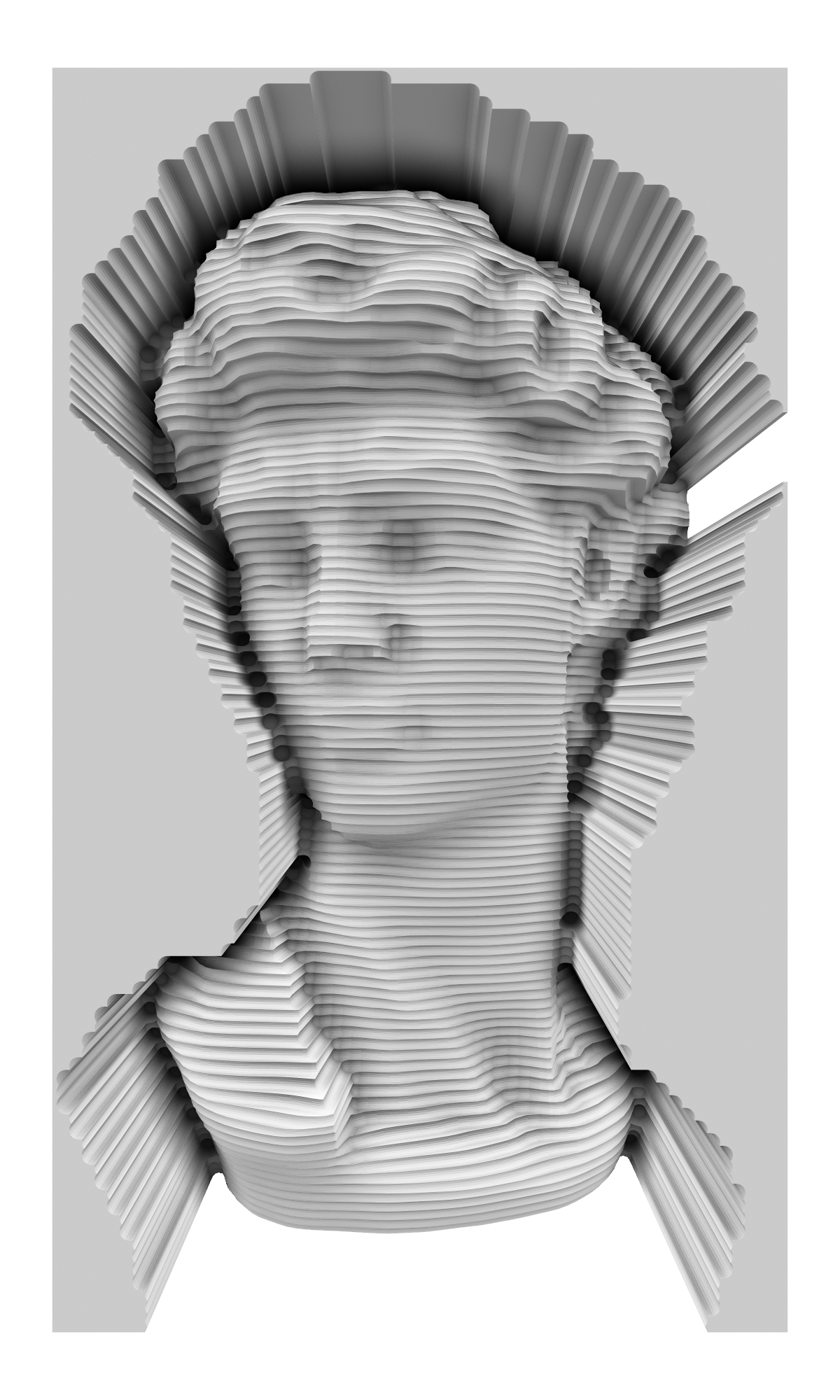}
    }
    \caption
    {
        The replication of experiments from \cite{Bernstein09} and \cite{Zhou16} using our approach:
        (a) Subtraction of \num{10000} dodecahedra from a block to create the bunny.
        (b) Subtraction of the sweep volumes of a tool along a path to mill David from a solid block.
    }
    \label{fig:app:bunny-david}
\end{figure}

Our approach excels at iteratively performing a sequence of CSG operations in an asymmetrical setting: 
Highly complex geometry (the ``workpiece'') in the Octree-BSP is subject to many iterative Booleans with a low-complexity \textit{Tool}-BSP.
With that in mind, we implemented applications for sweep-volume calculations (Figure~\ref{fig:app:sweep-volume}) and a milling simulation (Figure~\ref{fig:app:milling:images} and Figure~\ref{fig:app:milling:graphs}), as well as reproduced the bunny carving experiment from \cite{Bernstein09} and the David milling experiment from \cite{Zhou16} (Figure~\ref{fig:app:bunny-david}).

For the sweep-volume we took a series of discreet time-steps. 
Each step, the sweep prism (the \emph{tool}) of each individual triangle of the input shape is added to the resulting output sweep (the \emph{workpiece}).
A single triangle sweep is computed as the convex hull of the triangles vertices at the current step and the next one.

The milling simulation was performed in a similar manner.
A rotating drill-bit is moved through material with a \num{30} degree skew.
There are 90 four degree steps to achieve a full \num{360} degree rotation.
For each step, the sweep volume is precalculated.
After a full rotation, the sweeps can be translated according to the feed rate.
In this case, the \emph{tool} consists of one of the \num{90}, possibly translated, sweep volumes.
Then, each of the \num{90} sweep-volumes are iteratively subtracted from the \emph{workpice}, which starts as a cube.
This was done for three full rotations, resulting in a total of \num{270} tool subtractions.
We also implemented this simulation with other state-of-the-art methods.
Besides our approach, only CGAL with Nef-complexes~\citep{Hachenberger07} using exact construction and exact kernel successfully finished the simulation.
The results are shown in Figure~\ref{fig:app:milling:images}, the timings and number of resulting faces in Figure~\ref{fig:app:milling:graphs}.
Note that we used the per-step outputs from \cite{Zhou16} as inputs for \cite{CLSA20} because at the time of writing this, the published source code from \cite{CLSA20} only allows the computation of mesh-arrangements but not the resulting Boolean.
In order to compute Booleans a classification to determine whether a cell is part of the output is missing.
Therefore the computation of the arrangement gives a lower bound to the computation effort.

%% file: figures/milling-face-count.pgf
%% Creator: Matplotlib, PGF backend
%%
%% To include the figure in your LaTeX document, write
%%   \input{<filename>.pgf}
%%
%% Make sure the required packages are loaded in your preamble
%%   \usepackage{pgf}
%%
%% and, on pdftex
%%   \usepackage[utf8]{inputenc}\DeclareUnicodeCharacter{2212}{-}
%%
%% or, on luatex and xetex
%%   \usepackage{unicode-math}
%%
%% Figures using additional raster images can only be included by \input if
%% they are in the same directory as the main LaTeX file. For loading figures
%% from other directories you can use the `import` package
%% and then include the figures with
%%   \import{<path to file>}{<filename>.pgf}
%%
%% Matplotlib used the following preamble
%%   \usepackage{fontspec}
%%
\begingroup%
\makeatletter%
\begin{pgfpicture}%
\pgfpathrectangle{\pgfpointorigin}{\pgfqpoint{5.000000in}{3.000000in}}%
\pgfusepath{use as bounding box, clip}%
\begin{pgfscope}%
\pgfsetbuttcap%
\pgfsetmiterjoin%
\definecolor{currentfill}{rgb}{1.000000,1.000000,1.000000}%
\pgfsetfillcolor{currentfill}%
\pgfsetlinewidth{0.000000pt}%
\definecolor{currentstroke}{rgb}{1.000000,1.000000,1.000000}%
\pgfsetstrokecolor{currentstroke}%
\pgfsetdash{}{0pt}%
\pgfpathmoveto{\pgfqpoint{0.000000in}{0.000000in}}%
\pgfpathlineto{\pgfqpoint{5.000000in}{0.000000in}}%
\pgfpathlineto{\pgfqpoint{5.000000in}{3.000000in}}%
\pgfpathlineto{\pgfqpoint{0.000000in}{3.000000in}}%
\pgfpathclose%
\pgfusepath{fill}%
\end{pgfscope}%
\begin{pgfscope}%
\pgfsetbuttcap%
\pgfsetmiterjoin%
\definecolor{currentfill}{rgb}{1.000000,1.000000,1.000000}%
\pgfsetfillcolor{currentfill}%
\pgfsetlinewidth{0.000000pt}%
\definecolor{currentstroke}{rgb}{0.000000,0.000000,0.000000}%
\pgfsetstrokecolor{currentstroke}%
\pgfsetstrokeopacity{0.000000}%
\pgfsetdash{}{0pt}%
\pgfpathmoveto{\pgfqpoint{0.750000in}{0.630000in}}%
\pgfpathlineto{\pgfqpoint{5.000000in}{0.630000in}}%
\pgfpathlineto{\pgfqpoint{5.000000in}{2.850000in}}%
\pgfpathlineto{\pgfqpoint{0.750000in}{2.850000in}}%
\pgfpathclose%
\pgfusepath{fill}%
\end{pgfscope}%
\begin{pgfscope}%
\pgfpathrectangle{\pgfqpoint{0.750000in}{0.630000in}}{\pgfqpoint{4.250000in}{2.220000in}}%
\pgfusepath{clip}%
\pgfsetroundcap%
\pgfsetroundjoin%
\pgfsetlinewidth{1.003750pt}%
\definecolor{currentstroke}{rgb}{0.800000,0.800000,0.800000}%
\pgfsetstrokecolor{currentstroke}%
\pgfsetdash{}{0pt}%
\pgfpathmoveto{\pgfqpoint{0.914773in}{0.630000in}}%
\pgfpathlineto{\pgfqpoint{0.914773in}{2.850000in}}%
\pgfusepath{stroke}%
\end{pgfscope}%
\begin{pgfscope}%
\definecolor{textcolor}{rgb}{0.150000,0.150000,0.150000}%
\pgfsetstrokecolor{textcolor}%
\pgfsetfillcolor{textcolor}%
\pgftext[x=0.914773in,y=0.498056in,,top]{\color{textcolor}\sffamily\fontsize{11.000000}{13.200000}\selectfont 0}%
\end{pgfscope}%
\begin{pgfscope}%
\pgfpathrectangle{\pgfqpoint{0.750000in}{0.630000in}}{\pgfqpoint{4.250000in}{2.220000in}}%
\pgfusepath{clip}%
\pgfsetroundcap%
\pgfsetroundjoin%
\pgfsetlinewidth{1.003750pt}%
\definecolor{currentstroke}{rgb}{0.800000,0.800000,0.800000}%
\pgfsetstrokecolor{currentstroke}%
\pgfsetdash{}{0pt}%
\pgfpathmoveto{\pgfqpoint{1.625000in}{0.630000in}}%
\pgfpathlineto{\pgfqpoint{1.625000in}{2.850000in}}%
\pgfusepath{stroke}%
\end{pgfscope}%
\begin{pgfscope}%
\definecolor{textcolor}{rgb}{0.150000,0.150000,0.150000}%
\pgfsetstrokecolor{textcolor}%
\pgfsetfillcolor{textcolor}%
\pgftext[x=1.625000in,y=0.498056in,,top]{\color{textcolor}\sffamily\fontsize{11.000000}{13.200000}\selectfont 50}%
\end{pgfscope}%
\begin{pgfscope}%
\pgfpathrectangle{\pgfqpoint{0.750000in}{0.630000in}}{\pgfqpoint{4.250000in}{2.220000in}}%
\pgfusepath{clip}%
\pgfsetroundcap%
\pgfsetroundjoin%
\pgfsetlinewidth{1.003750pt}%
\definecolor{currentstroke}{rgb}{0.800000,0.800000,0.800000}%
\pgfsetstrokecolor{currentstroke}%
\pgfsetdash{}{0pt}%
\pgfpathmoveto{\pgfqpoint{2.335227in}{0.630000in}}%
\pgfpathlineto{\pgfqpoint{2.335227in}{2.850000in}}%
\pgfusepath{stroke}%
\end{pgfscope}%
\begin{pgfscope}%
\definecolor{textcolor}{rgb}{0.150000,0.150000,0.150000}%
\pgfsetstrokecolor{textcolor}%
\pgfsetfillcolor{textcolor}%
\pgftext[x=2.335227in,y=0.498056in,,top]{\color{textcolor}\sffamily\fontsize{11.000000}{13.200000}\selectfont 100}%
\end{pgfscope}%
\begin{pgfscope}%
\pgfpathrectangle{\pgfqpoint{0.750000in}{0.630000in}}{\pgfqpoint{4.250000in}{2.220000in}}%
\pgfusepath{clip}%
\pgfsetroundcap%
\pgfsetroundjoin%
\pgfsetlinewidth{1.003750pt}%
\definecolor{currentstroke}{rgb}{0.800000,0.800000,0.800000}%
\pgfsetstrokecolor{currentstroke}%
\pgfsetdash{}{0pt}%
\pgfpathmoveto{\pgfqpoint{3.045455in}{0.630000in}}%
\pgfpathlineto{\pgfqpoint{3.045455in}{2.850000in}}%
\pgfusepath{stroke}%
\end{pgfscope}%
\begin{pgfscope}%
\definecolor{textcolor}{rgb}{0.150000,0.150000,0.150000}%
\pgfsetstrokecolor{textcolor}%
\pgfsetfillcolor{textcolor}%
\pgftext[x=3.045455in,y=0.498056in,,top]{\color{textcolor}\sffamily\fontsize{11.000000}{13.200000}\selectfont 150}%
\end{pgfscope}%
\begin{pgfscope}%
\pgfpathrectangle{\pgfqpoint{0.750000in}{0.630000in}}{\pgfqpoint{4.250000in}{2.220000in}}%
\pgfusepath{clip}%
\pgfsetroundcap%
\pgfsetroundjoin%
\pgfsetlinewidth{1.003750pt}%
\definecolor{currentstroke}{rgb}{0.800000,0.800000,0.800000}%
\pgfsetstrokecolor{currentstroke}%
\pgfsetdash{}{0pt}%
\pgfpathmoveto{\pgfqpoint{3.755682in}{0.630000in}}%
\pgfpathlineto{\pgfqpoint{3.755682in}{2.850000in}}%
\pgfusepath{stroke}%
\end{pgfscope}%
\begin{pgfscope}%
\definecolor{textcolor}{rgb}{0.150000,0.150000,0.150000}%
\pgfsetstrokecolor{textcolor}%
\pgfsetfillcolor{textcolor}%
\pgftext[x=3.755682in,y=0.498056in,,top]{\color{textcolor}\sffamily\fontsize{11.000000}{13.200000}\selectfont 200}%
\end{pgfscope}%
\begin{pgfscope}%
\pgfpathrectangle{\pgfqpoint{0.750000in}{0.630000in}}{\pgfqpoint{4.250000in}{2.220000in}}%
\pgfusepath{clip}%
\pgfsetroundcap%
\pgfsetroundjoin%
\pgfsetlinewidth{1.003750pt}%
\definecolor{currentstroke}{rgb}{0.800000,0.800000,0.800000}%
\pgfsetstrokecolor{currentstroke}%
\pgfsetdash{}{0pt}%
\pgfpathmoveto{\pgfqpoint{4.465909in}{0.630000in}}%
\pgfpathlineto{\pgfqpoint{4.465909in}{2.850000in}}%
\pgfusepath{stroke}%
\end{pgfscope}%
\begin{pgfscope}%
\definecolor{textcolor}{rgb}{0.150000,0.150000,0.150000}%
\pgfsetstrokecolor{textcolor}%
\pgfsetfillcolor{textcolor}%
\pgftext[x=4.465909in,y=0.498056in,,top]{\color{textcolor}\sffamily\fontsize{11.000000}{13.200000}\selectfont 250}%
\end{pgfscope}%
\begin{pgfscope}%
\definecolor{textcolor}{rgb}{0.150000,0.150000,0.150000}%
\pgfsetstrokecolor{textcolor}%
\pgfsetfillcolor{textcolor}%
\pgftext[x=2.875000in,y=0.306833in,,top]{\color{textcolor}\sffamily\fontsize{12.000000}{14.400000}\selectfont step}%
\end{pgfscope}%
\begin{pgfscope}%
\pgfpathrectangle{\pgfqpoint{0.750000in}{0.630000in}}{\pgfqpoint{4.250000in}{2.220000in}}%
\pgfusepath{clip}%
\pgfsetroundcap%
\pgfsetroundjoin%
\pgfsetlinewidth{1.003750pt}%
\definecolor{currentstroke}{rgb}{0.800000,0.800000,0.800000}%
\pgfsetstrokecolor{currentstroke}%
\pgfsetdash{}{0pt}%
\pgfpathmoveto{\pgfqpoint{0.750000in}{0.838660in}}%
\pgfpathlineto{\pgfqpoint{5.000000in}{0.838660in}}%
\pgfusepath{stroke}%
\end{pgfscope}%
\begin{pgfscope}%
\definecolor{textcolor}{rgb}{0.150000,0.150000,0.150000}%
\pgfsetstrokecolor{textcolor}%
\pgfsetfillcolor{textcolor}%
\pgftext[x=0.399999in, y=0.785646in, left, base]{\color{textcolor}\sffamily\fontsize{11.000000}{13.200000}\selectfont \(\displaystyle {10^{1}}\)}%
\end{pgfscope}%
\begin{pgfscope}%
\pgfpathrectangle{\pgfqpoint{0.750000in}{0.630000in}}{\pgfqpoint{4.250000in}{2.220000in}}%
\pgfusepath{clip}%
\pgfsetroundcap%
\pgfsetroundjoin%
\pgfsetlinewidth{1.003750pt}%
\definecolor{currentstroke}{rgb}{0.800000,0.800000,0.800000}%
\pgfsetstrokecolor{currentstroke}%
\pgfsetdash{}{0pt}%
\pgfpathmoveto{\pgfqpoint{0.750000in}{1.324354in}}%
\pgfpathlineto{\pgfqpoint{5.000000in}{1.324354in}}%
\pgfusepath{stroke}%
\end{pgfscope}%
\begin{pgfscope}%
\definecolor{textcolor}{rgb}{0.150000,0.150000,0.150000}%
\pgfsetstrokecolor{textcolor}%
\pgfsetfillcolor{textcolor}%
\pgftext[x=0.399999in, y=1.271340in, left, base]{\color{textcolor}\sffamily\fontsize{11.000000}{13.200000}\selectfont \(\displaystyle {10^{2}}\)}%
\end{pgfscope}%
\begin{pgfscope}%
\pgfpathrectangle{\pgfqpoint{0.750000in}{0.630000in}}{\pgfqpoint{4.250000in}{2.220000in}}%
\pgfusepath{clip}%
\pgfsetroundcap%
\pgfsetroundjoin%
\pgfsetlinewidth{1.003750pt}%
\definecolor{currentstroke}{rgb}{0.800000,0.800000,0.800000}%
\pgfsetstrokecolor{currentstroke}%
\pgfsetdash{}{0pt}%
\pgfpathmoveto{\pgfqpoint{0.750000in}{1.810047in}}%
\pgfpathlineto{\pgfqpoint{5.000000in}{1.810047in}}%
\pgfusepath{stroke}%
\end{pgfscope}%
\begin{pgfscope}%
\definecolor{textcolor}{rgb}{0.150000,0.150000,0.150000}%
\pgfsetstrokecolor{textcolor}%
\pgfsetfillcolor{textcolor}%
\pgftext[x=0.399999in, y=1.757033in, left, base]{\color{textcolor}\sffamily\fontsize{11.000000}{13.200000}\selectfont \(\displaystyle {10^{3}}\)}%
\end{pgfscope}%
\begin{pgfscope}%
\pgfpathrectangle{\pgfqpoint{0.750000in}{0.630000in}}{\pgfqpoint{4.250000in}{2.220000in}}%
\pgfusepath{clip}%
\pgfsetroundcap%
\pgfsetroundjoin%
\pgfsetlinewidth{1.003750pt}%
\definecolor{currentstroke}{rgb}{0.800000,0.800000,0.800000}%
\pgfsetstrokecolor{currentstroke}%
\pgfsetdash{}{0pt}%
\pgfpathmoveto{\pgfqpoint{0.750000in}{2.295741in}}%
\pgfpathlineto{\pgfqpoint{5.000000in}{2.295741in}}%
\pgfusepath{stroke}%
\end{pgfscope}%
\begin{pgfscope}%
\definecolor{textcolor}{rgb}{0.150000,0.150000,0.150000}%
\pgfsetstrokecolor{textcolor}%
\pgfsetfillcolor{textcolor}%
\pgftext[x=0.399999in, y=2.242727in, left, base]{\color{textcolor}\sffamily\fontsize{11.000000}{13.200000}\selectfont \(\displaystyle {10^{4}}\)}%
\end{pgfscope}%
\begin{pgfscope}%
\pgfpathrectangle{\pgfqpoint{0.750000in}{0.630000in}}{\pgfqpoint{4.250000in}{2.220000in}}%
\pgfusepath{clip}%
\pgfsetroundcap%
\pgfsetroundjoin%
\pgfsetlinewidth{1.003750pt}%
\definecolor{currentstroke}{rgb}{0.800000,0.800000,0.800000}%
\pgfsetstrokecolor{currentstroke}%
\pgfsetdash{}{0pt}%
\pgfpathmoveto{\pgfqpoint{0.750000in}{2.781435in}}%
\pgfpathlineto{\pgfqpoint{5.000000in}{2.781435in}}%
\pgfusepath{stroke}%
\end{pgfscope}%
\begin{pgfscope}%
\definecolor{textcolor}{rgb}{0.150000,0.150000,0.150000}%
\pgfsetstrokecolor{textcolor}%
\pgfsetfillcolor{textcolor}%
\pgftext[x=0.399999in, y=2.728421in, left, base]{\color{textcolor}\sffamily\fontsize{11.000000}{13.200000}\selectfont \(\displaystyle {10^{5}}\)}%
\end{pgfscope}%
\begin{pgfscope}%
\definecolor{textcolor}{rgb}{0.150000,0.150000,0.150000}%
\pgfsetstrokecolor{textcolor}%
\pgfsetfillcolor{textcolor}%
\pgftext[x=0.344443in,y=1.740000in,,bottom,rotate=90.000000]{\color{textcolor}\sffamily\fontsize{12.000000}{14.400000}\selectfont faces}%
\end{pgfscope}%
\begin{pgfscope}%
\pgfpathrectangle{\pgfqpoint{0.750000in}{0.630000in}}{\pgfqpoint{4.250000in}{2.220000in}}%
\pgfusepath{clip}%
\pgfsetroundcap%
\pgfsetroundjoin%
\pgfsetlinewidth{1.505625pt}%
\definecolor{currentstroke}{rgb}{0.129412,0.588235,0.952941}%
\pgfsetstrokecolor{currentstroke}%
\pgfsetdash{}{0pt}%
\pgfpathmoveto{\pgfqpoint{0.943182in}{0.877118in}}%
\pgfpathlineto{\pgfqpoint{0.957386in}{0.877118in}}%
\pgfpathlineto{\pgfqpoint{0.971591in}{0.877118in}}%
\pgfpathlineto{\pgfqpoint{0.985795in}{0.877118in}}%
\pgfpathlineto{\pgfqpoint{1.000000in}{0.877118in}}%
\pgfpathlineto{\pgfqpoint{1.014205in}{1.023326in}}%
\pgfpathlineto{\pgfqpoint{1.028409in}{1.169534in}}%
\pgfpathlineto{\pgfqpoint{1.042614in}{1.255061in}}%
\pgfpathlineto{\pgfqpoint{1.056818in}{1.355660in}}%
\pgfpathlineto{\pgfqpoint{1.071023in}{1.452974in}}%
\pgfpathlineto{\pgfqpoint{1.085227in}{1.466300in}}%
\pgfpathlineto{\pgfqpoint{1.099432in}{1.541535in}}%
\pgfpathlineto{\pgfqpoint{1.113636in}{1.583727in}}%
\pgfpathlineto{\pgfqpoint{1.127841in}{1.635914in}}%
\pgfpathlineto{\pgfqpoint{1.142045in}{1.726829in}}%
\pgfpathlineto{\pgfqpoint{1.156250in}{1.809625in}}%
\pgfpathlineto{\pgfqpoint{1.170455in}{2.749091in}}%
\pgfusepath{stroke}%
\end{pgfscope}%
\begin{pgfscope}%
\pgfpathrectangle{\pgfqpoint{0.750000in}{0.630000in}}{\pgfqpoint{4.250000in}{2.220000in}}%
\pgfusepath{clip}%
\pgfsetroundcap%
\pgfsetroundjoin%
\pgfsetlinewidth{1.505625pt}%
\definecolor{currentstroke}{rgb}{0.188235,0.309804,0.996078}%
\pgfsetstrokecolor{currentstroke}%
\pgfsetdash{}{0pt}%
\pgfpathmoveto{\pgfqpoint{0.971591in}{1.336644in}}%
\pgfpathlineto{\pgfqpoint{0.985795in}{1.464137in}}%
\pgfpathlineto{\pgfqpoint{1.000000in}{1.584957in}}%
\pgfpathlineto{\pgfqpoint{1.014205in}{1.715250in}}%
\pgfpathlineto{\pgfqpoint{1.028409in}{1.859799in}}%
\pgfpathlineto{\pgfqpoint{1.042614in}{1.995766in}}%
\pgfpathlineto{\pgfqpoint{1.056818in}{2.109313in}}%
\pgfpathlineto{\pgfqpoint{1.071023in}{2.172532in}}%
\pgfpathlineto{\pgfqpoint{1.085227in}{2.239677in}}%
\pgfusepath{stroke}%
\end{pgfscope}%
\begin{pgfscope}%
\pgfpathrectangle{\pgfqpoint{0.750000in}{0.630000in}}{\pgfqpoint{4.250000in}{2.220000in}}%
\pgfusepath{clip}%
\pgfsetroundcap%
\pgfsetroundjoin%
\pgfsetlinewidth{1.505625pt}%
\definecolor{currentstroke}{rgb}{1.000000,0.341176,0.133333}%
\pgfsetstrokecolor{currentstroke}%
\pgfsetdash{}{0pt}%
\pgfpathmoveto{\pgfqpoint{0.943182in}{0.877118in}}%
\pgfpathlineto{\pgfqpoint{0.957386in}{0.877118in}}%
\pgfpathlineto{\pgfqpoint{0.971591in}{0.877118in}}%
\pgfpathlineto{\pgfqpoint{0.985795in}{0.877118in}}%
\pgfpathlineto{\pgfqpoint{1.000000in}{0.877118in}}%
\pgfpathlineto{\pgfqpoint{1.014205in}{1.023326in}}%
\pgfpathlineto{\pgfqpoint{1.028409in}{1.169534in}}%
\pgfpathlineto{\pgfqpoint{1.042614in}{1.260840in}}%
\pgfpathlineto{\pgfqpoint{1.056818in}{1.362811in}}%
\pgfpathlineto{\pgfqpoint{1.071023in}{1.426114in}}%
\pgfpathlineto{\pgfqpoint{1.085227in}{1.474739in}}%
\pgfpathlineto{\pgfqpoint{1.099432in}{1.532296in}}%
\pgfpathlineto{\pgfqpoint{1.113636in}{1.567048in}}%
\pgfpathlineto{\pgfqpoint{1.127841in}{1.618869in}}%
\pgfpathlineto{\pgfqpoint{1.142045in}{1.657849in}}%
\pgfpathlineto{\pgfqpoint{1.156250in}{1.690736in}}%
\pgfpathlineto{\pgfqpoint{1.170455in}{1.719181in}}%
\pgfpathlineto{\pgfqpoint{1.184659in}{1.758717in}}%
\pgfpathlineto{\pgfqpoint{1.198864in}{1.768701in}}%
\pgfpathlineto{\pgfqpoint{1.213068in}{1.800997in}}%
\pgfpathlineto{\pgfqpoint{1.227273in}{1.840990in}}%
\pgfpathlineto{\pgfqpoint{1.241477in}{1.876144in}}%
\pgfpathlineto{\pgfqpoint{1.255682in}{1.884013in}}%
\pgfpathlineto{\pgfqpoint{1.269886in}{1.907865in}}%
\pgfpathlineto{\pgfqpoint{1.284091in}{1.941401in}}%
\pgfpathlineto{\pgfqpoint{1.298295in}{1.962491in}}%
\pgfpathlineto{\pgfqpoint{1.312500in}{1.991704in}}%
\pgfpathlineto{\pgfqpoint{1.326705in}{2.003662in}}%
\pgfpathlineto{\pgfqpoint{1.340909in}{2.024194in}}%
\pgfpathlineto{\pgfqpoint{1.355114in}{2.032144in}}%
\pgfpathlineto{\pgfqpoint{1.369318in}{2.056185in}}%
\pgfpathlineto{\pgfqpoint{1.383523in}{2.084532in}}%
\pgfpathlineto{\pgfqpoint{1.397727in}{2.116440in}}%
\pgfpathlineto{\pgfqpoint{1.411932in}{2.151548in}}%
\pgfpathlineto{\pgfqpoint{1.426136in}{2.186792in}}%
\pgfpathlineto{\pgfqpoint{1.440341in}{2.203768in}}%
\pgfpathlineto{\pgfqpoint{1.454545in}{2.219721in}}%
\pgfpathlineto{\pgfqpoint{1.468750in}{2.237130in}}%
\pgfpathlineto{\pgfqpoint{1.482955in}{2.256946in}}%
\pgfpathlineto{\pgfqpoint{1.497159in}{2.273095in}}%
\pgfpathlineto{\pgfqpoint{1.511364in}{2.291221in}}%
\pgfpathlineto{\pgfqpoint{1.525568in}{2.307354in}}%
\pgfpathlineto{\pgfqpoint{1.539773in}{2.323194in}}%
\pgfpathlineto{\pgfqpoint{1.553977in}{2.342404in}}%
\pgfpathlineto{\pgfqpoint{1.568182in}{2.362515in}}%
\pgfpathlineto{\pgfqpoint{1.582386in}{2.382641in}}%
\pgfpathlineto{\pgfqpoint{1.596591in}{2.406449in}}%
\pgfpathlineto{\pgfqpoint{1.610795in}{2.423742in}}%
\pgfpathlineto{\pgfqpoint{1.625000in}{2.432635in}}%
\pgfpathlineto{\pgfqpoint{1.639205in}{2.440807in}}%
\pgfpathlineto{\pgfqpoint{1.653409in}{2.451497in}}%
\pgfpathlineto{\pgfqpoint{1.667614in}{2.457028in}}%
\pgfpathlineto{\pgfqpoint{1.681818in}{2.462226in}}%
\pgfpathlineto{\pgfqpoint{1.696023in}{2.474109in}}%
\pgfpathlineto{\pgfqpoint{1.710227in}{2.477576in}}%
\pgfpathlineto{\pgfqpoint{1.724432in}{2.476773in}}%
\pgfpathlineto{\pgfqpoint{1.738636in}{2.480372in}}%
\pgfpathlineto{\pgfqpoint{1.752841in}{2.485496in}}%
\pgfpathlineto{\pgfqpoint{1.767045in}{2.489944in}}%
\pgfpathlineto{\pgfqpoint{1.781250in}{2.492433in}}%
\pgfpathlineto{\pgfqpoint{1.795455in}{2.497161in}}%
\pgfpathlineto{\pgfqpoint{1.809659in}{2.501723in}}%
\pgfpathlineto{\pgfqpoint{1.823864in}{2.505736in}}%
\pgfpathlineto{\pgfqpoint{1.838068in}{2.511001in}}%
\pgfpathlineto{\pgfqpoint{1.852273in}{2.518293in}}%
\pgfpathlineto{\pgfqpoint{1.866477in}{2.525768in}}%
\pgfpathlineto{\pgfqpoint{1.880682in}{2.539992in}}%
\pgfpathlineto{\pgfqpoint{1.894886in}{2.561349in}}%
\pgfpathlineto{\pgfqpoint{1.909091in}{2.617684in}}%
\pgfpathlineto{\pgfqpoint{1.923295in}{2.675526in}}%
\pgfusepath{stroke}%
\end{pgfscope}%
\begin{pgfscope}%
\pgfpathrectangle{\pgfqpoint{0.750000in}{0.630000in}}{\pgfqpoint{4.250000in}{2.220000in}}%
\pgfusepath{clip}%
\pgfsetroundcap%
\pgfsetroundjoin%
\pgfsetlinewidth{1.505625pt}%
\definecolor{currentstroke}{rgb}{0.545098,0.764706,0.290196}%
\pgfsetstrokecolor{currentstroke}%
\pgfsetdash{}{0pt}%
\pgfpathmoveto{\pgfqpoint{0.943182in}{0.730909in}}%
\pgfpathlineto{\pgfqpoint{1.000000in}{0.730909in}}%
\pgfpathlineto{\pgfqpoint{1.014205in}{0.984868in}}%
\pgfpathlineto{\pgfqpoint{1.028409in}{1.077311in}}%
\pgfpathlineto{\pgfqpoint{1.042614in}{1.063244in}}%
\pgfpathlineto{\pgfqpoint{1.056818in}{1.182322in}}%
\pgfpathlineto{\pgfqpoint{1.071023in}{1.226894in}}%
\pgfpathlineto{\pgfqpoint{1.085227in}{1.246084in}}%
\pgfpathlineto{\pgfqpoint{1.099432in}{1.285050in}}%
\pgfpathlineto{\pgfqpoint{1.113636in}{1.299773in}}%
\pgfpathlineto{\pgfqpoint{1.127841in}{1.379695in}}%
\pgfpathlineto{\pgfqpoint{1.142045in}{1.426114in}}%
\pgfpathlineto{\pgfqpoint{1.156250in}{1.353834in}}%
\pgfpathlineto{\pgfqpoint{1.170455in}{1.387656in}}%
\pgfpathlineto{\pgfqpoint{1.184659in}{1.473702in}}%
\pgfpathlineto{\pgfqpoint{1.198864in}{1.515936in}}%
\pgfpathlineto{\pgfqpoint{1.213068in}{1.535421in}}%
\pgfpathlineto{\pgfqpoint{1.241477in}{1.608708in}}%
\pgfpathlineto{\pgfqpoint{1.255682in}{1.613046in}}%
\pgfpathlineto{\pgfqpoint{1.269886in}{1.621979in}}%
\pgfpathlineto{\pgfqpoint{1.284091in}{1.680463in}}%
\pgfpathlineto{\pgfqpoint{1.298295in}{1.678110in}}%
\pgfpathlineto{\pgfqpoint{1.312500in}{1.695872in}}%
\pgfpathlineto{\pgfqpoint{1.355114in}{1.766638in}}%
\pgfpathlineto{\pgfqpoint{1.369318in}{1.780915in}}%
\pgfpathlineto{\pgfqpoint{1.383523in}{1.815872in}}%
\pgfpathlineto{\pgfqpoint{1.397727in}{1.855592in}}%
\pgfpathlineto{\pgfqpoint{1.411932in}{1.889149in}}%
\pgfpathlineto{\pgfqpoint{1.426136in}{1.912068in}}%
\pgfpathlineto{\pgfqpoint{1.440341in}{1.921727in}}%
\pgfpathlineto{\pgfqpoint{1.454545in}{1.956888in}}%
\pgfpathlineto{\pgfqpoint{1.468750in}{1.975303in}}%
\pgfpathlineto{\pgfqpoint{1.482955in}{2.003998in}}%
\pgfpathlineto{\pgfqpoint{1.511364in}{2.032144in}}%
\pgfpathlineto{\pgfqpoint{1.525568in}{2.060218in}}%
\pgfpathlineto{\pgfqpoint{1.539773in}{2.070834in}}%
\pgfpathlineto{\pgfqpoint{1.568182in}{2.101672in}}%
\pgfpathlineto{\pgfqpoint{1.582386in}{2.107054in}}%
\pgfpathlineto{\pgfqpoint{1.596591in}{2.124477in}}%
\pgfpathlineto{\pgfqpoint{1.625000in}{2.143108in}}%
\pgfpathlineto{\pgfqpoint{1.639205in}{2.152923in}}%
\pgfpathlineto{\pgfqpoint{1.653409in}{2.156585in}}%
\pgfpathlineto{\pgfqpoint{1.667614in}{2.166741in}}%
\pgfpathlineto{\pgfqpoint{1.681818in}{2.174640in}}%
\pgfpathlineto{\pgfqpoint{1.696023in}{2.178168in}}%
\pgfpathlineto{\pgfqpoint{1.710227in}{2.187815in}}%
\pgfpathlineto{\pgfqpoint{1.724432in}{2.191512in}}%
\pgfpathlineto{\pgfqpoint{1.738636in}{2.189985in}}%
\pgfpathlineto{\pgfqpoint{1.752841in}{2.198449in}}%
\pgfpathlineto{\pgfqpoint{1.767045in}{2.203180in}}%
\pgfpathlineto{\pgfqpoint{1.781250in}{2.212021in}}%
\pgfpathlineto{\pgfqpoint{1.809659in}{2.222665in}}%
\pgfpathlineto{\pgfqpoint{1.838068in}{2.237992in}}%
\pgfpathlineto{\pgfqpoint{1.852273in}{2.248541in}}%
\pgfpathlineto{\pgfqpoint{1.880682in}{2.254318in}}%
\pgfpathlineto{\pgfqpoint{1.909091in}{2.266971in}}%
\pgfpathlineto{\pgfqpoint{1.951705in}{2.287548in}}%
\pgfpathlineto{\pgfqpoint{1.965909in}{2.292660in}}%
\pgfpathlineto{\pgfqpoint{1.980114in}{2.301114in}}%
\pgfpathlineto{\pgfqpoint{1.994318in}{2.304116in}}%
\pgfpathlineto{\pgfqpoint{2.051136in}{2.322601in}}%
\pgfpathlineto{\pgfqpoint{2.093750in}{2.330493in}}%
\pgfpathlineto{\pgfqpoint{2.107955in}{2.333389in}}%
\pgfpathlineto{\pgfqpoint{2.136364in}{2.342404in}}%
\pgfpathlineto{\pgfqpoint{2.150568in}{2.344088in}}%
\pgfpathlineto{\pgfqpoint{2.164773in}{2.344021in}}%
\pgfpathlineto{\pgfqpoint{2.178977in}{2.346490in}}%
\pgfpathlineto{\pgfqpoint{2.193182in}{2.351180in}}%
\pgfpathlineto{\pgfqpoint{2.207386in}{2.349176in}}%
\pgfpathlineto{\pgfqpoint{2.221591in}{2.359590in}}%
\pgfpathlineto{\pgfqpoint{2.235795in}{2.359668in}}%
\pgfpathlineto{\pgfqpoint{2.264205in}{2.368873in}}%
\pgfpathlineto{\pgfqpoint{2.278409in}{2.369156in}}%
\pgfpathlineto{\pgfqpoint{2.292614in}{2.375581in}}%
\pgfpathlineto{\pgfqpoint{2.306818in}{2.376733in}}%
\pgfpathlineto{\pgfqpoint{2.321023in}{2.380535in}}%
\pgfpathlineto{\pgfqpoint{2.349432in}{2.390256in}}%
\pgfpathlineto{\pgfqpoint{2.363636in}{2.388958in}}%
\pgfpathlineto{\pgfqpoint{2.377841in}{2.397619in}}%
\pgfpathlineto{\pgfqpoint{2.392045in}{2.401257in}}%
\pgfpathlineto{\pgfqpoint{2.406250in}{2.400205in}}%
\pgfpathlineto{\pgfqpoint{2.420455in}{2.406374in}}%
\pgfpathlineto{\pgfqpoint{2.505682in}{2.421302in}}%
\pgfpathlineto{\pgfqpoint{2.519886in}{2.420906in}}%
\pgfpathlineto{\pgfqpoint{2.534091in}{2.424740in}}%
\pgfpathlineto{\pgfqpoint{2.548295in}{2.425025in}}%
\pgfpathlineto{\pgfqpoint{2.562500in}{2.430597in}}%
\pgfpathlineto{\pgfqpoint{2.576705in}{2.430719in}}%
\pgfpathlineto{\pgfqpoint{2.590909in}{2.433240in}}%
\pgfpathlineto{\pgfqpoint{2.619318in}{2.436133in}}%
\pgfpathlineto{\pgfqpoint{2.647727in}{2.440064in}}%
\pgfpathlineto{\pgfqpoint{2.676136in}{2.444809in}}%
\pgfpathlineto{\pgfqpoint{2.747159in}{2.455668in}}%
\pgfpathlineto{\pgfqpoint{2.761364in}{2.459789in}}%
\pgfpathlineto{\pgfqpoint{2.775568in}{2.461064in}}%
\pgfpathlineto{\pgfqpoint{2.803977in}{2.466400in}}%
\pgfpathlineto{\pgfqpoint{2.875000in}{2.479783in}}%
\pgfpathlineto{\pgfqpoint{2.889205in}{2.480873in}}%
\pgfpathlineto{\pgfqpoint{2.903409in}{2.484136in}}%
\pgfpathlineto{\pgfqpoint{2.917614in}{2.483132in}}%
\pgfpathlineto{\pgfqpoint{2.931818in}{2.487077in}}%
\pgfpathlineto{\pgfqpoint{2.946023in}{2.485367in}}%
\pgfpathlineto{\pgfqpoint{2.960227in}{2.489507in}}%
\pgfpathlineto{\pgfqpoint{3.031250in}{2.494663in}}%
\pgfpathlineto{\pgfqpoint{3.088068in}{2.500615in}}%
\pgfpathlineto{\pgfqpoint{3.116477in}{2.505969in}}%
\pgfpathlineto{\pgfqpoint{3.130682in}{2.505533in}}%
\pgfpathlineto{\pgfqpoint{3.144886in}{2.509299in}}%
\pgfpathlineto{\pgfqpoint{3.159091in}{2.509620in}}%
\pgfpathlineto{\pgfqpoint{3.187500in}{2.514716in}}%
\pgfpathlineto{\pgfqpoint{3.201705in}{2.515670in}}%
\pgfpathlineto{\pgfqpoint{3.215909in}{2.519078in}}%
\pgfpathlineto{\pgfqpoint{3.258523in}{2.523006in}}%
\pgfpathlineto{\pgfqpoint{3.286932in}{2.527560in}}%
\pgfpathlineto{\pgfqpoint{3.301136in}{2.528339in}}%
\pgfpathlineto{\pgfqpoint{3.315341in}{2.532025in}}%
\pgfpathlineto{\pgfqpoint{3.343750in}{2.534909in}}%
\pgfpathlineto{\pgfqpoint{3.357955in}{2.534025in}}%
\pgfpathlineto{\pgfqpoint{3.386364in}{2.537164in}}%
\pgfpathlineto{\pgfqpoint{3.471591in}{2.543619in}}%
\pgfpathlineto{\pgfqpoint{3.500000in}{2.546053in}}%
\pgfpathlineto{\pgfqpoint{3.514205in}{2.549692in}}%
\pgfpathlineto{\pgfqpoint{3.528409in}{2.549571in}}%
\pgfpathlineto{\pgfqpoint{3.556818in}{2.553038in}}%
\pgfpathlineto{\pgfqpoint{3.571023in}{2.552564in}}%
\pgfpathlineto{\pgfqpoint{3.585227in}{2.556148in}}%
\pgfpathlineto{\pgfqpoint{3.656250in}{2.563725in}}%
\pgfpathlineto{\pgfqpoint{3.670455in}{2.566396in}}%
\pgfpathlineto{\pgfqpoint{3.698864in}{2.567365in}}%
\pgfpathlineto{\pgfqpoint{3.713068in}{2.570145in}}%
\pgfpathlineto{\pgfqpoint{3.727273in}{2.570013in}}%
\pgfpathlineto{\pgfqpoint{3.741477in}{2.572572in}}%
\pgfpathlineto{\pgfqpoint{3.826705in}{2.578744in}}%
\pgfpathlineto{\pgfqpoint{3.869318in}{2.581298in}}%
\pgfpathlineto{\pgfqpoint{3.883523in}{2.580840in}}%
\pgfpathlineto{\pgfqpoint{3.911932in}{2.583762in}}%
\pgfpathlineto{\pgfqpoint{3.997159in}{2.590241in}}%
\pgfpathlineto{\pgfqpoint{4.011364in}{2.593305in}}%
\pgfpathlineto{\pgfqpoint{4.196023in}{2.608051in}}%
\pgfpathlineto{\pgfqpoint{4.224432in}{2.609696in}}%
\pgfpathlineto{\pgfqpoint{4.295455in}{2.613209in}}%
\pgfpathlineto{\pgfqpoint{4.409091in}{2.620404in}}%
\pgfpathlineto{\pgfqpoint{4.437500in}{2.622793in}}%
\pgfpathlineto{\pgfqpoint{4.480114in}{2.625019in}}%
\pgfpathlineto{\pgfqpoint{4.508523in}{2.627305in}}%
\pgfpathlineto{\pgfqpoint{4.678977in}{2.634487in}}%
\pgfpathlineto{\pgfqpoint{4.764205in}{2.635125in}}%
\pgfpathlineto{\pgfqpoint{4.764205in}{2.635125in}}%
\pgfusepath{stroke}%
\end{pgfscope}%
\begin{pgfscope}%
\pgfpathrectangle{\pgfqpoint{0.750000in}{0.630000in}}{\pgfqpoint{4.250000in}{2.220000in}}%
\pgfusepath{clip}%
\pgfsetroundcap%
\pgfsetroundjoin%
\pgfsetlinewidth{1.505625pt}%
\definecolor{currentstroke}{rgb}{0.611765,0.152941,0.690196}%
\pgfsetstrokecolor{currentstroke}%
\pgfsetdash{}{0pt}%
\pgfpathmoveto{\pgfqpoint{0.943182in}{0.877118in}}%
\pgfpathlineto{\pgfqpoint{1.000000in}{0.877118in}}%
\pgfpathlineto{\pgfqpoint{1.014205in}{1.023326in}}%
\pgfpathlineto{\pgfqpoint{1.028409in}{1.077311in}}%
\pgfpathlineto{\pgfqpoint{1.042614in}{1.102910in}}%
\pgfpathlineto{\pgfqpoint{1.056818in}{1.198249in}}%
\pgfpathlineto{\pgfqpoint{1.071023in}{1.194379in}}%
\pgfpathlineto{\pgfqpoint{1.085227in}{1.216603in}}%
\pgfpathlineto{\pgfqpoint{1.099432in}{1.260840in}}%
\pgfpathlineto{\pgfqpoint{1.113636in}{1.271945in}}%
\pgfpathlineto{\pgfqpoint{1.127841in}{1.313534in}}%
\pgfpathlineto{\pgfqpoint{1.142045in}{1.344458in}}%
\pgfpathlineto{\pgfqpoint{1.156250in}{1.336644in}}%
\pgfpathlineto{\pgfqpoint{1.170455in}{1.344458in}}%
\pgfpathlineto{\pgfqpoint{1.184659in}{1.412674in}}%
\pgfpathlineto{\pgfqpoint{1.198864in}{1.435037in}}%
\pgfpathlineto{\pgfqpoint{1.213068in}{1.448338in}}%
\pgfpathlineto{\pgfqpoint{1.241477in}{1.525091in}}%
\pgfpathlineto{\pgfqpoint{1.255682in}{1.519311in}}%
\pgfpathlineto{\pgfqpoint{1.284091in}{1.567048in}}%
\pgfpathlineto{\pgfqpoint{1.298295in}{1.594546in}}%
\pgfpathlineto{\pgfqpoint{1.326705in}{1.617822in}}%
\pgfpathlineto{\pgfqpoint{1.340909in}{1.640675in}}%
\pgfpathlineto{\pgfqpoint{1.369318in}{1.667602in}}%
\pgfpathlineto{\pgfqpoint{1.383523in}{1.688871in}}%
\pgfpathlineto{\pgfqpoint{1.397727in}{1.720797in}}%
\pgfpathlineto{\pgfqpoint{1.411932in}{1.747387in}}%
\pgfpathlineto{\pgfqpoint{1.426136in}{1.769214in}}%
\pgfpathlineto{\pgfqpoint{1.454545in}{1.806431in}}%
\pgfpathlineto{\pgfqpoint{1.468750in}{1.828225in}}%
\pgfpathlineto{\pgfqpoint{1.482955in}{1.847094in}}%
\pgfpathlineto{\pgfqpoint{1.497159in}{1.862119in}}%
\pgfpathlineto{\pgfqpoint{1.511364in}{1.881322in}}%
\pgfpathlineto{\pgfqpoint{1.525568in}{1.896136in}}%
\pgfpathlineto{\pgfqpoint{1.553977in}{1.921478in}}%
\pgfpathlineto{\pgfqpoint{1.582386in}{1.949940in}}%
\pgfpathlineto{\pgfqpoint{1.610795in}{1.975784in}}%
\pgfpathlineto{\pgfqpoint{1.625000in}{1.987290in}}%
\pgfpathlineto{\pgfqpoint{1.653409in}{1.997768in}}%
\pgfpathlineto{\pgfqpoint{1.681818in}{2.012003in}}%
\pgfpathlineto{\pgfqpoint{1.724432in}{2.030592in}}%
\pgfpathlineto{\pgfqpoint{1.738636in}{2.033171in}}%
\pgfpathlineto{\pgfqpoint{1.781250in}{2.049242in}}%
\pgfpathlineto{\pgfqpoint{1.809659in}{2.058146in}}%
\pgfpathlineto{\pgfqpoint{1.838068in}{2.071568in}}%
\pgfpathlineto{\pgfqpoint{1.880682in}{2.087383in}}%
\pgfpathlineto{\pgfqpoint{1.909091in}{2.099276in}}%
\pgfpathlineto{\pgfqpoint{1.937500in}{2.109211in}}%
\pgfpathlineto{\pgfqpoint{1.951705in}{2.115748in}}%
\pgfpathlineto{\pgfqpoint{2.022727in}{2.140834in}}%
\pgfpathlineto{\pgfqpoint{2.051136in}{2.151423in}}%
\pgfpathlineto{\pgfqpoint{2.107955in}{2.161824in}}%
\pgfpathlineto{\pgfqpoint{2.136364in}{2.167672in}}%
\pgfpathlineto{\pgfqpoint{2.150568in}{2.172001in}}%
\pgfpathlineto{\pgfqpoint{2.164773in}{2.172077in}}%
\pgfpathlineto{\pgfqpoint{2.250000in}{2.190298in}}%
\pgfpathlineto{\pgfqpoint{2.392045in}{2.222187in}}%
\pgfpathlineto{\pgfqpoint{2.434659in}{2.231314in}}%
\pgfpathlineto{\pgfqpoint{2.505682in}{2.245806in}}%
\pgfpathlineto{\pgfqpoint{2.590909in}{2.257427in}}%
\pgfpathlineto{\pgfqpoint{2.605114in}{2.258083in}}%
\pgfpathlineto{\pgfqpoint{2.647727in}{2.264148in}}%
\pgfpathlineto{\pgfqpoint{2.690341in}{2.270091in}}%
\pgfpathlineto{\pgfqpoint{2.775568in}{2.283451in}}%
\pgfpathlineto{\pgfqpoint{2.860795in}{2.296940in}}%
\pgfpathlineto{\pgfqpoint{2.931818in}{2.308072in}}%
\pgfpathlineto{\pgfqpoint{3.116477in}{2.326884in}}%
\pgfpathlineto{\pgfqpoint{3.357955in}{2.356481in}}%
\pgfpathlineto{\pgfqpoint{3.414773in}{2.360724in}}%
\pgfpathlineto{\pgfqpoint{3.656250in}{2.382655in}}%
\pgfpathlineto{\pgfqpoint{3.826705in}{2.398476in}}%
\pgfpathlineto{\pgfqpoint{4.053977in}{2.414613in}}%
\pgfpathlineto{\pgfqpoint{4.281250in}{2.432822in}}%
\pgfpathlineto{\pgfqpoint{4.380682in}{2.438387in}}%
\pgfpathlineto{\pgfqpoint{4.451705in}{2.443002in}}%
\pgfpathlineto{\pgfqpoint{4.664773in}{2.455372in}}%
\pgfpathlineto{\pgfqpoint{4.806818in}{2.458369in}}%
\pgfpathlineto{\pgfqpoint{4.806818in}{2.458369in}}%
\pgfusepath{stroke}%
\end{pgfscope}%
\begin{pgfscope}%
\pgfsetrectcap%
\pgfsetmiterjoin%
\pgfsetlinewidth{1.254687pt}%
\definecolor{currentstroke}{rgb}{0.800000,0.800000,0.800000}%
\pgfsetstrokecolor{currentstroke}%
\pgfsetdash{}{0pt}%
\pgfpathmoveto{\pgfqpoint{0.750000in}{0.630000in}}%
\pgfpathlineto{\pgfqpoint{0.750000in}{2.850000in}}%
\pgfusepath{stroke}%
\end{pgfscope}%
\begin{pgfscope}%
\pgfsetrectcap%
\pgfsetmiterjoin%
\pgfsetlinewidth{1.254687pt}%
\definecolor{currentstroke}{rgb}{0.800000,0.800000,0.800000}%
\pgfsetstrokecolor{currentstroke}%
\pgfsetdash{}{0pt}%
\pgfpathmoveto{\pgfqpoint{5.000000in}{0.630000in}}%
\pgfpathlineto{\pgfqpoint{5.000000in}{2.850000in}}%
\pgfusepath{stroke}%
\end{pgfscope}%
\begin{pgfscope}%
\pgfsetrectcap%
\pgfsetmiterjoin%
\pgfsetlinewidth{1.254687pt}%
\definecolor{currentstroke}{rgb}{0.800000,0.800000,0.800000}%
\pgfsetstrokecolor{currentstroke}%
\pgfsetdash{}{0pt}%
\pgfpathmoveto{\pgfqpoint{0.750000in}{0.630000in}}%
\pgfpathlineto{\pgfqpoint{5.000000in}{0.630000in}}%
\pgfusepath{stroke}%
\end{pgfscope}%
\begin{pgfscope}%
\pgfsetrectcap%
\pgfsetmiterjoin%
\pgfsetlinewidth{1.254687pt}%
\definecolor{currentstroke}{rgb}{0.800000,0.800000,0.800000}%
\pgfsetstrokecolor{currentstroke}%
\pgfsetdash{}{0pt}%
\pgfpathmoveto{\pgfqpoint{0.750000in}{2.850000in}}%
\pgfpathlineto{\pgfqpoint{5.000000in}{2.850000in}}%
\pgfusepath{stroke}%
\end{pgfscope}%
\begin{pgfscope}%
\pgfsetbuttcap%
\pgfsetmiterjoin%
\definecolor{currentfill}{rgb}{1.000000,1.000000,1.000000}%
\pgfsetfillcolor{currentfill}%
\pgfsetfillopacity{0.800000}%
\pgfsetlinewidth{1.003750pt}%
\definecolor{currentstroke}{rgb}{0.800000,0.800000,0.800000}%
\pgfsetstrokecolor{currentstroke}%
\pgfsetstrokeopacity{0.800000}%
\pgfsetdash{}{0pt}%
\pgfpathmoveto{\pgfqpoint{3.919625in}{0.692500in}}%
\pgfpathlineto{\pgfqpoint{4.912500in}{0.692500in}}%
\pgfpathquadraticcurveto{\pgfqpoint{4.937500in}{0.692500in}}{\pgfqpoint{4.937500in}{0.717500in}}%
\pgfpathlineto{\pgfqpoint{4.937500in}{1.578374in}}%
\pgfpathquadraticcurveto{\pgfqpoint{4.937500in}{1.603374in}}{\pgfqpoint{4.912500in}{1.603374in}}%
\pgfpathlineto{\pgfqpoint{3.919625in}{1.603374in}}%
\pgfpathquadraticcurveto{\pgfqpoint{3.894625in}{1.603374in}}{\pgfqpoint{3.894625in}{1.578374in}}%
\pgfpathlineto{\pgfqpoint{3.894625in}{0.717500in}}%
\pgfpathquadraticcurveto{\pgfqpoint{3.894625in}{0.692500in}}{\pgfqpoint{3.919625in}{0.692500in}}%
\pgfpathclose%
\pgfusepath{stroke,fill}%
\end{pgfscope}%
\begin{pgfscope}%
\pgfsetroundcap%
\pgfsetroundjoin%
\pgfsetlinewidth{1.505625pt}%
\definecolor{currentstroke}{rgb}{0.188235,0.309804,0.996078}%
\pgfsetstrokecolor{currentstroke}%
\pgfsetdash{}{0pt}%
\pgfpathmoveto{\pgfqpoint{3.944625in}{1.509624in}}%
\pgfpathlineto{\pgfqpoint{4.194625in}{1.509624in}}%
\pgfusepath{stroke}%
\end{pgfscope}%
\begin{pgfscope}%
\definecolor{textcolor}{rgb}{0.150000,0.150000,0.150000}%
\pgfsetstrokecolor{textcolor}%
\pgfsetfillcolor{textcolor}%
\pgftext[x=4.294625in,y=1.465874in,left,base]{\color{textcolor}\sffamily\fontsize{9.000000}{10.800000}\selectfont CK10}%
\end{pgfscope}%
\begin{pgfscope}%
\pgfsetroundcap%
\pgfsetroundjoin%
\pgfsetlinewidth{1.505625pt}%
\definecolor{currentstroke}{rgb}{1.000000,0.341176,0.133333}%
\pgfsetstrokecolor{currentstroke}%
\pgfsetdash{}{0pt}%
\pgfpathmoveto{\pgfqpoint{3.944625in}{1.335374in}}%
\pgfpathlineto{\pgfqpoint{4.194625in}{1.335374in}}%
\pgfusepath{stroke}%
\end{pgfscope}%
\begin{pgfscope}%
\definecolor{textcolor}{rgb}{0.150000,0.150000,0.150000}%
\pgfsetstrokecolor{textcolor}%
\pgfsetfillcolor{textcolor}%
\pgftext[x=4.294625in,y=1.291624in,left,base]{\color{textcolor}\sffamily\fontsize{9.000000}{10.800000}\selectfont Zhou et al.}%
\end{pgfscope}%
\begin{pgfscope}%
\pgfsetroundcap%
\pgfsetroundjoin%
\pgfsetlinewidth{1.505625pt}%
\definecolor{currentstroke}{rgb}{0.611765,0.152941,0.690196}%
\pgfsetstrokecolor{currentstroke}%
\pgfsetdash{}{0pt}%
\pgfpathmoveto{\pgfqpoint{3.944625in}{1.160500in}}%
\pgfpathlineto{\pgfqpoint{4.194625in}{1.160500in}}%
\pgfusepath{stroke}%
\end{pgfscope}%
\begin{pgfscope}%
\definecolor{textcolor}{rgb}{0.150000,0.150000,0.150000}%
\pgfsetstrokecolor{textcolor}%
\pgfsetfillcolor{textcolor}%
\pgftext[x=4.294625in,y=1.116750in,left,base]{\color{textcolor}\sffamily\fontsize{9.000000}{10.800000}\selectfont cgal-nef}%
\end{pgfscope}%
\begin{pgfscope}%
\pgfsetroundcap%
\pgfsetroundjoin%
\pgfsetlinewidth{1.505625pt}%
\definecolor{currentstroke}{rgb}{0.129412,0.588235,0.952941}%
\pgfsetstrokecolor{currentstroke}%
\pgfsetdash{}{0pt}%
\pgfpathmoveto{\pgfqpoint{3.944625in}{0.984750in}}%
\pgfpathlineto{\pgfqpoint{4.194625in}{0.984750in}}%
\pgfusepath{stroke}%
\end{pgfscope}%
\begin{pgfscope}%
\definecolor{textcolor}{rgb}{0.150000,0.150000,0.150000}%
\pgfsetstrokecolor{textcolor}%
\pgfsetfillcolor{textcolor}%
\pgftext[x=4.294625in,y=0.941000in,left,base]{\color{textcolor}\sffamily\fontsize{9.000000}{10.800000}\selectfont cork}%
\end{pgfscope}%
\begin{pgfscope}%
\pgfsetroundcap%
\pgfsetroundjoin%
\pgfsetlinewidth{1.505625pt}%
\definecolor{currentstroke}{rgb}{0.545098,0.764706,0.290196}%
\pgfsetstrokecolor{currentstroke}%
\pgfsetdash{}{0pt}%
\pgfpathmoveto{\pgfqpoint{3.944625in}{0.810500in}}%
\pgfpathlineto{\pgfqpoint{4.194625in}{0.810500in}}%
\pgfusepath{stroke}%
\end{pgfscope}%
\begin{pgfscope}%
\definecolor{textcolor}{rgb}{0.150000,0.150000,0.150000}%
\pgfsetstrokecolor{textcolor}%
\pgfsetfillcolor{textcolor}%
\pgftext[x=4.294625in,y=0.766750in,left,base]{\color{textcolor}\sffamily\fontsize{9.000000}{10.800000}\selectfont ours}%
\end{pgfscope}%
\end{pgfpicture}%
\makeatother%
\endgroup%

%% file: paper-06-future-work.tex
% !TEX root paper.tex
\section{Limitations and Future Work}

Input meshes in floating point formats have to be rounded to suitable integer coordinates before they can be used in our system.
While our \SI{256}{\bit} arithmetic presented in Section~\ref{sec:method:int} provides nanometer resolution and works for all practical meshes we tested, this is technically a limitation.
Replacing our custom-tailored fixed-precision arithmetic with \texttt{gmp} is always possible but causes at least a $10 \times$ slowdown.
A potential future avenue is to track the precision requirements not per elementary operation (as arbitrary-precision libraries do) but per BSP plane and branch into different fixed-precision routines when constructing or classifying vertices.
This will not only remove any precision limitation but might also provide a further speedup when operations can be performed with less bit depth.

Currently, input meshes are required to be watertight and self-intersection free.
While not needed for our use cases, this limitation is easy to lift.
\citet{Campen10} showed how to treat self-intersections as BSP merges and \citet{Zhou16} use winding numbers to correctly classify cells.

The focus of this work is iterated CSG where a complex object is repeatedly merged with small, relatively simple objects.
In such a setting, keeping the octree-embedded BSP as a persistent data structure is essential to achieve high performance:
In each step, we only have to compute a series of BSP-BSP booleans where both BSPs have limited complexity.
The first is limited due to the subdivision criterion of the octree, the second due to the assumption that the ``tool BSP'' is simple.
Even with our optimizations, large BSP-BSP booleans are still expensive, exhibiting superlinear time complexity.
If the task is just to merge two complex objects once, an architecture closer to \cite{Campen10} might be more appropriate:
Build an octree containing triangles of the two meshes and only perform the BSP conversion and merging in octree cells containing geometry from both meshes.
This would still benefit significantly from our integer computation and BSP merging.

Finally, our current implementation is single-threaded.
Especially the octree is perfectly suited for parallelization where each octree cell can be merged in parallel with only minimal synchronization for the occasional octree subdivision and combination.
We expect an almost linear speedup in number of cores.

%% file: paper-07-conclusion.tex
\section{Conclusion}

In many applications, such as simulations of manufacturing processes, Boolean operations are not only performed once, but repeatedly on a progressively more complex object.
For these use cases, we designed a new persistent data structure that we call octree-embedded BSPs.
In each cell of a global octree we embed BSP trees with leaves labelled \textbf{in} and \textbf{out}, representing piecewise linear geometry.
Inner nodes of the BSP are defined by planes with homogeneous integer coordinates.

We present an exact algorithm for cutting a convex mesh against a plane that remains efficient even for cells in a deep BSP due to traversing only a fraction of the mesh and cutting it in-place.
Instead of floating point coordinates we use a plane-based geometry representation and homogeneous 4D integer coordinates with custom-tailored fixed-precision arithmetic to achieve up to \num{2.5} million mesh-plane cuts per second on a single CPU core.
This is used to implement equally exact and efficient CSG and mesh extraction.
While the BSPs are used to represent the geometry, the octree is used as an acceleration structure to keep the impact of modifications local and bound the BSP complexity.

In combination, this results in an unconditionally stable high-performance system for performing iterated CSG of a complex object against many small objects.
We demonstrate this by computing sweep volumes and simulating milling where our method outperforms the state of the art both in terms of robustness and performance.